\documentclass[12pt]{article}
\usepackage{graphicx}
\graphicspath{ {./images/} }
\usepackage{comment}
\usepackage{multicol}
\usepackage{caption}
\usepackage{mathrsfs}
\usepackage{float}
\usepackage{subcaption}
\usepackage[utf8]{inputenc}
\usepackage{longtable}
\usepackage{float}
\usepackage{booktabs}
\usepackage[english]{babel}
\usepackage[dvipsnames]{xcolor}
\usepackage{booktabs}
\usepackage{dcolumn}
\usepackage{multirow}
\usepackage{ragged2e}
\usepackage{hyperref}
\definecolor{darkblue}{RGB}{0,0,120}
\hypersetup{
      colorlinks=true,
      linkcolor=darkblue,
      filecolor=darkblue,
      citecolor=darkblue,      
      urlcolor=darkblue,
      }
\usepackage{multibib}
\newcites{supp}{Supplementary References}

\usepackage[justification=centering]{caption}

\usepackage[toc,page]{appendix}
\usepackage{adjustbox}
\usepackage{multirow}
\usepackage{threeparttable, tablefootnote,booktabs, stackengine}
\usepackage{array}
\usepackage{caption}
\usepackage{placeins}

\usepackage{indentfirst}
\usepackage[margin=1in]{geometry}
\usepackage{subcaption}

\usepackage{amsmath,amsthm,amssymb,scrextend}
\usepackage{cleveref}
\usepackage{bbm}
\usepackage{mathrsfs}
\usepackage{blindtext}

\usepackage{pdflscape}
\usepackage{longtable}
\usepackage{vcell}
\usepackage{booktabs}
\usepackage{natbib}
 \bibpunct[, ]{(}{)}{;}{a}{}{,}%

\DeclareCaptionLabelFormat{andtable}{#1~#2  \&  \tablename~\thetable}

\usepackage{setspace}
\usepackage{lipsum}
\usepackage{footmisc}

 \usepackage{spverbatim}
\usepackage{float}
\usepackage{graphicx}
\usepackage{titlesec}
\usepackage{nopageno}

\newcolumntype{x}[1]{%
>{\centering\hspace{0pt}}p{#1}}%

\titleformat{\section}
{\normalfont\fontsize{13}{13}\bfseries}
{}{-0.5em}{
\centering
}
[
] 

\titleformat{\subsection}
{\normalfont\fontsize{13}{13}\bfseries\itshape}
{}{-0.5em}{
}
[
] 

\titleformat{\subsubsection}
{\normalfont\fontsize{13}{13}\itshape}
{}{-0.5em}{
}
[
] 

 \title{Who Shares Fake News? Uncovering Insights from Social Media Users' Post Histories}
\date{}

\begin{document}

\maketitle

\begin{center}
Verena Schoenmuller, Simon J. Blanchard, Gita V. Johar    
\end{center} 

\begin{abstract}
\singlespacing
\noindent We propose that social-media users’ own post histories are an underused yet valuable resource for studying fake-news sharing. By extracting textual cues from their prior posts, and contrasting their prevalence against random social-media users and others (e.g., those with similar socio-demographics, political news-sharers, and fact-check sharers), researchers can identify cues that distinguish fake-news sharers, predict those most likely to share fake news, and identify promising constructs to build interventions. Our research includes studies along these lines. In Study 1, we explore the distinctive language patterns of fake-news sharers, highlighting elements such as their higher use of anger and power-related words. In Study 2, we show that adding textual cues into predictive models enhances their accuracy in predicting fake-news sharers. In Study 3, we explore the contrasting role of trait and situational anger, and show trait anger is associated with a greater propensity to share both true and fake news. In Study 4, we introduce a way to authenticate Twitter accounts in surveys, before using it to explore how crafting an ad copy that resonates with users' sense of power encourages the adoption of fact-checking tools. We hope to encourage the use of novel research methods for marketers and misinformation researchers.\\
\footnotesize
\\
\vspace{5mm}\noindent

\end{abstract}

\thispagestyle{empty}

\clearpage 
\doublespacing

\pagenumbering{arabic} 
	\newpage

\noindent Misinformation, the dissemination of claims contradicting or distorting verifiable facts \citep{guess2020misinformation}, poses significant risks across various domains, including politics, public health, climate change, and advertising \citep{albaOn}. It can manifest in different forms, including disinformation, which involves intentionally spreading false claims \citep{pennycook2021psychology}. Misinformation can also take the form of news, rumors, or conspiracy theories \citep[][]{del2016spreading}, and may be propagated not only by individuals but also by fake-news publishers \citep{rao2022deceptive} and bots \citep[][]{shao2018spread}. In this article, we study social media users who share misinformation in the form of news (i.e., fake news). The focus on the users' role in propagating fake news is crucial as such content can go from friends to strangers within seconds on social media \citep{gallotti2020assessing}. 

Research efforts to mitigate the re-sharing of fake news on social media were initially focused on accurately identifying fake-news content. For example, non-partisan organizations like Hoaxy, Media Bias Fact Check, and NewsGuard began to label publishing domains as either legitimate or primarily known for generating fake news. Scholars have also contributed to these efforts by creating lists of fake-news outlets \citep[e.g.,][]{grinberg2019fake, guess2020exposure}. While these efforts on the generators of fake news are useful, labeling publishing domains as fake-news sources is unlikely to be sufficient to eliminate fake-news sharing entirely. Indeed, many users share content without verifying source credibility \citep{kim2019says}. Additionally, not all misinformation originates from disinformation publishers (i.e., fake-news domains), and significant misinformation can spread through non-fake-news publishers \citep{scheufele2019science}.

In addition to labeling publishing domains, non-partisan organizations have also taken up manual fact-checking to verify the accuracy of information provided in individual news articles (e.g., Snopes, Politifact, Factcheck.org) in the hope that others can use the fact checks to correct false beliefs and prevent the spread of misinformation. However, labeling individual articles as misinformation is also insufficient because fact checks are often published well after the spread of fake news. Further, social media users often share without considering the content's veracity \citep{pennycook2021shifting}. Resources for fact-checking individual articles and claims within articles are also limited \citep{guo2022survey}, and scholars have found mixed results in debunking misinformation by posting fact-check links \citep{mosleh2021perverse}.

In another attempt to stem the spread of misinformation, non-partisan organizations, and researchers have started identifying individuals with the highest propensity to create or share fake news based on the idea that such identification is necessary to target mitigation efforts appropriately. For example, Politifact offers politicians scorecards based on the accuracy of their statements across media. Similarly, organizations such as the Center for Countering Digital Hate and Avaaz.org have published lists of the top fake news publishers in domains such as climate change. Rather than relying on publishers, researchers have focused on the socio-demographics, motivation, and personality of social media users to characterize and identify those most likely to engage in misinformation propagation \citep[e.g.,][]{arin2023ability, grinberg2019fake, kim2019says, lazer2018science, pennycook2020fighting, osmundsen2021partisan, shu2018understanding, jun2022social}. Using surveys and automated text analysis of the fake-news posts shared, this research generally finds that fake-news sharers are more likely to be male, older, have conservative leanings, and rely on emotions. However, using surveys and text analyses of fake-news sharing posts, can be limiting. Surveys rely on the assumption that sharing intentions reflect actual sharing behaviors, and automated text analyses on fake-news sharing posts often have insufficient accompanying text. 

Aligned with the research tradition of "conceptual contributions via non-deductive routes" described by \cite{lynch2012knowledge}, our research stems from the realization that the literature on misinformation largely neglects a valuable source of individual-level data, namely, that found in users' past social media post histories \footnote{Post history refers to all of the users' existing posts.}. These post histories can provide a comprehensive dataset and offer a window into the daily language used by news sharers; they enable the extraction of textual cues at the individual level using both dictionary-based (e.g., LIWC) and non-dictionary-based methods (e.g., pre-trained classifiers). Building on this observation, we hypothesize that specific textual cues extracted from post histories, particularly those related to users' chronic emotions and personality, could serve as valuable indicators to differentiate fake-news sharers from other social media users. In addition to providing actionable input to those involved in efforts to mitigate the spread of misinformation, our research can provide a more comprehensive set of characteristics of fake-news sharers over and above socio-demographics and political interest, thus, stimulating research into this critical topic of misinformation. This is particularly important given the lack of attention to this available data source and the growing significance of the misinformation domain within the marketing literature. 

In Study 1, we gather two datasets to explore the unique insights that users' post histories can provide into fake-news sharers. Our findings indicate that fake-news sharers' post histories contain textual cues different from others in the social-media ecosystem: not only random Twitter users, but also users matched on demographics, politically active users, conservative users, and fact-check sharers. Specifically, we find that fake-news sharers differ in their reliance on words about specific emotions, emphasizing high-arousal negative emotions like anger rather than low-arousal emotions. Additionally, we find that fake-news sharers' post histories are characterized by a higher usage of words related to power and existential concerns such as death and religion. 

In Study 2, we employ the socio-demographic and textual cues data extracted from post histories of fake-news sharers and random Twitter users to examine whether textual cues from these histories provide incremental predictive value to identify fake-news sharers. After we train a machine-learning classifier and a logistic regression model, we find that models that add textual cues from post histories achieve excellent predictive accuracy not only on withheld observations from the same training period (out-of-sample holdout) but also in a conservative prediction task for users whose data were collected two years later (out-of-time holdout) and with a different evaluation metric. 

In Studies 3 and 4, we leverage the descriptive findings on the textual cues that discriminate between fake-news sharers and others we uncovered in Study 1 to explore possible fake-news sharing mitigation strategies through experiments and surveys. Study 3 uses insights generated in Study 1 to explore how anger affects the intentions to share fake news focusing on the joint role of situational and trait anger. Study 4 investigates how power, a construct understudied in the context of fake-news sharing, but one that Study 1 finds differentiates such sharers, could be used to encourage fake-news mitigation behaviors. The experiment also addresses a limitation of relying on data from post histories -- it is difficult to safely and anonymously augment controlled experiment responses (e.g., randomized experiments within surveys) with data on post histories from accounts verified to belong to the survey respondent. To address this issue, we developed a tool for Study 4 that utilizes the Twitter API within a Qualtrics survey to anonymously and securely link participants' survey responses to their Twitter post histories. 

Across our multi-method investigation aimed at a substantive contribution \cite[c.f.,][]{blanchard2022promoting}, we make the following main contributions. First, we highlight a valuable, but often overlooked, source of information that can aid in understanding and reducing fake-news sharing among social media users: their post histories, which reflect their daily language, regardless of what is shared. We use these data to help uncover novel characteristics that are unique to fake-news sharers, not only compared to random social media users (the implicit comparison in most prior research), but also to others in the fake-news ecosystem (i.e., fact-check sharers, users matched on socio-demographics, those who share articles from right-leaning and left-leaning media). 

Second, we emphasize the importance of addressing participant misrepresentation in popular survey panels, particularly when dealing with low-incidence subpopulations like Twitter users \citep{baumgartner2022critical,sharpe2017mturk}. For Study 4, we integrate a social media account authentication within the survey flow (i.e., verify the Twitter account provided belongs to the respondents), while ensuring the protection of their privacy. We hope our approach will inspire others to design sampling systems prioritizing accuracy and participant privacy. Our replication files and analyses are at \url{https://osf.io/mp54w/}. 

\section*{Research Approaches Used to Study Fake-News Sharing} \label{review}

The dissemination of misinformation in news form, often referred to as fake news, involves various channels and actors, including the original creators, misinformation outlets, legitimate news sources, and social media accounts, encompassing not only individuals but also bots and companies. To understand how individuals share misinformation, researchers have utilized diverse data sources and methodologies to define and measure key concepts.

Initially, a significant number of studies relied on surveys and controlled experiments to explore what influences people's ability to differentiate between true and false news headlines and their subsequent sharing behaviors \citep[for instance,][]{allcott2017social, bronstein2019belief, guess2020digital}. In these studies, participants are typically presented with a mix of news headlines, pre-validated as either authentic or fabricated by reputable third parties or manipulated for the study. Researchers have then examined how various factors such as demographics, political leanings, personality traits, and emotional states affect people's discernment of news and their intentions to share fake news \citep[see, for example, cognitive reflection in][]{pennycook2021psychology} and how certain manipulations might influence these outcomes \citep[as seen in][]{bago2020fake, martel2020reliance}.

Despite the valuable insights gained from such studies, their controlled environments often limit the generalizability of the findings to the field (i.e., external validity). While some have suggested a correlation between the intention to share fake news in a survey and actual sharing behavior on social media \citep{mosleh2020self}, reliance on a survey from the general population may not accurately reflect the users most likely to encounter and share misinformation on social media platforms. While it is possible to carefully construct survey screeners to minimize the risk of false positives (e.g., including participants who would never be in the position to share fake news on social media), the risk of misrepresentation can remain high if participants have incentives to qualify for surveys \citep{sharpe2017mturk}. To the extent that those most likely to be exposed to fake news are different from those who are not, the inclusion of participants unlikely to be exposed to fake news on social media limits generalizability \citep{baumgartner2022critical}.\footnote{While some researchers ask survey respondents for their social media usernames to measure the effect of in-survey interventions on fake-news sharing behavior \citep[e.g.,][]{osmundsen2021partisan}, such researchers can only verify that the account is real -- not that it actually belongs to the survey taker. In our attempts, several implausible accounts were self-reported.} 

The limitations of laboratory experiments and surveys in establishing strong, real-world causal relationships have led researchers to employ field experiments to examine the sharing of fake news. Although innovative methods have been used to introduce randomization in these studies \citep[e.g.,][]{mosleh2021perverse, pennycook2021shifting}, conducting experiments on social media platforms presents significant challenges. One key issue is that researchers cannot observe the news items that users are exposed to but decide not to share. Additionally, the interplay between a user's actions and the platform's algorithms influences what content is presented to them. This phenomenon, known as divergent delivery bias \citep{eckles2018field,johnson2023inferno}, greatly limits the ability to use advertising campaigns as true experiments \citep{braun2024leveraging,braun2023ab}.

Given the challenges with lab and field experiments, many researchers have utilized observational data. Typically, they collect data by identifying individuals who share content from news domains labeled as fake by independent organizations and examine available information about these users or the language used in the posts sharing fake news \citep[e.g.,][]{grinberg2019fake, guess2019less, guess2020exposure, osmundsen2021partisan, shu2018understanding, verma2022examining}. One advantage of relying on observational data is that it is easy to gather behavioral data (e.g., their actual sharing behavior and their network of followers) for many users. Researchers investigate not only the action of sharing fake news, but also use the text provided along with the fake-news article link to glean insights into fake-news sharers' motivations \citep[e.g.,][]{ghanem2020emotional, jiang2018linguistic}. However, such analyses are fraught with challenges. First, automatically extracting a link of users who shared fake news from a pre-determined list of fake-news publishers neglects users who shared fake news from other sources (e.g., a generally legitimate news organization). Second, measuring theoretically and substantively relevant predictors can be challenging, as platforms do not usually have complete data on socio-demographics, political affiliation, or psychological states and traits. As such, researchers must rely on the users' texts to make inferences on such measures. 

Moreover, across these observational data approaches, challenges to causal inference abound. Say one finds that the language in posts with fake-news links contains a high proportion of power-related words (e.g., powerful, control). Can one assume that fake-news sharing is due to the feelings of powerfulness elicited by the post? Most likely not. Those who use more power-related words in a post may do so to compensate for their feelings of lacking power instead. Focusing only on the language of the post where fake news is shared may obscure gaining rich insights into the user because this post may not reflect the users' underlying traits, but rather reflect a situational response to the content of the fake news being shared. It is also likely that several observable (socio-demographics, social media usage) or unobservable factors (interests in politics) can explain the high use of power-related words in these users' posts in general. For example, those who share political news may use more power-related words in their posts, regardless of whether they contain a link to a fake-news post. As such, it is important to be careful with comparison groups. 

Across all these research approaches described above, some factors have been consistently shown to be important. First, socio-demographics and political affiliation  matter \citep[e.g.,][]{grinberg2019fake, lazer2018science, osmundsen2021partisan, shu2018understanding}. Second, past research has shown that the reliance on emotions tends to be heightened for fake-news sharers \citep[e.g.,][]{martel2020reliance, pennycook2021psychology} and that personality traits, notably the Big Five \citep{calvillo2023personality, shu2019beyond}, influence an individual's propensity to share fake news. We elaborate on these findings and compare them to ours at the end of Study 1. 

Building on prior research that has utilized dictionary-based analyses of fake-news promoting posts to understand the personalities of fake-news sharers \cite[e.g.,][]{giachanou2020role, shu2019beyond}, we argue that social media post histories are a valuable, yet underused, resource when it comes to understanding fake-news sharers across socio-demographics, social media activities, emotions, and personality. Moreover, we also argue that collecting such data for fake-news sharers and non-fake-news sharers is insufficient. Specifically, to isolate promising textual cues, it is crucial to compare their language not just with that of the average social media user, but also with that of individuals who share similar demographics, online behaviors, and exposure to both fake news and general political content. By doing so, we can better identify language patterns distinctly associated with fake-news sharers, enhance our ability to predict such behavior, and develop theories about the underlying psychological or situational factors that can inform effective interventions. 

The rest of the paper is structured as follows. In our initial study, we embark on an exploratory investigation into the textual cue differences in language used by fake-news sharers and various other groups within the fake-news ecosystem, such as random users, demographically similar users, fact-checkers, and political news sharers. By investigating the language used across posts rather than just in the post sharing fake news, we can infer differentiating characteristics of the fake-news sharer. We discuss our findings in light of prior research and present summary tables integrating our findings with past findings. In the second study, we use the textual cues extracted from post histories to predict who is more likely to share fake news and find a significant increase in predictive ability when we include textual markers of the post histories, compared to simply using socio-demographics. Studies 3 and 4 then use the insights generated from Study 1 regarding the language used by fake-news sharers to explore whether the textual cues uncovered in the first study can help to design interventions that potentially reduce fake-news sharing and increase fact-checking. 

\section*{Study 1: Exploring How Post Histories Can Help Distinguish Fake-News Sharers}

In this section, we outline our approach to exploring the use of post history data to study fake-news sharers on social media. First, we must define the criteria for identifying fake-news sharers (i.e., a classification rule) and select user comparison groups to help us analyze and understand what textual cues could be unique to this group. Once we have established these groups, we must determine which textual cues to extract. The following sections describe our choices for each step.

\subsection{Sampling Framework}

We employ two sampling approaches to identify fake-news sharers (Table \ref{tab:datadescription}). The first approach follows prior research that has relied on behavioral data to classify users as fake-news sharers. Specifically, we consider users who have shared articles verified as false by Snopes \citep[e.g.,][]{bronstein2019belief, mosleh2021perverse}. This method enables us to capture users who have engaged with specific fake news on social media. The second approach follows prior research that has classified users as fake-news sharers based on their sharing of any links from low-credibility domains \citep[e.g.,][]{grinberg2019fake, shu2018understanding}. By including users who share content from questionable sources, we can capture a broader range of fake-news sharers -- even those who might not have shared articles flagged by fact-checkers.

\begin{table}[htpb]
\caption{Sampling framework for our two datasets}
\label{tab:datadescription}
\footnotesize
\begin{tabular}{p{0.27\linewidth} | p{0.26\linewidth}  p{0.05\linewidth} p{0.26\linewidth} p{0.03\linewidth}}
\hline
\textbf{Identification}            & \textbf{ Dataset 1   (Snopes)}                                                                                                                               & \textbf{N} & \textbf{Dataset 2   (Hoaxy)}                                                          & \textbf{N} \\ \hline
\textbf{Fake-news sharers}         & Shared  1+ article identified as fake by Snopes.com                                                                                              & 1,727       & Shared 1+ fake-news publisher by Hoaxy                             & 235        \\
\textbf{Comparison}            &                                                                                                                                                             &            &                                                                                       &            \\
-- Fact-check sharers              & Shared   1+ Snopes.com                                                                                                                    & 671        & Shared   1+ fact-check publisher by Hoaxy                                                & 313        \\
-- Random Twitter users            & Random followers of one of 641 entities & 1,820       &                                                                                       &            \\
-- Matched sample           & Matched to fake-news sharers on socio-demographics (see Web Appendix C)                                                           & 1,725       &                                                                                       &            \\
-- Right-leaning news sharers         &                                                                                                                                                             &            & Shared: right-leaning  & 200        \\
-- Left-leaning news sharers          &                                                                                                                                                             &            & Shared: left-leaning & 284        \\ \hline
\textbf{Period}                    & 2015 to 2017                                                                                                                                                &            & Late 2018                                                                             &            \\
\textbf{Fake-news   articles}      & 66 mostly apolitical articles                                                                                                                  &             &          46 political articles                                                                             &            \\ \hline
\hline
\end{tabular}
\end{table}

\emph{Snopes.}\label{sec:snopesdata} To begin, we gathered 66 articles identified by Snopes as misinformation between 2015 and 2017. These articles covered both political and non-political fake news (see Web Appendix A). The presence of nonpolitical fake-news articles is important because it helps us get around any supply-side effects in determining who shares fake news, such as the fact that fake news tends to lean pro-conservative and is therefore more likely to be shared by conservatives \citep{allcott2017social,guess2019less}. For each article, we collected the Twitter usernames of users who shared them, and for whom we could access location and gender, forming the fake-news sharer group. Next, we formed three comparison groups. First, the fact-check sharer group comprised users who linked to Snopes.com fact-checking pages for the same articles (see Web Appendix A). Second, to approximate the behavior of an average Twitter user not engaged in misinformation sharing or fact-checking, we randomly sampled followers of brands, media sources, and non-profits, while considering their location and gender (random group)\footnote{See Web Appendix A for the list of brands.}. Third, we used propensity score matching to find users with similar characteristics for the matched comparison group to control for unique socio-demographics of fake-news sharers (see Web Appendix B). We gathered each user's last 3,200 tweets -- Twitter API's limit at that time.

\noindent \emph{Hoaxy.} Hoaxy classifies articles based solely on their source (publishing domain). It identifies "low-credibility" sources as more likely to publish inaccurate claims than reputable news. We label users as fake-news sharers (fact-check sharers) if they shared at least one link from fake-news (fact-checking) publishers in our observation window (October 2018) \citep {shao2018spread, shao2018anatomy, shao2016hoaxy}. Additionally, we collected data for users sharing left-leaning and right-leaning news from media identified as such by mediabiasfactcheck.com during the same observation window. For each user, we collected their last 3,200 Tweets.

\noindent \emph{Exclusions and robustness.} First, we only collect the tweets of users who shared either fake-news and fact-check articles but not both to avoid potential biases. Second, we exclude users with inaccessible profiles, such as those deleted or set to private within a year of our initial user ID collection, because we cannot retrieve complete information from these accounts. Third, we exclude re-tweets since we focus on analyzing users' original language. Fourth, we identified and excluded bots using the Botometer tool with a threshold of 50\% to enhance data accuracy. Fifth, we excluded individuals with fewer than 100 words in their post histories. In Web Appendix D, we present supplementary analyses about sensitivity. Our results are relatively insensitive to the amount of fake news that needed to be shared for classification, and to the inclusion of users without geographic information and of users who shared both fake-news and fact-checked articles. 

\subsection*{Measures and Variables}

\noindent \emph{Socio-demographics and social media usage} 

To explore how socio-demographics, user characteristics, and textual cues extractable from post histories relate to fake-news sharing, we must determine which socio-demographics and user characteristics to include. To do so, we built on previous research related to fake-news sharing. Fake-news sharers are often associated with being conservative, male, older, and more active on social media platforms \citep[e.g.,][]{grinberg2019fake, lazer2018science, osmundsen2021partisan, shu2018understanding}. To assess demographics, we used information from users' Twitter profiles. We accessed gender by building on the Python gender-guesser package.\footnote{https://pypi.org/project/gender-guesser/} We used self-declared information on the user's Twitter page to access geographical information. To assess political interest, we followed \cite{schoenmueller2023frontiers} and labeled users as Democrat (Republican) if they followed Democratic (Republican) reference accounts. For social media activity, we calculated activity per day (we use the status counts divided by the days a user has been on the platform), status counts, followers, and friends \citep[c.f.,][]{shu2018understanding}. We added followership of fake-news outlets, fact-checking outlets, and prominent media sources. Our measures are in Table \ref{tab:AllMeasures}.

\begin{table}[!htpb]
\caption{Summary of measures}
\label{tab:AllMeasures}
\begin{scriptsize}
\begin{tabular}{m{0.30\textwidth}p{0.65\textwidth}}
\hline 
\textbf{Gender} &
  Gender guesser API (https://pypi.org/project/gender-guesser/) \\ \hline
\textbf{Age} &
  Magic Sauce   (applymagicsauce) \\ \hline
\textbf{Geography} &
  Self-reported   on Twitter profile page \\ \hline
{\textbf{Political interest}} &
  - Democrat: 1 if follows Hillary Clinton’s Twitter account or the Democratic Party Twitter account and no Republican account; 0 otherwise. \\
 &
  - Republican: 1 if follows Donald Trump’s Twitter account or the GOP Twitter account and no Democratic account; 0 otherwise. \\ \hline
\textbf{Social media activity} &
  Activity per day (status counts/days on platform), number of followers, number of friends, status counts \\ \hline
\textbf{Media interest (followership of media Twitter accounts)} &  Salon, Wikileaks, Dailybeast, The New Yorker, Economist, Wall Street Journal, Washington Post, The New York Times, CNN, LA Times, USA Today, Buzzfeed, Aljazeera English, Fox News, Drudge Report \\ \hline
\textbf{Fake-news and Fact-check outlets followed} &  The number of fake-news and fact-check publishers followed (see Web Appendix I)\\ \hline
  
  {\textbf{Textual cues -- dictionary-based}} &
  Linguistic Inquiry and Word Count (LIWC) based on users’ tweets \\
   &
  Validations: NRC Lexica (Valence and arousal, anger, and fearfulness as well as their intensity), human-coded user tweets (anger, anxiety, and negative emotions), BERT (anger, anxiety, and negative emotions)\\
  \hline
{\textbf{Textual cues -- non-dictionary-based}} &
Big Five personality traits (openness to experience, conscientiousness, extraversion, agreeableness, neuroticism); Magic Sauce.\\ 
  & 
  Validation: personality traits model of \cite{park2015automatic}. \\ \hline
\end{tabular}
\end{scriptsize}
\end{table}

\noindent \emph{Textual cues from post histories} 

Next, we needed to determine the specific textual cues to extract from users' post histories. To accomplish this, we reviewed past research examining the factors influencing beliefs, accuracy assessments, and sharing behavior related to fake news. Previous studies have emphasized the significance of emotions, thinking styles, and even personality traits like the Big Five (openness, conscientiousness, extraversion, agreeableness, neuroticism) in shaping beliefs and behaviors concerning fake-news sharing. Specifically, research has highlighted the relevance of overall positive and negative emotional responses to fake-news exposure, as well as specific emotions such as happiness, anger, or anxiety \citep[e.g.,][]{bago2022emotion, rosenzweig2021happiness}. Moreover, in the more general context of information sharing, the arousal level of a specific emotion has been shown to impact the willingness to share information with others \citep{berger2012makes}. Apart from emotions, thinking styles relating to analytical thinking and cognitive processes have been shown to impact the sharing of fake news \citep[e.g.,][]{pennycook2019lazy, ross2021beyond}. Finally, past research has found that the underlying personality traits of a person are related to fake-news sharing \citep[e.g.,][]{lai2020falls, shu2018understanding} and has also explored textual cues related to writing style like the use of pronouns or temporal foci in language as indicators of fake-news sharing \citep[e.g.,][]{ghanem2020emotional, perez2018automatic}. Tables \ref{tab:comparison1}--\ref{tab:comparison3} summarize the specific findings on these different factors and compare them to ours. 

Researchers can rely on dictionary-based or non-dictionary-based approaches to extract these textual cues from post histories and gain insights into the characteristics, emotions, and traits that differentiate fake-news sharers from others. In our work, we use both. 

Dictionary-based approaches, such as Linguistic Inquiry and Word Count (LIWC) \citep{pennebaker2015development}, are commonly used to analyze consumers' writing samples \citep{berger2020uniting}. In our case, we can utilize LIWC on each user's entire post history to count words from pre-determined categories, including standard language categories (e.g., articles, prepositions), psychological processes (e.g., emotions, social), and specific content areas (e.g., school, work). LIWC then calculates the percentage of words in each language category out of the total number of words used in a document, which, in our case, would be all the past tweets accessible for a Twitter user. While LIWC is easy to implement and interpret, it does have limitations in scope. For example, differences regarding the level of arousal, which constitutes an important part of the quality of emotions, cannot be accessed via LIWC \citep{reisenzein1994pleasure}. As arousal has been shown to impact willingness to share information \citep{berger2012makes}, we also included four additional dictionaries for emotional valence (positive/negative) and arousal (high/low) \citep{villarroel2017unveiling}. These dictionaries enable us to extract numerous textual cues that may be useful for discriminating fake-news sharers in ways easily implementable by others.

Regarding personality traits, we chose a non-dictionary-based approach. In contrast to previous work that has relied on LIWC categories to infer psychological traits of fake-news propagators \citep[e.g.,][]{shu2018understanding}, we used the trait prediction engine Magic Sauce (applymagicsauce) to estimate the Big Five personality traits (openness to experience, conscientiousness, extraversion, agreeableness, neuroticism). Magic Sauce is a pre-trained classifier that utilizes data from participants who had both social media texts and psychometric survey results available. Extensively validated for personality and some socio-demographics \citep{youyou2015computer}, Magic Sauce provides a reliable, non-dictionary-based approach to infer personality traits from post histories. We also use alternative operationalizations and measures, as described in Table \ref{tab:AllMeasures}.

\subsection*{Results: Differences Between Groups} \label{descriptiveresults}

\begin{figure}[htbp]
    \caption{Snopes dataset: Values that positively discriminate fake-news sharers}
    \centering
    \includegraphics[width=0.80\textwidth]{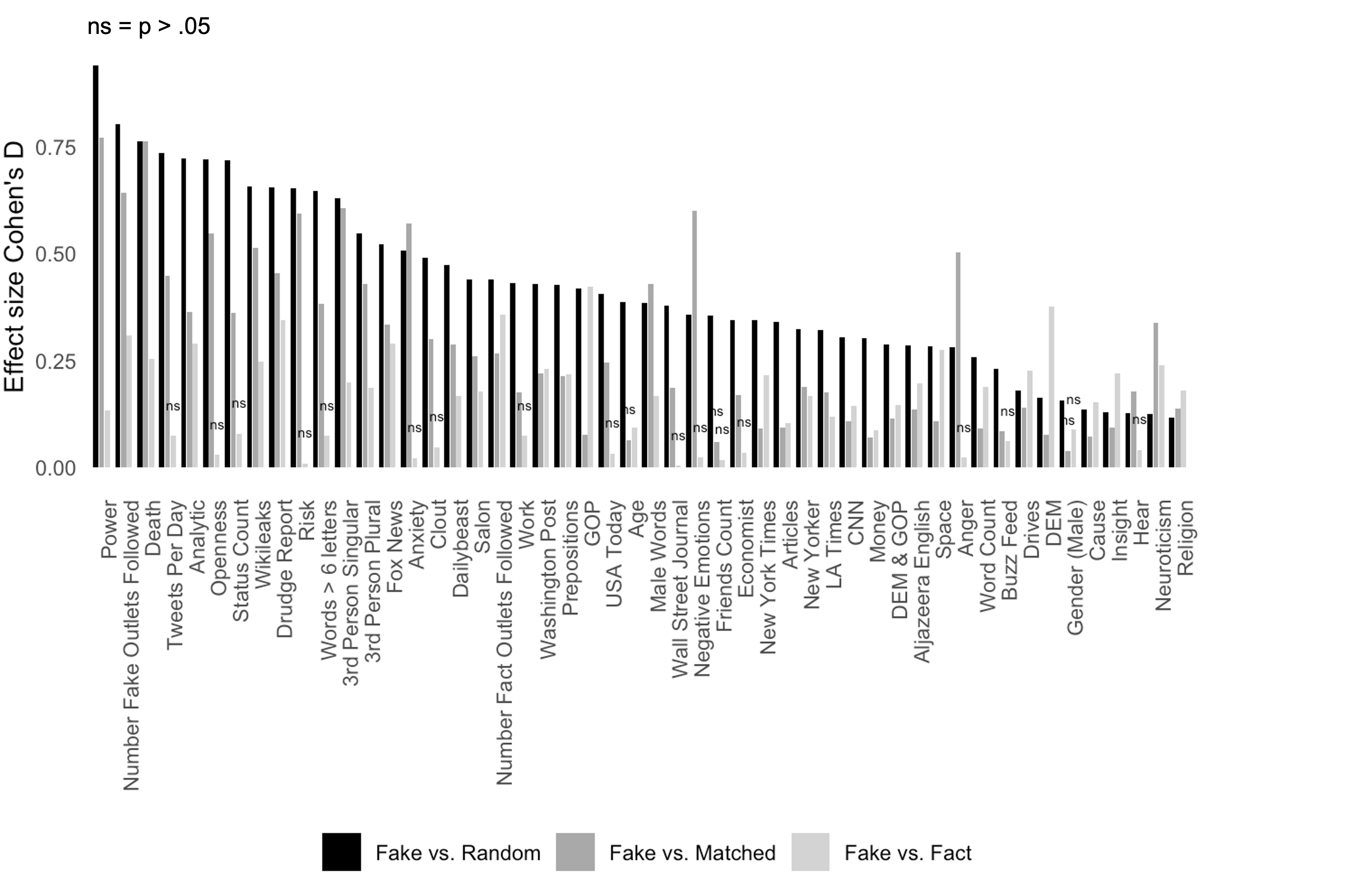}
    \label{fig:Figure_Snopes_CohensD_Pos}
\end{figure}

\begin{figure}[htbp]
    \caption{Snopes dataset: Values that negatively discriminate fake-news sharers}
    \centering
    \includegraphics[width=0.80\textwidth]{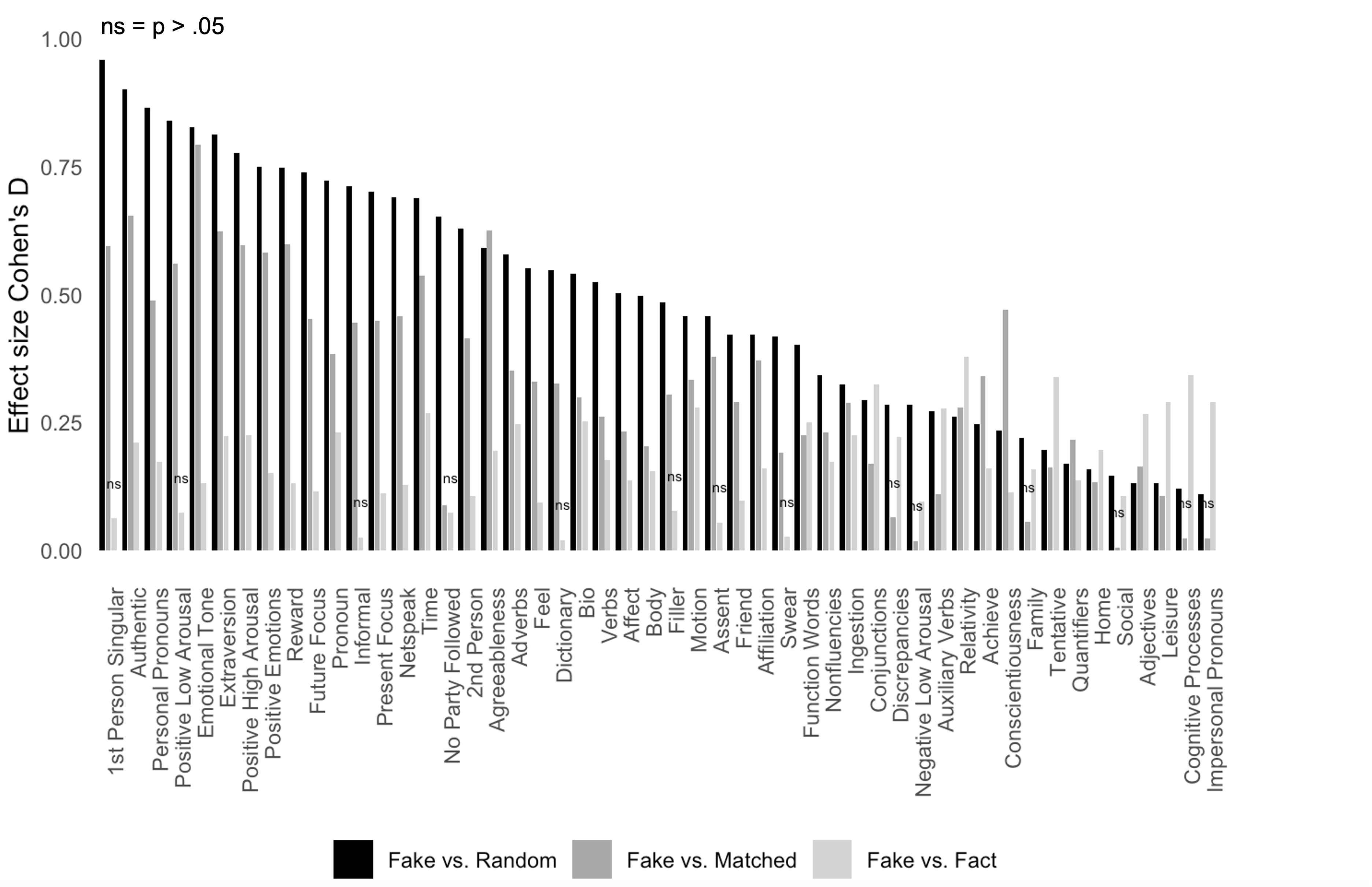}
    \label{fig:Figure_Snopes_CohensD_Neg}
\end{figure}

To explore differences between fake-news sharers and others, we present two figures: Figures \ref{fig:Figure_Snopes_CohensD_Pos} and \ref{fig:Figure_Snopes_CohensD_Neg}. Figure \ref{fig:Figure_Snopes_CohensD_Pos} compares the Snopes sample of fake-news sharers with non-sharers (random and matched samples) on characteristics where the fake-news sharers score significantly \emph{higher} than the random social media user group. We also include fact-check sharers as a reference group. Figure \ref{fig:Figure_Snopes_CohensD_Neg} shows the differences in the Snopes sample for characteristics where fake-news sharers score \emph{lower} than the random social media user group. Both charts show the 50 largest effect sizes, sorted from largest to smallest. Each figure has three data series. The first (in black) represents the observed effect size when comparing fake-news sharers to random Twitter users in the Snopes dataset. For instance, for the variable "power," the value ($d = .94$, a large effect size) indicates how much more fake-news sharers use power-related words than random Twitter users. The second data series (dark gray) shows the observed effect size comparing fake-news sharers to our matched Twitter users sample. For "power," the value ($d = .77$) indicates that fake-news sharers use more power-related words than Twitter users with similar socio-demographics. Notably, this significant difference in the frequency of using power-related words exists not only when compared to random Twitter users, but also when compared to users with similar socio-demographics. The third data series (in light gray) compares fake-news sharers to fact-check sharers ($d = .13$). Details for all variables and groups are in Web Appendix E.

\noindent \emph{Socio-demographics and social media activity}

Our results align with previous research regarding socio-demographics and political affiliation \citep[e.g.,][]{bovet2019influence, grinberg2019fake, lazer2018science, osmundsen2021partisan, shu2018understanding}. Fake-news sharers are notably older, more active on social media, and exhibit a higher level of interest in conservative Twitter accounts. This is evident from their followership of political accounts and their preference for following more conservative media outlets. However, it is worth mentioning that some of these distinctive characteristics also apply to fact-check sharers. For example, fact-check sharers exhibit similarity in their daily activity and the total number of tweets. That is, fake-news sharers and fact-check sharers have fairly similar social media usage. This is a novel finding in the misinformation literature with implications for uniquely predicting and targeting fake-news sharers.

\noindent \emph{Textual cues from post histories} 

We anticipated finding differences between fake-news sharers and other groups on emotions. We indeed find fake-news sharers tend to use words related to high-arousal negative emotions (e.g., anger and anxiety) more frequently than other Twitter users. The differences in high-arousal negative emotions also hold when we compare against users matched on socio-demographics. Moreover, fake-news sharers use fewer words related to positive emotions and low-arousal negative emotions. Regarding personality traits, they score higher on openness and neuroticism but are also \emph{less} agreeable, extroverted, and conscientious. Fact-check sharers are similar to fake-news sharers on high-arousal negative emotions and some personality traits. 

Beyond differences in the use of emotion-related words and personality traits, there are distinct language patterns in the post histories of fake-news sharers. The main distinguishing textual cue between them, and random as well as matched Twitter users is the usage of power-related words. Additionally, we note that fake-news sharers use language related to existential concerns more frequently, such as words associated with death and religion, compared to random Twitter users. They also use fewer words related to friends and family, and more words related to money. Regarding linguistic style, we found numerous differences between fake-news sharers and others. Fake-news sharers use fewer second-person singular words, future tense, filler words, swearing, and impersonal pronouns. Additionally, they utilize more clout, longer words (more than six letters), and articles. Furthermore, they are more inclined towards male-oriented language and display less assent.

\noindent \emph{Second dataset: Hoaxy.} Our identification of fake-news sharers through Hoaxy allows us to compare socio-demographics and textual cues against alternative comparison groups, specifically those who share political news including both left-leaning and right-leaning media users. Detailed results are in Web Appendix F.

Consistent with our findings in the Snopes sample, we observe that fake-news sharers are more likely to be conservative than fact-check sharers (and right-, as well as left-leaning news sharers). They tend to follow the GOP and conservative mainstream media outlets such as Fox News. However, compared to right-leaning news sharers, they tend to follow more accounts associated with fake-news publishers and are more active on social media. Fake-news sharers also use more words related to high-arousal negative emotions, anger, and negation, and their language is less analytic and comparative compared to right-leaning news sharers. Despite some similarities, there are notable differences between fake-news sharers and right-leaning news sharers. 

Similarities between fake-news and fact-check sharers were also observed, with both groups, showing lower levels of conscientiousness, agreeableness, and extraversion compared to right- and left-leaning news sharers. Additionally, both groups use more negative emotions, in particular negative high-arousal emotions; more words related to anger; certainty-related words; and swearing compared to right- and left-leaning sharers. These findings suggest that some prior observations of heightened emotionality in fake-news sharers may be partially attributed to their exposure and involvement to fake news, rather than solely to their sharing behavior. Nonetheless, some characteristics differentiate fake-news sharers from all other groups, such as their greater use of words related to death and religion. Overall, we find notable consistency between the fake-news sharer characteristics observed in the Hoaxy sample and those observed in the Snopes sample. 

\subsection*{Integrating our Exploratory Findings with Existing Literature on Fake News}

In examining the relationship between our exploratory findings and the existing literature on fake news, we categorize our key insights into three distinct constructs: emotions (Table \ref{tab:comparison1}), personality traits (Table \ref{tab:comparison2}), and others (Table \ref{tab:comparison3}). The tables present our findings in parallel with prior studies, highlighting both the constructs studied and the findings reported.

\noindent \emph{Emotions.} Research has consistently shown a link between intense emotional states (both positive and negative) and an impaired ability to distinguish real from fake news, which, in turn, influences the likelihood of sharing fake news. This phenomenon is observable in the general population and is consistent irrespective of whether the emotional states are naturally occurring or experimentally induced. Our study supports these findings, demonstrating that individuals who disseminate fake news on Twitter are more likely to use language indicative of negative high-arousal emotional states, such as anger and anxiety, compared to a control group of random and matched users. However, our findings also reveal a nuanced aspect of emotional expression: fake-news sharers tend to use fewer words associated with low-arousal negative emotions and fewer words related to positive emotions. Our study, therefore, contributes to the existing literature by highlighting the significance of the arousal level in emotional responses and how it relates to the language users use in social media posts. This finding underscores the importance of considering both emotion type and arousal level when examining the relationship between emotional states and the dissemination of fake news. Moreover, in line with \cite{giachanou2020role}, our study found no significant difference in the usage of negative emotion-related words between individuals sharing fake news and those engaged in fact-checking activities. This suggests that the heightened use of negative emotional language by fake-news sharers may be influenced by the emotionally charged nature of fake-news content, which fact-check sharers are also exposed to. 
\begin{table}[htpb]
\caption{Highlights: Negative emotions, positive emotions, anger, and anxiety}
\label{tab:comparison1}
\scriptsize
\resizebox{\textwidth}{!}{%
\begin{tabular}{p{\dimexpr \textwidth-2\tabcolsep}}
\hline
\textbf{Negative Emotions} \\ \hline
• \cite{martel2020reliance}: Those with higher states of negative emotions have a lower ability to discern true headlines. \\
• \cite{giachanou2020role}: There is no significant difference between fact-check sharers and fake-news sharers using negative emotion-related words in their post histories. \\
• \cite{ghanem2020emotional}: Fake-news stories use more negative emotion-related words than true-news stories. \\
• \cite{calvillo2021individual}: Negative emotionality is associated with a lower ability to discern fake news. \\
• \textbf{Study 1:} Fake-news sharers use more negative emotion-related words in their post histories (random, matched, left-, and right-leaning). While fake-news sharers use more high-arousal negative emotion-related words in their post histories (matched, left-, and right-leaning), this difference is not significant when compared to the random group. However, we do not find differences with fact-check sharers regarding negative high-arousal emotions and negative emotions in general. Fake-news sharers use fewer low-arousal negative emotion-related words compared to the random group, whereas we find no differences for the matched, left-leaning, and right-leaning media sharers. \\ \hline
\textbf{Positive Emotions} \\ \hline
• \cite{martel2020reliance}: Those with higher states of positive emotions are less able to discern fake news. \\
• \cite{bago2022emotion}: People who are happy after reading news headlines have lower discernment of fake news. \\
• \cite{rosenzweig2021happiness}: Those who are happy after reading COVID-19 headlines are more likely to believe and share fake news. \\
• \cite{giachanou2020role}: There is no significant difference between fact-check sharers and fake-news sharers regarding using positive emotion words in their post histories. \\
• \cite{ghanem2020emotional}: There is no significant difference between fake news and true news using positive emotion words. \\
• \textbf{Study 1:} Fake-news sharers use a smaller proportion of positive emotion-related words and positive high- and low-arousal emotion-related words in their post histories (random and matched). However, we do not find consistent evidence of differences between fact-check sharers or left- and right-leaning media sharers. \\ \hline
\textbf{Anger} \\ \hline
•  \cite{martel2020reliance}: Those with higher states of hostility or irritability are more likely to believe false headlines. \\
• \cite{rosenzweig2021happiness}: Those angered after reading COVID-19 headlines are more likely to believe, click, and share any news. \\
• \cite{greenstein2020anger}: Inducing anger increases susceptibility to misinformation. \\
• \cite{weeks2015emotions}: Those made angry by writing are more likely to believe fake news that aligns with their partisanship. \\
• \cite{bago2022emotion}: Those angered by news headlines have higher discernment for discordant headlines. \\
• \cite{ghanem2020emotional}: There is no significant difference between fake news and true news in using anger-related words. \\
• \cite{giachanou2020role}: There is no significant difference between fact-check sharers and fake-news sharers regarding using anger-related words in their post histories. \\
• \textbf{Study 1:} Fake-news sharers use a greater proportion of anger-related words in their post histories (random, matched, left-, and right-leaning). However, we do not find differences with fact-check sharers. \\
• \textbf{Study 3:} Those with higher trait anger share more fake news and true news, and reducing situational anger reduces the motivation of individuals with high trait anger to share news that is surprising. \\ \hline
\textbf{Anxiety} \\ \hline
• \cite{martel2020reliance}: Those with higher states of fear and distress are more likely to believe in false headlines, and have a lower ability to discern fake news. \\
• \cite{freiling2023believing,huang2022perceived, rosenzweig2021happiness}: Those anxious after reading COVID-19 headlines are more likely to believe and share any news. \\
• \cite{verma2022examining}: Twitter users who shared COVID-19 misinformation have higher anxiety when compared to similar users who did not share COVID-19 misinformation. \\
• \cite{bago2022emotion}: Those triggered with fear from headlines are not more or less likely to discern fake news. \\
• \cite{weeks2015emotions}: Those made anxious by writing are more likely to correct erroneous beliefs in line with their partisanship after being exposed to a correction. \\
• \cite{ghanem2020emotional}: There is no significant difference between fake news and true news in using fear-related words. \\
• \cite{giachanou2020role}: Compared to fact-check sharers, fake-news sharers use fewer anxiety-related words in their post histories. \\
• \textbf{Study 1:} Fake-news sharers use a greater proportion of anxiety-related words in their post histories (random, matched). However, we do not find consistent evidence of a difference for fact-check sharers, and no difference for right- or left-leaning media sharers. \\
• \textbf{Web Appendix G:} Increasing situational anxiety through a post on social media timelines did not significantly affect fake-news sharing. \\ \hline
\end{tabular}%
}
\end{table}

\emph{Anger and anxiety.} Several studies have established a connection between increased anger and a diminished capacity to discern fake news \citep{greenstein2020anger, martel2020reliance}, although this effect may vary depending on whether the news aligns with the individual's pre-existing beliefs \citep{bago2022emotion}. Our findings generally concur with these studies, showing that Twitter users who share fake news tend to use more anger-related words in their posts than other users. However, similar to \cite{giachanou2020role}, we did not find a significant difference in anger-related language usage between fake-news sharers and fact-checkers - a high use of anger-related words seems to characterize both groups. Regarding anxiety, past studies have found that anxiety lowers the ability to discern fake news and increases the likelihood of believing and sharing it \citep{martel2020reliance,verma2022examining}. While we find that fake-news sharers use more anxiety-related words than the random and the matched sample of Twitter users, our study did not find consistent evidence of higher anxiety-related word usage among fact-check sharers compared to fake-news sharers or political news sharing users (both left and right-leaning news). 

\noindent \emph{Personality traits.} Recent research has investigated into how the Big Five personality traits correlate with the ability to discern fake news and the propensity to share it. As indicated in Table \ref{tab:comparison2} and reviewed comprehensively by \cite{calvillo2023personality}, findings show a lack of consistency in which the Big Five personality traits are linked to higher or lower discernment of fake news and to increased or decreased sharing intentions. In contrast, our research uncovers a consistent pattern differentiating fake-news sharers from both randomly selected Twitter users and those matched on socio-demographics. Across our datasets, including Snopes (varied news) and Hoaxy (political news only), individuals who share fake news exhibit higher levels of neuroticism and openness but lower levels of extraversion, agreeableness, and conscientiousness.

\begin{table}[htpb]
\caption{Highlights: The Big Five personality traits}
\label{tab:comparison2}
\scriptsize
\resizebox{\textwidth}{!}{%
\begin{tabular}{p{\dimexpr \textwidth-2\tabcolsep}}
\hline
\textbf{Openness} \\ \hline
• \cite{shu2018understanding}: Across two datasets of Twitter users who share news from Buzzfeed and Politifact, there is no significant relationship between openness and the proportion of shared content that is fake news. \\
• \cite{ahmed2022personality, buchanan2021trust,sindermann2021evaluation}: Openness is not significantly correlated with fake-news discernment or sharing intentions.\\
• \cite{calvillo2021personality}: Those with higher openness have better fake-news discernment.\\
• \cite{ahmed2022social}: Those with higher openness have lower COVID-19 fake-news discernment, but no difference for sharing intentions.\\
• \cite{lawson2022pandemics}: Those with higher openness have higher sharing intentions for fake news. \\
• \textbf{Study 1:} Fake-news sharers have a higher predicted openness score than random and matched users. However, we do not find consistent evidence of differences between fact-check sharers, right-leaning and left-leaning media sharers.\\ \hline
\textbf{Neuroticism} \\ \hline
• \cite{shu2018understanding}: Among Twitter users who share news from Buzzfeed, those with a higher neuroticism score have a lower proportion of shared content that is fake news. However, the authors did not find a similar result in their Politifact dataset. \\
• \cite{lawson2022pandemics,shephard2023everyday}: Those with higher neuroticism scores have lower fake-news discernment and have higher sharing intentions for fake news. \\
• \cite{ahmed2022personality,ahmed2022social,buchanan2021trust,sindermann2021evaluation}: Neuroticism from a survey is not significantly correlated with fake-news discernment or fake-news sharing intentions. \\ 
• \textbf{Study 1:} Fake-news sharers have a higher predicted neuroticism score than random users, matched users, and left-/right-wing media sharers, while they seem to have a lower neuroticism score than fact-check sharers. \\ \hline 
\textbf{Extraversion} \\ \hline
• \cite{shu2018understanding}: Among Twitter users who share news from Buzzfeed, those with a higher extraversion score have a higher proportion of shared content that is fake news. However, the authors did not find the same result in their Politifact dataset. \\
• \cite{ahmed2022social,sindermann2021evaluation}: Those with higher extraversion have lower fake-news discernment. \\ 
• \cite{calvillo2021personality}: Those with higher extraversion have better fake-news discernment.\\
• \cite{shephard2023everyday}: Extraversion is not correlated with sharing fake news. \\
• \cite{ahmed2022social,lawson2022pandemics}: Those with higher extraversion have higher intentions to share fake news. \\
• \cite{ahmed2022personality}: Liberals high in extraversion have higher fake-news discernment, but neither liberals nor conservatives high in extraversion are less likely to share fake news. \\ 
• \textbf{Study 1:} Fake-news sharers have a lower predicted extraversion score than random users, matched users, and right-wing media sharers. We find no difference compared to left-leaning media sharers. Moreover, they also seem to have a higher extraversion score than fact-check sharers (but the difference is not significant in the Hoaxy data). \\ \hline
\textbf{Agreeableness} \\ \hline
• \cite{shu2018understanding}: Across two datasets of Twitter users who share news from Buzzfeed and Politifact, those with a higher agreeableness score have a greater proportion of shared content that is fake news. \\
• \cite{ahmed2022social,shephard2023everyday,sindermann2021evaluation}: Agreeableness is not significantly correlated with fake-news discernment or fake-news sharing intentions. \\ 
• \cite{calvillo2021personality}: Those with higher agreeableness have poorer fake-news discernment.\\
• \cite{lawson2022pandemics}: Those with higher agreeableness have higher fake-news sharing intentions.\\
• \cite{ahmed2022personality}: Liberals high in agreeableness have lower fake-news discernment. However, both liberals and conservatives high in agreeableness have lower intentions to share fake news.\\ 
• \textbf{Study 1:} Fake-news sharers have a lower predicted agreeableness score than random users, matched users, and left-/right-wing media sharers. They also seem to have a higher agreeableness score than fact-check sharers (we find no significant difference in the Hoaxy data). \\
\hline
\textbf{Conscientiousness} \\ \hline
• \cite{shu2018understanding}: Across two datasets of Twitter users who share news from Buzzfeed and Politifact, there is no significant relationship between conscientiousness and the proportion of shared content that is fake news. \\
• \cite{ahmed2022social,calvillo2021personality}: Those with higher conscientiousness have better fake-news discernment and lower fake-news sharing intentions \citep{ahmed2022social}.\\
• \cite{ahmed2022personality,buchanan2021trust,sindermann2021evaluation}: Conscientiousness is not significantly correlated with fake-news discernment. However, it can predict lower fake-news sharing intentions among both liberals and conservatives (Ahmed and Tan 2022).\\
• \cite{lawson2022pandemics}: Among those with high conscientiousness, conservatives are more likely to share fake news than liberals. \\
• \cite{lin2023conscientiousness}: Among those with high conscientiousness, conservatives are more likely than liberals to share both true and fake news. \\
• \textbf{Study 1:} Fake-news sharers have a lower predicted conscientiousness score than random users, matched users, and left-/right-wing media sharers. Moreover, they also seem to have a conscientiousness score higher than that of fact-check sharers (but the difference is not significant in the Hoaxy data).\\ \hline
\end{tabular}%

}
\end{table}

Our study also contributes to the understanding of personality's impact on fake-news sharing by emphasizing the need to consider not only the general population, but also those more likely to encounter misinformation. For example, while fake-news sharers demonstrate more openness than random or matched users, this trend is less evident when compared to individuals who share fact checks or content from politically-leaning media. This observation suggests that prior differences in openness may reflect broader social-media news sharing patterns, rather than being unique to fake-news sharers. Moreover, despite fake-news sharers exhibiting more neuroticism and less agreeableness, extraversion, and conscientiousness than randomly selected and matched Twitter users, they show less neuroticism and more agreeableness, extraversion, and conscientiousness than fact-check sharers. This finding indicates Big Five traits might be a source of self-selection in who chooses to share content related to misinformation, be it fake news or corrections to fake news.

\noindent \emph{Power, death, and religion.} Our research extends the understanding of fake-news sharing by examining existential elements like power, religion, and mortality (death), alongside emotions and personality traits. Scholars have increasingly focused on these existential factors to comprehend why individuals share fake news.

In the realm of fake news, the interplay of power and status-seeking (seen as a facet of power) with fake-news sharing has been under-researched. Studies exploring this relationship, particularly concerning COVID-19 fake news, have yielded inconsistent results \citep[e.g.,][]{apuke2020modelling, apuke2021social, jiang2018linguistic, jun2022social}. Our first study shows that fake-news sharers' post histories contain more power-related words than random users, matched users, and fact-check sharers. However, there is no significant difference in power-word usage between fake-news sharers and those sharing content from politically-leaning news sources. One could then infer that the influence of power and status-seeking for fake-news sharing may depend on the content's subject matter (e.g., whether it is political or not). However, it is crucial to note that these studies, including our Study 1, do not directly address the motivations behind using power-related words -- whether it stems from a sense of power or a desire for it. Our exploratory Study 4 reveals that individuals using more power-related words report lower personal power, suggesting a compensatory motive in their word choice. 

Regarding the prevalence of death-related words, our findings align with existing research \citep{giachanou2020role} showing that fake-news sharers use these words more frequently than other comparison groups. Prior studies have demonstrated that exposure to fake news, as opposed to true news, can heighten the awareness of death-related thoughts, which correlates with sharing fake news \citep{lim2021infodemic}. However, this specific aspect of fake-news sharing has been scarcely examined.

As for religion, the relationship between religiosity and fake-news sharing is complex and appears to vary based on how religion is operationalized. While high religious fundamentalism is linked to lower accuracy in discerning truth \citep{bronstein2019belief}, and regular religious service attendees show less discernment with COVID-19 fake news \citep{druckman2021role}, high self-reported religiosity does not necessarily correlate with increased fake-news sharing \citep{stefanone2019news}. Adding to the complexity of these findings, we find in Study 4 that using religion-related words correlates with self-reported religiosity. Finally, in Web Appendix H, we provide a study showing that activating religious values via a priming task can decrease the desire to share surprising content, but it does not seem to directly affect fake-news sharing. 

\begin{table}[htpb]
\caption{Highlights: Power, death, and religion}
\label{tab:comparison3}
\scriptsize
\resizebox{\textwidth}{!}{%
\begin{tabular}{p{\dimexpr \textwidth-2\tabcolsep}}
\hline
\textbf{Power} \\ \hline
• \cite{jiang2018linguistic}: Posts containing fake-news links contain a high proportion of power-related words.\\
• \cite{apuke2020modelling}: Those seeking status are not more likely to report having shared fake news about COVID-19. \\
• \cite{apuke2021social}: Those seeking status report sharing less fake news about COVID-19. \\
• \cite{kumar2023should}: Social status-seeking moderates the effect of perceived believability on sharing intentions of fake news.\\
• \textbf{Study 1:} Fake-news sharers use a greater proportion of power-related words in their post histories than random users, matched users, and fact-check sharers. However, we do not find differences between left- and right-leaning media sharers. \\
• \textbf{Study 4:} Adding power-related words in an ad copy for a fact-checking browser extension increases intentions to click and download the extension. \\ \hline
\textbf{Death} \\ \hline
• \cite{jiang2018linguistic}: Posts containing fake-news links include a high proportion of death-related words.\\
• \cite{lim2021infodemic}: Participants who read fake news had higher death-thought accessibility than those who read true news, and death-thought accessibility correlates with increased fake-news sharing.\\
• \cite{giachanou2020role}: Fact-check sharers use fewer death-related words than fake-news sharers in their post histories.\\
• \textbf{Study 1:} Fake-news sharers use a greater proportion of death-related words in their post histories than random users, matched users, fact-check sharers, and left- or right-leaning media sharers. \\ \hline
\textbf{Religion} \\ \hline
• \cite{bronstein2019belief}: Those scoring high on religious fundamentalism have lower discernment of true headlines. \\
• \cite{stefanone2019news}: Those self-reporting high in religiosity are not significantly more likely to share political fake news. \\
• \cite{druckman2021role}: Those who report attending religious services have lower discernment for COVID-19 fake news. \\
•  \cite{jiang2018linguistic}: Posts containing fake-news links contain a high proportion of religion-related words.\\
• \textbf{Study 1:} Fake-news sharers use a greater proportion of religion-related words in their post histories than random users, matched users, fact-check sharers, and left- or right-leaning media sharers.\\
• \textbf{Web Appendix H:}  Activating religious values decreases the motivation to share surprising news, affecting both fake and true news. However, we do not find a main effect of activating religious values on sharing fake news.\\ \hline
\end{tabular}%
}
\end{table}

\subsection*{Additional Analyses and Robustness Tests}

\noindent \emph{Using other tools to extract textual cues from post histories.} Several factors could explain the uncovered differences between fake-news sharers and others: a) our decision to classify fake-news sharers based on sharing a single piece of misinformation for the Snopes dataset, b) the use of LIWC, and c) the use of Magic Sauce to proxy personality traits. In Web Appendix I, we show that those we classify as fake-news sharers have incrementally more fake-news mentions in their post histories, justifying our classification procedure. In Web Appendix J, we document similar results on emotions using other lexicons (NRC-VAD, NRC Emotion, and NRC Emotion Intensity) and a BERT model based on tweets validated by blind judges through mTurk tasks to measure emotions. In Web Appendix K, we reach similar conclusions for personality traits using another pre-trained classifier instead of MagicSauce.

\noindent \emph{Similarities in textual cues in fake-news content to sharers' post histories.} In Web Appendix L, we explore whether users post fake news aligned with their linguistic profile. The textual cues of the fake-news articles shared are more similar to those who share them than to our random and matched samples. This could result from fake-news sharers being drawn to articles that match their characteristics. Engaging more with these articles also results in these users being served more such articles by the platform, resulting in a cycle of greater exposure and sharing of fake news. 

\noindent \emph{Political and non-political fake-news sharers.} One could be curious about the degree to which the observed differences are due to the articles we used as a starting point; these were mainly non-political articles in our Snopes data. In Web Appendix M, we examine the Snopes data to determine how much our findings differ based on whether the fake news shared was political or non-political. Although most articles in our Snopes sample do not pertain to political news, a subset covers political topics. We find that focusing solely on political fake-news sharers would, if anything, amplify the differences between the fake-news sharers and most comparison groups. However, it would be interesting to collect other samples of topically focused fake-news sharers, such as those focusing on COVID-19 or climate change.

\section*{Study 2: The Predictive Value of Post Histories}\label{prediction} 
In Study 1, we presented two sets of descriptive analyses of the socio-demographics, social media activities, and textual cues from post histories of those who share fake news. Although we found evidence that those who share such news differ from random Twitter users, we found that in some ways, they are similar to others in the fake-news ecosystem, particularly fact-check sharers, but also matched users, left-leaning news sharers, and right-leaning news sharers. Although one should be careful not to misinterpret such observational differences as causal, one should also remember that observational data can be particularly useful for prediction problems where causal inference is not necessary \cite[c.f.,][]{kleinberg2015prediction}. Identifying who is likely to share fake news can help with targeting efforts meant to ameliorate the sharing of fake news and increase fact-checking behaviors. Hence, in our second study, we test whether we can better predict fake-news sharers (vs. random users and others in the misinformation ecosystem) by adding textual cues from post histories to socio-demographic and social media usage and followership variables. 

\subsection*{Nested Comparison Evaluative Framework}

To quantify the extent to which the addition of textual cues can help distinguish fake-news sharers from both fact-check sharers and other Twitter users, we compare models trained solely on baseline variables (socio-demographics, followership of fake-news and fact-check outlets, media outlet followership, social-media activity) to a model that uses only textual cues, and one that includes both sets of variables.\footnote{We pre-train our classifier using the Snopes dataset because a) it is larger, and b) our measure of sharing is based on article-level misinformation sharing and is not restricted to only disinformation domains (i.e., fake-news publishers).} First, we divide our data into in-sample (80\%) and out-of-sample (20\%) test sets. Second, we use 10-fold cross-validation within the training set to determine tuning parameters and focus our evaluations on out-of-sample evaluations. Third, given that the prior probability of successful prediction (i.e., of being a fake-news sharer) is both imbalanced and unknown, we use the “area under the curve” (AUC) of the Receiver Operating Characteristic (ROC) curve, that is commonly used to evaluate predictive classifiers in marketing research \citep[e.g.,][]{netzer2019words}. Finally, because we do not wish to make an a priori judgment concerning the relative costs of misclassification for false positives and false negatives (i.e., false-positive: misclassifying a non-sharer of fake news as a fake-news sharer; false-negative: misclassifying a fake-news sharer as a non-sharer of fake news), we opted to select classification thresholds using Youden's Index, which selects the point that maximizes specificity (true positives) and sensitivity (true negatives) equally.

Numerous alternative machine-learning classifiers could be used when choosing a predictive model. We chose to use logistic regression and Bayesian Additive Regression Trees \citep[BART; ][]{chipman2010bart}. We chose BARTs as a second model because they not only have the ability to accommodate unanticipated non-linear main effects and multi-way interactions and provide similar performance to other models, but do not require using many researcher-degrees of freedom \citep[see][]{blanchard2023extraction}. 

\subsection*{Results}

Table \ref{tab:mainresults} presents prediction accuracy results, with ROC curves in Web Appendix N. Both logistic regressions and BARTs improve AUC and accuracy when we add textual cues from post histories to baseline variables. Out-of-sample BART models predict better overall when it comes to AUC (e.g., out-of-sample AUC$_{BART}$ = .9206 vs. AUC$_{LR}$ = .9074)\, but perform similarly to logistic regression for accuracy. In short, our predictive accuracy results suggest that including textual cues consistently improves accuracy across both logistic regression and BART models. Even using only textual cues from post histories performs better than baseline variables alone. 

\begin{table}[htbp]
\caption{Predictive value: results across in-sample, out-of-sample, and out-of-time}
\footnotesize
\label{tab:mainresults}
\resizebox{\textwidth}{!}{%
\begin{tabular}{lllllll}
 & \multicolumn{2}{c}{In-sample} & \multicolumn{2}{c}{Out-of-sample} & \multicolumn{2}{c}{Out-of-Y-and-time} \\
Model & AUC & Accuracy & AUC & Accuracy & AUC & Accuracy \\ \hline
LR: Baseline & .846 & 76.68\% & .839 & 75.70\% & .759 & 74.40\% \\
LR: Textual cues & .908 & 83.28\% & .878 & 80.90\% & .763 & 78.74\% \\
LR: Baseline + Textual cues & .932 & 86.46\% & .907 & 83.15\% & .747 & 78.26\% \\
BART: Baseline & .895 & 81.62\% & .876 & 80.76\% & .809 & 73.91\% \\
BART: Textual cues & .939 & 86.42\% & .901 & 82.72\% & .824 & 83.82\% \\
BART: Baseline + Textual cues & .946 & 87.76\% & .921 & 83.15\% & .846 & 80.92\% \\ \hline
\end{tabular}%
}
\end{table}

Incorporating textual cues extracted from post histories can significantly enhance the identification of fake-news sharers. Although our logistic regressions do not implement regularization, BARTs do. One could then be concerned that the observed performance improvements from adding textual cues in BARTs are merely due to providing a regularization-enabled model with more features to learn from. To address this concern, we collected an out-of-classfication-and-time sample with users from a time that was two years later, and classified users as fake-news sharers using a different mechanism (linked to a fake-news publisher instead of a fake-news article from Snopes).\footnote{Specifically, we sampled one user's follower list, selected one follower from this list, accessed the followers of this focal follower, and repeated this process six times. Six degrees of separation connects 97.91\% of Twitter users \citep{Sysomos}. We then added one user with a public profile to our sample. Our final sample included 35 users who shared at least one link from a fake-news publisher (9\%), and 379 who had not (91\%).} To understand why regularization benefits might diminish on this second test set that altered time and outcome for prediction, consider how BART's regularization works. BART uses Bayesian priors on model parameters, including tree structures, to manage complexity based on in-sample patterns. This aims to prevent overfitting by promoting simpler tree structures. However, this regularization assumes data generation is consistent between training and out-of-time datasets. Changes in population dynamics (and outcome measures) in the out-of-time dataset could alter predictor--outcome relationships. As these changes were not considered during training, they can easily lead to reduced predictive performance when applying a regularized model to out-of-time samples.

As shown in the last two columns of Table \ref{tab:mainresults}, both logistic regression models achieved approximately .75 AUC (much above chance), while BARTs had an AUC of .85 when textual cues were used. Importantly, these results help reduce concerns that the improved performance observed in contrasting nested variable sets is solely due to regularization. Moreover, these findings provide evidence of correspondence between different classifications of fake-news sharers (source vs. individual article) and underscore the predictive benefits of using textual cues from post histories. 

\section*{Overview of Exploratory Experiments (Studies 3 and 4)}
The findings of Study 2 add to the value of the exploratory findings in Study 1 by showing that a generally neglected data source (i.e., textual cues from post histories) can be used to better predict who is likely to share fake news. From an understanding and mitigation standpoint, several interesting findings emerge from the exploratory analyses in Study 1, regarding emotions, personality, and other textual cues that differentiate fake-news sharers from others. However, such observational data is limiting when it comes to theory testing. 

One limitation of such correlational analyses is that one should not assume that they capture underlying causal relationships between psychological states and fake-news sharing. For example, in Study 1, we found that fake-news sharers have a greater proportion of anger-related words in their post histories than all of the other comparison groups (except fact-check sharers). Fake-news sharers may use such words due to both situational (state) and trait (enduring) factors. Specifically, whereas situational anger refers to the experience of anger in a particular moment or state, trait anger refers to a relatively stable personality disposition in which people are more prone to experience anger in response to a variety of triggers, with reactions that may be intense and frequent \citep{deffenbacher1996state}. To the extent that state anger tends to be directed (e.g., toward others on social media), interventions that reduce anger could reduce actions based on anger (e.g., reduce fake-news sharing). However, suppose anger is a trait of fake-news sharers -- a trait that may be further perpetuated by high exposure to social media \citep{milli2023twitter}. In that case, it may be that trait-anger is more predictive of fake-news sharing and that reducing situational anger will have little effect. In fact, trait and state anger may additively predict fake-news sharing but if trait anger is fairly high, state anger may have little effect. As anger has been shown to increase social media sharing in general \citep[e.g.,][]{berger2012makes}, it may also be that previously uncovered effects of anger on fake-news sharing \citep[e.g.,][]{bago2022emotion, greenstein2020anger, rosenzweig2021happiness} can be explained by a higher propensity of high trait anger individuals to share news of all kinds. We investigate the role of state and trait anger in Study 3.

In addition to the difficulty of addressing causality using observational data, a second limitation of relying on observational data is that researchers often cannot establish a direct connection between the textual cues extracted from post histories (Study 1) and the survey responses from samples utilized in their causal research (i.e., self-reports from Twitter users on a panel). Given these constraints, we designed Study 4 to illustrate an alternative approach to using post histories in the context of controlled experiments: augmenting survey responses with aggregated data of Twitter post histories. We move from studying anger to studying power to explore another construct of potential relevance to mitigating fake-news sharing. Our choice to focus on power is based on the findings of Study 1 regarding the ability of power-related word usage in post histories to differentiate fake-news sharers. We found that fake-news and fact-check sharers use more power-related words than random Twitter users. The elevated usage of power-related words by these two groups, with presumably opposing underlying motivations regarding fake news, may suggest that the usage of power-related words in their texts reflects a low sense of power, which they try to compensate for or regain by sharing either fake news or fact checks. We test this assumption in the study by examining the relationship between textual cues in participants' post histories and their responses to questionnaires measuring desire for power and feelings of power. Besides testing this relationship, the main goal of this study is to examine the role of empowering language in ads in mitigating the spread of misinformation. Combined with the findings of previous research that power activation motivates individuals and increases their willingness to take action \citep{galinsky2003power,guinote2017power}, we explore if messages using empowering language can engage both fake-news sharers and fact-check sharers by encouraging behaviors that can help mitigate fake-news sharing. Specifically, we explore whether using more power-related words in ad copy can enhance beneficial actions such as clicking an ad and downloading a browser extension to fact-check news.

To summarize, our primary objective with these exploratory experiments is not to test hypotheses, but rather to explore potential mitigation strategies hinted at by our exploratory data, and to stimulate research ideas, especially considering the use of this available data source and the significance of the misinformation domain. Although our exploratory study guided us to focus on anger and power, we acknowledge that other emotions and factors (e.g., using first-person pronouns or using male/female words) likely also play a role. We hope other scholars will develop interventions using other findings from Study 1 that can serve as grounds for theory development and intervention building.

\section*{Study 3: Reducing Situational Anger}

This study explores the joint role of situational and trait anger. Particularly, we are interested in exploring whether the heightened use of anger-related words in the post histories of fake-news sharers might be due more to heightened trait anger than situational anger. This would potentially limit the effectiveness of situational anger mitigation interventions for reducing fake-news sharing. To examine the joint role of trait and state anger, we designed an experimental study in which we manipulated situational anger while measuring trait anger. 

\subsection*{Participants and Design}
The experiment was composed of multiple tasks. We first induced high situational anger in all of the participants. Specifically, they all began by describing an article that made them feel extremely angry. Then we randomly assigned half of the participants to an anger mitigation condition. Immediately following the anger induction, these participants were asked to think of how they would calm a supportive friend who felt angry after reading the same article \citep[see][for a similar manipulation]{reilly2019anger}.\footnote{A pre-test with 203 participants showed that participants in the no situational anger mitigation reduction condition reported higher anger levels ($M = 3.83$) than those in the mitigation condition ($M = 3.22, t(202) = 3.71, p<.01$), and we found no difference in trait anger, which was measured later in the survey ($M_1 = 2.22, M_2 = 2.24; t(202) = .16, p = .87$).} 

In the second task, participants were given 14 news headlines (seven true, seven false) from the news in March 2023 that were ambiguous in terms of veracity.\footnote{For example, "After the shootings of Dallas policemen, nearly 500 people applied to the Dallas Police Department in 12 days" is true, whereas "The last time that a Senate of a different party than the president in the White House confirmed a Supreme Court nominee was 1888." is false.} On the first screen, they indicated their willingness to share each article (1-extremely unlikely, 2-moderately unlikely, 3-slightly unlikely, 4-slightly likely, 5-moderately likely, 6-extremely likely). On a second screen, they indicated their belief about the accuracy (true or false). Finally, as in \cite{pennycook2021shifting}, we asked five questions about the importance of news content dimensions when making decisions about what to share on social media: "When deciding whether to share a piece of content on social media, how important is it to you that the content is..." accurate/surprising/interesting/aligned with your politics/funny on a five point-scale (1-not all, 2-slightly, 3-moderately, 4-very, and 5-extremely). Using the anger-arousal trait measurement scale from \cite{siegel1986multidimensional}, we then measured trait anger.

We aimed to recruit 400 US participants from CloudResearch Connect who self-reported as Twitter users in exchange for \$1. The final sample consisted of 398 participants with an average age of 37 (SD = 12). The sample was 52\% female, 60\% white, 14\% black, and 11\% Asian. Also, 49\% self-reported as Democrat, 19\% as Republican, and 32\% as Independent.

\subsection*{Results}

Given that we cannot run analyses at the individual-claim level \citep[as recommended by][]{pennycook2021practical}, that we do not want to conflate effects of the sharing of fake news with the sharing of true-news \citep{lin2023conscientiousness}, and that most fake-news sharing by individuals is done unintentionally \citep{arin2023ability, pennycook2019fighting}, we present alternative sharing outcomes that tease out sharing intentions for the subjectively believed versus actual nature of the headline as true or false. Table \ref{Study3summary} presents the summary statistics regarding sharing behaviors. 

\begin{table}[htpb]
\centering
\caption{Summary statistics of news-sharing behavior}
\label{Study3summary}
\label{table:news_sharing_summary}
\scriptsize
\begin{tabular}{l r}
\hline
\textbf{Variable} & \textbf{Mean (SD)} \\
\hline
Average share rating (1 to 6) & 1.46 (0.18) \\
Number shared ($>=4$; out of 14) & 4.48 (3.85) \\
Number shared believed to be false (out of 14) & 1.32 (1.74) \\
Number shared believed to be true (out of 14) & 3.16 (3.00) \\
Number shared actually false (out of 7)& 2.18 (2.16) \\
Number shared actually true (out of 7) & 2.30 (1.95) \\
Number of intentional fake-news items shared & .66 (1.04)\\
Number of intentional true-news items shared & 1.29 (1.49)\\
Number of unintentional fake-news items shared & 1.52 (1.72)\\
Number of unintentional true-news items shared & 1.01 (1.14)\\
\hline
 \multicolumn{2}{p{.8\textwidth}}{Notes. Number of intentional (unintentional) fake-news items shared: Number of fake-news items (i.e., false news headlines) shared while believing them to be false (true).} 
\end{tabular}
\end{table}

Consistent with past literature that has investigated the sharing intentions of news headlines, we find that the average intention to share news on social media is low on our rating scale \citep[M = 1.48, SD = .18, see][]{jun2022social}. If we classify headlines as to be shared when the rating is above the midpoint \citep[e.g.,][]{pennycook2021shifting}, then the average participant would share 4.48 news articles (SD = 3.85) out of the 14 headlines. On average, people want to share news that they believe to be true more than news that they think is false (71\% vs. 29\% of headlines shared), but given our use of ambiguous headlines, discernment between true and false is particularly challenging\footnote{Accuracy was 51.83\% across all claims, with an ability to detect fake-news headlines at 52.00\%.} such that participants end up sharing roughly the same amount of actual true news (51\%) as fake news (49\%). 

In Table \ref{table:combined_regression_results}, we present the results of regressions of each of these alternative sharing outcomes, with our manipulation of reduced situational anger (mean-centered), trait anger (mean-centered), and their interaction. We also control for expected socio-demographic determinants of fake-news sharing, such as religiosity, age, gender, race, and political ideology. The results presented do not materially change if they are excluded. 

\begin{table}[htbp] \centering 
  \caption{Situational and trait anger effects on news-sharing outcomes} 
  \label{table:combined_regression_results} 
\resizebox{\textwidth}{!}{%
\begin{tabular}{@{}l *{30}{r} @{}} 
\\[-1.8ex]\hline 
\hline \\[-1.8ex] 
&  \multicolumn{3}{c}{\textbf{Sharing Likelihood}} &  \multicolumn{3}{c}{\textbf{Total Shared ($>=4$)}} & \multicolumn{3}{c}{\textbf{Shared Believed False}} & \multicolumn{3}{c}{\textbf{Shared Believed True}} & \multicolumn{3}{l}{\textbf{Shared Actually False}}\\
\hline \\[-1.8ex] 
\multicolumn{1}{c}{} & \multicolumn{1}{c}{Est.} & \multicolumn{1}{c}{SE} & \multicolumn{1}{c}{Pval} & \multicolumn{1}{c}{Est.} & \multicolumn{1}{c}{SE} & \multicolumn{1}{c}{Pval} & \multicolumn{1}{c}{Est.} & \multicolumn{1}{c}{SE} & \multicolumn{1}{c}{Pval} & \multicolumn{1}{c}{Est.} & \multicolumn{1}{c}{SE} & \multicolumn{1}{c}{Pval} & \multicolumn{1}{c}{Est.} & \multicolumn{1}{c}{SE} & \multicolumn{1}{c}{Pval} \\ 
\hline \\[-1.8ex] 
\multicolumn{1}{l}{(Intercept)} & 2.835 & .254 & 0.000 & 4.909 & .899 & .000 & 3.290 & .717 & .00001 & 1.619 & .410 & .0001 & 2.476 & .455 & .000 \\ 
\multicolumn{1}{l}{Reduced Situational Anger Condition} & .064 & .106 & .546 & .071 & .374 & .850 & -.053 & .298 & .860 & .123 & .171 & .470 & -.012 & .189 & .951 \\ 
\multicolumn{1}{l}{Trait Anger} & .300 & .059 & .000& 1.100 & .208 & .000 & .659 & .166 & .0001 & .441 & .095 & .000 & .546 & .105 & .000 \\ 
\multicolumn{1}{l}{Reduced Situational Anger Condition X Trait Anger} & -.124 & .116 & .284 & -.222 & .411 & .589 & -.080 & .327 & .807 & -.142 & .187 & .449 & .052 & .208 & .801 \\ 
\multicolumn{1}{l}{Religiosity} & .045 & .041 & .278 & .210 & .147 & .152 & .173 & .117 & .138 & .037 & .067 & .582 & .059 & .074 & .431 \\ 
\multicolumn{1}{l}{Age} & -.008 & .005 & .091 & -.027 & .017 & .110 & -.015 & .014 & .285 & -.013 & .008 & .102 & -.014 & .009 & .102 \\ 
\multicolumn{1}{l}{Male} & .207 & .107 & .054 & .815 & .380 & .033 & .438 & .303 & .149 & .377 & .173 & .030 & .543 & .192 & .005 \\ 
\multicolumn{1}{l}{Asian-Pacific} & .025 & .176 & .887 & -.253 & .622 & .685 & .061 & .496 & .902 & -.314 & .284 & .269 & -.081 & .315 & .797 \\ 
\multicolumn{1}{l}{Black} & .191 & .164 & .243 & .657 & .579 & .257 & .291 & .462 & .529 & .366 & .264 & .167 & .400 & .293 & .173 \\ 
\multicolumn{1}{l}{Hispanic} & -.132 & .200 & .509 & -.416 & .708 & .557 & -.105 & .564 & .852 & -.311 & .323 & .336 & -.273 & .358 & .447 \\ 
\multicolumn{1}{l}{Democrat} & -.168 & .153 & .273 & -.378 & .542 & .486 & -.295 & .432 & .495 & -.082 & .247 & .740 & -.059 & .274 & .829 \\ 
\multicolumn{1}{l}{Independent} & -.139 & .160 & .384 & -.464 & .565 & .411 & -.328 & .450 & .467 & -.136 & .258 & .597 & -.165 & .286 & .564 \\ 
\hline \\[-1.8ex] 
\end{tabular}%
}

\vspace{.25cm} 

\resizebox{\textwidth}{!}{%
\begin{tabular}{@{}l *{30}{r} @{}} 
\\[-1.8ex]\hline 
\hline \\[-1.8ex] 
&  \multicolumn{3}{c}{\textbf{Shared Actually True}} &  \multicolumn{3}{c}{\textbf{Intentional Fake}} & \multicolumn{3}{c}{\textbf{Unintentional Fake}} & \multicolumn{3}{c}{\textbf{Unintentional True}} & \multicolumn{3}{l}{\textbf{Intentional True}}\\
\hline \\[-1.8ex] 
\multicolumn{1}{c}{} & \multicolumn{1}{c}{Est.} & \multicolumn{1}{c}{SE} & \multicolumn{1}{c}{Pval} & \multicolumn{1}{c}{Est.} & \multicolumn{1}{c}{SE} & \multicolumn{1}{c}{Pval} & \multicolumn{1}{c}{Est.} & \multicolumn{1}{c}{SE} & \multicolumn{1}{c}{Pval} & \multicolumn{1}{c}{Est.} & \multicolumn{1}{c}{SE} & \multicolumn{1}{c}{Pval} & \multicolumn{1}{c}{Est.} & \multicolumn{1}{c}{SE} & \multicolumn{1}{c}{Pval} \\ 
\hline \\[-1.8ex] 
\multicolumn{1}{l}{(Intercept)} & 2.433 & .509 & .000 & .716 & .248 & .004 & 1.717 & .410 & .00003 & 1.075 & .273 & .0001 & 1.401 & .346 & .0001 \\ 
\multicolumn{1}{l}{Reduced Situational Anger Condition} & .082 & .212 & .698 & .057 & .103 & .580 & .025 & .170 & .883 & .016 & .113 & .889 & -.028 & .144 & .848 \\ 
\multicolumn{1}{l}{Trait Anger} & .554 & .118 & .000 & .204 & .057 & .0004 & .350 & .095 & .0003 & .140 & .063 & .028 & .406 & .080 & .000 \\ 
\multicolumn{1}{l}{Reduced Situational Anger Condition X Trait Anger} & -.275 & .232 & .238 & -.236 & .113 & .038 & -.039 & .187 & .836 & .023 & .125 & .852 & .029 & .158 & .854 \\ 
\multicolumn{1}{l}{Religiosity} & .152 & .083 & .068 & -.004 & .040 & .916 & .156 & .067 & .020 & -.096 & .044 & .032 & .154 & .056 & .007 \\ 
\multicolumn{1}{l}{Age} & -.013 & .010 & .174 & -.005 & .005 & .305 & -.008 & .008 & .286 & -.002 & .005 & .711 & -.012 & .007 & .063 \\ 
\multicolumn{1}{l}{Male} & .271 & .215 & .208 & .230 & .105 & .029 & .041 & .173 & .812 & .443 & .115 & .0001 & .101 & .146 & .491 \\ 
\multicolumn{1}{l}{Asian Pacific} & -.172 & .352 & .626 & -.234 & .172 & .175 & .062 & .284 & .828 & -.202 & .189 & .285 & .121 & .240 & .614 \\ 
\multicolumn{1}{l}{Black} & .256 & .328 & .435 & .280 & .160 & .081 & -.024 & .264 & .929 & .111 & .176 & .528 & .289 & .223 & .195 \\ 
\multicolumn{1}{l}{Hispanic} & -.144 & .401 & .720 & -.100 & .195 & .607 & -.043 & .323 & .894 & -.204 & .215 & .342 & -.068 & .273 & .802 \\ 
\multicolumn{1}{l}{Democrat} & -.318 & .307 & .300 & .049 & .150 & .744 & -.367 & .247 & .138 & .107 & .164 & .514 & -.166 & .209 & .426 \\ 
\multicolumn{1}{l}{Independent} & -.299 & .320 & .350 & -.012 & .156 & .937 & -.287 & .257 & .266 & -.060 & .171 & .727 & -.105 & .217 & .628 \\ 
\hline \\[-1.8ex] 
\end{tabular}%
}
\end{table}

Given that both the situational anger manipulation and trait anger have been mean-centered, we can interpret their coefficients as main effects even in the presence of the interaction term. Consistent across all ten measures of sharing, and as found in past literature on social media sharing \citep{berger2012makes}, we find that trait anger is an important predictor of sharing news in general. Specifically, we find that those with higher trait anger express higher intentions of sharing more news headlines, regardless of whether they believe the news to be true or false, and regardless of the actual truth state of the news (all $p<.05$; all but one $p<.01$). Moreover, we do not find evidence that our situational anger reduction manipulation had any direct effect on sharing outcomes (all $p's >.50$). As such, interventions to reduce situational anger may be ineffective at reducing fake-news sharing. More generally, past results correlating fake-news sharing and higher anger may reflect high anger individuals' propensity to share news regardless of its subjective or actual veracity. 

While we do not find evidence that reducing situational anger affected sharing intentions directly, or that its effect differed much by trait anger,\footnote{While we do find an interaction for intentional fake-news sharing, there is no significant level of trait anger at which the manipulation affects sharing outcomes.} the lack of significant interaction effects could be due to our specific operationalization of situational anger reduction. Alternatively, it may be the case that the situational anger reduction manipulation has an effect outside of our specific stimuli by affecting the type of news that individuals want to share, for example. In Web Appendix O, we investigate how the situational anger reduction manipulation and trait anger affect the desire to share news that is accurate, surprising, interesting, aligned with one's politics, or funny \citep[][]{pennycook2021shifting}. We find that reducing situational anger helps decrease the desire to share surprising news among those with high-trait anger (an interaction effect). However, this effect is not sufficient to have a material effect (be it direct or indirect) on sharing intentions in our study.  

\subsection*{Discussion}
In Study 3, we built on exploratory findings from Study 1 about the heightened use of anger-related words by fake-news sharers, and discrepant findings in the literature about the role of anger in fake-news sharing, to explore how situational and trait anger influence such sharing. While we found evidence that trait anger is correlated with higher fake-news sharing, we also found evidence that trait anger is correlated with higher sharing more generally. Moreover, we did not find evidence that reducing situational anger reduces fake-news sharing. These findings suggest that some past findings may be due more to trait anger differences. Further, although our specific manipulation of reducing situational anger was ineffective, we cannot rule out the possibility that other anger-reducing methods would be effective. In line with this notion, we do find evidence that reducing situational anger reduces the desire of high-trait anger individuals to share surprising news. 

\section*{Study 4: Authenticated Twitter Users and Ads Using Power-related Words}

Study 1 found that words related to power are used by those who share misinformation and those who seek to correct it (fact-check sharers). Both groups use more power-related words than random and matched users. As discussed in the overview of studies 3 and 4, one could assume that using power-related words in their general language reflects a low sense of power when posting on Twitter. If this holds true, one could hypothesize that a Twitter ad suggesting that a fact-checking product (e.g., a browser extension) restores power should increase its appeal to Twitter users. Unlike in Study 3 when we rely on behavioral screeners (and self-reports) to identify a sample of Twitter users, here we use a different sampling strategy to establish with a greater certainty that our results hold with verified Twitter users and how the effect is influenced by unanticipated heterogeneity in post histories. 

\subsection*{Participants and Design}
\begin{figure}[htbp]
     \centering
     \caption{Twitter API authentication procedure}
     \begin{subfigure}[b]{0.32\textwidth}
         \centering
         \includegraphics[width=\textwidth]{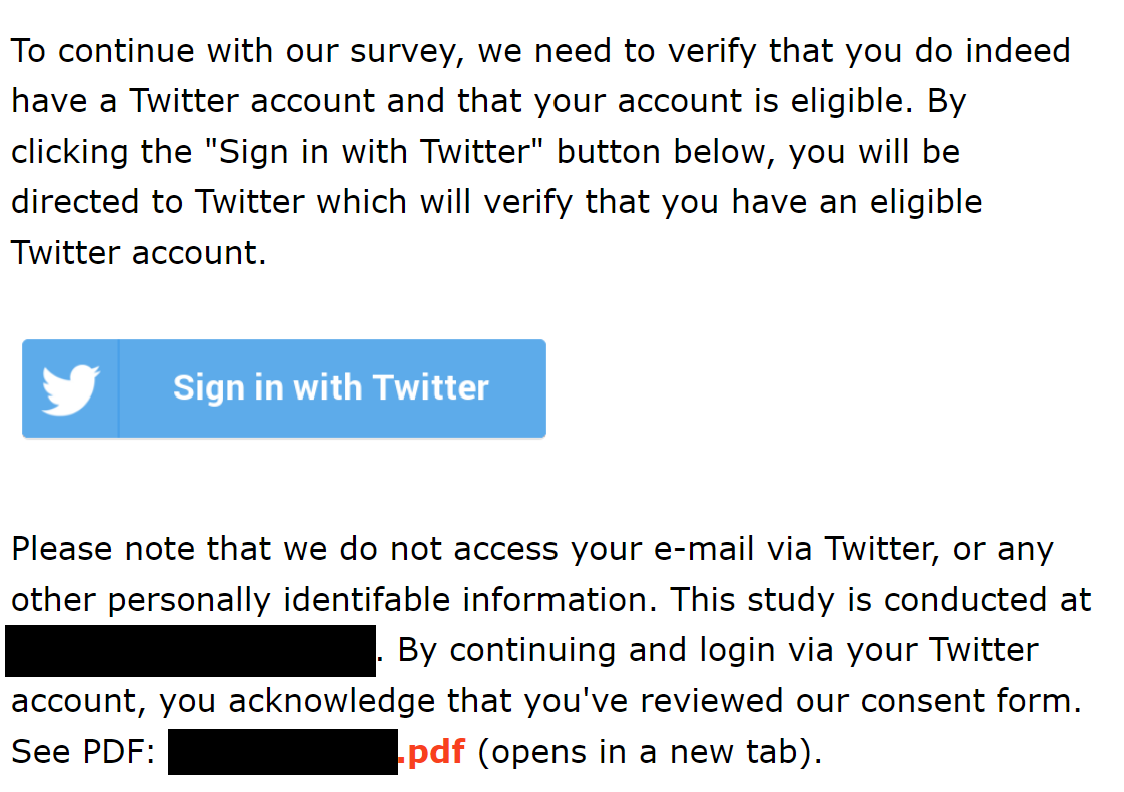}
         \caption{Instructions and consent form}
         \label{TwitterLoginA}
     \end{subfigure}
     \hfill
     \begin{subfigure}[b]{0.32\textwidth}
         \centering
         \includegraphics[width=\textwidth]{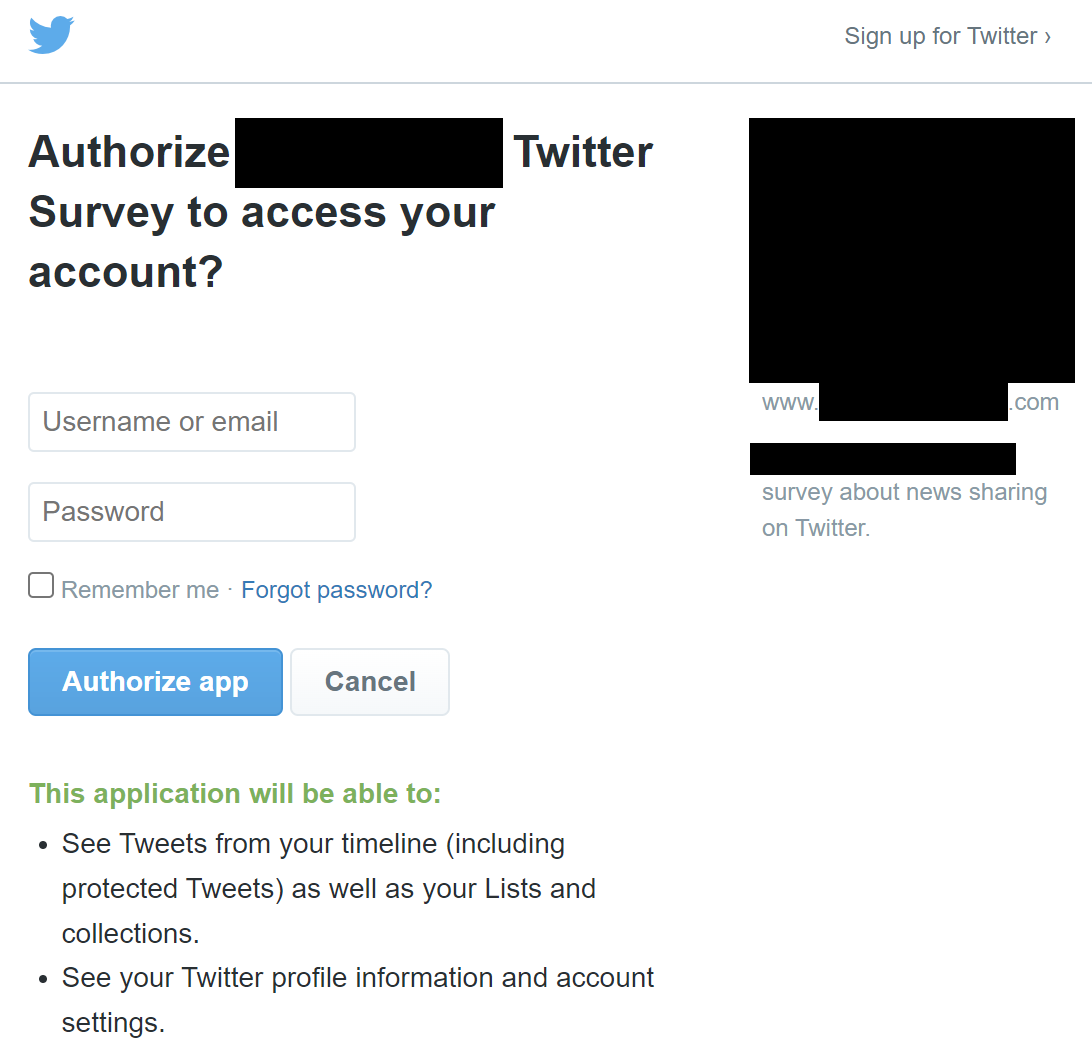}
         \caption{Twitter API login screen}
         \label{TwitterLoginB}
     \end{subfigure}
     \label{fig:TwitterLogin}
\end{figure}

We contracted Qualtrics Research Services to recruit 600 qualified US participants for a 10-minute study in exchange for \$9.31 per completed response. To obtain these 600 complete responses, Qualtrics invited 1,947 survey takers to participate in our study. After the screening portion of the questionnaire, participants proceeded to indicate which of several statements applied to them. They were asked questions about religion ("How important is religion in your life?" I am not religious, not important at all although I consider myself religious, moderately important, very important, center of my life), and to report their age, political orientation (very liberal, liberal, middle of the road, conservative, very conservative), and gender. Participants were then redirected to a website where they were asked to authenticate their Twitter accounts using the Twitter API via OAuth (Figure \ref{fig:TwitterLogin}). For those who did, we stored post history data and associated textual markers in an encrypted file on a separate drive before removing the tweets and Twitter ID to merge with the Qualtrics file. We merged scored Twitter data with Qualtrics survey responses without enabling the possibility of participant identification in the final data. The raw post history data was never held in a single file along with the surveys. The final sample consisted of 481 survey participants who had written over 100 words on Twitter. All exclusions are documented in Web Appendix P.\\

\begin{figure}[htbp]
     \centering
     \caption{Study 4: Addition of power-related words in ad copy}
    \begin{subfigure}[b]{0.35\textwidth}
         \centering
         \includegraphics[width=\textwidth]{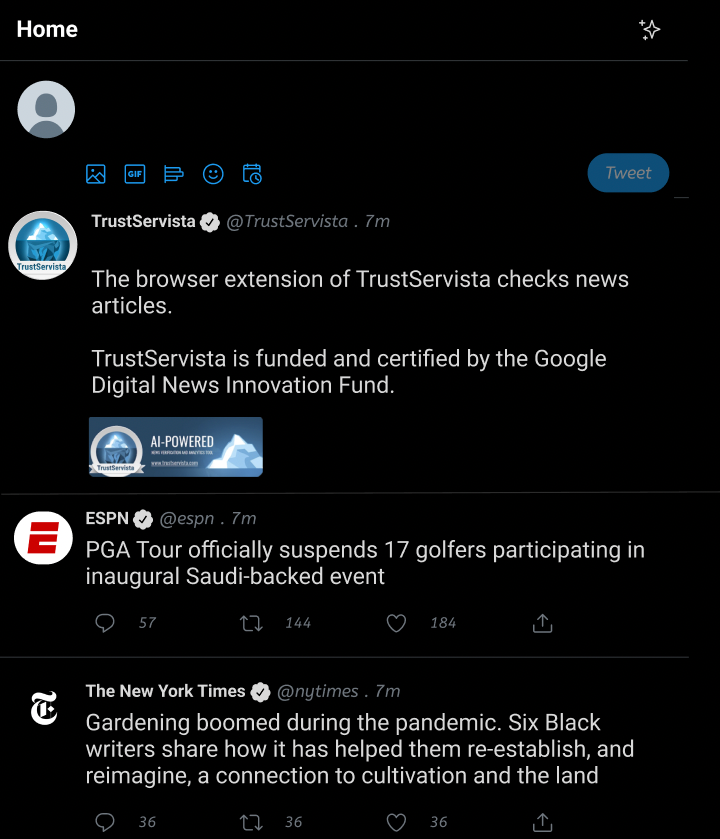}
         \caption{Power-related words: Low}
         \label{fig:lowpower}
     \end{subfigure}
     \begin{subfigure}[b]{0.32\textwidth}
         \centering
         \includegraphics[width=\textwidth]{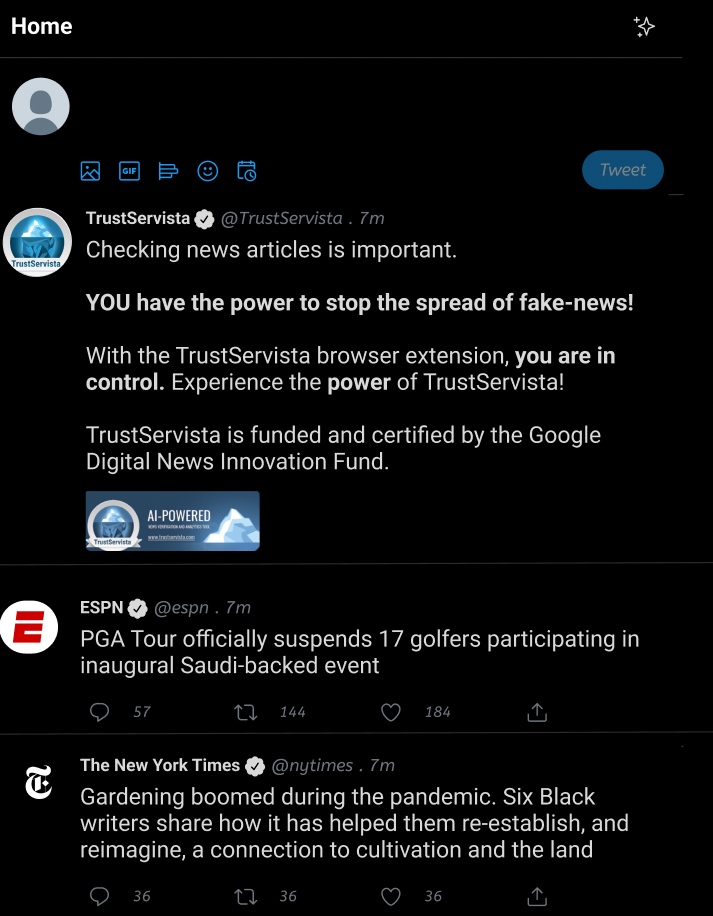}
         \caption{Power-related words: High}
         \label{fig:highpower}
     \end{subfigure}
     \label{fig:powermanipulation}
\end{figure}

Upon completing the screener and authenticating their Twitter accounts, participants were presented with a timeline, as shown in Figure \ref{fig:powermanipulation}. Half of them were assigned to the control advertisement for TrustServista, a browser extension for fact-checking articles (Figure \ref{fig:lowpower}), followed by two unrelated tweets (ESPN, New York Times). The remainder saw the same tweets, but the TrustServista message was modified to emphasize increases in control and power. 

Afterwards, all participants indicated their likelihood to click on the ads to learn more (1-extremely unlikely to 9-extremely likely). On the following page, all rated their likelihood to download the extension and join a fact-checking group (an act unrelated to the empowering message about TrustServista) on the same 9-point scale. Finally, participants completed the "desire for control" \citep{burger1984desire} and the "personal sense of power" scales \citep{anderson2012personal} to examine whether the use of more power-related words in their post histories is indicative of a desire for power or of feeling powerful. As described below, the use of power-related words reflects a low personal sense of power, thus justifying our empowering language to instigate action.

\subsection*{Results}

In Table \ref{tab:powerresults}, we show regressions that include the power manipulation (1: high power, 0 control; then mean-centered). The manipulation increased intentions to click  ($B = .94; t(468) = 3.746, p < .01)$ and to download the browser extension ($B = 1.13; t(468) = 4.448, p < .01$). We note that our ad-specific power manipulation did not induce participants to comply with just any action; it did not affect the likelihood of joining a fact-checking group, thus, showing the targeted effect of the intervention on the behavior endorsed in the power-enhancing message (i.e., to download the fact-checking browser extension). 

\begin{table}[htbp] \centering 
\tiny
  \caption{Results power -- click, download, join fact-sharer group} 
  \label{} 
\begin{tabular}{@{\extracolsep{5pt}}lD{.}{.}{-3} D{.}{.}{-3} D{.}{.}{-3} D{.}{.}{-3} D{.}{.}{-3} D{.}{.}{-3} } 
\\[-1.8ex]\hline 
\hline \\[-1.8ex] 
 & \multicolumn{6}{c}{\textit{Dependent variable:}} \\ 
\cline{2-7} 
\\[-1.8ex] & \multicolumn{2}{c}{Click} & \multicolumn{2}{c}{Download} & \multicolumn{2}{c}{Join}  \\ 
\\[-1.8ex] & \multicolumn{1}{c}{coef (se)} & \multicolumn{1}{c}{p-value} & \multicolumn{1}{c}{coef (se)} & \multicolumn{1}{c}{p-value} & \multicolumn{1}{c}{coef (se)} & \multicolumn{1}{c}{p-value}\\ 
\hline \\[-1.8ex] 
Power Dummy & .943 (0.251) & .0002 & 1.131 (0.255) & .00001 & .129 (0.108) & .233 \\ 
Religion & .246 (0.098) & .013 & .195 (0.100) & .052 & .090 (0.042) & .035 \\ 
Age & -.001 (0.008) & .903 & -.006 (0.009) & .487 & -.008 (0.004) & .037 \\
Political Ideology (baseline = very liberal) & & &  &  &  &   \\ 
Liberal & .448 (0.389) & .250 & .321 (0.396) & .418 & -.066 (0.168) & .694 \\ 
Middle of the Road & .136 (0.357) & .703 & .020 (0.363) & .956 & -.147 (0.154) & .340 \\ 
Conservative & .062 (0.449) & .891 & -.453 (0.457) & .321 & -.400 (0.194) & .039 \\ 
Very Conservative & -.489 (0.530) & .357 & -.637 (0.539) & .238 & -.268 (0.229) & .242 \\ 
Gender (baseline = male) & & &  &  &  &  \\ 
Female & -.118 (0.252) & .639 & -.271 (0.257) & .292 & -.123 (0.109) & .259 \\ 
Non-binary/Third Gender & -1.727 (2.722) & .526 & -1.951 (2.767) & .481 & -2.979 (1.174) & .012 \\ 
Constant & 4.320 (0.536) & .000 & 4.586 (0.545) & .000 & 3.867 (0.231) & .000 \\
N & \multicolumn{2}{c}{481} & \multicolumn{2}{c}{481} & \multicolumn{2}{c}{481} \\ 
R$^{2}$ & \multicolumn{2}{c}{.055} & \multicolumn{2}{c}{.065} & \multicolumn{2}{c}{.047} \\ 
Adjusted R$^{2}$ & \multicolumn{2}{c}{.037} & \multicolumn{2}{c}{.043} & \multicolumn{2}{c}{.025} \\ 
\hline 
\hline \\[-1.8ex] 
\end{tabular} 
\label{tab:powerresults}
\end{table}

\subsection*{Accounting for Unexpected Heterogeneity}

The results presented in Table \ref{tab:powerresults} assume that random assignment of the power manipulation worked perfectly and that no pre-treatment differences between the power conditions could explain the differential response to outcomes. However, our Twitter authentication allows us to see if unanticipated differences in post histories across conditions could account for our results. In Web Appendix Q, we show that despite randomization, those who received the ad copy with power-related words differ on several pre-treatment covariates, which could be related to how they react to power-related words. For example, those in the treatment condition use fewer power-related and anger-related words in their post histories.

Given this unexpected heterogeneity, we use causal random forests to estimate individual-level nonparametric treatment effects without restricting the number of covariates \citep{chen2021treatment}. This allows us to estimate individual-specific treatment effects of the power manipulation on clicking, downloading the browser extension, and joining a fact-checking community. Once we have estimated these individual-level treatment effects and adjusted the estimates to be doubly robust \citep{semenova2021debiased}, we find little evidence that the effect depends on self-reported background variables, except for the use of power-related words being less effective at encouraging the download of the browser extension among those who self-identify as conservative. We find the effect stronger for those with fewer friends, and stronger for those who use comparative language. Importantly, the effects remain large on both intentions to click to learn more and to download the extension. Details are in Web Appendix Q.

\subsection{Additional Analyses}

\subsubsection*{Sampling framework considerations}

One could question the adequacy of using a commercial survey panel to sample active users on Twitter and whether our requirement to link Twitter accounts (which resulted in a nearly 60\% dropout) resulted in a sample whose decision to opt-in could be responsible for our results. We find no meaningful differences between participants who refused to authenticate their Twitter accounts and those who did regarding self-reported variables (see Web Appendix P). Additionally, when comparing the final sample from our survey to the random sample of Twitter users (Study 1), we find few significant differences -- most importantly, none regarding power-related words. 

\subsubsection*{Adequacy of textual cues for capturing emotions, personality, and socio-demographics}

Using public dictionary-based APIs to extract textual cues raises considerations about whether the use of certain words directly implies the corresponding emotional state or trait. For example, using power-related words may indicate either a desire for power or a sense of already having power. With authenticated Twitter accounts, our sampling approach allows us to shed light on the relationship between textual cues and underlying psychological traits. We compare self-reported data on the desire for control and personal sense of power with the extracted proxies from Twitter texts. The correlations are in Table \ref{tab:selfreports}.

\begin{table}[htbp]
\centering
\caption{Comparison of self-reports to proxies extracted via textual cues}
\label{tab:selfreports}
\begin{footnotesize}    
\begin{tabular}{lll}
\hline
\textbf{Self-report} & \multicolumn{1}{c}{\textbf{Correlation (r)}} & \multicolumn{1}{c}{\textbf{p}} \\ \hline
Age  vs. Age (Magic Sauce) & .360 & \textless{}0.001 \\ 
Gender Female vs. Gender Female (genderGuesser)$^a$ & .812 & \textless{}0.001 \\ 
Political Affiliation$^b$ vs. Following Democrat Twitter Accounts & -.252 & \textless{}0.001 \\ 
Political Affiliation$^b$ vs. Following Republican Twitter Accounts & .316 & \textless{}0.001 \\ 
Religiosity vs. Religion LIWC & .118 & .0098 \\ \hline
\textbf{Personality Scales} &  &  \\ 
Desire for Control Scale Survey vs. Power LIWC & -.001 & .9817 \\ 
Personal Sense of Power Survey vs. Power LIWC & -.147 & .0012 \\ \hline
\multicolumn{3}{l}{$^a$ Excluded were the 171 users for whom gender was predicted as unknown.} \\ 
\multicolumn{3}{l}{$^b$ Political affiliation was captured on a 5-point scale from Liberal to Conservative.} \\ 
\end{tabular}
\end{footnotesize}
\end{table}

The results indicate that the proxies extracted from post histories are surprisingly good at approximating socio-demographics. We also find those using more power-releated words have a lower sense of personal power. This aligns with our reasoning to explore an empowerment manipulation to mitigate fake-news sharing via downloading a fact-checking browser extension. This highlights a critical point of caution: one should not equate frequent word usage with possessing the trait implied by the word (e.g., power). 

\section*{General Discussion}

This article explored the usefulness of analyzing users' social media post histories to study fake-news sharing. To do so, we conducted two studies exploring the language in the post histories of actual Twitter users, a study showing the utility of adding textual cues to socio-demographics to better predict fake-news sharers, and two exploratory experiments that build on the identified promising textual cues. 

Study 1 involved an exploratory analysis of textual cues in Twitter post histories in differentiating between users who share fake news and several types of users who do not. The analyses revealed that fake-news sharers use distinct textual cues that we then used to reconcile and expand empirical findings from past research. Study 2 showed that incorporating textual cues from post histories improves the classification of fake-news sharers. Studies 3 and 4 built on the first study's results to explore how textual cues from fake-news users' post histories could be further investigated. In Study 3, we evaluated how situational anger and trait anger jointly relate to fake-news sharing intentions. We found that individuals high in trait anger tend to share more fake news, but also more news in general, and that our attempts to reduce situational anger did not affect fake-news sharing intentions. In Study 4, we authenticated participants' Twitter accounts and extracted textual cues from their post histories that were subsequently linked to their survey responses. The study suggests that messages using empowering language can facilitate the desired behavior, increasing the clicking and downloading intentions of a browser extension for fact-checking in this case. The effects are robust to accounting for unanticipated heterogeneity observable from post histories. Results also support that those with more power-related words in their post histories feel like they have less power, as assessed by the psychometric scale in our survey.

We also report two additional exploratory studies that investigate two textual cues identified in Study 1 and their relevance to mitigate fake-news sharing: anxiety and religion. Due to space concerns, we moved their discussions to Web Appendix G and H.  For anxiety, we do not find significant evidence that an anxiety-inducing post on a user's timeline would affect fake-news sharing. For religion, we also do not find that activating religious values affects fake-news sharing, but like situational anger, it seems to reduce a general motivation to share surprising content. We encourage future research to build on our exploratory studies.

\subsection*{Substantive Contribution}

Numerous calls have been released for academic work to help reduce the effects of misinformation \citep{lazer2018science,pennycook2019fighting, johar2022untangling}. Since then, academics and policymakers have taken many complementary steps. Some have tackled the reduction of misinformation at the source (e.g., identifying low-credibility domains and identifying and labeling articles containing misinformation). In contrast, others have focused on using surveys to understand the beliefs and motivations of those who share misinformation. Others, like us, have attempted to use fake-news sharers' user-generated social media to better understand their sharing behavior. Our work extends prior studies in the context of fake-news sharers' social media sharing, which has been mostly confined to two-group comparisons: those who share fake news to those who do not or to those who share fact checks of articles. Our findings challenge the assumption that features distinguishing fake-news sharers from random Twitter users, like age and political affiliation, are exclusive to those who propagate misinformation. Additionally, discovering commonalities between fake-news and fact-check sharers encourages further the study of fact-check sharers \citep{giachanou2020role, mosleh2021perverse}. Beyond the platforms who can use the rich textual post data they can access to a priori identify and target potential fake news sharers, numerous stakeholders should be interested in harvesting the unique information in users' post histories to describe and predict who is likely to share fake news. For example, those marketing products related to misinformation should benefit from collecting post histories' textual cues, as doing so may help avoid targeting based on reductive and unproductive generalizations (e.g., target conservative males). 

\subsection*{Methodological Contribution}
Conducting experiments focused on misinformation within the realm of social media users presents inherent challenges. On the one hand, research has established a correlation between sharing intentions as measured in surveys and actual sharing behavior of misinformation on social media \citep{mosleh2020self}. However, surveys encounter difficulties in maintaining ecological validity - such as simulating misinformation within the context of a genuine timeline - or effectively uncovering real social media activity. For instance, requesting participants to provide their Twitter handles for verification remains prone to misrepresentation. On the other hand, conducting internally valid social-media experiments is challenging because of the interplay between exposure to potential interventions, platform dynamics, and users' behaviors \citep{johnson2023inferno}. Here, we introduce an innovative approach: integrating an account authentication step within the research survey. Within the ethical confines of panel policies, this helps reduce misrepresentation concerns and facilitate the extraction of textual cues from post histories. For example, we showed that those who use more power-related words perceive a lack of power (measured by a psychometric instrument) as opposed to feeling powerful. Thus, we encourage researchers to proceed cautiously and verify their assumptions about the relationship between frequent word usage in dictionary-based approaches (e.g., power-related words) and the corresponding emotional states (e.g., individuals using power-related words feel more powerful). More generally, we hope to motivate a focus on sampling quality \citep{baumgartner2022critical}.

\subsection*{Limitations and Future Research}

A limitation of post history data is that they do not capture intentions directly. Our position is that any resharing of fake news, regardless of deliberateness, contributes to its spread. In fact, intentional fake-news sharing is relatively rare: approximately 16\% in \cite{pennycook2019fighting} and 12\% in \cite{arin2023ability}. Nevertheless, it is important to recognize underlying personality traits and emotions may differ between those who unintentionally share fake news versus those who intentionally do so. We encourage future research to pay attention to intentionality. We also encourage future research to build on our findings and investigate how textual cues relate to the motivations for fake-news sharing. Additionally, studying more personality traits, such as the Dark Triad \citep[narcissism, psychopathy, machievalism, see][]{calvillo2023personality} or cognitive reflection \citep[cf.][]{pennycook2019lazy} is important.

\subsection*{Concluding remarks}

Our study highlights users' post histories on social media as an overlooked data source for formulating strategies to mitigate the spread of fake news. While we are optimistic about others building upon our work, we recognize the challenges of relying on public APIs to obtain post histories. Notably, the data in our research was obtained using the first version of the Twitter API. Since then, Twitter (X) has restructured its pricing and data accessibility via the free tier. As such, studies like ours are now unfeasible for scholars who cannot afford to pay the considerable monthly subscription for a higher tier. We hope that our findings not only contribute to the scientific community, but also remind both funding bodies and social media platforms that facilitating reasonable access to social-media APIs is important. 

\setstretch{1.0}
	\restoregeometry
	\newgeometry{margin=1in}
	\singlespacing
	\bibliographystyle{jmr.bst}
	\bibliography{bibliography}

\newpage

\begin{table}[htbp]
\begin{tabular}{l}
\multicolumn{1}{c}{\textbf{Table of Web Appendices}} \\ \hline
\begin{tabular}[c]{@{}l@{}}Web Appendix A -- Data Sources for Snopes Dataset\end{tabular} \\ \hline
\begin{tabular}[c]{@{}l@{}}Web Appendix B -- Demographics and Media Followership Snopes and Hoaxy Sample\end{tabular} \\ \hline
Web Appendix C -- Propensity Score Matching \\ \hline
Web Appendix D -- Robustness Snopes Sample Operationalization \\ \hline
Web Appendix E -- Exploratory Analyses: Snopes All Variables \\ \hline
Web Appendix F -- Exploratory Analyses: Fact-Check Sharers, Left/Right-Leaning Media Sharers \\ \hline
Web Appendix G -- Exploratory Anxiety Field Experiment \\ \hline
Web Appendix H -- Exploratory Religion Experiment \\ \hline
Web Appendix I -- Robustness Analyses: Followerhsip and Mentions in Tweets \\ \hline
\begin{tabular}[c]{@{}l@{}}Web Appendix J -- Robustness Alternative Measures\\ -- Alternative Emotions Measures via NRC\\ -- Alternative Emotions Measures via MTurk Coded Emotions.\end{tabular} \\ \hline
Web Appendix K -- Comparison of MagicSauce to Park et al. (2005) \\ \hline
Web Appendix L -- Fake-News Article Content Analysis \\ \hline
Web Appendix M -- Political vs. Non-Political Fake-News Sharers \\ \hline
Web Appendix N -- Details on Fake-News Sharer Predictions \\ \hline
Web Appendix O -- Moderation of Content Characteristics Study 3\\ \hline
Web Appendix P -- Exclusion Field Study\\ \hline
Web Appendix Q -- Heterogeneous Effect Analyses \\ \hline
\end{tabular}
\end{table}

\setcounter{table}{0}
\renewcommand{\thetable}{WA\arabic{table}}

\section*{Web Appendix A - Data Sources for Snopes Dataset}
\vspace{8pt}

\vspace*{-\baselineskip}
\begin{table}[htbp]
\caption{List of Fake-News Articles Used}
\label{tab:SnopesArticle}
\begin{scriptsize}
\begin{multicols}{2}
snapchat-shutting-end-2017-prank/ \\
indian-ocean-earthquake-tsunami-prediction-2017/ \\
colorado-infant-become-first-pot-overdose-death/ \\
bulldog-saves-sleeping-children/ \\
ryan-zinke-dead-elephant/ \\
trump-turkey-pardons-reversed/ \\
actor-harry-shearer-leaving-simpsons/ \\
did-elderly-woman-train-cats-to-steal/ \\
has-tony-podesta-been-arrested-in-the-mueller-investigation/ \\
was-grand-juror-clinton-probe-dead/ \\
iceland-mandates-mental-health-warnings-on-all-bibles/ \\
jay-z-said-satan-is-our-true-lord-and-only-idiots-believe-in-jesus/ \\
hpv-kissing-aids/ \\
did-canada-legalize-euthanasia-children-disabilities/ \\
morgue-employee-cremated-nap/ \\
did-nfl-team-declare-bankruptcy-protests/ \\
did-germany-release-statement-denouncing-u-s/ \\
bob-marley-cia/ \\
media/notnews/imaginarycrime.asp \\
bbc-pedophilia-journalist-dead/ \\
media/iftrue/anthony.asp \\
media/notnews/elvishomeless.asp \\
seven-witnesses-to-vegas-shooting-died-suspicious/ \\
child-abduction-kansas/ \\
muslim-self-defense/ \\
roy-moores-accusers-arrested-charged-falsification/ \\
hpv-kissing-aids/ \\
tiger-woods-paternity/ \\
girl-disconnect-life-support-plug-in-charger/ \\
jared-kushner-go-saudi-arabia-doesnt-extradition-treaty-us/ \\
morgan-freeman-jailing-hillary-restore-faith/ \\
george-washington-historic-church-plaque/ \\
sandra-bullock-trump-clinton/ \\
425-pound-teacher-suspended-sitting-student-farting-mouth/ \\
ghost-on-camera-hallways-ireland-school/ \\
john-mccain-fire-uss-forrestal/ \\
frederica-wilson-veterans/ \\
weinstein-fbi-names/ \\
creepy-clowns-halloween-night-purge/ \\
drag-queen-library/ \\
chad-nishimura-mandalay-bay-las-vegas/ \\
creepy-clown-haunted-house-massacre/ \\
trump-mocks-trudeau-celebrating-thanksgiving-six-weeks-early/ \\
new-research-yellowstone-eruption-occur-sooner/ \\
kaepernick-anthemk-stand-if-signed/ \\
is-antifa-planning-a-civil-war/ \\
surgeon-clinton-foundation-corruption-haiti/ \\
birmingham-caught-voting-multiple-times/ \\
keaton-jones-beaten-robbed-gofundme-bullying/ \\
judi-dench-harvey-weinstein-tattoo/ \\
democrat-principal-defecates-in-front-of-students-during-pledge-of-allegiance/ \\
morgan-freeman-death-hoax/ \\
does-arctic-ice-doubt-reality-climate-change/ \\
nike-cowboys/ \\
jared-kushner-register-vote-new-york-woman/ \\
fbi-seizes-3000-penises-raid-morgue-employees-home/ \\
is-the-trump-administration-puerto-rico-evacuees/ \\
steelers-villanueva-police-brutality-protest/ \\
is-it-illegal-florida-power-home-solar-storm/ \\
nfl-refunds/ \\
pedophiles-decapitated-corpse-found-on-judges-doorstep-after-bail-hearing/ \\
ashley-judd-middle-east-women-rights/ \\
hillary-clinton-hurricane-irma-florida-hillbillies/ \\
alec-baldwin-arrested-trump/ \\
man-head-stuck-wife/ \\
media/notnews/humanmeat.asp
\end{multicols}
\vspace*{-\baselineskip}
Note: 62 articles were published in 2017, one of these refers to a story from 2008 but was published in 2017. Four articles are from 2015.
\end{scriptsize}
\end{table}

\newpage
\vspace{8pt}

\begin{table}[]
\caption{Twitter Brand Accounts for Random Sample}
\tiny
\begin{tabular}{llll}
\multicolumn{4}{c}{\textbf{Brands}}                                                                                     \\
\hline\hline
3M                               & Boost Mobile             & CW                                & FX Network                    \\
7eleven                          & Bose                     & Daily Beast                       & GameStop                      \\
A \& E Network                   & Boston Consulting Group  & Dallas Cowboys                    & Gap                           \\
ABC News                         & Boston Globe             & Dallas Mavericks                  & GE                            \\
Abercrombie                      & Breaking News            & DAVHQ                             & General Motors                \\
Academy Sports + Outdoors        & British Airways          & DC Shoes                          & General Nutrition Corporation \\
Ace Hardware                     & Broncos                  & DC Skateboarding                  & Gilt                          \\
ACLU                             & Brooklyn Nets            & Dell                              & GitHub                        \\
Acura                            & Bud Light                & Delta                             & Glassdoor                     \\
Adidas                           & Budweiser                & Denny's                           & Gmail                         \\
Adobe                            & Buffalo Wild Wings       & Detroit Lions                     & Golden State Warriors         \\
Adobe InDesign                   & Buick                    & Detroit Pistons                   & Goldman Sachs                 \\
Aeropostale                      & Burberry                 & Deutsche Bank                     & Golf Channel                  \\
Air Canada                       & Burger King              & Dick's Sporting Goods             & Golf.com                      \\
Airbnb                           & Burt's Bees              & Diesel                            & Google                        \\
Airbus                           & Burton                   & Dior                              & Google Chrome                 \\
Al Jazeera English               & BuzzFeed                 & Discovery                         & Google Earth Outreach         \\
Alaska Airlines                  & C-SPAN                   & Disney                            & Google Maps                   \\
alzassociation                   & CA Technologies          & DocuSign                          & Google News                   \\
Amazon                           & Cabelas                  & Dodge                             & Google Talk                   \\
AMD                              & Cadillac                 & Dogfish Head                      & Green Bay Packers             \\
American Airlines                & Callaway Golf            & Dolce \& Gabbana                  & Greenpeace                    \\
American Apparel                 & Calvin Klein             & Dollar Tree                       & Groupon                       \\
American Cancer Society          & Capital One              & Dominos                           & Gucci                         \\
American Eagle Outfitters        & CARE                     & Donna Karan                       & Guess                         \\
American Heart Association       & Carolina Panthers        & dosomething                       & H\&M                          \\
American Red Cross               & Caterpillar              & Dr Pepper                         & Habitat for Humanity          \\
AMNH                             & CBS TV                   & Dreamworks                        & Harper's Magazine             \\
Anthropologie                    & Celtics                  & Dropbox                           & Hautelook                     \\
AOL                              & Century 21               & Drudge Report                     & HBO                           \\
App Store                        & Champs Sports            & DucksUnlimited                    & Heineken                      \\
Apple                            & Chanel                   & Dunkin' Donuts                    & HGTV                          \\
Apple News                       & Chargers                 & E! News                           & Hilton Hotels                 \\
Applebees                        & Chase                    & Ebay                              & History                       \\
Arbys                            & Cheesecake Factory       & Ebony Magazine                    & Holiday Inn                   \\
Arizona Cardinals                & Chevrolet                & Economist                         & Hollister                     \\
Armani                           & Chevron                  & Electronic Arts                   & Home Depot                    \\
artinstitutechi                  & Chicago Bears            & Emirates                          & HomeGoods                     \\
ASPCA                            & Chicago Bulls            & EnvDefenseFund                    & Honda                         \\
Aston Martin                     & Chicago Tribune          & ESPN                              & Hotmail                       \\
AT\&T                            & Chick-fil-A              & Estee Lauder                      & Houston Rockets               \\
Atlanta Falcons                  & Chili's                  & Etihad Airways                    & Houston Texans                \\
Atlanta Hawks                    & Chipotle                 & Etsy                              & HP                            \\
Atlantic                         & Chloe                    & Evernote                          & hrw                           \\
Audi                             & Chobani                  & Expedia                           & HSBC                          \\
Audible                          & Chrysler                 & Express                           & Hugo Boss                     \\
Avon                             & Cincinnati Bengals       & ExxonMobil                        & Hulu                          \\
Babies'R'Us                      & CISCO                    & EY News                           & Hyundai                       \\
Bain \& Company                  & Citibank                 & Facebook                          & IBM                           \\
Banana Republic                  & Citigroup                & Fairmont                          & IHOP                          \\
Bank of America                  & Citrix                   & Fedex                             & IMAX                          \\
Barclays                         & Cleveland Browns         & FeedingAmerica                    & IMDb                          \\
bareMinerals                     & Cleveland Cavaliers      & feedthechildren                   & Infiniti                      \\
Barnes \& Nobles                 & Clinique                 & Fendi                             & Instagram                     \\
Barneys New York                 & ClintonFdn               & Ferragamo                         & Intel                         \\
BBC AMERICA                      & CNBC                     & Ferrari                           & InterContinental              \\
BBC News                         & CNET                     & Fiat Chrysler                     & Invesco                       \\
BCBG Max Azria                   & CNN                      & Fidelity Investments              & iShares                       \\
Bed Bath \& Beyond               & CNN Money                & Financial Times (@FinancialTimes) & iTunes                        \\
Benefit                          & Coach                    & Financial Times (@FT)             & J.Crew                        \\
Bentley Motors                   & Cobra Golf               & Finish Line Inc.                  & Jack Daniels                  \\
Bergdorf Goodman                 & Coca Cola                & Firefox                           & Jacksonville Jaguars          \\
Best Buy                         & Cognizant                & Flickr                            & Jaegermeister                 \\
BGEA                             & Columbia Sportswear      & Foot Locker                       & Jaguar                        \\
Bill \& Melinda Gates Foundation & Comedy Central           & Footaction                        & Java                          \\
Bing                             & compassion               & Forbes                            & JBL                           \\
Bitcoin                          & ConocoPhillips           & Ford                              & JC Penney                     \\
Black Entertainment Television   & ConservationOrg          & Four Seasons Hotels and Resorts   & Jetblue                       \\
BlackRock                        & Consumer Reports         & FOX Business                      & Jim Beam                      \\
Bloomberg                        & Converse                 & Fox News                          & Jimmy Choo                    \\
Bloomberg Businessweek           & Coors Light              & FOX Sports                        & Jo-Ann Stores                 \\
Bloomingdale's                   & Country Music Television & Fox Sports 1                      & John Deere                    \\
BMW                              & Coursera                 & Fox Sports Motor                  & JP Morgan                     \\
Bobbi Brown                      & Covergirl                & Fox TV                            & Juicy Couture                 \\
Boeing                           & creativecommons          & Franklin Templeton Investments    & Karl Lagerfeld                \\
Booking.com                      & Credit Suisse            & Free People                       & kate spade         \\
\hline
\end{tabular}
\end{table}

\begin{table}[]
\caption*{Twitter Brand Accounts for Random Sample - Continued}
\tiny
\begin{tabular}{llll}
\multicolumn{4}{c}{\textbf{Brands}}                                                                                     \\
\hline\hline
Kayak                       & Monster                          & Pinterest                   & Stand Up to Cancer        \\
Keen                        & MontereyAq                       & Pizza Hut                   & Staples                   \\
Kenneth Cole                & Morgan Stanley                   & Polo Ralph Lauren           & Star Alliance             \\
Keurig                      & Mountain Hardwear                & Porsche                     & Starbucks                 \\
KFC                         & Mozilla                          & Pottery Barn                & Starwood Hotels           \\
Kia Motors                  & MSF                              & PPact                       & Steelers                  \\
Kickstarter                 & MSF USA                          & PPFA                        & StJude                    \\
Kmart                       & MSN                              & Prada                       & Subaru                    \\
Kohls                       & MSN Money                        & Procter \& Gamble           & Subway                    \\
Korean Air                  & MSNBC                            & Public Broadcasting Service & Susan G. Komen            \\
Kroger                      & MTV                              & Publix                      & Syfy                      \\
LA Clippers                 & MuseumModernArt                  & Puma                        & Symantec                  \\
Lacoste                     & Myspace                          & Qualcomm                    & T-Mobile                  \\
Lakers                      & National Geographic (@NatGeo)    & Quiznos                     & T.J.Maxx                  \\
Lancome                     & National Geographic (@NatGeoMag) & Quora                       & Taco Bell                 \\
Lands' End                  & National Rifle Association       & QVC                         & Target                    \\
Last.fm                     & nature\_org                      & RadioShack                  & Tata Consultancy Services \\
Leading Hotels of the World & Nautica                          & Raiders                     & Tata Group                \\
Lenovo                      & NBC                              & Ravens                      & TaylorMade                \\
Levi's                      & NBC News                         & Raytheon Co                 & TeachForAmerica           \\
Lexus                       & Neiman Marcus                    & Red                         & Telemundo                 \\
LG Mobile                   & Nest                             & Red Hat                     & Tennis Channel            \\
Lifetime Television         & Net-A-Porter                     & Red Lobster                 & Tesla                     \\
Linkedin                    & Netflix                          & Redbox                      & TGI Fridays               \\
Linux                       & New Balance                      & Reddit                      & The Week                  \\
LivingSocial                & New Belgium Brewing Company      & Reebok                      & theIRC                    \\
Lockheed Martin             & New England Patriots             & Rei                         & TheStreet                 \\
Logitech                    & New Orleans Saints               & Renaissance Hotels          & Thomson Reuters           \\
Los Angeles Rams            & New York Giants                  & Revlon                      & Threadless                \\
Los Angeles Times           & New York Jets                    & Ritz-Carlton                & Ticketmaster              \\
Louis Vuitton               & New York Knicks                  & Roberto Cavalli             & Tiffany \& Co.            \\
Lowe's                      & New York Times                   & Roche                       & Timberland                \\
Lufthansa                   & New Yorker                       & Rolling Stone               & TIME                      \\
Lululemon                   & Newsweek                         & Rolls Royce                 & Titleist                  \\
Lush                        & NFL Network                      & Rotary                      & TNT Drama                 \\
Lyft                        & ngadc                            & Rotten Tomatoes             & Tommy Hilfiger            \\
MAC Cosmetics               & NHL Network                      & Roxy                        & Toronto Raptors           \\
Macy's                      & NHL on NBC                       & Saks Fifth Avenue           & Toshiba                   \\
Make-A-Wish-Foundation      & Nickelodeon                      & Salesforce.com              & Toyota                    \\
Maker's Mark                & Nike                             & Salon                       & Toys'R'Us                 \\
Mandarin Oriental           & Nintendo                         & Sam's Club                  & Travelocity               \\
Market Watch                & Nissan                           & SamaritansPurse             & Travelzoo                 \\
Marks \& Spencer            & Nordstrom                        & Samsung Mobile              & Tripadvisor               \\
Marmot                      & North Face                       & San Francisco 49ers         & truTV                     \\
Marriott International      & Northrop Grumman                 & SanDisk                     & Tumblr                    \\
Marshalls                   & Norton Antivirus Security        & SAP                         & TWCO                      \\
Martha Stewart              & Novartis                         & Save the Children           & Twitter                   \\
Mary Kay                    & NPR                              & save\_children              & U.S. News                 \\
Maserati                    & nypl                             & Seagate Hotel \& Spa        & U.S. Open (Golf)          \\
Masters Tournament          & Old Navy                         & Seamless                    & Uber                      \\
Maybelline                  & Olive Garden                     & Sears                       & UBS                       \\
MayoClinic                  & operationbless                   & Seattle Seahawks            & Udacity                   \\
Mazda                       & Oracle                           & Sephora                     & UGG                       \\
Mc Donalds                  & Orbitz                           & Sheetz                      & Under Armour              \\
McKinsey                    & Oxfam America                    & Shell                       & UNICEF                    \\
Medium                      & Oxygen                           & Sherwin Williams            & Unilever                  \\
MentalHealthAm              & Pacers                           & Showtime                    & United Way                \\
Mercedes Benz               & PacSun                           & Siemens                     & Universal Music Group     \\
Merck                       & Panera Bread                     & Sierra Club                 & UPS                       \\
mercycorps                  & Papa John's Pizza                & Sierra Nevada               & Urban Decay               \\
Merrell                     & Paramount Pictures               & Singapore Airlines          & Urban Outfitters          \\
Merrill Lynch               & Patagonia                        & Sixers                      & USA Today                 \\
metmuseum                   & Patron                           & Skype                       & Valentino                 \\
MetOpera                    & Paypal                           & Smarty Pig                  & Vanguard Group            \\
Miami Dolphins              & Pelicans                         & smithsonian                 & Vanity Fair               \\
Miami Heat                  & Pepsi                            & Snapchat                    & Vans                      \\
Michael Kors                & PETA                             & Sony                        & Vera Wang                 \\
Michaels Stores             & PETCO                            & Southwest Airlines          & Verizon Wireless          \\
Microsoft                   & PetSmart                         & SpaceX                      & Versace Official          \\
Miller Lite                 & Pfizer                           & Speedway                    & VH1                       \\
Mint                        & Philadelphia Eagles              & Spike                       & Viber                     \\
MLB Network                 & Philips                          & Spotify                     & Victoria's Secret         \\
ModCloth                    & Pier 1 Imports                   & Sprint                      & Vikings                   \\
Money                       & PIMCO                            & Spurs                       & Vimeo                     \\
MoneyGram                   & Ping                             & Square                      & Virgin Atlantic          
\\
\hline
\end{tabular}
\end{table}

\clearpage
\begin{table}[!ht]
\caption*{Twitter Brand Accounts for Random Sample - Continued}
\tiny
\begin{tabular}{llll}
\multicolumn{4}{c}{\textbf{Brands}}                                                                                     \\
\hline\hline
vmware              & WE (www.we.org)    & Wikipedia            & Yahoo! News        \\
Volvo               & Wells Fargo        & Women's British Open & Yelp               \\
Vox                 & Wendy's            & World Vision         & YouTube            \\
Walgreens           & Western Union      & WWF                  & Yves Saint Laurent \\
Wall Street Journal & WFP                & WWFUS                & Zumiez             \\
Walmart             & WFPUSA             & wwp                  & Zynga              \\
Washington Post     & WhatsApp           & Yahoo                &                    \\
Washington Redskins & Whole Foods Market & Yahoo! Finance       &                    \\
Washington Wizards  & Wikileaks          & Yahoo! Music         &        
\\
\hline
\end{tabular}
\end{table} 
\setcounter{table}{0}
\setcounter{figure}{0}
\renewcommand{\thetable}{WB\arabic{table}}
\renewcommand{\thefigure}{WB-\arabic{figure}}

\section*{Web Appendix B - Demographics and Media Followership Snopes and Hoaxy Sample}

\begin{singlespace*}
\scriptsize

\begin{table}[ht]
\caption{Profile/Followership Information Snopes Sample}
\centering
\tiny
\begin{tabular}{lccc}
  \hline
& \textbf{Fake-news Sharers} & \textbf{Fact-check Sharers} & \textbf{Random Sample} \\ 
  \hline\hline
No. of Users & 1,727 &  671 & 1,820 \\ 
\hline
  \textbf{Political followership} &  &  &  \\ 
  Democrats & 16.91\% & 32.94\% & 11.26\% \\ 
  Republicans & 33.99\% & 15.95\% & 16.1\% \\ 
  Both & 14.36\% & 19.82\% & 5.88\% \\ 
  None & 34.74\% & 31.3\% & 66.76\% \\ 
  \textbf{Demographics} &  &  &  \\ 
  Age & 32.60 & 32.19 & 30.42 \\ 
  Proportion Male & 63.29\% & 59.02\% & 55.66\% \\ 
  \textbf{Social Media Usage Behavior} &  &  &  \\ 
  Friends (median) & 977 & 991 &  409 \\ 
  Followers (median) &  681 & 846 &  320 \\ 
  Ratio Followers Friends &  3.79 & 36.47 & 11.34 \\ 
  Statuses (median) & 14,194 & 15,748 &  3,406 \\ 
  Tweets per day & 10.55 & 11.68 &  2.45 \\ 
  \textbf{Location} &  &  &  \\ 
  Blue States & 43.49\% & 53.8\% & 41.76\% \\ 
  Red States & 28.26\% & 22.35\% & 34.12\% \\ 
  Swing States & 28.26\% & 23.85\% & 24.12\% \\ 
  \textbf{News/Media Outlets Followed} &  &  &  \\ 
  Salon & 11.12\% & 17.29\% & 1.43\% \\ 
  Wikileaks & 27.74\% & 17.44\% & 5.05\% \\ 
  Dailybeast & 15.34\% & 21.76\% & 2.75\% \\ 
  New Yorker & 19.8\% & 26.83\% & 8.68\% \\ 
  Economist & 16.44\% & 17.73\% & 5.88\% \\ 
  Wall Stree Journal & 24.72\% & 24.89\% & 10.55\% \\ 
  Washington Post & 25.88\% & 36.51\% & 9.84\% \\ 
  The New York Times & 31.56\% & 41.88\% & 16.92\% \\ 
  CNN & 26.11\% & 32.64\% & 14.01\% \\ 
  LA Times & 11.06\% & 15.05\% & 3.13\% \\ 
  USA Today & 16.97\% & 18.18\% & 4.84\% \\ 
  Buzzfeed & 13.09\% & 15.2\% & 6.37\% \\ 
  Aljazeera English & 8.05\% & 14.16\% & 2.09\% \\ 
  Fox News & 27.62\% & 15.8\% & 8.3\% \\ 
  Drudge Report & 21.6\% & 9.39\% & 2.42\% \\ 
   \hline
\end{tabular}
\end{table}

\begin{table}[ht]
\caption{Profile/Followership Information Hoaxy Sample}
\centering
\tiny
\begin{tabular}{lcccc}
  \hline
 & \textbf{Fake-news} & \textbf{Fact-check} & \textbf{Left Leaning} & \textbf{Right Leaning} \\ 
  & \textbf{Sharers} & \textbf{Sharers} & \textbf{Media Sharers} & \textbf{Media Sharers} \\ 
  \hline\hline
No. of Users & 235 & 313 & 284 & 200 \\ 
\hline
  \textbf{Political followership} &  &  &  &  \\ 
  Democrats & 9.36\% & 23.64\% & 25.7\% & 9.5\% \\ 
  Republicans & 53.62\% & 24.6\% & 11.97\% & 26.5\% \\ 
  Both & 20.85\% & 25.88\% & 25\% & 27\% \\ 
  None & 16.17\% & 25.88\% & 37.32\% & 37\% \\ 
  \textbf{Demographics} &  &  &  &  \\ 
  Age & 32.70 & 32.50 & 34.77 & 35.42 \\ 
  Proportion Male & 70.21\% & 57.19\% & 66.2\% & 69\% \\ 
  \textbf{Social Media Usage Behavior} &  &  &  &  \\ 
  Friends (median) & 1,944 &  890 & 1,192 & 1,181 \\ 
  Followers (median) &  2,228 &  637 & 4,655 & 6,779  \\ 
  Ratio Followers Friends &  19.02 & 133.82 & 468.88 & 161.63 \\ 
  Statuses (median) & 22,488 &  8,534 & 14,005 &  8,864\\ 
  Tweets per day & 23.35 & 13.97 & 14.21 & 10.05 \\ 
  \textbf{Location} &  &  &  &  \\ 
  Blue States & 46.81\% & 53.99\% & 66.2\% & 76\% \\ 
  Red States & 28.94\% & 21.41\% & 16.2\% & 11.5\% \\ 
  Swing States & 24.26\% & 24.6\% & 17.61\% & 12.5\% \\ 
  \textbf{News/Media Outlets Followed} &  &  &  &  \\ 
  Salon & 10.64\% & 12.14\% & 17.61\% & 9\% \\ 
  Wikileaks & 42.13\% & 21.73\% & 15.85\% & 15\% \\ 
  Dailybeast & 17.02\% & 14.7\% & 20.07\% & 20\% \\ 
  New Yorker & 20.85\% & 20.45\% & 35.21\% & 30.5\% \\ 
  Economist & 21.7\% & 16.93\% & 26.41\% & 34\% \\ 
  Wall Stree Journal & 28.94\% & 23\% & 35.56\% & 56\% \\ 
  Washington Post & 29.79\% & 35.46\% & 44.37\% & 44.5\% \\ 
  The New York Times & 33.62\% & 38.98\% & 52.46\% & 47\% \\ 
  CNN & 27.23\% & 38.98\% & 35.21\% & 29.5\% \\ 
  LA Times & 16.17\% & 15.02\% & 21.83\% & 15\% \\ 
  USA Today & 19.15\% & 19.49\% & 24.65\% & 16.5\% \\ 
  Buzzfeed & 13.19\% & 14.7\% & 19.37\% & 16.5\% \\ 
  Aljazeera English & 14.89\% & 7.67\% & 15.85\% & 11\% \\ 
  Fox News & 47.66\% & 29.07\% & 16.55\% & 28\% \\ 
  Drudge Report & 45.53\% & 15.34\% & 12.68\% & 22\% \\ 
   \hline
\end{tabular}
\end{table}

\end{singlespace*} 
\setcounter{table}{0}
\setcounter{figure}{0}
\renewcommand{\thetable}{WC\arabic{table}}
\renewcommand{\thefigure}{WC-\arabic{figure}}

\section*{Web Appendix C - Propensity Score Matching}
\begin{singlespace*}

For creating a group of matched users (i.e., similar socio-demographics, political affiliation, and social-media activity to fake-news sharers), we first sampled an additional set of users following one of the Twitter brand accounts (n = 1,320), combined it with the random user sample (n = 1,820) and used propensity score matching (matchit package in R). This procedure resulted in 1,725 users in the matched user group. As shown in Figure \ref{fig:PropensityScoreMatching}, the samples match well except for the number of tweets per day (all p's except for \# tweets per day $>$ 0.05). Even after the matching, fake-news sharers are more active on social media (p$<$0.01). 

\begin{figure}[htbp]
    \caption{Covariate Balance Matched Sample Snopes dataset}
    \centering
    \includegraphics[width=\textwidth]{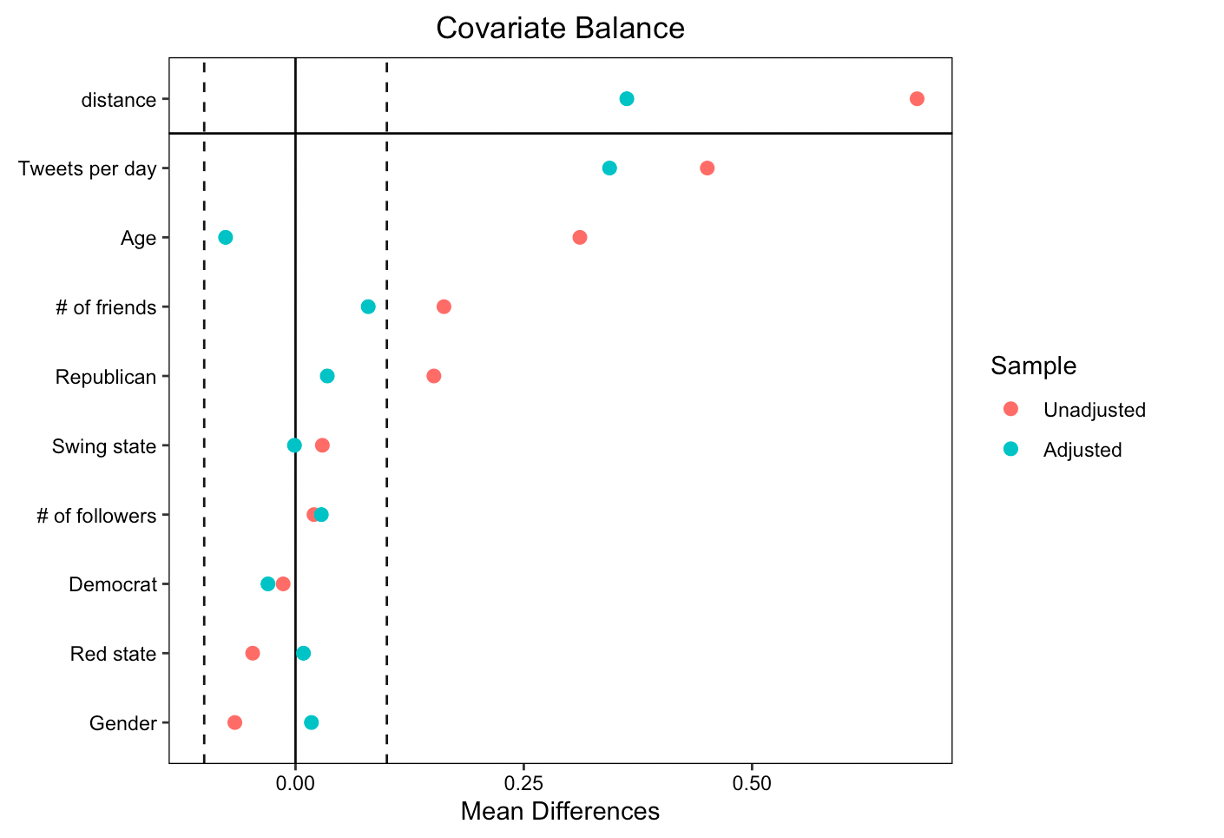}
    \label{fig:PropensityScoreMatching}
\end{figure}

\end{singlespace*} 
\setcounter{table}{0}
\renewcommand{\thetable}{WD\arabic{table}}

\section*{Web Appendix D - Robustness Snopes Sample Operationalization}

\begin{singlespace*}

\singlespacing

Below, we present the results of multiple robustness tests of our operationalization of fake-news sharers. We compare the results from our exploratory analysis of the language of fake-news sharers to two different groups of fake-news sharers that were excluded from our Snopes sample: 1) fake-news sharers that did not provide their location on their Twitter profile and 2) users that shared both a fake-news article and a fact-check article. Additionally, we also test the robustness of our results by comparing two sub-samples of fake-news sharers: fake-news sharers that shared only one fake-news article vs. fake-news sharers that shared more than one fake-news article from our Snopes sample of fake-news articles.

\singlespacing
\noindent \emph{Robustness Fake-News Sharers that Indicate No Location}

We only included fake news sharers in our sample for whom we could identify the location, who were accessed via the self-provided information on Twitter. We did so to be 1) able to compare fake-news sharers to other groups based on this information (red vs. blue vs. swing states) and 2) use the location as a matching criterion to create our matched sample.

To verify whether our results are robust to the focus on users that provide their location on their Twitter profile, we collected the tweets for a random sample of fake-news sharers that did not provide their location and were thus excluded from the original Snopes sample of fake-news sharers. We compare the sub-sample of fake-news sharers that did not provide their location (842 users) to the Snopes sample of fake-news sharers that did provide their location and to the random sample. Table B1 below shows all the differences between the two fake-news sharer groups and the random group. We find that the differences between the fake-news sharers and the random group persist when comparing fake-news sharers that did not indicate their location. Focusing on the key textual cues we identified in our exploratory analysis (power, anger) and comparing the fake-news sharer group that did not indicate their location (NFN), the fake-news sharer group that indicated their location (FN), and the random sharer group, we find:

\begin{enumerate}
\item	Power-related words: FN $=$ NFN $>$ Random users (similar finding when focusing on fake-news sharers who do not share location)
\item	Anger-related words: FN $>$ NFN $>$ Random users (similar finding when focusing on fake-news sharers who do not share location)
\end{enumerate}

Overall, we can conclude that there is no systematic evidence that the exclusion of fake-news sharers who did not indicate their location biases our results.

\begin{center}
\tiny
\begin{longtable}{lllllllll}
\caption{Comparison Snopes Fake-News Sharers No Location} \\
\hline
 & Fake  & Fake  & t-value & Random & t-value & Matched & t-value & Fact  \\ 
  & News  & News & (p-value) &  Group & (p-value) & Group & (p-value) & Checkers  \\ 
    & Snopes (1) & No Location (2)& 1 vs. 2 & (3) & 3 vs. 2  & (4)& 4 vs. 2 \\
\hline
\endfirsthead
\multicolumn{9}{c}%
{{\bfseries \tablename\ \thetable{} -- Continued from previous page}} \\
\hline
 & Fake  & Fake  & t-value & Random & t-value & Matched & t-value & Fact  \\ 
  & News  & News & (p-value) &  Group & (p-value) & Group & (p-value) & Checkers  \\ 
    & Snopes (1) & No Location (2)& 1 vs. 2 & (3) & 3 vs. 2  & (4)& 4 vs. 2 \\
\hline
\endhead
\hline \multicolumn{9}{r}{{Continued on next page}} \\ 
\endfoot
\hline \hline
\endlastfoot
Word Count & 21280.361 & 38249.417 & -19.675 ( 0.000 ) & 18401.074 & -23.222 ( 0.000 ) & 22328.980 & -18.444 ( 0.000 ) & 23477.387 \\ 
  Analytic &    80.682 &    83.340 &  -5.275 ( 0.000 ) &    69.411 & -24.059 ( 0.000 ) &    75.493 & -14.323 ( 0.000 ) &    76.995 \\ 
  Clout &    70.011 &    71.469 &  -3.483 ( 0.001 ) &    63.849 & -15.703 ( 0.000 ) &    66.450 & -10.613 ( 0.000 ) &    69.547 \\ 
  Authentic &    25.840 &    22.947 &   4.490 ( 0.000 ) &    42.021 &  28.031 ( 0.000 ) &    37.226 &  21.214 ( 0.000 ) &    22.447 \\ 
  Emotional Tone &    51.260 &    47.068 &   3.840 ( 0.000 ) &    71.868 &  24.002 ( 0.000 ) &    71.198 &  23.097 ( 0.000 ) &    47.859 \\ 
  Words $>$ 6 letters &    24.198 &    23.657 &   2.556 ( 0.011 ) &    20.866 & -13.127 ( 0.000 ) &    22.279 &  -6.536 ( 0.000 ) &    24.552 \\ 
  Dictionary &    71.690 &    69.482 &   4.372 ( 0.000 ) &    75.331 &  11.569 ( 0.000 ) &    73.861 &   8.637 ( 0.000 ) &    71.562 \\ 
  Function Words &    39.017 &    37.632 &   3.881 ( 0.000 ) &    41.335 &  10.405 ( 0.000 ) &    40.323 &   7.522 ( 0.000 ) &    40.416 \\ 
  Pronoun &    10.094 &     9.422 &   4.624 ( 0.000 ) &    12.597 &  21.197 ( 0.000 ) &    11.381 &  13.205 ( 0.000 ) &    10.825 \\ 
  Personal Pronouns &     6.512 &     6.037 &   4.585 ( 0.000 ) &     8.893 &  25.302 ( 0.000 ) &     7.773 &  15.832 ( 0.000 ) &     6.908 \\ 
  1st Person Singluar &     2.600 &     2.146 &   6.179 ( 0.000 ) &     4.752 &  30.474 ( 0.000 ) &     3.814 &  20.584 ( 0.000 ) &     2.709 \\ 
  1st Person Plural &     0.660 &     0.756 &  -5.351 ( 0.000 ) &     0.670 &  -4.259 ( 0.000 ) &     0.702 &  -2.649 ( 0.008 ) &     0.678 \\ 
  2nd Person &     1.743 &     1.556 &   4.619 ( 0.000 ) &     2.440 &  19.968 ( 0.000 ) &     2.194 &  14.386 ( 0.000 ) &     1.840 \\ 
  3rd Personl Singular &     0.924 &     0.902 &   0.941 ( 0.347 ) &     0.609 & -13.334 ( 0.000 ) &     0.612 & -12.940 ( 0.000 ) &     1.037 \\ 
  3rd Person Plural &     0.584 &     0.676 &  -5.810 ( 0.000 ) &     0.428 & -16.913 ( 0.000 ) &     0.457 & -14.649 ( 0.000 ) &     0.644 \\ 
  Impersonal Pronouns &     3.576 &     3.382 &   3.615 ( 0.000 ) &     3.698 &   6.077 ( 0.000 ) &     3.603 &   4.176 ( 0.000 ) &     3.912 \\ 
  Articles &     5.175 &     5.327 &  -2.491 ( 0.013 ) &     4.750 &  -9.414 ( 0.000 ) &     5.059 &  -4.331 ( 0.000 ) &     5.295 \\ 
  Prepositions &    11.648 &    11.382 &   2.243 ( 0.025 ) &    10.712 &  -5.628 ( 0.000 ) &    11.187 &  -1.635 ( 0.102 ) &    11.194 \\ 
  Auxiliary Verbs &     5.703 &     5.403 &   4.087 ( 0.000 ) &     6.125 &   9.913 ( 0.000 ) &     5.872 &   6.381 ( 0.000 ) &     6.122 \\ 
  Adverbs &     3.702 &     3.280 &   8.107 ( 0.000 ) &     4.424 &  21.642 ( 0.000 ) &     4.134 &  16.154 ( 0.000 ) &     3.985 \\ 
  Conjunctions &     3.507 &     3.557 &  -0.994 ( 0.321 ) &     3.814 &   5.126 ( 0.000 ) &     3.685 &   2.534 ( 0.011 ) &     3.838 \\ 
  Negate &     0.896 &     0.869 &   1.632 ( 0.103 ) &     0.940 &   4.139 ( 0.000 ) &     0.852 &  -0.988 ( 0.324 ) &     1.007 \\ 
  Verbs &    12.215 &    11.384 &   6.671 ( 0.000 ) &    13.520 &  16.923 ( 0.000 ) &    12.860 &  11.648 ( 0.000 ) &    12.620 \\ 
  Adjectives &     4.390 &     4.196 &   3.798 ( 0.000 ) &     4.517 &   6.268 ( 0.000 ) &     4.556 &   6.821 ( 0.000 ) &     4.155 \\ 
  Compare &     1.719 &     1.723 &  -0.185 ( 0.854 ) &     1.702 &  -0.808 ( 0.419 ) &     1.734 &   0.379 ( 0.704 ) &     1.745 \\ 
  Interrogatives &     1.273 &     1.180 &   4.937 ( 0.000 ) &     1.288 &   5.535 ( 0.000 ) &     1.224 &   2.244 ( 0.025 ) &     1.356 \\ 
  Number &     0.682 &     0.684 &  -0.247 ( 0.805 ) &     0.657 &  -2.089 ( 0.037 ) &     0.671 &  -1.070 ( 0.285 ) &     0.677 \\ 
  Quantifiers &     1.560 &     1.535 &   1.040 ( 0.298 ) &     1.661 &   4.948 ( 0.000 ) &     1.692 &   5.990 ( 0.000 ) &     1.628 \\ 
  Affect &     6.452 &     5.803 &   8.461 ( 0.000 ) &     7.275 &  18.343 ( 0.000 ) &     6.829 &  12.750 ( 0.000 ) &     6.263 \\ 
  Positive Emotions &     3.941 &     3.520 &   5.923 ( 0.000 ) &     5.096 &  21.801 ( 0.000 ) &     4.860 &  18.173 ( 0.000 ) &     3.729 \\ 
  Negative Emotions &     2.464 &     2.237 &   5.972 ( 0.000 ) &     2.128 &  -2.705 ( 0.007 ) &     1.917 &  -8.030 ( 0.000 ) &     2.483 \\ 
  Anxiety &     0.286 &     0.272 &   1.813 ( 0.070 ) &     0.217 &  -7.545 ( 0.000 ) &     0.209 &  -8.691 ( 0.000 ) &     0.283 \\ 
  Anger &     1.021 &     0.923 &   4.536 ( 0.000 ) &     0.852 &  -2.986 ( 0.003 ) &     0.736 &  -8.073 ( 0.000 ) &     1.033 \\ 
  Sadness &     0.421 &     0.379 &   5.395 ( 0.000 ) &     0.435 &   6.951 ( 0.000 ) &     0.385 &   0.830 ( 0.407 ) &     0.424 \\ 
  Social &     8.775 &     8.684 &   0.917 ( 0.359 ) &     9.088 &   3.916 ( 0.000 ) &     8.765 &   0.780 ( 0.436 ) &     8.978 \\ 
  Family &     0.402 &     0.395 &   0.675 ( 0.500 ) &     0.479 &   6.317 ( 0.000 ) &     0.423 &   1.986 ( 0.047 ) &     0.360 \\ 
  Friend &     0.325 &     0.284 &   4.409 ( 0.000 ) &     0.454 &  15.022 ( 0.000 ) &     0.403 &  11.440 ( 0.000 ) &     0.304 \\ 
  Female Words &     0.615 &     0.585 &   1.797 ( 0.073 ) &     0.580 &  -0.273 ( 0.785 ) &     0.512 &  -4.250 ( 0.000 ) &     0.603 \\ 
  Male Words &     1.094 &     1.050 &   1.975 ( 0.048 ) &     0.893 &  -7.007 ( 0.000 ) &     0.867 &  -8.037 ( 0.000 ) &     1.182 \\ 
  Cognitive Processes &     8.096 &     7.626 &   4.822 ( 0.000 ) &     8.342 &   7.229 ( 0.000 ) &     8.146 &   5.180 ( 0.000 ) &     8.753 \\ 
  Insight &     1.688 &     1.555 &   5.748 ( 0.000 ) &     1.619 &   2.656 ( 0.008 ) &     1.636 &   3.283 ( 0.001 ) &     1.798 \\ 
  Cause &     1.340 &     1.331 &   0.452 ( 0.651 ) &     1.281 &  -2.471 ( 0.014 ) &     1.308 &  -1.110 ( 0.267 ) &     1.395 \\ 
  Discrepancies &     1.251 &     1.130 &   6.167 ( 0.000 ) &     1.382 &  12.488 ( 0.000 ) &     1.279 &   7.477 ( 0.000 ) &     1.345 \\ 
  Tentative &     1.761 &     1.620 &   5.119 ( 0.000 ) &     1.886 &   9.717 ( 0.000 ) &     1.865 &   8.834 ( 0.000 ) &     1.975 \\ 
  Certain &     1.316 &     1.253 &   3.033 ( 0.002 ) &     1.355 &   5.014 ( 0.000 ) &     1.292 &   1.899 ( 0.058 ) &     1.367 \\ 
  Differentiation &     1.811 &     1.728 &   2.796 ( 0.005 ) &     1.873 &   4.895 ( 0.000 ) &     1.817 &   2.929 ( 0.003 ) &     2.065 \\ 
  Perceptional Processes &     2.737 &     2.523 &   4.545 ( 0.000 ) &     2.838 &   6.646 ( 0.000 ) &     2.725 &   4.274 ( 0.000 ) &     2.482 \\ 
  See &     1.402 &     1.304 &   2.604 ( 0.009 ) &     1.393 &   2.322 ( 0.020 ) &     1.360 &   1.490 ( 0.136 ) &     1.191 \\ 
  Hear &     0.802 &     0.712 &   5.613 ( 0.000 ) &     0.746 &   2.094 ( 0.036 ) &     0.722 &   0.606 ( 0.544 ) &     0.786 \\ 
  Feel &     0.401 &     0.379 &   1.688 ( 0.092 ) &     0.530 &  11.593 ( 0.000 ) &     0.476 &   7.490 ( 0.000 ) &     0.380 \\ 
  Bio &     2.010 &     1.985 &   0.579 ( 0.563 ) &     2.616 &  12.989 ( 0.000 ) &     2.356 &   7.396 ( 0.000 ) &     1.799 \\ 
  Body &     0.559 &     0.543 &   1.107 ( 0.268 ) &     0.764 &  12.522 ( 0.000 ) &     0.636 &   5.453 ( 0.000 ) &     0.517 \\ 
  Health &     0.593 &     0.690 &  -4.551 ( 0.000 ) &     0.592 &  -4.347 ( 0.000 ) &     0.573 &  -4.950 ( 0.000 ) &     0.530 \\ 
  Sexual &     0.226 &     0.219 &   0.609 ( 0.542 ) &     0.232 &   1.062 ( 0.288 ) &     0.194 &  -1.997 ( 0.046 ) &     0.229 \\ 
  Ingestion &     0.554 &     0.488 &   3.088 ( 0.002 ) &     0.763 &  10.914 ( 0.000 ) &     0.762 &   9.791 ( 0.000 ) &     0.452 \\ 
  Drives &     7.907 &     7.921 &  -0.168 ( 0.867 ) &     7.616 &  -3.629 ( 0.000 ) &     7.676 &  -2.883 ( 0.004 ) &     7.575 \\ 
  Affiliation &     2.143 &     2.265 &  -2.925 ( 0.004 ) &     2.553 &   6.227 ( 0.000 ) &     2.490 &   4.889 ( 0.000 ) &     2.026 \\ 
  Achieve &     1.358 &     1.362 &  -0.194 ( 0.847 ) &     1.510 &   6.083 ( 0.000 ) &     1.574 &   8.422 ( 0.000 ) &     1.283 \\ 
  Power &     3.345 &     3.343 &   0.039 ( 0.969 ) &     2.311 & -19.986 ( 0.000 ) &     2.468 & -16.602 ( 0.000 ) &     3.180 \\ 
  Reward &     1.430 &     1.298 &   4.735 ( 0.000 ) &     1.905 &  21.768 ( 0.000 ) &     1.828 &  18.326 ( 0.000 ) &     1.353 \\ 
  Risk &     0.616 &     0.610 &   0.537 ( 0.591 ) &     0.438 & -15.276 ( 0.000 ) &     0.452 & -13.841 ( 0.000 ) &     0.619 \\ 
  Past Focus &     2.465 &     2.555 &  -2.110 ( 0.035 ) &     2.463 &  -2.136 ( 0.033 ) &     2.380 &  -4.008 ( 0.000 ) &     2.579 \\ 
  Present Focus &     8.376 &     7.615 &   8.182 ( 0.000 ) &     9.756 &  22.678 ( 0.000 ) &     9.233 &  17.266 ( 0.000 ) &     8.574 \\ 
  Future Focus &     1.015 &     0.987 &   1.767 ( 0.077 ) &     1.312 &  18.539 ( 0.000 ) &     1.199 &  11.833 ( 0.000 ) &     0.980 \\ 
  Relativity &    12.637 &    12.317 &   2.463 ( 0.014 ) &    13.325 &   7.849 ( 0.000 ) &    13.381 &   8.170 ( 0.000 ) &    11.645 \\ 
  Motion &     1.695 &     1.585 &   4.962 ( 0.000 ) &     1.963 &  16.143 ( 0.000 ) &     1.899 &  12.856 ( 0.000 ) &     1.552 \\ 
  Space &     6.507 &     6.183 &   3.833 ( 0.000 ) &     6.002 &  -2.207 ( 0.027 ) &     6.312 &   1.541 ( 0.123 ) &     6.005 \\ 
  Time &     4.562 &     4.668 &  -1.559 ( 0.119 ) &     5.566 &  13.038 ( 0.000 ) &     5.355 &   9.838 ( 0.000 ) &     4.200 \\ 
  Work &     2.702 &     2.758 &  -0.976 ( 0.329 ) &     2.082 & -11.256 ( 0.000 ) &     2.430 &  -5.150 ( 0.000 ) &     2.611 \\ 
  Leisure &     2.013 &     1.887 &   1.586 ( 0.113 ) &     2.208 &   4.141 ( 0.000 ) &     2.165 &   3.645 ( 0.000 ) &     1.595 \\ 
  Home &     0.397 &     0.407 &  -0.862 ( 0.389 ) &     0.467 &   3.754 ( 0.000 ) &     0.460 &   3.086 ( 0.002 ) &     0.349 \\ 
  Money &     0.977 &     1.046 &  -2.320 ( 0.020 ) &     0.748 &  -9.459 ( 0.000 ) &     0.917 &  -3.701 ( 0.000 ) &     0.920 \\ 
  Religion &     0.528 &     0.468 &   2.319 ( 0.021 ) &     0.446 &  -0.865 ( 0.387 ) &     0.431 &  -1.438 ( 0.150 ) &     0.414 \\ 
  Death &     0.341 &     0.373 &  -2.461 ( 0.014 ) &     0.181 & -16.186 ( 0.000 ) &     0.179 & -16.308 ( 0.000 ) &     0.279 \\ 
  Informal &     1.895 &     1.839 &   0.714 ( 0.475 ) &     3.262 &  15.550 ( 0.000 ) &     2.636 &   9.182 ( 0.000 ) &     1.865 \\ 
  Swear &     0.392 &     0.342 &   2.524 ( 0.012 ) &     0.683 &  12.773 ( 0.000 ) &     0.509 &   6.704 ( 0.000 ) &     0.403 \\ 
  Netspeak &     0.949 &     1.086 &  -2.368 ( 0.018 ) &     1.865 &  11.528 ( 0.000 ) &     1.453 &   5.761 ( 0.000 ) &     0.853 \\ 
  Assent &     0.382 &     0.322 &   3.317 ( 0.001 ) &     0.571 &  13.070 ( 0.000 ) &     0.540 &  11.287 ( 0.000 ) &     0.400 \\ 
  Nonfluencies &     0.216 &     0.191 &   1.621 ( 0.105 ) &     0.281 &   5.672 ( 0.000 ) &     0.262 &   4.428 ( 0.000 ) &     0.254 \\ 
  Filler &     0.020 &     0.018 &   1.731 ( 0.084 ) &     0.039 &  13.661 ( 0.000 ) &     0.031 &   8.837 ( 0.000 ) &     0.023 \\ 
  Openness &     0.620 &     0.616 &   1.325 ( 0.185 ) &     0.554 & -17.212 ( 0.000 ) &     0.570 & -12.614 ( 0.000 ) &     0.623 \\ 
  Conscientuousness &     0.520 &     0.523 &  -0.972 ( 0.331 ) &     0.539 &   4.753 ( 0.000 ) &     0.558 &  10.127 ( 0.000 ) &     0.512 \\ 
  Extraversion &     0.414 &     0.402 &   3.582 ( 0.000 ) &     0.498 &  24.077 ( 0.000 ) &     0.475 &  18.784 ( 0.000 ) &     0.396 \\ 
  Agreeableness &     0.443 &     0.422 &   5.884 ( 0.000 ) &     0.501 &  21.210 ( 0.000 ) &     0.502 &  21.826 ( 0.000 ) &     0.426 \\ 
  Neuroticism &     0.421 &     0.425 &  -1.314 ( 0.189 ) &     0.410 &  -4.379 ( 0.000 ) &     0.392 &  -9.564 ( 0.000 ) &     0.439 \\ 
  Negative Low Arousal &     0.379 &     0.367 &   1.690 ( 0.091 ) &     0.430 &   8.057 ( 0.000 ) &     0.382 &   1.989 ( 0.047 ) &     0.394 \\ 
  Negative High Arousal &     0.873 &     0.735 &   7.628 ( 0.000 ) &     0.859 &   5.963 ( 0.000 ) &     0.735 &  -0.042 ( 0.967 ) &     0.899 \\ 
  Positive Low Arousal &     1.467 &     1.322 &   4.005 ( 0.000 ) &     2.055 &  19.667 ( 0.000 ) &     1.861 &  14.330 ( 0.000 ) &     1.422 \\ 
  Positive High Arousal &     0.823 &     0.699 &   5.275 ( 0.000 ) &     1.297 &  23.536 ( 0.000 ) &     1.164 &  18.843 ( 0.000 ) &     0.711 \\ 
   \hline
\end{longtable}
\end{center}

\noindent \emph{Robustness Sharers that Shared Both Fake-News and Fact-Checks}

We excluded Twitter users who shared both a fake news article and a fact-check article in our sample as these users could not be clearly assigned to one group. We compare the excluded Twitter users based on this criterion to fake-news sharers and fact-check sharers. To do so, we collected the users' information that shared both fake-news and fact-check article(s). From the 253 excluded users that shared both, we could access 150 users in July 2023 who still had an open and active Twitter profile. We compare the group that shared both fake-news and fact-check articles to the fake-news sharer group (FN) and fact-check sharer group (FC) of our Snopes sample. Overall, we find no systematic pattern of differences in the users that share both fake-news sharers and fact-check articles with the other groups. Focusing only on the key descriptive findings related to textual cues (power, anger), we find:

\begin{enumerate}
\item Power-related words: FN $<$ FNBoth $>$ FC (more power-related words when focusing on fake-news sharers that also share fact-checks)
\item	Anger-related words: FN $>$ FNBoth $<$ FC (less anger-related words to fake-news sharers and fact-check sharers)
\end{enumerate}

\begin{center}
\tiny
\begin{longtable}{lllllllll}
\caption{Snopes Users Sharing Fake-News and Fact Check} \\
\hline
 & Fake  & Fake and  & t-value & Fact & t-value & Random & t-value & Matched  \\ 
  & News  & Fact & (p-value) &  Checkers & (p-value) & Group & (p-value) & Group  \\ 
    & Snopes (1) & Both (2)& 1 vs. 2 & (3) & 3 vs. 2  & (4)& 4 vs. 2 \\
\hline
\endfirsthead
\multicolumn{9}{c}%
{{\bfseries \tablename\ \thetable{} -- Continued from previous page}} \\
\hline
 & Fake  & Fake and  & t-value & Fact & t-value & Random & t-value & Matched  \\ 
  & News  & Fact & (p-value) &  Checkers & (p-value) & Group & (p-value) & Group  \\ 
    & Snopes (1) & Both (2)& 1 vs. 2 & (3) & 3 vs. 2  & (4)& 4 vs. 2 \\
\hline
\endhead
\hline \multicolumn{9}{r}{{Continued on next page}} \\ 
\endfoot
\hline \hline
\endlastfoot
Word Count & 21280.361 & 53481.095 & -16.114 ( 0.000 ) & 23477.387 & -14.770 ( 0.000 ) & 18401.074 & -17.584 ( 0.000 ) & 22328.980 \\ 
  Analytic &    80.682 &    84.969 &  -5.383 ( 0.000 ) &    76.995 &  -9.057 ( 0.000 ) &    69.411 & -18.392 ( 0.000 ) &    75.493 \\ 
  Clout &    70.011 &    72.731 &  -4.316 ( 0.000 ) &    69.547 &  -4.611 ( 0.000 ) &    63.849 & -13.133 ( 0.000 ) &    66.450 \\ 
  Authentic &    25.840 &    23.452 &   2.098 ( 0.037 ) &    22.447 &  -0.825 ( 0.410 ) &    42.021 &  16.020 ( 0.000 ) &    37.226 \\ 
  Emotional Tone &    51.260 &    44.530 &   3.452 ( 0.001 ) &    47.859 &   1.617 ( 0.107 ) &    71.868 &  14.256 ( 0.000 ) &    71.198 \\ 
  Words $>$ 6 letters &    24.198 &    24.540 &  -0.973 ( 0.332 ) &    24.552 &   0.031 ( 0.975 ) &    20.866 & -10.422 ( 0.000 ) &    22.279 \\ 
  Dictionary &    71.690 &    73.365 &  -3.320 ( 0.001 ) &    71.562 &  -3.426 ( 0.001 ) &    75.331 &   3.893 ( 0.000 ) &    73.861 \\ 
  Function Words &    39.017 &    40.774 &  -3.774 ( 0.000 ) &    40.416 &  -0.736 ( 0.463 ) &    41.335 &   1.206 ( 0.229 ) &    40.323 \\ 
  Pronoun &    10.094 &     9.582 &   2.343 ( 0.020 ) &    10.825 &   5.305 ( 0.000 ) &    12.597 &  13.607 ( 0.000 ) &    11.381 \\ 
  Personal Pronouns &     6.512 &     5.842 &   4.258 ( 0.000 ) &     6.908 &   6.326 ( 0.000 ) &     8.893 &  18.656 ( 0.000 ) &     7.773 \\ 
  1st Person Singluar &     2.600 &     1.783 &   7.966 ( 0.000 ) &     2.709 &   8.195 ( 0.000 ) &     4.752 &  26.611 ( 0.000 ) &     3.814 \\ 
  1st Person Plural &     0.660 &     0.909 &  -6.020 ( 0.000 ) &     0.678 &  -5.446 ( 0.000 ) &     0.670 &  -5.643 ( 0.000 ) &     0.702 \\ 
  2nd Person &     1.743 &     1.332 &   7.385 ( 0.000 ) &     1.840 &   8.449 ( 0.000 ) &     2.440 &  18.951 ( 0.000 ) &     2.194 \\ 
  3rd Personl Singular &     0.924 &     1.019 &  -2.059 ( 0.041 ) &     1.037 &   0.360 ( 0.719 ) &     0.609 &  -9.071 ( 0.000 ) &     0.612 \\ 
  3rd Person Plural &     0.584 &     0.798 &  -7.716 ( 0.000 ) &     0.644 &  -5.323 ( 0.000 ) &     0.428 & -13.654 ( 0.000 ) &     0.457 \\ 
  Impersonal Pronouns &     3.576 &     3.737 &  -1.964 ( 0.051 ) &     3.912 &   1.983 ( 0.048 ) &     3.698 &  -0.480 ( 0.632 ) &     3.603 \\ 
  Articles &     5.175 &     5.957 &  -7.960 ( 0.000 ) &     5.295 &  -6.512 ( 0.000 ) &     4.750 & -12.261 ( 0.000 ) &     5.059 \\ 
  Prepositions &    11.648 &    12.661 &  -6.660 ( 0.000 ) &    11.194 &  -9.081 ( 0.000 ) &    10.712 & -12.789 ( 0.000 ) &    11.187 \\ 
  Auxiliary Verbs &     5.703 &     5.856 &  -1.435 ( 0.153 ) &     6.122 &   2.345 ( 0.020 ) &     6.125 &   2.540 ( 0.012 ) &     5.872 \\ 
  Adverbs &     3.702 &     3.479 &   2.798 ( 0.006 ) &     3.985 &   6.006 ( 0.000 ) &     4.424 &  11.770 ( 0.000 ) &     4.134 \\ 
  Conjunctions &     3.507 &     4.115 &  -7.749 ( 0.000 ) &     3.838 &  -3.343 ( 0.001 ) &     3.814 &  -3.854 ( 0.000 ) &     3.685 \\ 
  Negate &     0.896 &     0.884 &   0.406 ( 0.685 ) &     1.007 &   3.991 ( 0.000 ) &     0.940 &   1.919 ( 0.057 ) &     0.852 \\ 
  Verbs &    12.215 &    12.003 &   1.299 ( 0.196 ) &    12.620 &   3.577 ( 0.000 ) &    13.520 &   9.217 ( 0.000 ) &    12.860 \\ 
  Adjectives &     4.390 &     4.370 &   0.323 ( 0.747 ) &     4.155 &  -3.465 ( 0.001 ) &     4.517 &   2.416 ( 0.017 ) &     4.556 \\ 
  Compare &     1.719 &     1.998 &  -8.116 ( 0.000 ) &     1.745 &  -7.132 ( 0.000 ) &     1.702 &  -8.335 ( 0.000 ) &     1.734 \\ 
  Interrogatives &     1.273 &     1.244 &   1.139 ( 0.256 ) &     1.356 &   4.141 ( 0.000 ) &     1.288 &   1.704 ( 0.090 ) &     1.224 \\ 
  Number &     0.682 &     0.800 &  -6.727 ( 0.000 ) &     0.677 &  -6.486 ( 0.000 ) &     0.657 &  -7.536 ( 0.000 ) &     0.671 \\ 
  Quantifiers &     1.560 &     1.718 &  -4.517 ( 0.000 ) &     1.628 &  -2.474 ( 0.014 ) &     1.661 &  -1.577 ( 0.116 ) &     1.692 \\ 
  Affect &     6.452 &     5.749 &   6.652 ( 0.000 ) &     6.263 &   4.640 ( 0.000 ) &     7.275 &  14.089 ( 0.000 ) &     6.829 \\ 
  Positive Emotions &     3.941 &     3.384 &   5.395 ( 0.000 ) &     3.729 &   3.187 ( 0.002 ) &     5.096 &  16.438 ( 0.000 ) &     4.860 \\ 
  Negative Emotions &     2.464 &     2.316 &   2.585 ( 0.010 ) &     2.483 &   2.733 ( 0.007 ) &     2.128 &  -3.189 ( 0.002 ) &     1.917 \\ 
  Anxiety &     0.286 &     0.301 &  -1.749 ( 0.082 ) &     0.283 &  -1.994 ( 0.047 ) &     0.217 &  -9.850 ( 0.000 ) &     0.209 \\ 
  Anger &     1.021 &     0.943 &   2.001 ( 0.047 ) &     1.033 &   2.172 ( 0.031 ) &     0.852 &  -2.258 ( 0.025 ) &     0.736 \\ 
  Sadness &     0.421 &     0.410 &   1.009 ( 0.314 ) &     0.424 &   1.158 ( 0.248 ) &     0.435 &   2.274 ( 0.024 ) &     0.385 \\ 
  Social &     8.775 &     8.854 &  -0.517 ( 0.606 ) &     8.978 &   0.786 ( 0.433 ) &     9.088 &   1.522 ( 0.130 ) &     8.765 \\ 
  Family &     0.402 &     0.370 &   1.734 ( 0.084 ) &     0.360 &  -0.530 ( 0.597 ) &     0.479 &   5.511 ( 0.000 ) &     0.423 \\ 
  Friend &     0.325 &     0.251 &   5.942 ( 0.000 ) &     0.304 &   3.938 ( 0.000 ) &     0.454 &  14.504 ( 0.000 ) &     0.403 \\ 
  Female Words &     0.615 &     0.527 &   3.317 ( 0.001 ) &     0.603 &   2.612 ( 0.010 ) &     0.580 &   2.004 ( 0.046 ) &     0.512 \\ 
  Male Words &     1.094 &     1.101 &  -0.138 ( 0.890 ) &     1.182 &   1.689 ( 0.093 ) &     0.893 &  -4.614 ( 0.000 ) &     0.867 \\ 
  Cognitive Processes &     8.096 &     8.498 &  -2.247 ( 0.026 ) &     8.753 &   1.369 ( 0.172 ) &     8.342 &  -0.867 ( 0.387 ) &     8.146 \\ 
  Insight &     1.688 &     1.764 &  -0.772 ( 0.441 ) &     1.798 &   0.341 ( 0.734 ) &     1.619 &  -1.477 ( 0.142 ) &     1.636 \\ 
  Cause &     1.340 &     1.459 &  -4.453 ( 0.000 ) &     1.395 &  -2.284 ( 0.023 ) &     1.281 &  -6.443 ( 0.000 ) &     1.308 \\ 
  Discrepancies &     1.251 &     1.233 &   0.617 ( 0.538 ) &     1.345 &   3.506 ( 0.001 ) &     1.382 &   4.998 ( 0.000 ) &     1.279 \\ 
  Tentative &     1.761 &     1.801 &  -0.760 ( 0.448 ) &     1.975 &   3.162 ( 0.002 ) &     1.886 &   1.644 ( 0.102 ) &     1.865 \\ 
  Certain &     1.316 &     1.348 &  -0.766 ( 0.445 ) &     1.367 &   0.450 ( 0.653 ) &     1.355 &   0.180 ( 0.857 ) &     1.292 \\ 
  Differentiation &     1.811 &     2.009 &  -3.896 ( 0.000 ) &     2.065 &   1.041 ( 0.299 ) &     1.873 &  -2.672 ( 0.008 ) &     1.817 \\ 
  Perceptional Processes &     2.737 &     2.256 &   7.838 ( 0.000 ) &     2.482 &   3.632 ( 0.000 ) &     2.838 &   9.447 ( 0.000 ) &     2.725 \\ 
  See &     1.402 &     1.089 &   7.335 ( 0.000 ) &     1.191 &   2.388 ( 0.018 ) &     1.393 &   7.003 ( 0.000 ) &     1.360 \\ 
  Hear &     0.802 &     0.697 &   5.251 ( 0.000 ) &     0.786 &   4.399 ( 0.000 ) &     0.746 &   2.450 ( 0.015 ) &     0.722 \\ 
  Feel &     0.401 &     0.366 &   2.211 ( 0.028 ) &     0.380 &   0.877 ( 0.382 ) &     0.530 &  10.503 ( 0.000 ) &     0.476 \\ 
  Bio &     2.010 &     1.668 &   6.009 ( 0.000 ) &     1.799 &   2.175 ( 0.031 ) &     2.616 &  15.513 ( 0.000 ) &     2.356 \\ 
  Body &     0.559 &     0.426 &   7.672 ( 0.000 ) &     0.517 &   4.932 ( 0.000 ) &     0.764 &  16.916 ( 0.000 ) &     0.636 \\ 
  Health &     0.593 &     0.645 &  -2.316 ( 0.021 ) &     0.530 &  -4.512 ( 0.000 ) &     0.592 &  -2.237 ( 0.026 ) &     0.573 \\ 
  Sexual &     0.226 &     0.187 &   2.814 ( 0.005 ) &     0.229 &   2.807 ( 0.005 ) &     0.232 &   3.085 ( 0.002 ) &     0.194 \\ 
  Ingestion &     0.554 &     0.385 &   5.854 ( 0.000 ) &     0.452 &   2.328 ( 0.021 ) &     0.763 &  11.890 ( 0.000 ) &     0.762 \\ 
  Drives &     7.907 &     8.747 &  -6.101 ( 0.000 ) &     7.575 &  -8.192 ( 0.000 ) &     7.616 &  -8.135 ( 0.000 ) &     7.676 \\ 
  Affiliation &     2.143 &     2.320 &  -2.860 ( 0.005 ) &     2.026 &  -4.566 ( 0.000 ) &     2.553 &   3.591 ( 0.000 ) &     2.490 \\ 
  Achieve &     1.358 &     1.545 &  -5.188 ( 0.000 ) &     1.283 &  -7.014 ( 0.000 ) &     1.510 &  -0.921 ( 0.358 ) &     1.574 \\ 
  Power &     3.345 &     3.903 &  -5.143 ( 0.000 ) &     3.180 &  -6.394 ( 0.000 ) &     2.311 & -14.965 ( 0.000 ) &     2.468 \\ 
  Reward &     1.430 &     1.274 &   4.983 ( 0.000 ) &     1.353 &   2.395 ( 0.017 ) &     1.905 &  20.065 ( 0.000 ) &     1.828 \\ 
  Risk &     0.616 &     0.764 &  -5.507 ( 0.000 ) &     0.619 &  -5.281 ( 0.000 ) &     0.438 & -12.330 ( 0.000 ) &     0.452 \\ 
  Past Focus &     2.465 &     2.733 &  -4.208 ( 0.000 ) &     2.579 &  -2.320 ( 0.021 ) &     2.463 &  -4.219 ( 0.000 ) &     2.380 \\ 
  Present Focus &     8.376 &     7.939 &   3.714 ( 0.000 ) &     8.574 &   5.150 ( 0.000 ) &     9.756 &  15.320 ( 0.000 ) &     9.233 \\ 
  Future Focus &     1.015 &     1.026 &  -0.514 ( 0.608 ) &     0.980 &  -2.096 ( 0.037 ) &     1.312 &  12.750 ( 0.000 ) &     1.199 \\ 
  Relativity &    12.637 &    12.993 &  -1.967 ( 0.051 ) &    11.645 &  -6.889 ( 0.000 ) &    13.325 &   1.842 ( 0.067 ) &    13.381 \\ 
  Motion &     1.695 &     1.722 &  -0.622 ( 0.534 ) &     1.552 &  -3.868 ( 0.000 ) &     1.963 &   5.636 ( 0.000 ) &     1.899 \\ 
  Space &     6.507 &     6.668 &  -1.380 ( 0.169 ) &     6.005 &  -5.231 ( 0.000 ) &     6.002 &  -5.801 ( 0.000 ) &     6.312 \\ 
  Time &     4.562 &     4.738 &  -2.277 ( 0.024 ) &     4.200 &  -6.590 ( 0.000 ) &     5.566 &  10.670 ( 0.000 ) &     5.355 \\ 
  Work &     2.702 &     3.500 &  -6.239 ( 0.000 ) &     2.611 &  -6.802 ( 0.000 ) &     2.082 & -10.973 ( 0.000 ) &     2.430 \\ 
  Leisure &     2.013 &     1.327 &   8.879 ( 0.000 ) &     1.595 &   3.377 ( 0.001 ) &     2.208 &  11.694 ( 0.000 ) &     2.165 \\ 
  Home &     0.397 &     0.451 &  -2.499 ( 0.013 ) &     0.349 &  -4.620 ( 0.000 ) &     0.467 &   0.649 ( 0.517 ) &     0.460 \\ 
  Money &     0.977 &     1.307 &  -4.519 ( 0.000 ) &     0.920 &  -5.218 ( 0.000 ) &     0.748 &  -7.577 ( 0.000 ) &     0.917 \\ 
  Religion &     0.528 &     0.415 &   1.649 ( 0.101 ) &     0.414 &  -0.024 ( 0.981 ) &     0.446 &   0.456 ( 0.649 ) &     0.431 \\ 
  Death &     0.341 &     0.345 &  -0.212 ( 0.833 ) &     0.279 &  -3.408 ( 0.001 ) &     0.181 &  -9.039 ( 0.000 ) &     0.179 \\ 
  Informal &     1.895 &     1.130 &  13.197 ( 0.000 ) &     1.865 &  11.793 ( 0.000 ) &     3.262 &  28.741 ( 0.000 ) &     2.636 \\ 
  Swear &     0.392 &     0.238 &   5.780 ( 0.000 ) &     0.403 &   5.652 ( 0.000 ) &     0.683 &  13.871 ( 0.000 ) &     0.509 \\ 
  Netspeak &     0.949 &     0.553 &  13.420 ( 0.000 ) &     0.853 &   9.260 ( 0.000 ) &     1.865 &  28.967 ( 0.000 ) &     1.453 \\ 
  Assent &     0.382 &     0.244 &   9.475 ( 0.000 ) &     0.400 &  10.462 ( 0.000 ) &     0.571 &  20.807 ( 0.000 ) &     0.540 \\ 
  Nonfluencies &     0.216 &     0.128 &  10.717 ( 0.000 ) &     0.254 &   9.228 ( 0.000 ) &     0.281 &  18.019 ( 0.000 ) &     0.262 \\ 
  Filler &     0.020 &     0.017 &   2.195 ( 0.029 ) &     0.023 &   3.168 ( 0.002 ) &     0.039 &  11.425 ( 0.000 ) &     0.031 \\ 
  Openness &     0.620 &     0.635 &  -3.453 ( 0.001 ) &     0.623 &  -2.782 ( 0.006 ) &     0.554 & -17.683 ( 0.000 ) &     0.570 \\ 
  Conscientuousness &     0.520 &     0.557 &  -8.255 ( 0.000 ) &     0.512 &  -9.332 ( 0.000 ) &     0.539 &  -3.677 ( 0.000 ) &     0.558 \\ 
  Extraversion &     0.414 &     0.387 &   4.950 ( 0.000 ) &     0.396 &   1.557 ( 0.121 ) &     0.498 &  19.280 ( 0.000 ) &     0.475 \\ 
  Agreeableness &     0.443 &     0.420 &   3.733 ( 0.000 ) &     0.426 &   0.973 ( 0.331 ) &     0.501 &  12.887 ( 0.000 ) &     0.502 \\ 
  Neuroticism &     0.421 &     0.418 &   0.686 ( 0.494 ) &     0.439 &   3.785 ( 0.000 ) &     0.410 &  -1.436 ( 0.152 ) &     0.392 \\ 
  Negative Low Arousal &     0.379 &     0.378 &   0.074 ( 0.941 ) &     0.394 &   1.332 ( 0.184 ) &     0.430 &   4.686 ( 0.000 ) &     0.382 \\ 
  Negative High Arousal &     0.873 &     0.712 &   6.161 ( 0.000 ) &     0.899 &   6.583 ( 0.000 ) &     0.859 &   5.247 ( 0.000 ) &     0.735 \\ 
  Positive Low Arousal &     1.467 &     1.224 &   5.888 ( 0.000 ) &     1.422 &   4.566 ( 0.000 ) &     2.055 &  19.738 ( 0.000 ) &     1.861 \\ 
  Positive High Arousal &     0.823 &     0.586 &   8.048 ( 0.000 ) &     0.711 &   4.067 ( 0.000 ) &     1.297 &  22.882 ( 0.000 ) &     1.164 \\ 
  \hline
\end{longtable}
\end{center}

\noindent \emph{Robustness Number of Fake-News Shared}

In our original data, most users labeled as fake-news sharers had shared only 1 fake-news article in our sample of fake-news articles. Very few (92 users, 5\%) shared more than 1. To investigate whether there are systematic differences between fake-news sharers that shared only one vs. more than one article from our Snopes sample of articles, we compare those labeled as fake-news sharers after sharing exactly 1 fake-news to those labeled fake-news sharers after sharing 2+ fake-news. Table WD-3 shows the differences for all the variables included in our analysis. Our results demonstrate that by focusing on those who shared more fake news, our descriptive findings about differences in textual cues may have been amplified if anything. Focusing on the key textual cues (power, anger) and comparing the groups that shared only one fake-news article (1 FN) vs. those that shared more than one fake-news article (2+ FN), we again find that if anything further distinguishing the group of fake-news sharers amplifies the differences to the random group.

\begin{enumerate}
\item	Power-related words: 2+ FN $>$ 1 FN $>$ Random users (finding amplified when focusing on those with 2+ vs 1)
\item	Anger-related words: 2+ FN $>$ 1 FN $>$ Random users (finding amplified when focusing on those with 2+ vs 1)
\end{enumerate}

\begin{center}
\tiny
\begin{longtable}{llllllll}
\caption{One Time vs. Multiple Time Fake-News Sharers} \\
\hline
 & One Time  & Multiple Times  & t-value & Random  & t-value  & Matched  & Fact \\
  & Fake-News  & Fake-News  & (p-value) & Group & (p-value) &  Group & Checkers \\ 
    & Sharers (1) & Sharers (2)& 1 vs. 2 & (3) & 3 vs. 2 &  &  \\ 
\hline
\endfirsthead
\multicolumn{8}{c}%
{{\bfseries \tablename\ \thetable{} -- Continued from previous page}} \\
\hline
 & One Time  & Multiple Times  & t-value & Random  & t-value  & Matched  & Fact \\
  & Fake-News  & Fake-News  & (p-value) & Group & (p-value) &  Group & Checkers \\ 
    & Sharers (1) & Sharers (2)& 1 vs. 2 & (3) & 3 vs. 2 &  &  \\ 
\hline
\endhead
\hline
\multicolumn{8}{r}{{Continued on next page}} \\
\endfoot
\hline
\endlastfoot
Word Count & 21476.481 & 17794.957 &  3.315 ( 0.001 ) & 18401.074 &  -0.550 ( 0.584 ) & 22328.980 & 23477.387 \\ 
  Analytic &    80.501 &    83.900 & -2.862 ( 0.005 ) &    69.411 &  11.891 ( 0.000 ) &    75.493 &    76.995 \\ 
  Clout &    69.864 &    72.623 & -3.045 ( 0.003 ) &    63.849 &   9.368 ( 0.000 ) &    66.450 &    69.547 \\ 
  Authentic &    26.193 &    19.559 &  4.953 ( 0.000 ) &    42.021 & -16.601 ( 0.000 ) &    37.226 &    22.447 \\ 
  Emotional Tone &    52.089 &    36.542 &  6.038 ( 0.000 ) &    71.868 & -13.872 ( 0.000 ) &    71.198 &    47.859 \\ 
  Words $>$ 6 letters &    24.126 &    25.478 & -2.738 ( 0.007 ) &    20.866 &   9.353 ( 0.000 ) &    22.279 &    24.552 \\ 
  Dictionary &    71.788 &    69.944 &  2.544 ( 0.012 ) &    75.331 &  -7.434 ( 0.000 ) &    73.861 &    71.562 \\ 
  Function Words &    39.104 &    37.479 &  2.530 ( 0.013 ) &    41.335 &  -6.015 ( 0.000 ) &    40.323 &    40.416 \\ 
  Pronoun &    10.148 &     9.133 &  3.049 ( 0.003 ) &    12.597 & -10.362 ( 0.000 ) &    11.381 &    10.825 \\ 
  Personal Pronouns &     6.554 &     5.773 &  3.433 ( 0.001 ) &     8.893 & -13.491 ( 0.000 ) &     7.773 &     6.908 \\ 
  1st Person Singluar &     2.642 &     1.866 &  4.758 ( 0.000 ) &     4.752 & -17.139 ( 0.000 ) &     3.814 &     2.709 \\ 
  1st Person Plural &     0.657 &     0.719 & -1.559 ( 0.122 ) &     0.670 &   1.198 ( 0.233 ) &     0.702 &     0.678 \\ 
  2nd Person &     1.756 &     1.520 &  2.541 ( 0.013 ) &     2.440 &  -9.749 ( 0.000 ) &     2.194 &     1.840 \\ 
  3rd Personl Singular &     0.918 &     1.043 & -2.381 ( 0.019 ) &     0.609 &   8.406 ( 0.000 ) &     0.612 &     1.037 \\ 
  3rd Person Plural &     0.582 &     0.625 & -1.373 ( 0.173 ) &     0.428 &   6.462 ( 0.000 ) &     0.457 &     0.644 \\ 
  Impersonal Pronouns &     3.589 &     3.357 &  1.788 ( 0.077 ) &     3.698 &  -2.648 ( 0.009 ) &     3.603 &     3.912 \\ 
  Articles &     5.197 &     4.783 &  3.252 ( 0.002 ) &     4.750 &   0.255 ( 0.800 ) &     5.059 &     5.295 \\ 
  Prepositions &    11.625 &    12.043 & -2.105 ( 0.038 ) &    10.712 &   6.719 ( 0.000 ) &    11.187 &    11.194 \\ 
  Auxiliary Verbs &     5.715 &     5.499 &  1.341 ( 0.183 ) &     6.125 &  -3.908 ( 0.000 ) &     5.872 &     6.122 \\ 
  Adverbs &     3.720 &     3.377 &  3.027 ( 0.003 ) &     4.424 &  -9.225 ( 0.000 ) &     4.134 &     3.985 \\ 
  Conjunctions &     3.516 &     3.349 &  1.653 ( 0.101 ) &     3.814 &  -4.597 ( 0.000 ) &     3.685 &     3.838 \\ 
  Negate &     0.895 &     0.907 & -0.307 ( 0.759 ) &     0.940 &  -0.866 ( 0.388 ) &     0.852 &     1.007 \\ 
  Verbs &    12.239 &    11.780 &  1.895 ( 0.061 ) &    13.520 &  -7.172 ( 0.000 ) &    12.860 &    12.620 \\ 
  Adjectives &     4.395 &     4.309 &  1.181 ( 0.240 ) &     4.517 &  -2.873 ( 0.005 ) &     4.556 &     4.155 \\ 
  Compare &     1.715 &     1.776 & -1.689 ( 0.094 ) &     1.702 &   2.024 ( 0.045 ) &     1.734 &     1.745 \\ 
  Interrogatives &     1.274 &     1.247 &  0.719 ( 0.474 ) &     1.288 &  -1.073 ( 0.286 ) &     1.224 &     1.356 \\ 
  Number &     0.680 &     0.708 & -1.110 ( 0.269 ) &     0.657 &   1.971 ( 0.051 ) &     0.671 &     0.677 \\ 
  Quantifiers &     1.565 &     1.474 &  2.104 ( 0.038 ) &     1.661 &  -4.264 ( 0.000 ) &     1.692 &     1.628 \\ 
  Affect &     6.453 &     6.439 &  0.086 ( 0.931 ) &     7.275 &  -5.161 ( 0.000 ) &     6.829 &     6.263 \\ 
  Positive Emotions &     3.968 &     3.463 &  3.098 ( 0.003 ) &     5.096 & -10.001 ( 0.000 ) &     4.860 &     3.729 \\ 
  Negative Emotions &     2.438 &     2.930 & -5.624 ( 0.000 ) &     2.128 &   9.083 ( 0.000 ) &     1.917 &     2.483 \\ 
  Anxiety &     0.282 &     0.359 & -4.016 ( 0.000 ) &     0.217 &   7.446 ( 0.000 ) &     0.209 &     0.283 \\ 
  Anger &     1.006 &     1.294 & -5.247 ( 0.000 ) &     0.852 &   7.942 ( 0.000 ) &     0.736 &     1.033 \\ 
  Sadness &     0.417 &     0.489 & -2.260 ( 0.026 ) &     0.435 &   1.684 ( 0.095 ) &     0.385 &     0.424 \\ 
  Social &     8.765 &     8.958 & -1.030 ( 0.306 ) &     9.088 &  -0.691 ( 0.491 ) &     8.765 &     8.978 \\ 
  Family &     0.404 &     0.369 &  1.488 ( 0.140 ) &     0.479 &  -4.456 ( 0.000 ) &     0.423 &     0.360 \\ 
  Friend &     0.326 &     0.323 &  0.098 ( 0.922 ) &     0.454 &  -5.772 ( 0.000 ) &     0.403 &     0.304 \\ 
  Female Words &     0.612 &     0.672 & -1.797 ( 0.075 ) &     0.580 &   2.707 ( 0.008 ) &     0.512 &     0.603 \\ 
  Male Words &     1.091 &     1.155 & -1.333 ( 0.185 ) &     0.893 &   5.484 ( 0.000 ) &     0.867 &     1.182 \\ 
  Cognitive Processes &     8.108 &     7.899 &  1.087 ( 0.280 ) &     8.342 &  -2.304 ( 0.023 ) &     8.146 &     8.753 \\ 
  Insight &     1.689 &     1.681 &  0.175 ( 0.861 ) &     1.619 &   1.448 ( 0.150 ) &     1.636 &     1.798 \\ 
  Cause &     1.341 &     1.307 &  0.999 ( 0.320 ) &     1.281 &   0.759 ( 0.450 ) &     1.308 &     1.395 \\ 
  Discrepancies &     1.254 &     1.198 &  1.452 ( 0.149 ) &     1.382 &  -4.749 ( 0.000 ) &     1.279 &     1.345 \\ 
  Tentative &     1.766 &     1.678 &  1.380 ( 0.170 ) &     1.886 &  -3.263 ( 0.002 ) &     1.865 &     1.975 \\ 
  Certain &     1.316 &     1.319 & -0.076 ( 0.940 ) &     1.355 &  -0.738 ( 0.462 ) &     1.292 &     1.367 \\ 
  Differentiation &     1.815 &     1.750 &  1.041 ( 0.300 ) &     1.873 &  -1.999 ( 0.048 ) &     1.817 &     2.065 \\ 
  Perceptional Processes &     2.750 &     2.506 &  3.135 ( 0.002 ) &     2.838 &  -4.273 ( 0.000 ) &     2.725 &     2.482 \\ 
  See &     1.409 &     1.275 &  2.515 ( 0.013 ) &     1.393 &  -2.204 ( 0.029 ) &     1.360 &     1.191 \\ 
  Hear &     0.803 &     0.782 &  0.738 ( 0.462 ) &     0.746 &   1.231 ( 0.221 ) &     0.722 &     0.786 \\ 
  Feel &     0.404 &     0.342 &  3.286 ( 0.001 ) &     0.530 & -10.116 ( 0.000 ) &     0.476 &     0.380 \\ 
  Bio &     2.014 &     1.939 &  0.880 ( 0.381 ) &     2.616 &  -7.644 ( 0.000 ) &     2.356 &     1.799 \\ 
  Body &     0.561 &     0.525 &  1.355 ( 0.178 ) &     0.764 &  -8.522 ( 0.000 ) &     0.636 &     0.517 \\ 
  Health &     0.591 &     0.624 & -0.773 ( 0.441 ) &     0.592 &   0.740 ( 0.461 ) &     0.573 &     0.530 \\ 
  Sexual &     0.221 &     0.312 & -3.942 ( 0.000 ) &     0.232 &   3.398 ( 0.001 ) &     0.194 &     0.229 \\ 
  Ingestion &     0.559 &     0.457 &  2.691 ( 0.008 ) &     0.763 &  -7.648 ( 0.000 ) &     0.762 &     0.452 \\ 
  Drives &     7.873 &     8.523 & -3.886 ( 0.000 ) &     7.616 &   5.386 ( 0.000 ) &     7.676 &     7.575 \\ 
  Affiliation &     2.142 &     2.165 & -0.260 ( 0.795 ) &     2.553 &  -4.309 ( 0.000 ) &     2.490 &     2.026 \\ 
  Achieve &     1.363 &     1.267 &  3.379 ( 0.001 ) &     1.510 &  -7.927 ( 0.000 ) &     1.574 &     1.283 \\ 
  Power &     3.305 &     4.053 & -5.395 ( 0.000 ) &     2.311 &  12.731 ( 0.000 ) &     2.468 &     3.180 \\ 
  Reward &     1.440 &     1.260 &  3.501 ( 0.001 ) &     1.905 & -12.620 ( 0.000 ) &     1.828 &     1.353 \\ 
  Risk &     0.610 &     0.723 & -3.325 ( 0.001 ) &     0.438 &   8.494 ( 0.000 ) &     0.452 &     0.619 \\ 
  Past Focus &     2.466 &     2.449 &  0.245 ( 0.807 ) &     2.463 &  -0.196 ( 0.845 ) &     2.380 &     2.579 \\ 
  Present Focus &     8.397 &     8.000 &  2.026 ( 0.045 ) &     9.756 &  -8.949 ( 0.000 ) &     9.233 &     8.574 \\ 
  Future Focus &     1.020 &     0.928 &  3.601 ( 0.000 ) &     1.312 & -14.412 ( 0.000 ) &     1.199 &     0.980 \\ 
  Relativity &    12.653 &    12.351 &  1.318 ( 0.190 ) &    13.325 &  -4.279 ( 0.000 ) &    13.381 &    11.645 \\ 
  Motion &     1.697 &     1.671 &  0.498 ( 0.619 ) &     1.963 &  -5.520 ( 0.000 ) &     1.899 &     1.552 \\ 
  Space &     6.497 &     6.683 & -1.015 ( 0.313 ) &     6.002 &   3.757 ( 0.000 ) &     6.312 &     6.005 \\ 
  Time &     4.585 &     4.142 &  4.351 ( 0.000 ) &     5.566 & -14.028 ( 0.000 ) &     5.355 &     4.200 \\ 
  Work &     2.681 &     3.084 & -3.457 ( 0.001 ) &     2.082 &   8.505 ( 0.000 ) &     2.430 &     2.611 \\ 
  Leisure &     2.034 &     1.651 &  3.451 ( 0.001 ) &     2.208 &  -5.109 ( 0.000 ) &     2.165 &     1.595 \\ 
  Home &     0.398 &     0.379 &  0.956 ( 0.341 ) &     0.467 &  -3.827 ( 0.000 ) &     0.460 &     0.349 \\ 
  Money &     0.974 &     1.020 & -1.022 ( 0.309 ) &     0.748 &   6.009 ( 0.000 ) &     0.917 &     0.920 \\ 
  Religion &     0.524 &     0.600 & -0.896 ( 0.373 ) &     0.446 &   1.807 ( 0.074 ) &     0.431 &     0.414 \\ 
  Death &     0.335 &     0.453 & -3.549 ( 0.001 ) &     0.181 &   8.281 ( 0.000 ) &     0.179 &     0.279 \\ 
  Informal &     1.908 &     1.671 &  2.176 ( 0.032 ) &     3.262 & -13.479 ( 0.000 ) &     2.636 &     1.865 \\ 
  Swear &     0.390 &     0.434 & -0.871 ( 0.386 ) &     0.683 &  -4.608 ( 0.000 ) &     0.509 &     0.403 \\ 
  Netspeak &     0.955 &     0.836 &  1.682 ( 0.095 ) &     1.865 & -13.123 ( 0.000 ) &     1.453 &     0.853 \\ 
  Assent &     0.388 &     0.286 &  4.737 ( 0.000 ) &     0.571 & -12.890 ( 0.000 ) &     0.540 &     0.400 \\ 
  Nonfluencies &     0.219 &     0.169 &  3.631 ( 0.000 ) &     0.281 &  -8.054 ( 0.000 ) &     0.262 &     0.254 \\ 
  Filler &     0.020 &     0.022 & -0.641 ( 0.523 ) &     0.039 &  -5.529 ( 0.000 ) &     0.031 &     0.023 \\ 
  Openness &     0.619 &     0.641 & -3.544 ( 0.001 ) &     0.554 &  13.831 ( 0.000 ) &     0.570 &     0.623 \\ 
  Conscientuousness &     0.520 &     0.516 &  0.560 ( 0.577 ) &     0.539 &  -3.233 ( 0.002 ) &     0.558 &     0.512 \\ 
  Extraversion &     0.415 &     0.398 &  2.407 ( 0.018 ) &     0.498 & -13.558 ( 0.000 ) &     0.475 &     0.396 \\ 
  Agreeableness &     0.445 &     0.399 &  6.043 ( 0.000 ) &     0.501 & -13.144 ( 0.000 ) &     0.502 &     0.426 \\ 
  Neuroticism &     0.420 &     0.442 & -3.205 ( 0.002 ) &     0.410 &   4.607 ( 0.000 ) &     0.392 &     0.439 \\ 
  Negative Low Arousal &     0.379 &     0.375 &  0.193 ( 0.847 ) &     0.430 &  -2.659 ( 0.009 ) &     0.382 &     0.394 \\ 
  Negative High Arousal &     0.865 &     1.012 & -3.209 ( 0.002 ) &     0.859 &   3.249 ( 0.002 ) &     0.735 &     0.899 \\ 
  Positive Low Arousal &     1.480 &     1.239 &  4.551 ( 0.000 ) &     2.055 & -15.265 ( 0.000 ) &     1.861 &     1.422 \\ 
  Positive High Arousal &     0.833 &     0.645 &  3.826 ( 0.000 ) &     1.297 & -13.018 ( 0.000 ) &     1.164 &     0.711 \\ 
   \hline
\end{longtable}
\end{center}

\end{singlespace*} 
\setcounter{table}{0}
\setcounter{figure}{0}
\renewcommand{\thetable}{WE\arabic{table}}
\renewcommand{\thefigure}{WE-\arabic{figure}}

\section*{Web Appendix E - Exploratory Analyses: Snopes All Variables}
\begin{singlespace*}

\ \\

The tables on the following pages show the comparison of Fake-Sharers, the Random and Matched Sample as well as the Fact-Check Sharers. For factor variables the mean column indicates the proportion, the column of the t-value indicates the z-value and the columns of Cohen's D indicates Cohen's H.

\begin{landscape} 
\begin{scriptsize} 
 \begin{longtable}[c]{ l  c c  c c  c c  c c  c c  c c  c}
\caption{Snopes Positive Effect Sizes}\\
 \hline
 & {\begin{tabular}[c]{@{}c@{}}Fake-news\\ Sharers\end{tabular}} & \multicolumn{4}{c}{Random Sample} & \multicolumn{4}{c}{Matched Sample} & \multicolumn{4}{c}{Fact-Check Sharers}\\
 \hline\hline
Predictor	& Mean & 	Mean	& D	& t	& p & Mean	& D	& t	& p & Mean	& D	& t	& p\\
 \hline
Power & 3.345 & 2.311 & 0.940 & 27.767 & 0.000 & 2.468 & 0.771 & 22.658 & 0.000 & 3.180 & 0.134 & 3.087 & 0.002 \\ 
  Number Fake Outlets Followed & 2.010 & 0.238 & 0.802 & 23.324 & 0.000 & 0.506 & 0.641 & 18.835 & 0.000 & 1.148 & 0.309 & 8.335 & 0.000 \\ 
  Death & 0.341 & 0.181 & 0.762 & 22.384 & 0.000 & 0.179 & 0.762 & 22.402 & 0.000 & 0.279 & 0.254 & 6.110 & 0.000 \\ 
  Tweets Per Day & 10.547 & 2.452 & 0.735 & 21.385 & 0.000 & 5.176 & 0.447 & 13.141 & 0.000 & 11.678 & 0.073 & -1.566 & 0.118 \\ 
  Analytic & 80.682 & 69.411 & 0.722 & 21.670 & 0.000 & 75.493 & 0.364 & 10.694 & 0.000 & 76.995 & 0.290 & 6.418 & 0.000 \\ 
  Openness & 0.620 & 0.554 & 0.721 & 21.606 & 0.000 & 0.570 & 0.548 & 16.106 & 0.000 & 0.623 & 0.029 & -0.685 & 0.494 \\ 
  Status Count & 30071.969 & 7239.498 & 0.717 & 20.875 & 0.000 & 17264.187 & 0.361 & 10.617 & 0.000 & 33551.738 & 0.078 & -1.652 & 0.099 \\ 
  Wikileaks & 0.277 & 0.051 & 0.656 & 335.828 & 0.000 & 0.086 & 0.513 & 210.222 & 0.000 & 0.174 & 0.248 & 26.897 & 0.000 \\ 
  Drudge Report & 0.216 & 0.024 & 0.654 & 312.398 & 0.000 & 0.064 & 0.454 & 163.346 & 0.000 & 0.094 & 0.344 & 47.606 & 0.000 \\ 
  Risk & 0.616 & 0.438 & 0.653 & 19.315 & 0.000 & 0.452 & 0.593 & 17.416 & 0.000 & 0.619 & 0.009 & -0.221 & 0.825 \\ 
  Words $>$ 6 letters & 24.198 & 20.866 & 0.647 & 19.288 & 0.000 & 22.279 & 0.383 & 11.251 & 0.000 & 24.552 & 0.073 & -1.725 & 0.085 \\ 
  3rd Person Singular & 0.924 & 0.609 & 0.629 & 18.605 & 0.000 & 0.612 & 0.607 & 17.823 & 0.000 & 1.037 & 0.199 & -4.310 & 0.000 \\ 
  3rd Person Plural & 0.584 & 0.428 & 0.548 & 16.155 & 0.000 & 0.457 & 0.428 & 12.581 & 0.000 & 0.644 & 0.185 & -4.269 & 0.000 \\ 
  Fox News & 0.276 & 0.083 & 0.522 & 225.767 & 0.000 & 0.142 & 0.334 & 93.115 & 0.000 & 0.158 & 0.289 & 36.068 & 0.000 \\ 
  Anxiety & 0.286 & 0.217 & 0.506 & 15.001 & 0.000 & 0.209 & 0.571 & 16.784 & 0.000 & 0.283 & 0.022 & 0.534 & 0.593 \\ 
  Clout & 70.011 & 63.849 & 0.489 & 14.700 & 0.000 & 66.450 & 0.299 & 8.796 & 0.000 & 69.547 & 0.047 & 1.049 & 0.295 \\ 
  Dailybeast & 0.153 & 0.027 & 0.472 & 172.225 & 0.000 & 0.066 & 0.287 & 67.541 & 0.000 & 0.218 & 0.166 & 13.551 & 0.000 \\ 
  Salon & 0.111 & 0.014 & 0.440 & 142.540 & 0.000 & 0.043 & 0.260 & 54.476 & 0.000 & 0.173 & 0.178 & 15.888 & 0.000 \\ 
  Number Fact Outlets Followed & 0.241 & 0.043 & 0.439 & 12.819 & 0.000 & 0.107 & 0.267 & 7.856 & 0.000 & 0.492 & 0.357 & -6.579 & 0.000 \\ 
  Work & 2.702 & 2.082 & 0.430 & 12.847 & 0.000 & 2.430 & 0.176 & 5.161 & 0.000 & 2.611 & 0.073 & 1.745 & 0.081 \\ 
  Washington Post & 0.259 & 0.098 & 0.429 & 155.917 & 0.000 & 0.169 & 0.219 & 40.607 & 0.000 & 0.365 & 0.230 & 26.081 & 0.000 \\ 
  Prepositions & 11.648 & 10.712 & 0.427 & 12.721 & 0.000 & 11.187 & 0.213 & 6.246 & 0.000 & 11.194 & 0.217 & 4.988 & 0.000 \\ 
  GOP & 0.340 & 0.161 & 0.419 & 151.088 & 0.000 & 0.305 & 0.075 & 4.672 & 0.031 & 0.159 & 0.423 & 75.626 & 0.000 \\ 
  USA Today & 0.170 & 0.048 & 0.406 & 134.744 & 0.000 & 0.089 & 0.244 & 49.563 & 0.000 & 0.182 & 0.032 & 0.418 & 0.518 \\ 
  Age & 32.596 & 30.417 & 0.387 & 11.617 & 0.000 & 32.950 & 0.064 & -1.866 & 0.062 & 32.188 & 0.093 & 2.206 & 0.028 \\ 
  Male Words & 1.094 & 0.893 & 0.384 & 11.430 & 0.000 & 0.867 & 0.429 & 12.596 & 0.000 & 1.182 & 0.167 & -3.559 & 0.000 \\ 
  Wall Street Journal & 0.247 & 0.105 & 0.379 & 122.624 & 0.000 & 0.172 & 0.185 & 28.896 & 0.000 & 0.249 & 0.004 & 0.001 & 0.976 \\ 
  Negative Emotions & 2.464 & 2.128 & 0.357 & 10.689 & 0.000 & 1.917 & 0.599 & 17.586 & 0.000 & 2.483 & 0.023 & -0.536 & 0.592 \\ 
  Friends Count & 2293.471 & 763.677 & 0.355 & 10.404 & 0.000 & 1852.286 & 0.060 & 1.771 & 0.077 & 2391.864 & 0.017 & -0.352 & 0.725 \\ 
  Economist & 0.164 & 0.059 & 0.345 & 99.781 & 0.000 & 0.107 & 0.170 & 24.094 & 0.000 & 0.177 & 0.034 & 0.487 & 0.485 \\ 
  New York Times & 0.316 & 0.169 & 0.345 & 103.105 & 0.000 & 0.274 & 0.091 & 6.907 & 0.009 & 0.419 & 0.215 & 22.338 & 0.000 \\ 
  Articles & 5.175 & 4.750 & 0.341 & 10.157 & 0.000 & 5.059 & 0.092 & 2.716 & 0.007 & 5.295 & 0.104 & -2.476 & 0.013 \\ 
  New Yorker & 0.198 & 0.087 & 0.324 & 89.602 & 0.000 & 0.129 & 0.187 & 29.304 & 0.000 & 0.268 & 0.167 & 13.585 & 0.000 \\ 
  LA Times & 0.111 & 0.031 & 0.322 & 84.429 & 0.000 & 0.062 & 0.175 & 25.197 & 0.000 & 0.151 & 0.119 & 6.835 & 0.009 \\ 
  CNN & 0.261 & 0.140 & 0.305 & 80.672 & 0.000 & 0.215 & 0.108 & 9.846 & 0.002 & 0.326 & 0.143 & 9.892 & 0.002 \\ 
  Money & 0.977 & 0.748 & 0.302 & 9.034 & 0.000 & 0.917 & 0.069 & 2.021 & 0.043 & 0.920 & 0.087 & 2.130 & 0.033 \\ 
  DEM \& GOP & 0.144 & 0.059 & 0.287 & 69.829 & 0.000 & 0.106 & 0.114 & 10.775 & 0.001 & 0.198 & 0.145 & 10.379 & 0.001 \\ 
  Aljazeera English & 0.080 & 0.021 & 0.285 & 65.157 & 0.000 & 0.048 & 0.136 & 15.091 & 0.000 & 0.142 & 0.196 & 19.794 & 0.000 \\ 
  Space & 6.507 & 6.002 & 0.284 & 8.428 & 0.000 & 6.312 & 0.107 & 3.133 & 0.002 & 6.005 & 0.275 & 6.243 & 0.000 \\ 
  Anger & 1.021 & 0.852 & 0.281 & 8.416 & 0.000 & 0.736 & 0.502 & 14.733 & 0.000 & 1.033 & 0.023 & -0.529 & 0.597 \\ 
  Word Count & 21280.361 & 18401.074 & 0.257 & 7.630 & 0.000 & 22328.980 & 0.090 & -2.647 & 0.008 & 23477.387 & 0.188 & -4.088 & 0.000 \\ 
  Buzz Feed & 0.131 & 0.064 & 0.230 & 45.064 & 0.000 & 0.104 & 0.084 & 5.859 & 0.015 & 0.152 & 0.061 & 1.656 & 0.198 \\ 
  Drives & 7.907 & 7.616 & 0.179 & 5.338 & 0.000 & 7.676 & 0.140 & 4.121 & 0.000 & 7.575 & 0.227 & 5.174 & 0.000 \\ 
  DEM & 0.169 & 0.113 & 0.163 & 22.964 & 0.000 & 0.199 & 0.077 & 4.896 & 0.027 & 0.329 & 0.375 & 72.873 & 0.000 \\ 
  Gender (Male) & 0.633 & 0.557 & 0.156 & 21.070 & 0.000 & 0.651 & 0.038 & 1.156 & 0.282 & 0.590 & 0.088 & 3.569 & 0.059 \\ 
  Cause & 1.340 & 1.281 & 0.136 & 4.077 & 0.000 & 1.308 & 0.071 & 2.100 & 0.036 & 1.395 & 0.152 & -3.550 & 0.000 \\ 
  Insight & 1.688 & 1.619 & 0.128 & 3.821 & 0.000 & 1.636 & 0.093 & 2.721 & 0.007 & 1.798 & 0.220 & -4.974 & 0.000 \\ 
  Hear & 0.802 & 0.746 & 0.126 & 3.747 & 0.000 & 0.722 & 0.178 & 5.226 & 0.000 & 0.786 & 0.040 & 1.052 & 0.293 \\ 
  Neuroticism & 0.421 & 0.410 & 0.125 & 3.758 & 0.000 & 0.392 & 0.337 & 9.889 & 0.000 & 0.439 & 0.238 & -5.606 & 0.000 \\ 
  Religion & 0.528 & 0.446 & 0.117 & 3.475 & 0.001 & 0.431 & 0.138 & 4.048 & 0.000 & 0.414 & 0.180 & 4.950 & 0.000 \\ 
  Followers Count & 4449.825 & 1708.484 & 0.098 & 2.880 & 0.004 & 3432.604 & 0.037 & 1.092 & 0.275 & 10196.757 & 0.103 & -1.637 & 0.102 \\ 
  Female Words & 0.615 & 0.580 & 0.087 & 2.595 & 0.009 & 0.512 & 0.261 & 7.663 & 0.000 & 0.603 & 0.032 & 0.697 & 0.486 \\ 
  Number & 0.682 & 0.657 & 0.076 & 2.280 & 0.023 & 0.671 & 0.036 & 1.058 & 0.290 & 0.677 & 0.021 & 0.452 & 0.651 \\    \hline
\end{longtable}
\end{scriptsize} 

\pagebreak

\vspace{-4mm}
\begin{scriptsize} 
 \begin{longtable}[c]{ l  c c  c c  c c  c c  c c  c c  c}
  \caption{Snopes Negative Effect Sizes}\\
   \hline
 & {\begin{tabular}[c]{@{}c@{}}Fake-news\\ Sharers\end{tabular}} & \multicolumn{4}{c}{Random Sample} & \multicolumn{4}{c}{Matched Sample} & \multicolumn{4}{c}{Fact-Check Sharers}\\

 \hline\hline
Predictor	& Mean & 	Mean	& D	& t	& p & Mean	& D	& t	& p & Mean	& D	& t	& p\\
 \hline
1st Person Singular & 2.600 & 4.752 & 0.959 & -28.824 & 0.000 & 3.814 & 0.595 & -17.473 & 0.000 & 2.709 & 0.062 & -1.415 & 0.157 \\ 
  Authentic & 25.840 & 42.021 & 0.901 & -26.927 & 0.000 & 37.226 & 0.654 & -19.214 & 0.000 & 22.447 & 0.211 & 4.792 & 0.000 \\ 
  Personal Pronouns & 6.512 & 8.893 & 0.866 & -25.962 & 0.000 & 7.773 & 0.489 & -14.373 & 0.000 & 6.908 & 0.172 & -3.946 & 0.000 \\ 
  Positive Low Arousal & 1.467 & 2.055 & 0.841 & -25.137 & 0.000 & 1.861 & 0.561 & -16.477 & 0.000 & 1.422 & 0.073 & 1.726 & 0.085 \\ 
  Emotional Tone & 51.260 & 71.868 & 0.827 & -24.507 & 0.000 & 71.198 & 0.794 & -23.320 & 0.000 & 47.859 & 0.131 & 3.019 & 0.003 \\ 
  Extraversion & 0.414 & 0.498 & 0.814 & -24.430 & 0.000 & 0.475 & 0.623 & -18.307 & 0.000 & 0.396 & 0.223 & 5.420 & 0.000 \\ 
  Positive High Arousal & 0.823 & 1.297 & 0.777 & -23.271 & 0.000 & 1.164 & 0.597 & -17.536 & 0.000 & 0.711 & 0.226 & 5.555 & 0.000 \\ 
  Positive Emotions & 3.941 & 5.096 & 0.750 & -22.367 & 0.000 & 4.860 & 0.583 & -17.133 & 0.000 & 3.729 & 0.151 & 3.562 & 0.000 \\ 
  Reward & 1.430 & 1.905 & 0.748 & -22.276 & 0.000 & 1.828 & 0.598 & -17.567 & 0.000 & 1.353 & 0.131 & 3.242 & 0.001 \\ 
  Future Focus & 1.015 & 1.312 & 0.740 & -22.238 & 0.000 & 1.199 & 0.453 & -13.301 & 0.000 & 0.980 & 0.115 & 2.773 & 0.006 \\ 
  Pronoun & 10.094 & 12.597 & 0.723 & -21.592 & 0.000 & 11.381 & 0.384 & -11.283 & 0.000 & 10.825 & 0.231 & -5.268 & 0.000 \\ 
  Informal & 1.895 & 3.262 & 0.712 & -21.492 & 0.000 & 2.636 & 0.446 & -13.097 & 0.000 & 1.865 & 0.025 & 0.617 & 0.537 \\ 
  Present Focus & 8.376 & 9.756 & 0.701 & -20.920 & 0.000 & 9.233 & 0.449 & -13.196 & 0.000 & 8.574 & 0.111 & -2.683 & 0.007 \\ 
  Netspeak & 0.949 & 1.865 & 0.691 & -20.919 & 0.000 & 1.453 & 0.458 & -13.461 & 0.000 & 0.853 & 0.128 & 3.161 & 0.002 \\ 
  Time & 4.562 & 5.566 & 0.688 & -20.498 & 0.000 & 5.355 & 0.537 & -15.762 & 0.000 & 4.200 & 0.269 & 6.568 & 0.000 \\ 
  No Party Followed & 0.347 & 0.668 & 0.652 & 362.243 & 0.000 & 0.390 & 0.089 & 6.584 & 0.010 & 0.313 & 0.073 & 2.413 & 0.120 \\ 
  2nd Person & 1.743 & 2.440 & 0.630 & -18.887 & 0.000 & 2.194 & 0.415 & -12.189 & 0.000 & 1.840 & 0.106 & -2.456 & 0.014 \\ 
  Agreeableness & 0.443 & 0.501 & 0.592 & -17.698 & 0.000 & 0.502 & 0.626 & -18.379 & 0.000 & 0.426 & 0.195 & 4.610 & 0.000 \\ 
  Adverbs & 3.702 & 4.424 & 0.579 & -17.282 & 0.000 & 4.134 & 0.352 & -10.341 & 0.000 & 3.985 & 0.246 & -5.769 & 0.000 \\ 
  Feel & 0.401 & 0.530 & 0.551 & -16.383 & 0.000 & 0.476 & 0.330 & -9.703 & 0.000 & 0.380 & 0.094 & 2.480 & 0.013 \\ 
  Dictionary & 71.690 & 75.331 & 0.548 & -16.325 & 0.000 & 73.861 & 0.326 & -9.590 & 0.000 & 71.562 & 0.020 & 0.478 & 0.633 \\ 
  Bio & 2.010 & 2.616 & 0.541 & -16.280 & 0.000 & 2.356 & 0.300 & -8.808 & 0.000 & 1.799 & 0.252 & 5.976 & 0.000 \\ 
  Verbs & 12.215 & 13.520 & 0.524 & -15.638 & 0.000 & 12.860 & 0.261 & -7.663 & 0.000 & 12.620 & 0.177 & -4.127 & 0.000 \\ 
  Affect & 6.452 & 7.275 & 0.503 & -15.050 & 0.000 & 6.829 & 0.233 & -6.847 & 0.000 & 6.263 & 0.136 & 3.187 & 0.001 \\ 
  Body & 0.559 & 0.764 & 0.497 & -14.995 & 0.000 & 0.636 & 0.203 & -5.952 & 0.000 & 0.517 & 0.154 & 3.576 & 0.000 \\ 
  Filler & 0.020 & 0.039 & 0.485 & -14.575 & 0.000 & 0.031 & 0.304 & -8.921 & 0.000 & 0.023 & 0.077 & -1.770 & 0.077 \\ 
  Motion & 1.695 & 1.963 & 0.457 & -13.666 & 0.000 & 1.899 & 0.334 & -9.802 & 0.000 & 1.552 & 0.280 & 6.542 & 0.000 \\ 
  Assent & 0.382 & 0.571 & 0.457 & -13.673 & 0.000 & 0.540 & 0.379 & -11.124 & 0.000 & 0.400 & 0.054 & -1.412 & 0.158 \\ 
  Friend & 0.325 & 0.454 & 0.422 & -12.698 & 0.000 & 0.403 & 0.290 & -8.522 & 0.000 & 0.304 & 0.097 & 2.335 & 0.020 \\ 
  Affiliation & 2.143 & 2.553 & 0.422 & -12.701 & 0.000 & 2.490 & 0.371 & -10.884 & 0.000 & 2.026 & 0.160 & 3.747 & 0.000 \\ 
  Swear & 0.392 & 0.683 & 0.419 & -12.662 & 0.000 & 0.509 & 0.191 & -5.598 & 0.000 & 0.403 & 0.026 & -0.580 & 0.562 \\ 
  Function Words & 39.017 & 41.335 & 0.402 & -11.977 & 0.000 & 40.323 & 0.226 & -6.627 & 0.000 & 40.416 & 0.250 & -5.787 & 0.000 \\ 
  Nonfluencies & 0.216 & 0.281 & 0.342 & -10.224 & 0.000 & 0.262 & 0.230 & -6.747 & 0.000 & 0.254 & 0.173 & -3.042 & 0.002 \\ 
  Ingestion & 0.554 & 0.763 & 0.325 & -9.779 & 0.000 & 0.762 & 0.288 & -8.457 & 0.000 & 0.452 & 0.225 & 5.889 & 0.000 \\ 
  Conjunctions & 3.507 & 3.814 & 0.293 & -8.735 & 0.000 & 3.685 & 0.170 & -4.999 & 0.000 & 3.838 & 0.325 & -7.401 & 0.000 \\ 
  Discrepancies & 1.251 & 1.382 & 0.285 & -8.517 & 0.000 & 1.279 & 0.064 & -1.867 & 0.062 & 1.345 & 0.221 & -4.885 & 0.000 \\ 
  Negative Low Arousal & 0.379 & 0.430 & 0.284 & -8.524 & 0.000 & 0.382 & 0.018 & -0.523 & 0.601 & 0.394 & 0.096 & -2.031 & 0.043 \\ 
  Auxiliary Verbs & 5.703 & 6.125 & 0.272 & -8.100 & 0.000 & 5.872 & 0.109 & -3.188 & 0.001 & 6.122 & 0.277 & -6.376 & 0.000 \\ 
  Relativity & 12.637 & 13.325 & 0.262 & -7.803 & 0.000 & 13.381 & 0.279 & -8.196 & 0.000 & 11.645 & 0.379 & 8.495 & 0.000 \\ 
  Achieve & 1.358 & 1.510 & 0.246 & -7.400 & 0.000 & 1.574 & 0.341 & -10.016 & 0.000 & 1.283 & 0.161 & 3.865 & 0.000 \\ 
  Conscientiousness & 0.520 & 0.539 & 0.235 & -7.048 & 0.000 & 0.558 & 0.471 & -13.823 & 0.000 & 0.512 & 0.114 & 2.660 & 0.008 \\ 
  Family & 0.402 & 0.479 & 0.220 & -6.613 & 0.000 & 0.423 & 0.055 & -1.629 & 0.103 & 0.360 & 0.159 & 3.747 & 0.000 \\ 
  Tentative & 1.761 & 1.886 & 0.197 & -5.859 & 0.000 & 1.865 & 0.162 & -4.773 & 0.000 & 1.975 & 0.339 & -7.566 & 0.000 \\ 
  Quantifiers & 1.560 & 1.661 & 0.170 & -5.077 & 0.000 & 1.692 & 0.216 & -6.338 & 0.000 & 1.628 & 0.136 & -3.349 & 0.001 \\ 
  Home & 0.397 & 0.467 & 0.158 & -4.795 & 0.000 & 0.460 & 0.134 & -3.949 & 0.000 & 0.349 & 0.196 & 4.649 & 0.000 \\ 
  Social & 8.775 & 9.088 & 0.145 & -4.342 & 0.000 & 8.765 & 0.005 & 0.140 & 0.889 & 8.978 & 0.107 & -2.498 & 0.013 \\ 
  Adjectives & 4.390 & 4.517 & 0.132 & -3.933 & 0.000 & 4.556 & 0.163 & -4.784 & 0.000 & 4.155 & 0.267 & 6.734 & 0.000 \\ 
  Leisure & 2.013 & 2.208 & 0.131 & -3.877 & 0.000 & 2.165 & 0.107 & -3.143 & 0.002 & 1.595 & 0.291 & 7.492 & 0.000 \\ 
  Cognitive Processes & 8.096 & 8.342 & 0.121 & -3.603 & 0.000 & 8.146 & 0.024 & -0.707 & 0.480 & 8.753 & 0.343 & -7.708 & 0.000 \\ 
  Impersonal Pronouns & 3.576 & 3.698 & 0.110 & -3.272 & 0.001 & 3.603 & 0.024 & -0.693 & 0.488 & 3.912 & 0.290 & -6.436 & 0.000 \\ 
  Negate & 0.896 & 0.940 & 0.109 & -3.258 & 0.001 & 0.852 & 0.113 & 3.325 & 0.001 & 1.007 & 0.297 & -6.610 & 0.000 \\ 
  Perceptional Processes & 2.737 & 2.838 & 0.102 & -3.028 & 0.002 & 2.725 & 0.012 & 0.366 & 0.714 & 2.482 & 0.286 & 7.375 & 0.000 \\ 
  Differentiation & 1.811 & 1.873 & 0.093 & -2.784 & 0.005 & 1.817 & 0.008 & -0.228 & 0.819 & 2.065 & 0.388 & -8.622 & 0.000 \\ 
  Certain & 1.316 & 1.355 & 0.087 & -2.592 & 0.010 & 1.292 & 0.051 & 1.496 & 0.135 & 1.367 & 0.117 & -2.735 & 0.006 \\ 
  Sadness & 0.421 & 0.435 & 0.068 & -2.018 & 0.044 & 0.385 & 0.181 & 5.318 & 0.000 & 0.424 & 0.015 & -0.346 & 0.729 \\ 
\hline
\end{longtable}
\end{scriptsize}

\newpage
\vspace{-4mm}
\begin{scriptsize} 
 \begin{longtable}[c]{ l  c c  c c  c c  c c  c c  c c  c}
  \caption{Snopes Non Significant Effect Sizes}\\
   \hline
 & {Fake-news Sharers} & \multicolumn{4}{c}{Random Sample} & \multicolumn{4}{c}{Matched Sample} & \multicolumn{4}{c}{Fact-Check Sharers}\\

 \hline\hline
Predictor	& Mean & 	Mean	& D	& t	& p & Mean	& D	& t	& p & Mean	& D	& t	& p\\
 \hline
  Interrogatives & 1.273 & 1.288 & 0.035 & -1.047 & 0.295 & 1.224 & 0.117 & 3.425 & 0.001 & 1.356 & 0.217 & -4.995 & 0.000 \\ 
  Compare & 1.719 & 1.702 & 0.028 & 0.839 & 0.401 & 1.734 & 0.024 & -0.702 & 0.483 & 1.745 & 0.052 & -1.283 & 0.200 \\ 
  Negative High Arousal & 0.873 & 0.859 & 0.025 & 0.765 & 0.444 & 0.735 & 0.283 & 8.314 & 0.000 & 0.899 & 0.066 & -1.475 & 0.141 \\ 
  Sexual & 0.226 & 0.232 & 0.024 & -0.714 & 0.475 & 0.194 & 0.122 & 3.589 & 0.000 & 0.229 & 0.015 & -0.331 & 0.741 \\ 
  1st Person Plural & 0.660 & 0.670 & 0.021 & -0.624 & 0.533 & 0.702 & 0.086 & -2.536 & 0.011 & 0.678 & 0.049 & -1.124 & 0.261 \\ 
  See & 1.402 & 1.393 & 0.011 & 0.322 & 0.748 & 1.360 & 0.056 & 1.657 & 0.098 & 1.191 & 0.311 & 8.217 & 0.000 \\ 
  Past Focus & 2.465 & 2.463 & 0.003 & 0.094 & 0.925 & 2.380 & 0.107 & 3.153 & 0.002 & 2.579 & 0.158 & -3.640 & 0.000 \\ 
  Health & 0.593 & 0.592 & 0.002 & 0.047 & 0.963 & 0.573 & 0.040 & 1.165 & 0.244 & 0.530 & 0.156 & 3.432 & 0.001 \\ 
\hline
\end{longtable}
\end{scriptsize} 
\end{landscape}
\end{singlespace*}
\setcounter{table}{0}
\setcounter{figure}{0}
\renewcommand{\thetable}{WF\arabic{table}}
\renewcommand{\thefigure}{WF-\arabic{figure}}
\setlength{\baselineskip}{21pt plus.2pt}

\section*{Web Appendix F - Exploratory Analyses: Fact-Check Sharers, Left/Right-Leaning Media Sharers}

\begin{singlespace*}

\ \\

\noindent \emph{Distinguishing Fake-News Sharers from Fact-Check Sharers.}

The comparison of fake-news sharers and fact-check sharers shows consistent effects across both samples (Snopes and Hoaxy; see Figure \ref{Figure_Hoaxy_CohensD_Pos} and \ref{Figure_Hoaxy_CohensD_Neg} for Hoaxy). Notably, we find that the variables distinguishing fake-news sharers from the random or matched sample do not necessarily distinguish them well from fact-check sharers. Fake-news sharers and fact-check sharers have important commonalities in terms of their social media activity and certain textual cues. For example, both groups are very active on social media platforms. Moreover, the two groups show similarly high levels of negative emotions, particularly anger.

However, we also find variables that distinguish between the two groups well in terms of socio-demographics, political affiliation, and textual cues. As previously mentioned, fake-news sharers follow a considerably larger number of fake-news outlets, confirming the face validity of our approach to identifying fake-news and fact-check sharers. Compared to fact-check sharers, fake-news sharers are more conservative, as seen in higher exclusive followership of Republican accounts, and also of conservative media outlets such as Fox News.

We see traces in language that also distinguish fake-news sharers from fact-check sharers. Mentions of existentially based needs, such as death and religion, are considerably greater for fake-news sharers. Power, an important variable that distinguishes fake-news sharers from the random and the matched sample, is also considerably higher for fake-news sharers. Interestingly, the language of fake-news sharers is more analytical than that of fact-check sharers. Regarding differences in personality traits, we find that fake-news sharers are more agreeable and exhibit lower levels of neuroticism than fact-check sharers, although both groups have significantly higher levels of neuroticism compared to all other groups. \\

\noindent \emph{Distinguishing Fake-News Sharers from Left and Right-Leaning Media Sharers.}

Even though fake-news sharers and followers of right-leaning media outlets both tend to be conservative, they have notable differences. Compared to sharers of right-leaning media outlets, fake-news sharers have a greater exclusive followership of Republican accounts, and conservative media outlets such as Fox News. However, these differences are presumably largely driven by the greater social media activity of fake-news sharers as observed in the number of tweets per day and the number of status counts. To test the robustness of our results particularly given the large average follower count, we also test the robustness of our Hoaxy comparisons between the groups on a subset of users with less than 10,000 followers and find similar results.

In terms of emotions, we again find negative emotions and particularly the high arousal negative emotion of anger to be elevated for fake-news sharers compared to both right and left-leaning media outlet sharers. Additionally, mentions of existentially based needs such as death and religion are elevated for fake-news sharers compared to left and right-leaning media outlet sharers, similar to results presented when comparing fake-news sharers to the random and the matched sample of Twitter users. Fake-news sharers are more neurotic compared to both left and right-leaning media sharers. The results of all variables comparing fake-news sharers to the right-leaning, left-leaning, and fact-check sharers are displayed in the Tables below. 

Apart from the consistency of results across the two datasets, the Hoaxy sample allows us to gain insights into differences between fake-news sharers and other individuals with a similarly conservative-leaning political opinion through direct comparisons of fake-news sharers and right-leaning news sharers. This comparison enables us to go beyond a simple statement about the role of political ideology in misinformation. 

The figures and tables below show the comparison of Fake-Sharers, the right- and left-leaning media outlet sharer and the Fact-Check Sharers. In the tables, the mean column for factor variables indicates the proportion, the column of the t-value indicates the z-value and the columns of Cohen's D indicates Cohen's H.

\begin{figure}[htbp]
    \caption{Hoaxy dataset: Values that positively discriminate fake-news sharers}
    \includegraphics[width=\textwidth]{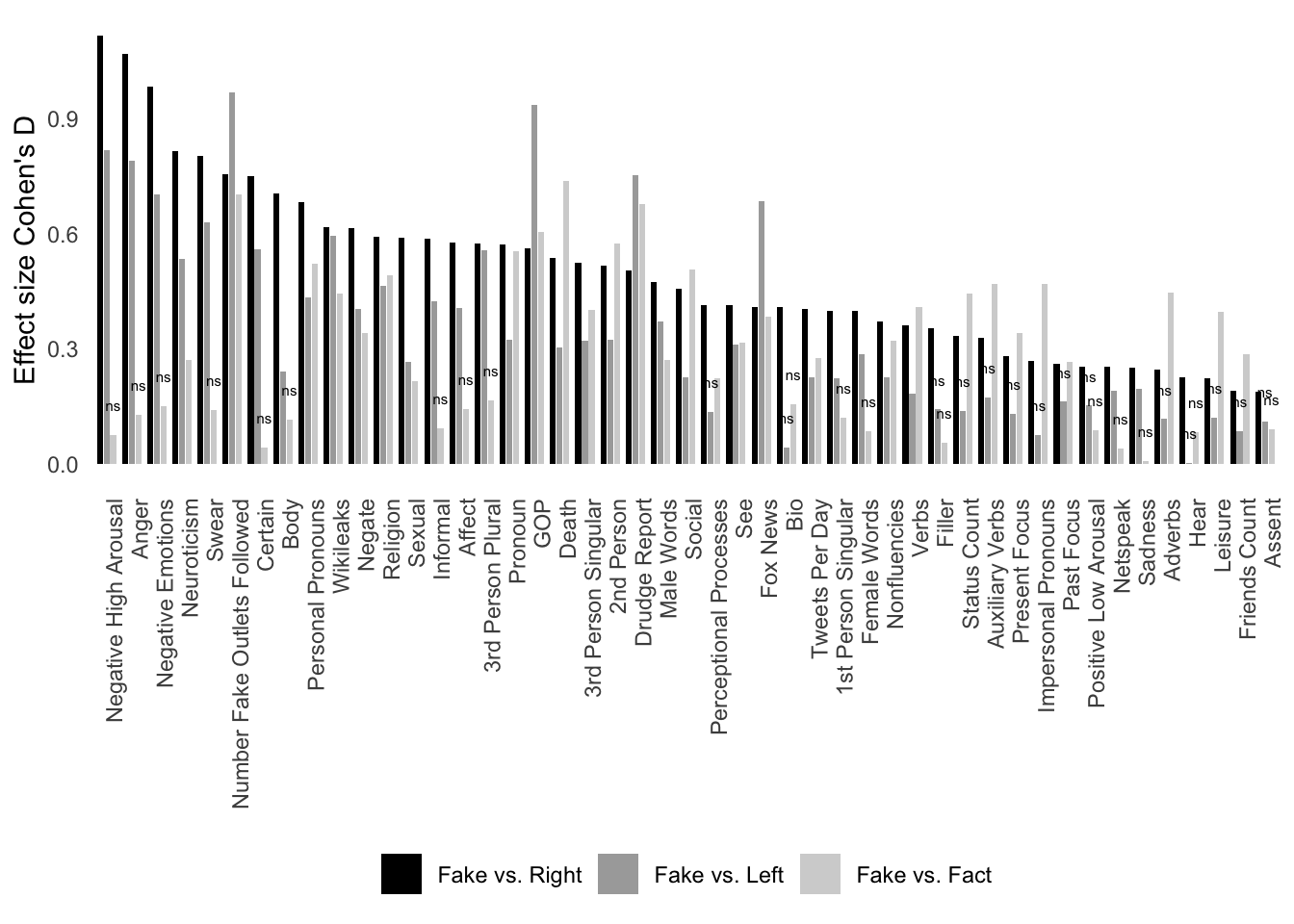}
    \label{Figure_Hoaxy_CohensD_Pos}
\end{figure}

\begin{figure}[htbp]
    \caption{Hoaxy dataset: Values that negatively discriminate fake-news sharers}
    \includegraphics[width=\textwidth]{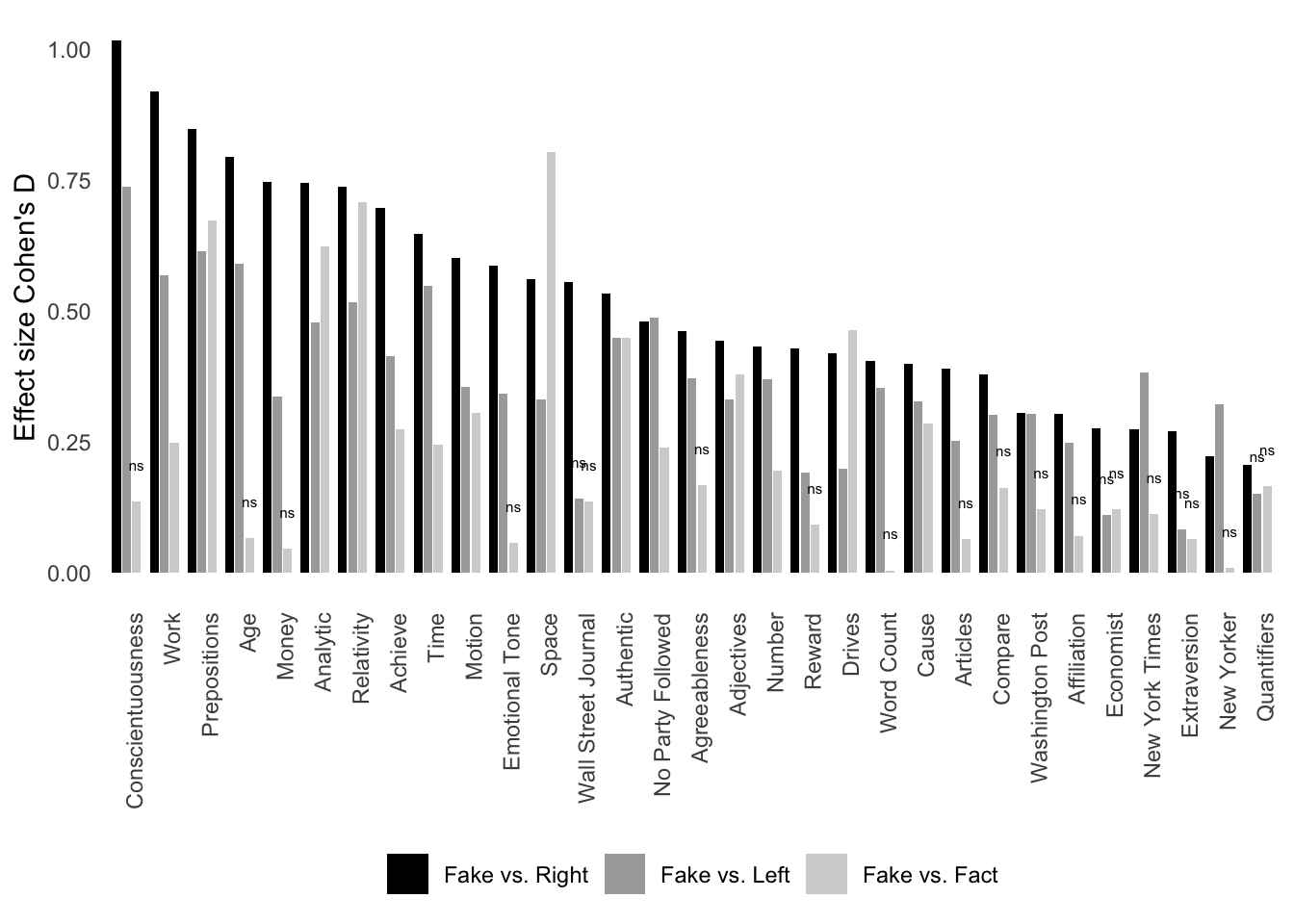}
    \label{Figure_Hoaxy_CohensD_Neg}
\end{figure}

\begin{landscape}
\vspace{-4mm}
\begin{scriptsize} 
 \begin{longtable}[c]{ l  c c  c c  c c  c c  c c  c c  c}
  \caption{Hoaxy Positive Effect Sizes}\\
   \hline
 & {F-N Sharers} & \multicolumn{4}{c}{Right-Leaning Media Sharers} & \multicolumn{4}{c}{Left-Leaning Media Sharers} & \multicolumn{4}{c}{Fact-Check Sharers}\\

 \hline\hline
Predictor	& Mean & 	Mean	& D	& t	& p & Mean	& D	& t	& p & Mean	& D	& t	& p\\
 \hline
Negative High Arousal & 0.991 & 0.575 & 1.117 & 11.821 & 0.000 & 0.679 & 0.818 & 9.160 & 0.000 & 1.025 & 0.075 & -0.890 & 0.374 \\ 
  Anger & 1.208 & 0.689 & 1.070 & 11.233 & 0.000 & 0.817 & 0.791 & 8.915 & 0.000 & 1.139 & 0.128 & 1.496 & 0.135 \\ 
  Negative Emotions & 2.746 & 1.956 & 0.983 & 10.234 & 0.000 & 2.171 & 0.703 & 7.980 & 0.000 & 2.613 & 0.151 & 1.779 & 0.076 \\ 
  Neuroticism & 0.446 & 0.386 & 0.815 & 8.407 & 0.000 & 0.408 & 0.535 & 6.084 & 0.000 & 0.464 & 0.271 & -3.094 & 0.002 \\ 
  Swear & 0.443 & 0.130 & 0.802 & 8.827 & 0.000 & 0.194 & 0.629 & 6.804 & 0.000 & 0.511 & 0.140 & -1.612 & 0.108 \\ 
  Number Fake Outlets Followed & 3.932 & 1.290 & 0.755 & 8.252 & 0.000 & 0.898 & 0.968 & 10.189 & 0.000 & 1.537 & 0.702 & 7.562 & 0.000 \\ 
  Certain & 1.459 & 1.113 & 0.750 & 7.929 & 0.000 & 1.196 & 0.560 & 6.272 & 0.000 & 1.478 & 0.043 & -0.485 & 0.628 \\ 
  Body & 0.483 & 0.329 & 0.706 & 7.431 & 0.000 & 0.424 & 0.240 & 2.747 & 0.006 & 0.513 & 0.115 & -1.363 & 0.173 \\ 
  Personal Pronouns & 6.412 & 4.958 & 0.683 & 7.225 & 0.000 & 5.465 & 0.435 & 4.873 & 0.000 & 7.489 & 0.522 & -5.851 & 0.000 \\ 
  Wikileaks & 0.421 & 0.150 & 0.617 & 36.826 & 0.000 & 0.158 & 0.594 & 43.008 & 0.000 & 0.217 & 0.443 & 25.416 & 0.000 \\ 
  Negate & 1.007 & 0.781 & 0.615 & 6.515 & 0.000 & 0.855 & 0.404 & 4.525 & 0.000 & 1.133 & 0.341 & -3.856 & 0.000 \\ 
  Religion & 0.752 & 0.275 & 0.593 & 6.611 & 0.000 & 0.367 & 0.464 & 4.991 & 0.000 & 0.373 & 0.493 & 5.157 & 0.000 \\ 
  Sexual & 0.249 & 0.123 & 0.590 & 6.291 & 0.000 & 0.180 & 0.266 & 3.055 & 0.002 & 0.204 & 0.216 & 2.390 & 0.017 \\ 
  Informal & 1.898 & 1.296 & 0.588 & 6.342 & 0.000 & 1.435 & 0.425 & 4.718 & 0.000 & 2.001 & 0.094 & -1.060 & 0.290 \\ 
  Affect & 6.477 & 5.772 & 0.578 & 6.029 & 0.000 & 5.921 & 0.407 & 4.679 & 0.000 & 6.291 & 0.143 & 1.676 & 0.094 \\ 
  3rd Person Plural & 0.689 & 0.506 & 0.574 & 6.171 & 0.000 & 0.512 & 0.558 & 6.121 & 0.000 & 0.747 & 0.166 & -1.892 & 0.059 \\ 
  Pronoun & 10.014 & 8.231 & 0.573 & 6.073 & 0.000 & 8.972 & 0.324 & 3.631 & 0.000 & 11.618 & 0.554 & -6.116 & 0.000 \\ 
  GOP & 0.536 & 0.265 & 0.562 & 31.698 & 0.000 & 0.120 & 0.937 & 102.641 & 0.000 & 0.246 & 0.605 & 47.222 & 0.000 \\ 
  Death & 0.360 & 0.246 & 0.538 & 5.614 & 0.000 & 0.294 & 0.304 & 3.451 & 0.001 & 0.221 & 0.739 & 8.235 & 0.000 \\ 
  3rd Person Singular & 0.972 & 0.709 & 0.525 & 5.434 & 0.000 & 0.806 & 0.322 & 3.683 & 0.000 & 1.191 & 0.401 & -4.772 & 0.000 \\ 
  2nd Person & 1.809 & 1.296 & 0.518 & 5.559 & 0.000 & 1.479 & 0.323 & 3.579 & 0.000 & 2.442 & 0.574 & -6.578 & 0.000 \\ 
  Drudge Report & 0.455 & 0.220 & 0.505 & 25.372 & 0.000 & 0.127 & 0.753 & 67.907 & 0.000 & 0.153 & 0.677 & 58.857 & 0.000 \\ 
  Male Words & 1.139 & 0.884 & 0.474 & 4.826 & 0.000 & 0.956 & 0.371 & 4.241 & 0.000 & 1.274 & 0.271 & -3.180 & 0.002 \\ 
  Social & 8.844 & 7.984 & 0.456 & 4.762 & 0.000 & 8.409 & 0.226 & 2.557 & 0.011 & 9.771 & 0.508 & -5.785 & 0.000 \\ 
  Perceptional Processes & 2.418 & 2.135 & 0.414 & 4.403 & 0.000 & 2.323 & 0.135 & 1.509 & 0.132 & 2.263 & 0.223 & 2.512 & 0.012 \\ 
  See & 1.213 & 0.995 & 0.413 & 4.479 & 0.000 & 1.056 & 0.312 & 3.365 & 0.001 & 1.042 & 0.317 & 3.499 & 0.001 \\ 
  Fox News & 0.477 & 0.280 & 0.409 & 16.797 & 0.000 & 0.165 & 0.686 & 57.113 & 0.000 & 0.291 & 0.385 & 19.093 & 0.000 \\ 
  Bio & 1.654 & 1.340 & 0.408 & 4.264 & 0.000 & 1.694 & 0.043 & -0.504 & 0.615 & 1.544 & 0.156 & 1.753 & 0.080 \\ 
  Tweets Per Day & 23.348 & 10.053 & 0.405 & 4.385 & 0.000 & 14.211 & 0.227 & 2.586 & 0.010 & 13.969 & 0.276 & 3.073 & 0.002 \\ 
  1st Person Singular & 2.229 & 1.670 & 0.399 & 4.260 & 0.000 & 1.901 & 0.224 & 2.499 & 0.013 & 2.404 & 0.120 & -1.359 & 0.175 \\ 
  Female Words & 0.555 & 0.411 & 0.398 & 3.991 & 0.000 & 0.465 & 0.287 & 3.322 & 0.001 & 0.583 & 0.086 & -1.041 & 0.299 \\ 
  Nonfluencies & 0.212 & 0.160 & 0.371 & 3.938 & 0.000 & 0.178 & 0.226 & 2.557 & 0.011 & 0.270 & 0.321 & -3.836 & 0.000 \\ 
  Verbs & 12.105 & 11.303 & 0.362 & 3.834 & 0.000 & 11.681 & 0.183 & 2.055 & 0.040 & 12.967 & 0.409 & -4.546 & 0.000 \\ 
  Filler & 0.022 & 0.013 & 0.354 & 3.810 & 0.000 & 0.018 & 0.142 & 1.591 & 0.112 & 0.024 & 0.056 & -0.649 & 0.517 \\ 
  Status Count & 60820.460 & 30681.875 & 0.334 & 3.561 & 0.000 & 43537.982 & 0.139 & 1.619 & 0.106 & 26349.895 & 0.445 & 4.726 & 0.000 \\ 
  Auxiliary Verbs & 5.874 & 5.362 & 0.329 & 3.483 & 0.001 & 5.595 & 0.173 & 1.942 & 0.053 & 6.561 & 0.470 & -5.218 & 0.000 \\ 
  Present Focus & 8.344 & 7.879 & 0.281 & 2.987 & 0.003 & 8.117 & 0.130 & 1.455 & 0.146 & 8.903 & 0.342 & -3.828 & 0.000 \\ 
  Impersonal Pronouns & 3.599 & 3.270 & 0.269 & 2.833 & 0.005 & 3.503 & 0.075 & 0.844 & 0.399 & 4.124 & 0.470 & -5.195 & 0.000 \\ 
  Past Focus & 2.382 & 2.194 & 0.262 & 2.733 & 0.007 & 2.260 & 0.162 & 1.847 & 0.065 & 2.566 & 0.265 & -3.023 & 0.003 \\ 
  Positive Low Arousal & 1.360 & 1.242 & 0.254 & 2.708 & 0.007 & 1.283 & 0.152 & 1.709 & 0.088 & 1.404 & 0.087 & -1.001 & 0.318 \\ 
  Netspeak & 0.891 & 0.728 & 0.253 & 2.738 & 0.006 & 0.752 & 0.191 & 2.140 & 0.033 & 0.864 & 0.041 & 0.458 & 0.647 \\ 
  Sadness & 0.417 & 0.375 & 0.251 & 2.593 & 0.010 & 0.390 & 0.196 & 2.169 & 0.031 & 0.415 & 0.008 & 0.100 & 0.920 \\ 
  Adverbs & 3.685 & 3.422 & 0.246 & 2.585 & 0.010 & 3.555 & 0.117 & 1.320 & 0.187 & 4.115 & 0.446 & -4.934 & 0.000 \\ 
  Hear & 0.771 & 0.702 & 0.226 & 2.374 & 0.018 & 0.772 & 0.002 & -0.021 & 0.983 & 0.798 & 0.083 & -0.957 & 0.339 \\ 
  Leisure & 1.680 & 1.407 & 0.224 & 2.436 & 0.015 & 1.524 & 0.121 & 1.338 & 0.182 & 1.192 & 0.396 & 4.359 & 0.000 \\ 
  Friends Count & 6304.357 & 3186.135 & 0.191 & 2.105 & 0.036 & 4656.158 & 0.086 & 0.961 & 0.337 & 2239.486 & 0.287 & 2.930 & 0.004 \\ 
  Assent & 0.386 & 0.327 & 0.189 & 2.028 & 0.043 & 0.350 & 0.111 & 1.236 & 0.217 & 0.413 & 0.090 & -0.998 & 0.319 \\ 
  \hline
\end{longtable}
\end{scriptsize}

\vspace{-4mm}
\begin{scriptsize} 
 \begin{longtable}[c]{ l  c c  c c  c c  c c  c c  c c  c}
  \caption{Hoaxy Negative Effect Sizes}\\
   \hline
 & {FN Sharers} & \multicolumn{4}{c}{Right-leaning Media Sharers} & \multicolumn{4}{c}{Left-leaning Media Sharers} & \multicolumn{4}{c}{Fact-Check Sharers}\\

 \hline\hline
Predictor	& Mean & 	Mean	& D	& t	& p & Mean	& D	& t	& p & Mean	& D	& t	& p\\
 \hline
Conscientuousness & 0.506 & 0.577 & 1.016 & -10.647 & 0.000 & 0.556 & 0.738 & -8.250 & 0.000 & 0.498 & 0.136 & 1.530 & 0.127 \\ 
  Work & 2.723 & 4.267 & 0.919 & -9.197 & 0.000 & 3.598 & 0.569 & -6.661 & 0.000 & 2.451 & 0.248 & 2.780 & 0.006 \\ 
  Prepositions & 11.227 & 13.095 & 0.847 & -8.863 & 0.000 & 12.654 & 0.614 & -6.982 & 0.000 & 9.813 & 0.672 & 7.610 & 0.000 \\ 
  Age & 32.702 & 35.415 & 0.794 & -8.187 & 0.000 & 34.767 & 0.590 & -6.762 & 0.000 & 32.501 & 0.066 & 0.747 & 0.456 \\ 
  Money & 0.882 & 1.662 & 0.747 & -7.311 & 0.000 & 1.148 & 0.337 & -4.010 & 0.000 & 0.906 & 0.047 & -0.535 & 0.593 \\ 
  Analytic & 79.036 & 87.827 & 0.744 & -7.929 & 0.000 & 84.855 & 0.478 & -5.330 & 0.000 & 71.438 & 0.623 & 7.049 & 0.000 \\ 
  Relativity & 11.514 & 13.481 & 0.738 & -7.579 & 0.000 & 12.970 & 0.516 & -5.971 & 0.000 & 9.856 & 0.708 & 8.078 & 0.000 \\ 
  Achieve & 1.312 & 1.742 & 0.696 & -7.071 & 0.000 & 1.601 & 0.414 & -4.878 & 0.000 & 1.181 & 0.274 & 3.094 & 0.002 \\ 
  Time & 3.881 & 4.606 & 0.647 & -6.609 & 0.000 & 4.468 & 0.548 & -6.287 & 0.000 & 3.616 & 0.244 & 2.880 & 0.004 \\ 
  Motion & 1.486 & 1.741 & 0.602 & -6.157 & 0.000 & 1.691 & 0.355 & -4.238 & 0.000 & 1.372 & 0.305 & 3.508 & 0.000 \\ 
  Emotional Tone & 43.212 & 57.129 & 0.587 & -6.127 & 0.000 & 51.775 & 0.343 & -3.910 & 0.000 & 44.570 & 0.058 & -0.669 & 0.504 \\ 
  Space & 6.252 & 7.303 & 0.561 & -5.874 & 0.000 & 6.932 & 0.331 & -3.787 & 0.000 & 4.968 & 0.803 & 8.780 & 0.000 \\ 
  Wall Street Journal & 0.289 & 0.560 & 0.555 & 31.519 & 0.000 & 0.356 & 0.142 & 2.279 & 0.131 & 0.230 & 0.136 & 2.182 & 0.140 \\ 
  Authentic & 18.867 & 25.754 & 0.534 & -5.532 & 0.000 & 25.096 & 0.449 & -5.172 & 0.000 & 13.535 & 0.449 & 5.112 & 0.000 \\ 
  No Party Followed & 0.162 & 0.370 & 0.480 & 23.443 & 0.000 & 0.373 & 0.487 & 27.658 & 0.000 & 0.259 & 0.240 & 6.882 & 0.009 \\ 
  Agreeableness & 0.413 & 0.449 & 0.461 & -4.828 & 0.000 & 0.442 & 0.372 & -4.169 & 0.000 & 0.400 & 0.168 & 1.877 & 0.061 \\ 
  Adjectives & 4.151 & 4.496 & 0.444 & -4.672 & 0.000 & 4.415 & 0.331 & -3.728 & 0.000 & 3.863 & 0.379 & 4.291 & 0.000 \\ 
  Number & 0.637 & 0.768 & 0.433 & -4.330 & 0.000 & 0.720 & 0.369 & -4.206 & 0.000 & 0.595 & 0.196 & 2.258 & 0.024 \\ 
  Reward & 1.285 & 1.483 & 0.428 & -4.394 & 0.000 & 1.375 & 0.191 & -2.195 & 0.029 & 1.248 & 0.092 & 1.044 & 0.297 \\ 
  Drives & 7.943 & 8.633 & 0.419 & -4.349 & 0.000 & 8.292 & 0.198 & -2.277 & 0.023 & 7.262 & 0.463 & 5.218 & 0.000 \\ 
  Word Count & 20634.089 & 25742.435 & 0.405 & -4.211 & 0.000 & 25258.479 & 0.354 & -4.040 & 0.000 & 20571.498 & 0.005 & 0.058 & 0.954 \\ 
  Cause & 1.320 & 1.496 & 0.399 & -4.117 & 0.000 & 1.460 & 0.327 & -3.718 & 0.000 & 1.428 & 0.286 & -3.214 & 0.001 \\ 
  Articles & 5.085 & 5.586 & 0.390 & -4.081 & 0.000 & 5.411 & 0.253 & -2.851 & 0.005 & 5.161 & 0.065 & -0.730 & 0.466 \\ 
  Compare & 1.739 & 1.943 & 0.379 & -4.034 & 0.000 & 1.894 & 0.302 & -3.316 & 0.001 & 1.661 & 0.163 & 1.761 & 0.079 \\ 
  Washington Post & 0.298 & 0.445 & 0.306 & 9.460 & 0.002 & 0.444 & 0.303 & 11.017 & 0.001 & 0.355 & 0.121 & 1.707 & 0.191 \\ 
  Affiliation & 1.977 & 2.238 & 0.304 & -3.117 & 0.002 & 2.180 & 0.248 & -2.833 & 0.005 & 1.926 & 0.071 & 0.804 & 0.422 \\ 
  Economist & 0.217 & 0.340 & 0.276 & 7.615 & 0.006 & 0.264 & 0.110 & 1.304 & 0.253 & 0.169 & 0.121 & 1.687 & 0.194 \\ 
  New York Times & 0.336 & 0.470 & 0.274 & 7.530 & 0.006 & 0.525 & 0.383 & 17.789 & 0.000 & 0.390 & 0.112 & 1.438 & 0.230 \\ 
  Extraversion & 0.402 & 0.422 & 0.270 & -2.794 & 0.005 & 0.408 & 0.083 & -0.939 & 0.348 & 0.397 & 0.065 & 0.741 & 0.459 \\ 
  New Yorker & 0.209 & 0.305 & 0.222 & 4.826 & 0.028 & 0.352 & 0.322 & 12.264 & 0.000 & 0.204 & 0.010 & 0.000 & 0.993 \\ 
  Quantifiers & 1.545 & 1.643 & 0.207 & -2.216 & 0.027 & 1.621 & 0.152 & -1.692 & 0.091 & 1.619 & 0.166 & -1.810 & 0.071 \\ 
\hline
\end{longtable}
\end{scriptsize}

\vspace{-4mm}
\begin{scriptsize} 
 \begin{longtable}[c]{ l  c c  c c  c c  c c  c c  c c  c}
  \caption{Hoaxy Non Significant Effect Sizes}\\
   \hline
 & {Fake-news Sharers} & \multicolumn{4}{c}{Right-leaning Media Sharers} & \multicolumn{4}{c}{Left-leaning Media Sharers} & \multicolumn{4}{c}{Fact-Check Sharers}\\

 \hline\hline
Predictor	& Mean & 	Mean	& D	& t	& p & Mean	& D	& t	& p & Mean	& D	& t	& p\\
 \hline
Openness & 0.626 & 0.612 & 0.189 & 1.956 & 0.051 & 0.633 & 0.103 & -1.167 & 0.244 & 0.611 & 0.239 & 2.697 & 0.007 \\ 
  Home & 0.329 & 0.377 & 0.181 & -1.878 & 0.061 & 0.363 & 0.149 & -1.646 & 0.101 & 0.278 & 0.220 & 2.470 & 0.014 \\ 
  Dictionary & 69.676 & 70.765 & 0.173 & -1.856 & 0.064 & 71.220 & 0.238 & -2.633 & 0.009 & 69.436 & 0.035 & 0.397 & 0.692 \\ 
  Interrogatives & 1.320 & 1.250 & 0.170 & 1.766 & 0.078 & 1.288 & 0.081 & 0.912 & 0.362 & 1.484 & 0.417 & -4.763 & 0.000 \\ 
  Followers Count & 20483.221 & 74520.130 & 0.164 & -1.574 & 0.117 & 96277.261 & 0.253 & -3.148 & 0.002 & 113227.994 & 0.086 & -1.156 & 0.249 \\ 
  Anxiety & 0.305 & 0.283 & 0.157 & 1.614 & 0.107 & 0.296 & 0.061 & 0.703 & 0.482 & 0.267 & 0.320 & 3.614 & 0.000 \\ 
  Discrepancies & 1.240 & 1.180 & 0.149 & 1.552 & 0.121 & 1.240 & 0.000 & -0.002 & 0.998 & 1.398 & 0.406 & -4.643 & 0.000 \\ 
  DEM \& GOP & 0.209 & 0.270 & 0.144 & 1.933 & 0.164 & 0.250 & 0.099 & 1.023 & 0.312 & 0.259 & 0.119 & 1.607 & 0.205 \\ 
  Negative Low Arousal & 0.369 & 0.348 & 0.142 & 1.462 & 0.145 & 0.356 & 0.100 & 1.112 & 0.267 & 0.408 & 0.197 & -2.433 & 0.015 \\ 
  1st Person Plural & 0.714 & 0.777 & 0.132 & -1.346 & 0.179 & 0.767 & 0.115 & -1.329 & 0.184 & 0.706 & 0.022 & 0.248 & 0.804 \\ 
  Aljazeera English & 0.149 & 0.110 & 0.116 & 1.117 & 0.291 & 0.158 & 0.026 & 0.031 & 0.860 & 0.077 & 0.231 & 6.562 & 0.010 \\ 
  Cognitive Processes & 8.159 & 7.963 & 0.099 & 1.033 & 0.302 & 8.277 & 0.059 & -0.667 & 0.505 & 8.959 & 0.446 & -5.035 & 0.000 \\ 
  Positive High Arousal & 0.701 & 0.745 & 0.099 & -1.038 & 0.300 & 0.726 & 0.046 & -0.537 & 0.592 & 0.676 & 0.058 & 0.654 & 0.513 \\ 
  Future Focus & 1.014 & 1.045 & 0.097 & -1.015 & 0.310 & 1.004 & 0.035 & 0.394 & 0.694 & 0.942 & 0.245 & 2.759 & 0.006 \\ 
  Buzz Feed & 0.132 & 0.165 & 0.093 & 0.697 & 0.404 & 0.194 & 0.168 & 3.114 & 0.078 & 0.147 & 0.043 & 0.143 & 0.706 \\ 
  Risk & 0.654 & 0.677 & 0.087 & -0.886 & 0.376 & 0.677 & 0.086 & -0.989 & 0.323 & 0.599 & 0.265 & 2.948 & 0.003 \\ 
  Family & 0.347 & 0.325 & 0.079 & 0.788 & 0.431 & 0.337 & 0.046 & 0.530 & 0.596 & 0.338 & 0.041 & 0.486 & 0.627 \\ 
  Ingestion & 0.344 & 0.373 & 0.079 & -0.817 & 0.414 & 0.440 & 0.200 & -2.355 & 0.019 & 0.332 & 0.036 & 0.399 & 0.690 \\ 
  Dailybeast & 0.170 & 0.200 & 0.077 & 0.456 & 0.500 & 0.201 & 0.079 & 0.599 & 0.439 & 0.147 & 0.064 & 0.387 & 0.534 \\ 
  Words $>$ 6 letters & 25.579 & 25.918 & 0.072 & -0.772 & 0.441 & 25.524 & 0.012 & 0.131 & 0.896 & 26.433 & 0.170 & -1.944 & 0.053 \\ 
  USA Today & 0.191 & 0.165 & 0.069 & 0.351 & 0.554 & 0.246 & 0.133 & 1.947 & 0.163 & 0.195 & 0.009 & 0.000 & 1.000 \\ 
  Health & 0.517 & 0.487 & 0.066 & 0.681 & 0.496 & 0.644 & 0.225 & -2.623 & 0.009 & 0.420 & 0.281 & 2.983 & 0.003 \\ 
  Function Words & 38.668 & 38.296 & 0.064 & 0.681 & 0.496 & 38.870 & 0.034 & -0.374 & 0.709 & 40.369 & 0.300 & -3.329 & 0.001 \\ 
  Positive Emotions & 3.685 & 3.759 & 0.064 & -0.667 & 0.505 & 3.693 & 0.006 & -0.068 & 0.946 & 3.629 & 0.047 & 0.540 & 0.590 \\ 
  Clout & 71.048 & 70.509 & 0.056 & 0.588 & 0.557 & 71.012 & 0.004 & 0.043 & 0.966 & 74.081 & 0.319 & -3.631 & 0.000 \\ 
  Salon & 0.106 & 0.090 & 0.055 & 0.168 & 0.682 & 0.176 & 0.202 & 4.502 & 0.034 & 0.121 & 0.047 & 0.168 & 0.682 \\ 
  Differentiation & 1.875 & 1.840 & 0.051 & 0.533 & 0.594 & 1.942 & 0.095 & -1.088 & 0.277 & 2.153 & 0.453 & -5.099 & 0.000 \\ 
  CNN & 0.272 & 0.295 & 0.050 & 0.173 & 0.677 & 0.352 & 0.172 & 3.426 & 0.064 & 0.390 & 0.250 & 7.741 & 0.005 \\ 
  Insight & 1.653 & 1.676 & 0.045 & -0.462 & 0.645 & 1.759 & 0.219 & -2.494 & 0.013 & 1.770 & 0.263 & -3.011 & 0.003 \\ 
  Number Fact Outlets Followed & 0.323 & 0.295 & 0.043 & 0.455 & 0.649 & 0.342 & 0.024 & -0.276 & 0.782 & 0.387 & 0.084 & -0.985 & 0.325 \\ 
  LA Times & 0.162 & 0.150 & 0.032 & 0.041 & 0.840 & 0.218 & 0.145 & 2.298 & 0.130 & 0.150 & 0.032 & 0.063 & 0.802 \\ 
  Tentative & 1.696 & 1.679 & 0.027 & 0.280 & 0.779 & 1.762 & 0.106 & -1.197 & 0.232 & 1.954 & 0.442 & -4.984 & 0.000 \\ 
  Gender (Male) & 0.702 & 0.690 & 0.026 & 0.029 & 0.865 & 0.662 & 0.086 & 0.778 & 0.378 & 0.572 & 0.272 & 9.194 & 0.002 \\ 
  Conjunctions & 3.540 & 3.518 & 0.022 & 0.228 & 0.820 & 3.595 & 0.052 & -0.584 & 0.559 & 3.887 & 0.362 & -3.992 & 0.000 \\ 
  Power & 3.648 & 3.673 & 0.022 & -0.234 & 0.815 & 3.573 & 0.061 & 0.697 & 0.486 & 3.121 & 0.467 & 5.325 & 0.000 \\ 
  Friend & 0.257 & 0.260 & 0.013 & -0.136 & 0.892 & 0.246 & 0.080 & 0.892 & 0.373 & 0.268 & 0.074 & -0.845 & 0.399 \\ 
  Feel & 0.342 & 0.345 & 0.012 & -0.121 & 0.904 & 0.365 & 0.116 & -1.250 & 0.212 & 0.323 & 0.098 & 1.037 & 0.301 \\ 
  DEM & 0.094 & 0.095 & 0.005 & 0.000 & 1.000 & 0.257 & 0.441 & 21.887 & 0.000 & 0.236 & 0.394 & 17.968 & 0.000 \\ 
  \hline
\end{longtable}
\end{scriptsize} 

\end{landscape}

\end{singlespace*} 
\setcounter{table}{0}
\setcounter{figure}{0}
\renewcommand{\thetable}{WG-\arabic{table}}
\renewcommand{\thefigure}{WG-\arabic{figure}}

\section*{Web Appendix G - Exploratory Anxiety Field Experiment}

\begin{singlespace*}
\singlespacing

As anxiety is one of the emotion that distinguishes fake-news sharers uncovered in exploratory study (Study 1), and as heightened anxiety is associated with frequent mentions of anxiety-related words on social media (Gruda and Hasan 2019) and exposure to misinformation is associated with heightened anxiety (Verma et al. 2019), we explore whether incidental exposure to anxiety-inducing content on one's timelines (e.g., other posts not containing misinformation) would increase intentions to share fake-news. We attached this second study to our power study (Study 4) in the same Qualtrics survey to collect it at the same cost. In this section, for completeness, we report the findings from this study, which did not find evidence that heightening anxiety in a post affected fake-news sharing intentions. 

\subsection*{Participants and Methods}

Upon completing the power study, participants were informed that we had a second unrelated study and that they should proceed to the next page. When they proceeded to the second study, participants were presented with a Twitter feed with three posts and asked to imagine that this was content on their own Twitter feed. The first tweet displayed a tweet of a fictitious person named Janet Newman, who was identified as a journalist via her Twitter handle ($"@Janet\_Journalist"$). The second and the third tweets on the Twitter timeline were the same across conditions. The second tweet was a tweet about Serena Williams by ESPN and served as a distractor. The third tweet, our fake-news article, was a real fake-news headline and picture taken from our Snopes dataset used in Study 1 above. 

Participants were randomly assigned to one of two tweets for the journalist, as presented in Figure \ref{fig:anxietymanipulation}. We varied two components of the tweet--the text of the tweet and the profile picture-- following Ryan (2012) who investigated the impact of angry and anxious political ads on click-through rates of individuals. We manipulated anxiety via facial expression (neutral vs. high anxiety) and a text related to the 2016 presidential election. Specifically, we used the facial expressions from Ryan (2012) as the profile picture for the journalist and created text related to the effects of a recession. 

\begin{figure}[htbp]
     \centering
     \caption{Study 3B: Anxiety manipulation}
     \begin{subfigure}[b]{0.40\textwidth}
         \centering
         \includegraphics[width=\textwidth]{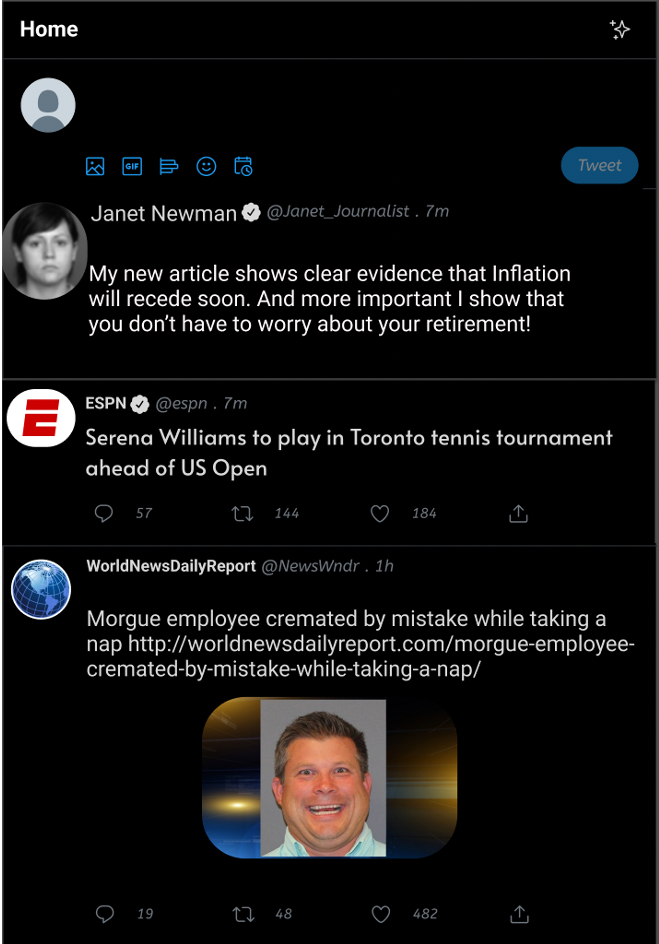}
         \caption{Neutral anxiety condition}
         \label{fig:highanxiety}
     \end{subfigure}
     \hfill
     \begin{subfigure}[b]{0.40\textwidth}
         \centering
         \includegraphics[width=\textwidth]{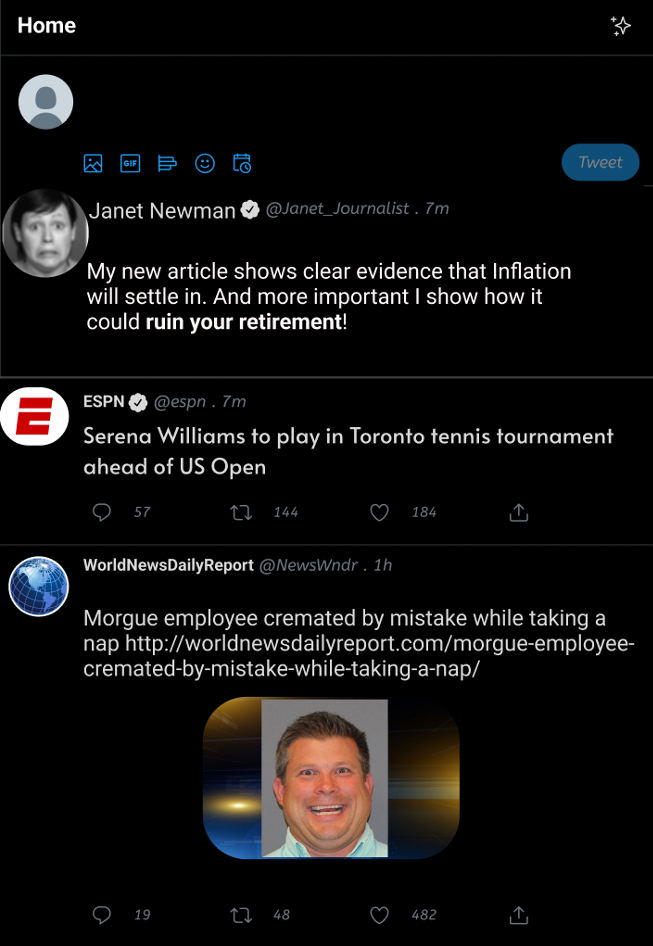}
         \caption{High anxiety condition}
         \label{fig:lowanxiety}
     \end{subfigure}
     \label{fig:anxietymanipulation}
\end{figure}

After seeing the manipulation, participants were asked to indicate how anxious the journalist's tweet made them (1 - not anxious at all; 7 - very anxious). Participants in the high anxiety condition reported feeling more anxious $M=3.83$ than those in the neutral anxiety condition ($M=3.48, t(479)=3.03,p <0.01$). Then, they completed the short form to the state-trait anxiety scale (Marteau and Bekker 1992). We did not find that participants in the high anxiety condition had higher state-trait anxiety ($M=2.28$) than those who were in the neutral condition ($M=2.29, t(479)=0.34, p=.74$). Then, we measured their intention to share the fake-news article by asking them how likely they were to retweet the fake-news article (1 -  extremely unlikely; 5 - extremely likely). This measure served as our focal dependent variable. We also asked participants how likely they would be to fact-check the fake-news article (1 -  extremely unlikely; 5 - extremely likely).

Table \ref{tab:expresults} shows the regressions in which we include the anxiety manipulation (1: high anxiety, 0 low anxiety), state-trait anxiety, along with control variables for both intentions to share the fake-news and for fact-checking it. First, we fail to find evidence that our anxiety manipulation affected intentions to share the fake-news ($B=-.089, SE=.132, p=.501$) or to fact-check it ($B=-.043, SE=.133, p=.747$). Second, we also fail to find evidence that state-trait anxiety is related to intentions to reshare the fake-news tweet ($B=-0.066, SE=.119, p=.578$) or to fact-check it ($B=.005, SE=.120, p=.967$).

\begin{table}[htbp] \centering 
  \caption{Anxiety Study Results} 
  \label{tab:expresults} 
  \footnotesize
\begin{tabular}{@{\extracolsep{5pt}} D{.}{.}{-3} D{.}{.}{-3} D{.}{.}{-3} D{.}{.}{-3} D{.}{.}{-3} D{.}{.}{-3} D{.}{.}{-3} D{.}{.}{-3} D{.}{.}{-3} } 
\\[-1.8ex]\hline 
\hline \\[-1.8ex] 
\multicolumn{1}{l}{} & \multicolumn{3}{c}{Retweet Fake} & \multicolumn{3}{c}{Fact-Check Claim} \\ 
\hline \\[-1.8ex] 
\multicolumn{1}{l}{} & \multicolumn{1}{c}{Est.} & \multicolumn{1}{c}{SE} & \multicolumn{1}{c}{p-val} & \multicolumn{1}{c}{Est.} & \multicolumn{1}{c}{SE} & \multicolumn{1}{c}{p-val} & \\ 
\hline \\[-1.8ex] 
\multicolumn{1}{l}{(Intercept)} & -0.088 & 0.249 & 0.725 & -0.482 & 0.251 & 0.056 \\ 
\multicolumn{1}{l}{Anxiety (1-high, 0-low)} & -0.089 & 0.132 & 0.501 & -0.043 & 0.133 & 0.747 \\ 
\multicolumn{1}{l}{Trait Anxiety} & -0.066 & 0.119 & 0.578 & 0.005 & 0.120 & 0.967 \\ 
\multicolumn{1}{l}{Religion} & 0.118 & 0.051 & 0.022 & 0.107 & 0.052 & 0.039 \\ 
\multicolumn{1}{l}{Age} & -0.025 & 0.004 & 0.000 & -0.014 & 0.005 & 0.002 \\
\multicolumn{1}{l}{Gender: Female} & -0.087 & 0.133 & 0.516 & 0.229 & 0.135 & 0.089 \\ 
\multicolumn{1}{l}{Gender: Non-binary} & -1.551 & 1.435 & 0.280 & 0.330 & 1.448 & 0.820 \\  
\multicolumn{1}{l}{Political: Very Liberal} & -0.145 & 0.188 & 0.442 & -0.189 & 0.190 & 0.320 \\
\multicolumn{1}{l}{Political: Liberal} & -0.030 & 0.179 & 0.865 & -0.269 & 0.180 & 0.137 \\ 
\multicolumn{1}{l}{Political: Conservative} & -0.150 & 0.205 & 0.466 & -0.428 & 0.207 & 0.039 \\
\multicolumn{1}{l}{Political: Very Conservative} & -0.088 & 0.249 & 0.725 & -0.482 & 0.251 & 0.056 \\  
\hline \\[-1.8ex] 
\end{tabular} 
\end{table} 

Although we were surprised not to find evidence that heightened anxiety through nearby posts affects intentions to retweet or fact-check a piece of fake-news, numerous factors could explain this result. For example, it may be that the source of the anxiety-inducing information (i.e., the journalist's post about inflation) is too disconnected from the topic of the fake-news we presented (i.e., a morgue employee being accidentally cremated). Another reason may be carry-over effects from our power study preceding the anxiety study in the survey flow. For example, we found that although there was no difference in intention to retweet between the high power condition from the first study ($M=2.21$) and those in the control condition in the first study ($M=2.32, t(479)=0.84, p=.40$), those in the high power condition were more likely to fact-check the fake-news tweet in this study ($M=3.79$) than those in the control power condition ($M=3.44, t(479)=2.62, p=.01$). Finally, we also find that the use of anxiety-related words (via LIWC) correlates with the state-trait anxiety scale ($r=.10, t(479)=2.23, p=.03$). This provides evidence that the use of anxiety-related words as measured by via LIWC can be a useful proxy for an individual's state-trait anxiety. 

\end{singlespace*}
\setcounter{table}{0}
\setcounter{figure}{0}
\renewcommand{\thetable}{WH\arabic{table}}
\renewcommand{\thefigure}{WH-\arabic{figure}}

\section*{Web Appendix H - Exploratory Religion Experiment}
\begin{singlespace*}
\singlespacing

Similar to Study 3 in the manuscript in which we studied the joint role of trait and situational anger, we explored the role of religion. We do so again based on the findings of Study 1 which reveals a higher usage of words related to religion by fake-news sharers. As we discuss, the relationship between religiosity and fake-news sharing is complex and appears to vary based on how religion is operationalized. While high religious fundamentalism is linked to lower accuracy in discerning truth (Bronstein et al. 2019), and regular religious service attendees show less discernment with COVID-19 fake news (Druckman et al. 2021), high self-reported religiosity does not necessarily correlate with increased fake-news sharing 
(Stefanone, Vollmer, and Covert 2019). Thus, we aimed to explore whether activating religious values would increase intentions to share fake news. 

Except for the manipulation, the study followed the same procedure as the anger study reported in the manuscript. To manipulate the activation of religious values, participants were told to complete a sentence unscrambling task following prior work which has primed religiosity through the same task (Casidy et al. 2021, Hyodo and Bolton 2021, Randolph-Seng and Nielsen 2007). Each participant had to choose four words for eight sets of five words to form eight proper sentences. For each of the eight sets, four words were the same across conditions (e.g.: I, the, overcoat, street). However, the fifth word changed based on whether the participant was in the religiosity prime condition (e.g., cross; where the sentence becomes "I cross the street") or neutral condition (e.g., wore; "I wore the overcoat").\footnote{The eight religious words prime were "religion, morality, heaven, church, divine, cross, worship, holy" while the neutral prime words were "water, cards, seagull, ruins, delicious, wore, follow, cold." These are the same words used in prior research (e.g., Hyodo and Bolton 2021).} Our pre-test with 200 separate participants shows that those in the religious words condition generate more religion-related words ($M=0.62$) than those in the neutral prime condition ($M=0.47; t(199)=1.79, p=.08$). 

The final sample consisted of 397 participants with an average age of 39 (SD=12). The sample was 53\% male, 45\% white, 32\% black or African American, and 11\% Asian. Regarding political affiliation, 50\% self-reported as Democrat, 22\% Republican, and 28\% independent. 

\subsection*{Results}

As in Study 3 on anger in the manuscript, we present alternative sharing outcomes that tease out sharing intentions from the believed versus actual nature of the headline as true or false. Table \ref{table:Study3summary_religion} presents the summary statistics regarding sharing behaviors. 

\begin{table}[htpb]
\centering
\caption{Summary Statistics of News Sharing Behavior}
\label{table:Study3summary_religion}
\footnotesize
\begin{tabular}{l l}
\hline
\textbf{Variable} & \textbf{Mean (SD)} \\
\hline
Average share rating (1 to 6) & 0.51 (0.19)\\
Number shared ($>=4$; out of 14) & 3.84 (3.94)\\
Number shared believed to be false (out of 14) & 1.07 (1.63)\\
Number shared believed to be true (out of 14) & 2.77 (2.98)\\
Number shared actually false (out of 7) & 1.87 (2.12)\\
Number shared actually true (out of 7) & 1.97 (2.11)\\
Number of intentional fake-news shared &  0.53 (0.93)\\
Number of intentional true-news shared &  1.34 (1.66)\\
Number of unintentional fake-news shared & 0.91 (1.33)\\
Number of unintentional true-news shared & 1.06 (1.46)\\
\hline
 \multicolumn{2}{p{.6\textwidth}}{Notes. Number of intentional (unintentional) fake-news shared: Number of fake-news (i.e., false news headlines) shared while believing them to be false (true).} 
\end{tabular}
\end{table}

In Table \ref{table:combined_regression_results_religion}, we present the results of regressions of each of these alternative sharing outcomes, with our manipulation of religion (mean-centered). We control for expected socio-demographic determinants of fake-news sharing, such as age, gender, race, and political ideology. The results presented do not materially change if they are excluded. 

\begin{table}[htbp] \centering 
  \caption{Effects of Priming Religion on News Sharing Outcomes} 
  \label{table:combined_regression_results_religion} 
\resizebox{\textwidth}{!}{%
\begin{tabular}{@{}l *{30}{r} @{}} 
\\[-1.8ex]\hline 
\hline \\[-1.8ex] 
&  \multicolumn{3}{c}{\textbf{Sharing Likelihood}} &  \multicolumn{3}{c}{\textbf{Total shared ($>=4$)}} & \multicolumn{3}{c}{\textbf{Shared Believed False}} & \multicolumn{3}{c}{\textbf{Shared Believed True}} & \multicolumn{3}{l}{\textbf{Shared Actually False}}\\
\hline \\[-1.8ex] 
\multicolumn{1}{c}{} & \multicolumn{1}{c}{Est.} & \multicolumn{1}{c}{SE} & \multicolumn{1}{c}{Pval} & \multicolumn{1}{c}{Est.} & \multicolumn{1}{c}{SE} & \multicolumn{1}{c}{Pval} & \multicolumn{1}{c}{Est.} & \multicolumn{1}{c}{SE} & \multicolumn{1}{c}{Pval} & \multicolumn{1}{c}{Est.} & \multicolumn{1}{c}{SE} & \multicolumn{1}{c}{Pval} & \multicolumn{1}{c}{Est..4} & \multicolumn{1}{c}{SE} & \multicolumn{1}{c}{Pval} \\ 
\hline \\[-1.8ex] 
\multicolumn{1}{l}{(Intercept)} & 2.805 & 0.273 & 0 & 5.012 & 0.957 & 0.00000 & 3.724 & 0.728 & 0.00000 & 1.288 & 0.400 & 0.001 & 2.595 & 0.504 & 0.00000 \\ 
\multicolumn{1}{l}{Religion Primed} & 0.013 & 0.111 & 0.906 & 0.034 & 0.390 & 0.930 & 0.012 & 0.297 & 0.967 & 0.022 & 0.163 & 0.892 & 0.112 & 0.206 & 0.586 \\ 
\multicolumn{1}{l}{Age} & -0.005 & 0.005 & 0.331 & -0.017 & 0.017 & 0.333 & -0.012 & 0.013 & 0.387 & -0.005 & 0.007 & 0.456 & -0.013 & 0.009 & 0.172 \\ 
\multicolumn{1}{l}{Male} & 0.233 & 0.113 & 0.041 & 0.776 & 0.397 & 0.051 & 0.348 & 0.302 & 0.250 & 0.428 & 0.166 & 0.010 & 0.380 & 0.209 & 0.070 \\ 
\multicolumn{1}{l}{Asian Pacific} & 0.022 & 0.184 & 0.907 & -0.151 & 0.646 & 0.815 & 0.015 & 0.491 & 0.975 & -0.167 & 0.270 & 0.538 & 0.008 & 0.340 & 0.980 \\ 
\multicolumn{1}{l}{Black} & 0.561 & 0.181 & 0.002 & 1.946 & 0.635 & 0.002 & 1.367 & 0.483 & 0.005 & 0.579 & 0.266 & 0.030 & 1.574 & 0.335 & 0.00000 \\ 
\multicolumn{1}{l}{Hispanic} & -0.129 & 0.198 & 0.516 & -0.635 & 0.694 & 0.361 & -0.661 & 0.528 & 0.212 & 0.026 & 0.290 & 0.930 & -0.335 & 0.366 & 0.361 \\ 
\multicolumn{1}{l}{Democrat} & -0.379 & 0.146 & 0.010 & -1.138 & 0.513 & 0.027 & -0.803 & 0.390 & 0.040 & -0.335 & 0.214 & 0.119 & -0.475 & 0.270 & 0.080 \\ 
\multicolumn{1}{l}{Independent} & -0.471 & 0.158 & 0.003 & -1.769 & 0.556 & 0.002 & -1.367 & 0.422 & 0.001 & -0.402 & 0.232 & 0.084 & -0.866 & 0.293 & 0.003 \\ 
\hline \\[-1.8ex] 
\end{tabular}%
}

\vspace{.25cm} 

\resizebox{\textwidth}{!}{%
\begin{tabular}{@{}l *{30}{r} @{}} 
\\[-1.8ex]\hline 
\hline \\[-1.8ex] 
&  \multicolumn{3}{c}{\textbf{Shared Actually True}} &  \multicolumn{3}{c}{\textbf{Intentional Fake}} & \multicolumn{3}{c}{\textbf{Unintentional Fake}} & \multicolumn{3}{c}{\textbf{Unintentional True}} & \multicolumn{3}{l}{\textbf{Intentional True}}\\
\hline \\[-1.8ex] 
\multicolumn{1}{c}{} & \multicolumn{1}{c}{Est.} & \multicolumn{1}{c}{SE} & \multicolumn{1}{c}{Pval} & \multicolumn{1}{c}{Est.} & \multicolumn{1}{c}{SE} & \multicolumn{1}{c}{Pval} & \multicolumn{1}{c}{Est.} & \multicolumn{1}{c}{SE} & \multicolumn{1}{c}{Pval} & \multicolumn{1}{c}{Est.} & \multicolumn{1}{c}{SE} & \multicolumn{1}{c}{Pval} & \multicolumn{1}{c}{Est..4} & \multicolumn{1}{c}{SE} & \multicolumn{1}{c}{Pval} \\ 
\hline \\[-1.8ex] 
\multicolumn{1}{l}{(Intercept))} & 2.416 & 0.521 & 0.00000 & 0.333 & 0.229 & 0.147 & 2.083 & 0.408 & 0.00000 & 0.820 & 0.315 & 0.010 & 1.776 & 0.359 & 0.00000 \\ 
\multicolumn{1}{l}{Religion primed} & -0.078 & 0.212 & 0.715 & -0.007 & 0.094 & 0.943 & -0.071 & 0.166 & 0.670 & 0.074 & 0.129 & 0.566 & 0.038 & 0.147 & 0.795 \\ 
\multicolumn{1}{l}{Age} & -0.004 & 0.010 & 0.649 & 0.002 & 0.004 & 0.697 & -0.006 & 0.007 & 0.424 & -0.005 & 0.006 & 0.428 & -0.008 & 0.007 & 0.220 \\ 
\multicolumn{1}{l}{Male} & 0.396 & 0.216 & 0.068 & 0.279 & 0.095 & 0.004 & 0.117 & 0.169 & 0.490 & 0.306 & 0.131 & 0.020 & 0.074 & 0.149 & 0.619 \\ 
\multicolumn{1}{l}{Asian Pacific} & -0.160 & 0.352 & 0.650 & -0.169 & 0.155 & 0.274 & 0.010 & 0.276 & 0.972 & -0.149 & 0.213 & 0.485 & 0.157 & 0.243 & 0.517 \\ 
\multicolumn{1}{l}{Black} & 0.372 & 0.346 & 0.283 & 0.271 & 0.152 & 0.075 & 0.100 & 0.271 & 0.712 & 1.067 & 0.209 & 0.00000 & 0.507 & 0.238 & 0.034 \\ 
\multicolumn{1}{l}{Hispanic} & -0.300 & 0.378 & 0.428 & 0.035 & 0.166 & 0.833 & -0.335 & 0.296 & 0.258 & -0.126 & 0.229 & 0.582 & -0.209 & 0.261 & 0.424 \\ 
\multicolumn{1}{l}{Democrat} & -0.663 & 0.279 & 0.018 & 0.010 & 0.123 & 0.938 & -0.673 & 0.219 & 0.002 & 0.153 & 0.169 & 0.366 & -0.628 & 0.192 & 0.001 \\ 
\multicolumn{1}{l}{Independent} & -0.903 & 0.302 & 0.003 & -0.124 & 0.133 & 0.354 & -0.779 & 0.237 & 0.001 & -0.226 & 0.183 & 0.218 & -0.640 & 0.209 & 0.002 \\ 
\hline \\[-1.8ex] 
\end{tabular}%
}
\end{table}

Overall, we do not find evidence that our activating religious values had any direct effect on sharing outcomes (all $p>.50$). Our findings are thus in line with (Stefanone, Vollmer, and Covert 2019) who did not find a correlation between high self-reported religiosity and increased fake-news sharing. The difference to studies that do reveal a relationship between religion and fake-news sharing might be attributed to the operationalization of religiosity across different studies (Bronstein et al. 2019, Druckman et al. 2021).

In line with the exploratory study on anger, we also investigate how activating religious values affects the desire to share news that is accurate, surprising, interesting, aligned with one's politics, or funny. We find that priming religiosity decreased the motivation to be surprising and funny, but had no effect on motivations to be accurate, interesting, or aligned with political leaning. We regressed our predictors and the five motivations on our individual-level outcome variables. We find that although the motivation to be funny does not influence sharing intentions, the motivation to be surprising is related to higher sharing intentions, greater number of believed-true and greater believed-false news shared. The indirect effect of activating religious values on sharing intentions, number of believed-true shared and number of believed-false shared suggest that priming religious activation reduces the sharing intentions of the claims in our sample through a lowered desire to be surprising regardless of the perceived veracity. 

 \begin{landscape}
 \begin{table}[!htbp] \centering 
 \tiny
   \caption{Regression Moderation Content Characteristics} 
   \label{tab:moderations_religion} 
 \begin{tabular}{@{\extracolsep{5pt}}lD{.}{.}{-3} D{.}{.}{-3} D{.}{.}{-3} D{.}{.}{-3} D{.}{.}{-3} D{.}{.}{-3} D{.}{.}{-3} D{.}{.}{-3} } 
 \\[-1.8ex]\hline 
 \hline \\[-1.8ex] 
  & \multicolumn{8}{c}{\textit{Dependent variable:}} \\ 
 \cline{2-9} 
  & \multicolumn{1}{c}{Accurate} & \multicolumn{1}{c}{Surprising} & \multicolumn{1}{c}{Interesting} & \multicolumn{1}{c}{Aligned with Politics} & \multicolumn{1}{c}{Funny} & \multicolumn{1}{c}{Sharing Intentions} & \multicolumn{1}{c}{Believed True Shared} & \multicolumn{1}{c}{Believed False Shared} \\ 
 \\[-1.8ex] & \multicolumn{1}{c}{(1)} & \multicolumn{1}{c}{(2)} & \multicolumn{1}{c}{(3)} & \multicolumn{1}{c}{(4)} & \multicolumn{1}{c}{(5)} & \multicolumn{1}{c}{(6)} & \multicolumn{1}{c}{(7)} & \multicolumn{1}{c}{(8)}\\ 
 \hline \\[-1.8ex] 
   Constant & 3.887^{***} & 3.312^{***} & 3.992^{***} & 3.200^{***} & 3.650^{***} & 1.517^{***} & 0.713 & 0.326 \\ 
   & (0.277) & (0.308) & (0.278) & (0.321) & (0.316) & (0.357) & (0.978) & (0.542) \\ 
   & & & & & & & & \\ 
  Religion Primed & 0.026 & -0.227^{*} & -0.117 & -0.013 & -0.265^{**} & 0.077 & 0.126 & 0.100 \\ 
   & (0.111) & (0.123) & (0.112) & (0.129) & (0.127) & (0.105) & (0.287) & (0.159) \\ 
   & & & & & & & & \\ 
  Age & 0.006 & -0.003 & 0.002 & -0.005 & -0.008 & -0.003 & -0.010 & -0.004 \\ 
   & (0.005) & (0.006) & (0.005) & (0.006) & (0.006) & (0.005) & (0.013) & (0.007) \\ 
   & & & & & & & & \\ 
  Male & -0.126 & -0.173 & -0.101 & -0.161 & -0.076 & 0.298^{***} & 0.499^{*} & 0.485^{***} \\ 
   & (0.113) & (0.125) & (0.113) & (0.131) & (0.129) & (0.106) & (0.291) & (0.161) \\ 
   & & & & & & & & \\ 
  Asian Pacific & -0.440^{**} & 0.052 & -0.551^{***} & -0.443^{**} & -0.037 & 0.076 & 0.192 & -0.177 \\ 
   & (0.184) & (0.204) & (0.185) & (0.213) & (0.210) & (0.176) & (0.481) & (0.267) \\ 
   & & & & & & & & \\ 
  Black & 0.025 & 0.400^{**} & 0.170 & 0.456^{**} & 0.091 & 0.392^{**} & 1.002^{**} & 0.407 \\ 
   & (0.181) & (0.201) & (0.182) & (0.209) & (0.206) & (0.171) & (0.468) & (0.260) \\ 
   & & & & & & & & \\ 
  Hispanic & -0.116 & 0.144 & 0.177 & -0.031 & -0.142 & -0.180 & -0.760 & -0.052 \\ 
   & (0.197) & (0.219) & (0.198) & (0.229) & (0.225) & (0.186) & (0.509) & (0.283) \\ 
   & & & & & & & & \\ 
  Democrat & 0.256^{*} & -0.513^{***} & -0.140 & -0.100 & -0.245 & -0.217 & -0.511 & -0.123 \\ 
   & (0.146) & (0.162) & (0.146) & (0.169) & (0.166) & (0.140) & (0.382) & (0.212) \\ 
   & & & & & & & & \\ 
  Independent & -0.014 & -0.189 & -0.117 & -0.595^{***} & -0.003 & -0.340^{**} & -1.075^{***} & -0.290 \\ 
   & (0.158) & (0.175) & (0.159) & (0.183) & (0.180) & (0.150) & (0.412) & (0.228) \\ 
   & & & & & & & & \\ 
  Accurate &  &  &  &  &  & -0.038 & 0.043 & -0.128 \\ 
   &  &  &  &  &  & (0.056) & (0.152) & (0.084) \\ 
   & & & & & & & & \\ 
  Surprising &  &  &  &  &  & 0.265^{***} & 0.552^{***} & 0.338^{***} \\ 
   &  &  &  &  &  & (0.051) & (0.140) & (0.078) \\ 
   & & & & & & & & \\ 
  Interesting &  &  &  &  &  & 0.055 & 0.110 & 0.060 \\ 
   &  &  &  &  &  & (0.062) & (0.170) & (0.094) \\ 
   & & & & & & & & \\ 
  Aligned with Politics &  &  &  &  &  & 0.125^{***} & 0.293^{**} & 0.072 \\ 
   &  &  &  &  &  & (0.044) & (0.121) & (0.067) \\ 
   & & & & & & & & \\ 
  Funny &  &  &  &  &  & -0.020 & -0.101 & -0.039 \\ 
   &  &  &  &  &  & (0.047) & (0.129) & (0.072) \\ 
   & & & & & & & & \\ 
 \hline \\[-1.8ex] 
 \hline 
 \hline \\[-1.8ex] 
 \textit{Note:}  & \multicolumn{8}{r}{$^{*}$p$<$0.1; $^{**}$p$<$0.05; $^{***}$p$<$0.01} \\ 
 \end{tabular} 
 \end{table}

\end{landscape}

\end{singlespace*}

\setcounter{table}{0}
\setcounter{figure}{0}
\renewcommand{\thetable}{WI\arabic{table}}
\renewcommand{\thefigure}{WI-\arabic{figure}}
\setlength{\baselineskip}{21pt plus.2pt}

\section*{Web Appendix I - Robustness Analyses: Followerhsip and Mentions in Tweets}

\begin{table}[htbp]
\caption{Fake-news and Fact-check Outlets' Twitter Accounts Used to Quantify Fake-News and Fact-Check Outlet Followership and Mentions of Fake-News and Fact-Check Outlets in Tweets}
\label{tab:HoaxyFollership}
\begin{scriptsize}
\begin{multicols}{4}
\href{https://twitter.com/gatewaypundit}{thegatewaypundit.com}\\
\href{https://twitter.com/conserv_tribune/media}{conservativetribune.com}\\
\href{https://twitter.com/zerohedge}{zerohedge.com}\\
\href{https://twitter.com/rickrwells?lang=it}{rickwells.us}\\
\href{https://twitter.com/TPInsidr}{thepoliticalinsider.com}\\
\href{https://twitter.com/therightscoop}{therightscoop.com}\\
\href{https://twitter.com/teapartyorg}{teaparty.org}\\
\href{https://twitter.com/thefederalist1?lang=it}{thefederalistpapers.org}\\
\href{https://twitter.com/21WIRE/}{21stcenturywire.com}\\
\href{https://twitter.com/ActivistPost}{activistpost.com}\\
\href{https://twitter.com/AddInfoOrg}{addictinginfo.org}\\
\href{https://twitter.com/beforeitsnews}{beforeitsnews.com}\\
\href{https://twitter.com/bipartisanism}{bipartisanreport.com}\\
\href{https://twitter.com/BlueNationRev}{bluenationreview.com}\\
\href{https://twitter.com/BreitbartNews}{breitbart.com}\\
\href{https://twitter.com/BurrardStreetJ}{burrardstreetjournal.com}\\
\href{https://twitter.com/Call_TheCops}{callthecops.net}\\
\href{https://twitter.com/thechristiantim?lang=en}{christiantimes.com}\\
\href{https://twitter.com/christwire}{christwire.org}\\
\href{https://twitter.com/chroniclesu}{chronicle.su}\\
\href{https://twitter.com/ClickHole}{clickhole.com}\\
\href{https://twitter.com/coasttocoastam}{coasttocoastam.com}\\
\href{https://twitter.com/collectiveevol}{collective-evolution.com}\\
\href{https://twitter.com/conscious_news}{consciouslifenews.com}\\
\href{https://twitter.com/counterpsyops?lang=en}{counterpsyops.com}\\
\href{https://twitter.com/dailybuzzlive}{dailybuzzlive.com	}\\
\href{https://twitter.com/thedailycurrant}{dailycurrant.com	}\\
\href{https://twitter.com/dcclothesline?lang=it}{dcclothesline.com	}\\
\href{https://twitter.com/DerfMagazine}{derfmagazine.com	}\\
\href{https://twitter.com/disclosetv}{disclose.tv	}\\
\href{https://twitter.com/duffelblog}{duffelblog.com	}\\
\href{https://twitter.com/duh_progressive}{duhprogressive.com	}\\
\href{https://twitter.com/empirenewsnet}{empirenews.net	}\\
\href{https://twitter.com/EmpireSportsNet}{empiresports.co	}\\
\href{https://twitter.com/enduringvision}{enduringvision.com	}\\
\href{https://twitter.com/flyheightvideos}{flyheight.com	}\\
\href{https://twitter.com/freewoodpost}{freewoodpost.com	}\\
\href{https://twitter.com/geoengwatch}{geoengineeringwatch.org	}\\
\href{https://twitter.com/CRG_CRM}{globalresearch.ca	}\\
\href{https://twitter.com/gomerblog}{gomerblog.com	}\\
\href{https://twitter.com/hangthebankers}{hangthebankers.com	}\\
\href{https://twitter.com/huzlers}{huzlers.com	}\\
\href{https://twitter.com/ifyou0nlynews?lang=it}{ifyouonlynews.com	}\\
\href{https://twitter.com/InfoWarsMedia?lang=en}{infowars.com	}\\
\href{https://twitter.com/intellihubnews}{intellihub.com	}\\
\href{https://twitter.com/itaglive}{itaglive.com	}\\
\href{https://twitter.com/lewrockwell}{lewrockwell.com	}\\
\href{https://twitter.com/LibAmericaOrg}{liberalamerica.org	}\\
\href{https://twitter.com/lt_fm?lang=it}{libertytalk.fm	}\\
\href{https://twitter.com/libertyvideos}{libertyvideos.org	}\\
\href{https://twitter.com/tnronline}{nationalreport.net	}\\
\href{https://twitter.com/natnewsfollow}{naturalnews.com	}\\
\href{https://twitter.com/newsbiscuit}{newsbiscuit.com	}\\
\href{https://twitter.com/thenewslo}{newslo.com	}\\
\href{https://twitter.com/nowtheendbegins}{nowtheendbegins.com	}\\
\href{https://twitter.com/OccupyDemocrats/}{occupydemocrats.com	}\\
\href{https://twitter.com/other98}{other98.com	}\\
\href{https://twitter.com/poliblindspot}{politicalblindspot.com	}\\
\href{https://twitter.com/politicususa}{politicususa.com	}\\
\href{https://twitter.com/prisonplanet}{prisonplanet.com	}\\
\href{https://twitter.com/realfarmacy}{realfarmacy.com	}\\
\href{https://twitter.com/rhobbusjd}{realnewsrightnow.com	}\\
\href{https://twitter.com/RedState}{redstate.com	}\\
\href{https://twitter.com/rockcitytimes}{rockcitytimes.com	}\\
\href{https://twitter.com/stuppidcom}{stuppid.com	}\\
\href{https://twitter.com/theblaze?lang=it}{theblaze.com	}\\
\href{https://twitter.com/TheDailySheeple}{thedailysheeple.com	}\\
\href{https://twitter.com/thedcgazette}{thedcgazette.com	}\\
\href{https://twitter.com/thelapine}{thelapine.ca	}\\
\href{https://twitter.com/thenewsnerd}{thenewsnerd.com	}\\
\href{https://twitter.com/TheOnion}{theonion.com	}\\
\href{https://twitter.com/rundownlive}{therundownlive.com	}\\
\href{https://twitter.com/thespoof}{thespoof.com	}\\
\href{https://twitter.com/truthfrequency}{truthfrequencyradio.com	}\\
\href{https://twitter.com/twitchyteam}{twitchy.com	}\\
\href{https://twitter.com/veteranstoday}{veteranstoday.com	}\\
\href{https://twitter.com/weeklyworldnews}{weeklyworldnews.com	}\\
\href{https://twitter.com/worldnetdaily}{wnd.com	}\\
\href{https://twitter.com/WorldTruthTV}{worldtruth.tv	}\\
\href{https://twitter.com/yournewswire?lang=it}{yournewswire.com}\\
\end{multicols}
Note: Fact-check outlets were \href{https://twitter.com/factcheckdotorg}{FactCheck.org}, \href{https://twitter.com/PolitiFact}{PolitiFact.com}, \href{https://twitter.com/snopes}{Snopes.com}, \href{https://twitter.com/erumors}{TruthOrFiction.com}, and \href{https://twitter.com/RealClearNews}{RealClearPolitics.com}. Clicking on any domain listed in this table will lead to the associated Twitter account. 
\end{scriptsize}
\end{table}

\begin{singlespace*}
\footnotesize

\begin{table}[ht]
\caption{Comparisons of Mentions in Tweets and Fake-News and Fact-Check Outlet Followership by Group}
\footnotesize
\tiny
\label{tab:mentionsbygroup}
\bigskip
\subcaption{Panel A: Snopes Sample*}
\begin{tabular}{lrrrrrr}
  \hline
Group & Avg Fake  & Avg Fact Check  & Avg \# Fake Outlets & Avg \# Fact Outlets  & Proportion Following  & Proportion Following  \\ 
 &  Mentions & Mentions & Followed & Followed & at least 1 Fake Outlet &  at least 1 Fact Outlet \\ 
  \hline
Fake-news Sharers & 3.83 & 0.14 & 2.01 & 0.24 & 0.57 & 0.17 \\ 
  Fact-check Sharers & 2.76 & 0.23 & 1.15 & 0.49 & 0.50 & 0.30 \\ 
  Random Group & 0.22 & 0.02 & 0.24 & 0.04 & 0.16 & 0.04 \\ 
   \hline
\end{tabular}
\medskip
\subcaption{Panel B: Hoaxy Sample*}
\begin{tabular}{lrrrrrr}
  \hline
Group & Avg Fake  & Avg Fact Check  & Avg \# Fake Outlets & Avg \# Fact Outlets  & Proportion Following  & Proportion Following  \\ 
 &  Mentions & Mentions & Followed & Followed & at least 1 Fake Outlet &  at least 1 Fact Outlet \\ 
  \hline
Fake-news Sharers & 2.57 & 0.20 & 3.93 & 0.32 & 0.79 & 0.21 \\ 
  Fact-check Sharers & 1.19 & 3.51 & 1.54 & 0.39 & 0.53 & 0.25 \\ 
  Left-leaning Media Sharers & 1.31 & 0.23 & 0.90 & 0.34 & 0.44 & 0.21 \\ 
  Right-leaning Media Sharers & 1.48 & 0.34 & 1.29 & 0.29 & 0.55 & 0.24 \\ 
   \hline
\end{tabular}
Notes:\\
* For the Snopes sample, the differences between fake-news sharers and all other groups regarding the fake-news outlets followed as well as the fake-news outlets mentioned are statistically significant except for the difference between the fake-news mentioned by fake-news sharers and fact-check sharers. \\
* For the Hoaxy sample, the differences between fake-news sharers and all other groups for fake mentions as well as fake-news outlets followership are statistically significant except for the difference of fake-news mentions between fake-news sharers and right-leaning news sharers (p = 0.055). None of the differences is significant for fact-check outlet mentions or followership. \\
\end{table}

\begin{table}[ht]
\caption{Comparisons of Mentions in Tweets by Group in 2023}
\footnotesize
\tiny
\label{tab:mentionsbygroup}
\begin{tabular}{lrrrrrr}
  \hline
Group & Avg Fake  & Avg Fact Check   \\ 
 &  Mentions & Mentions  \\ 
  \hline
Fake-news Sharers & 4.42 & 0.177  \\ 
  Fact-check Sharers & 2.43 & 0.289  \\ 
  Random Group & 0.242  & 0.0194   \\ 
   \hline
\end{tabular}
\end{table}

\end{singlespace*} 
\setcounter{table}{0}
\setcounter{figure}{0}
\renewcommand{\thetable}{WJ\arabic{table}}
\renewcommand{\thefigure}{WJ-\arabic{figure}}
\setlength{\baselineskip}{21pt plus.2pt}
\vspace{-6mm}
\section*{Web Appendix J - Robustness Alternative Measures -- Alternative Emotions Measures via NRC -- Alternative Emotions Measures via MTurk Coded Emotions.}

\begin{singlespace*}

\vspace{+3mm}
\subsection{Alternative emotions measures via NRC}
\label{Alternative emotions measures via NRC.}
We analyze the Snopes dataset using the three NRC lexica: NRC-VAD, NRC emotion, and NRC emotion intensity. As with the LIWC dictionary, we find that fake-news sharers write significantly more negatively and with higher arousal in their language compared to the random group. Moreover, in line with our previous findings we find that fake-news sharers and fact-check sharers use significantly more angry and fearful words compared to the random group. Additionally, the words used by fake-news sharers and fact-check sharers have a significantly higher intensity on both anger and fear compared to the random group.

\begin{table}[ht]
\centering
\small
\caption{NRC comparison fake-news sharers vs. other groups} 
\begin{tabular}{lllllllll}
  \hline
Variable & Mean  & Mean   &  Mean &  Fake vs.  & Fake vs.   & Fake vs.  & Fake vs.  \\ 
  &   Fake-news & Fact-check  & Random  & Fact  & Fact & Random & Random \\ 
      & Sharers & Sharers  & Sample  &  t-Value & p-value & t-Value & p-value  \\ 
  \hline  \hline
Positive & 0.042 & 0.040 & 0.041 & 6.182 & 0.000 & 3.479 & 0.001 \\ 
Negative & 0.031 & 0.031 & 0.022 & 1.293 & 0.196 & 27.890 & 0.000 \\ 
Anger & 0.016 & 0.015 & 0.011 & 1.610 & 0.108 & 24.530 & 0.000 \\ 
Anger intensity & 0.496 & 0.492 & 0.491 & 1.927 & 0.054 & 2.763 & 0.006 \\ 
Fear & 0.019 & 0.018 & 0.012 & 4.021 & 0.000 & 30.210 & 0.000 \\ 
Fear intensity & 0.393 & 0.393 & 0.348 & 0.217 & 0.828 & 24.055 & 0.000 \\ 
VAD valence & 0.160 & 0.154 & 0.161 & 5.700 & 0.000 & -0.311 & 0.756 \\ 
VAD arousal & 0.126 & 0.121 & 0.116 & 5.553 & 0.000 & 14.508 & 0.000 \\ 
   \hline
\end{tabular}
\label{tab:ttest_results}
\end{table}

\vspace{+3mm}
\subsection{Alternative emotions measures via MTurk coded emotions} \label{Alternative emotions measures via mTurk coded emotions}

To further test the robustness of our results on emotions, a subset of the tweets were rated by the human coders on MTurk. We conducted a survey in which MTurkers first read the following descriptions and instructions: "We would like to assess emotions in tweets people post on their Twitter page. A tweet can incorporate multiple emotions and we are interested in how much a tweet incorporates different emotions." We specified that their task would be to answer questions about multiple tweets and we asked them to look at an exemplary tweet before starting the survey. The participants then clicked on an embedded URL that displayed the first original tweet. Raters were then instructed to read the tweet and score it on negative emotions, anger, and anxiety on a 1-5 scale answering the question on whether the tweets contained the particular emotion. Survey participants could rate as many tweets as they wanted. In total we gathered 92,165 ratings from MTurk coders for 21,060 unique tweets of the Fake-News Sharer, the Fact-Check Sharers and the Random Sample. We compare the language differences across the three groups (fake-news sharers, fact-check sharers, random group) on a tweet level and a user level. On the tweet level, we compare the average emotion scores for each tweet assigned by the human coders across the groups. On the user level, we average the emotion scores of the tweets for each user and subsequently compare the users of the three different groups. We replicate our results both on a tweet as well as on a user level. Specifically, we find that fake-news sharers write overall more negatively and share more angry and anxious tweets compared to the random group (see Table \ref{tab:MTurk_User} and \ref{tab:MTurk_Tweets} for the results). 

Finally, we use the subset of coded tweets to fine-tune a BERT model and predict the remaining tweets that were not coded by humans. Comparing the full sample of tweets among the groups leads to the same results such that fake-news sharers use significantly more negative, angry and anxious language compared to the random group (See Table \ref{tab:BERT_Tweet} and Table \ref{tab:BERT_User} for the results).

\begin{table}[ht]
\centering
\small
\caption{MTurk Subsample - Tweet Level} 
\begin{tabular}{lllllllll}
  \hline
Variable & Mean  & Mean   &  Mean &  Fake vs.  & Fake vs.   & Fake vs.  & Fake vs.  \\ 
  &   Fake-news & Fact-check  & Random  & Fact  & Fact & Random & Random \\ 
      & Sharers & Sharers  & Sample  &  t-Value & p-value & t-Value & p-value  \\ 
      \hline \hline
Anger & 2.73 & 2.80 & 2.50 & -3.81 & 0.00 & 16.04 & 0.00 \\ 
 Anxiety & 2.74 & 2.76 & 2.68 & -1.42 & 0.16 & 4.70 & 0.00 \\ 
 Negative & 3.03 & 3.10 & 2.74 & -3.33 & 0.00 & 18.93 & 0.00 \\ 
   \hline
\end{tabular}
\label{tab:MTurk_Tweets}
\end{table}

\begin{table}[ht]
\centering
\small
\caption{MTurk Subsample - User Level} 
\begin{tabular}{lllllllll}
  \hline
Variable & Mean  & Mean   &  Mean &  Fake vs.  & Fake vs.   & Fake vs.  & Fake vs.  \\ 
  &   Fake-news & Fact-check  & Random  & Fact  & Fact & Random & Random \\ 
      & Sharers & Sharers  & Sample  &  t-Value & p-value & t-Value & p-value  \\ 
      \hline \hline
Anger & 2.73 & 2.80 & 2.50 & -3.36 & 0.00 & 14.31 & 0.00 \\ 
Anxiety & 2.74 & 2.76 & 2.68 & -1.39 & 0.17 & 4.45 & 0.00 \\ 
Negative & 3.03 & 3.10 & 2.74 & -2.90 & 0.00 & 16.49 & 0.00 \\ 
   \hline
\end{tabular}
\label{tab:MTurk_User}
\end{table}

\begin{table}[ht]
\centering
\small
\caption{BERT - Tweet Level} 
\begin{tabular}{lllllllll}
  \hline
Variable & Mean  & Mean   &  Mean &  Fake vs.  & Fake vs.   & Fake vs.  & Fake vs.  \\ 
  &   Fake-news & Fact-check  & Random  & Fact  & Fact & Random & Random \\ 
      & Sharers & Sharers  & Sample  &  t-Value & p-value & t-Value & p-value  \\ 
      \hline \hline
Anger & 2.65 & 2.73 & 2.45 & -107.72 & 0.00 & 357.25 & 0.00 \\ 
Anxiety & 2.73 & 2.77 & 2.64 & -52.92 & 0.00 & 174.39 & 0.00 \\ 
Negative & 2.74 & 2.79 & 2.64 & -63.88 & 0.00 & 189.56 & 0.00 \\ 
   \hline
\end{tabular}
\label{tab:BERT_Tweet}
\end{table}

\begin{table}[htpb]
\caption{BERT - User Level} 
\centering
\small
\begin{tabular}{lllllllll}
  \hline
Variable & Mean  & Mean   &  Mean &  Fake vs.  & Fake vs.   & Fake vs.  & Fake vs.  \\ 
  &   Fake-news & Fact-check  & Random  & Fact  & Fact & Random & Random \\ 
      & Sharers & Sharers  & Sample  &  t-Value & p-value & t-Value & p-value  \\ 
      \hline \hline
Anger& 2.67 & 2.75 & 2.45 & -5.83 & 0.00 & 25.66 & 0.00 \\ 
Anxiety & 2.74 & 2.78 & 2.63 & -4.80 & 0.00 & 18.83 & 0.00 \\ 
Negative & 2.76 & 2.80 & 2.63 & -4.93 & 0.00 & 19.71 & 0.00 \\ 
   \hline
\end{tabular}
\label{tab:BERT_User}
\end{table}
\end{singlespace*}

\setcounter{table}{0}
\setcounter{figure}{0}
\renewcommand{\thetable}{WK\arabic{table}}

\section*{Web Appendix K - Comparison of MagicSauce to Park et al. (2005) }

\begin{singlespace*}
\footnotesize 

\begin{longtable}{cc}
\caption{Correlations of Two Personality Models}
\label{tab:my-table}\\
\hline
\textbf{\begin{tabular}[c]{@{}c@{}} Correlation Personality Scores -  Snopes\end{tabular}} \\ \hline
\endfirsthead
\endhead
Agreeableness: 0.69 (p $<$ 0.01)  \\ \hline
Openness: 0.54 (p $<$ 0.01)\\ \hline
Extraversion:  0.35 (p $<$ 0.01)\\ \hline
Neuroticism:  0.58 (p $<$ 0.01)\\ \hline
Conscientuousness:  0.39 (p $<$ 0.01)\\ \hline
\end{longtable}

\begin{longtable}{lcccc}
\caption{Comparison Personality Models - Snopes}
\vspace*{-\baselineskip}
\endfirsthead
\endhead

 & \multicolumn{2}{c}{\textbf{Park et al. (2015)}} & \multicolumn{2}{c}{\textbf{Magic Sauce}} \\ \hline
\textbf{Openness} & \textit{\textbf{Cohen's   d}} & \textit{\textbf{p}} & \textit{\textbf{Cohen's d}} & \textit{\textbf{p}} \\ \hline
Fake-news Sharers vs. Fact-check Sharers   & 0.23 & 0.00 & 0.03 & 0.49 \\ \hline
Fake-news Sharers vs. Random Sample & 0.54 & 0.00 & 0.72 & 0.00 \\ \hline
\multicolumn{2}{l}{\textbf{Conscientuousness}} &  &  &  \\ \hline
Fake-news Sharers vs. Fact-check Sharers& 0.15 & 0.00 & 0.11 & 0.01\\ \hline
Fake-news Sharers vs. Random Sample & 0.36 & 0.00 & 0.23 & 0.00 \\ \hline
 \multicolumn{2}{l}{\textbf{Extraversion}} &  &  &  \\ \hline
Fake-news Sharers vs. Fact-check Sharers & 0.32 & 0.00 & 0.22 & 0.00 \\ \hline
Fake-news Sharers vs. Random Sample & 0.64 & 0.00 & 0.81 & 0.00 \\ \hline
\multicolumn{2}{l}{\textbf{Neuroticism}} &  &  &  \\ \hline
Fake-news Sharers vs. Fact-check Sharers & 0.15 & 0.00 & 0.24 & 0.00 \\ \hline
Fake-news Sharers vs. Random Sample & 0.35 & 0.00 & 0.13 & 0.00\\ \hline
\multicolumn{2}{l}{\textbf{Agreeableness}} &  &  &  \\ \hline
Fake-news Sharers vs. Fact-check Sharers & 0.29 & 0.00 & 0.20 & 0.00 \\ \hline
Fake-news Sharers vs. Random Sample & 0.56 & 0.00 & 0.59 & 0.00 \\ \hline
\end{longtable}

\end{singlespace*} 
\setcounter{table}{0}
\setcounter{figure}{0}
\renewcommand{\thetable}{WL\arabic{table}}
\renewcommand{\thefigure}{WL-\arabic{figure}}

\section*{Web Appendix L - Fake-news Article Content Analysis}

\begin{singlespace*}

\singlespacing

To analyze the article-user matching hypothesis we first assess for the fake-news articles in our sample their textual profile to compare it to the textual profile of fake-news and non-fake-news sharers. To assess the text of articles in our Snopes sample we use waybackmachine as many articles in our sample were no longer available. Of the 66 articles used, we could not find an archive of the text for nine. Our articles span various misinformation categories within Snopes, with "Junk News" being the most represented (28 articles), followed by politics (9 articles), uncategorized (4), science (3), and several with two articles (media matters, fauxtography, and entertainment) or just one article (humor, viral phenomena, military, crime and hurricane Katrina). For the set of 57 articles, we then created a textual profile for each of the articles. 

We assess all 79 LIWC dimensions (excluding word count, words per sentence, and punctuation). We then compute for each LIWC dimension the average value across the 57 articles. This creates a vector of 79 attributes, one for each LIWC dimension that reflects the average value of all fake-news articles.  We can compare this vector of the average linguistic profile of all fake-news articles to the same individual vector of each user in our sample. We compute the Euclidean distance between the fake-news article LIWC vector and each user's LIWC vector. We then average the distance across all users in each group (fake-news sharers vs. random social media users vs. matched sample) and compare the group differences. Figure \ref{EuclideanDistanceContentAnalysisLIWC} shows the process of creating the vectors representing the average fake-news article and the computation for the Euclidean distance of the linguistic profile of the fake-news articles and the three user groups (fake-news sharers, random and matched group).

\begin{figure}[htbp]
    \caption{Similarity Calculation Fake-News Articles and Fake-News Users' Text}
    \centering
    \includegraphics[width=\textwidth]{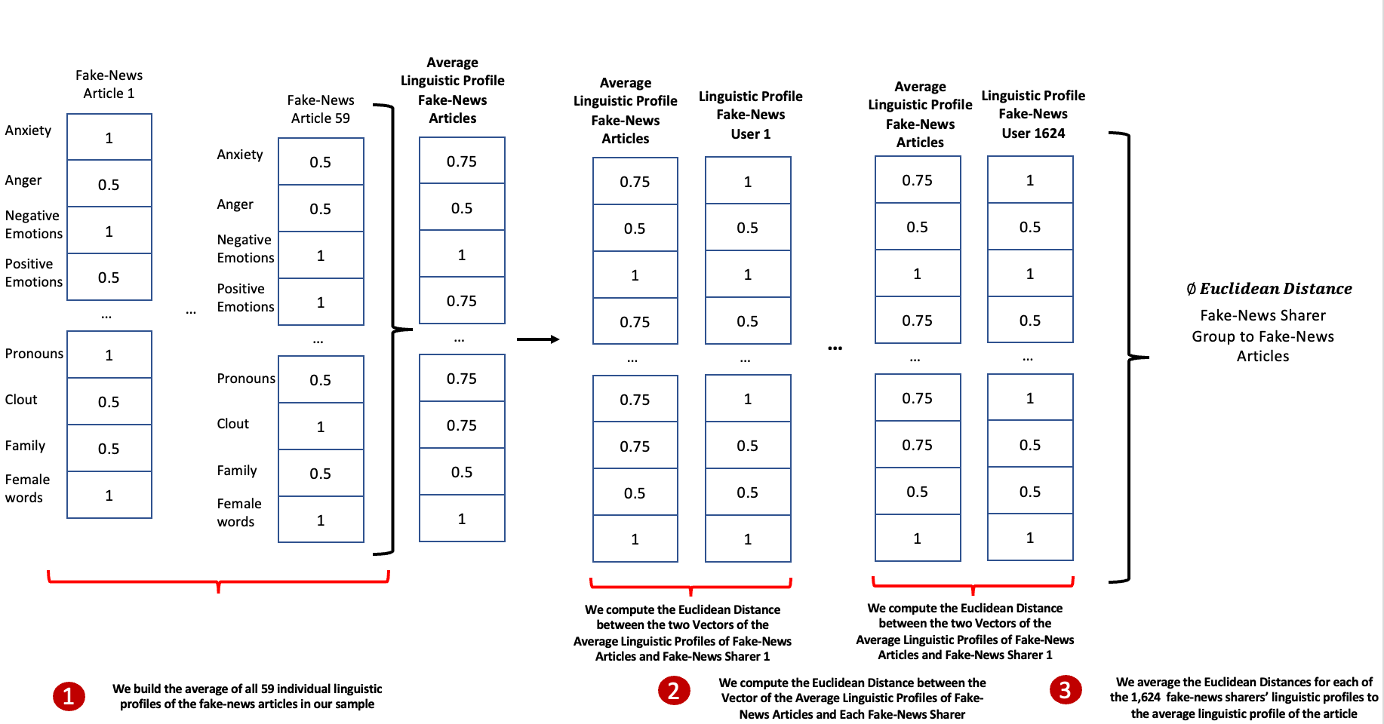}
    \label{EuclideanDistanceContentAnalysisLIWC}
\end{figure}

\end{singlespace*} 
\setcounter{table}{0}
\setcounter{figure}{0}
\renewcommand{\thetable}{WM\arabic{table}}
\renewcommand{\thefigure}{WM-\arabic{figure}}
\setlength{\baselineskip}{21pt plus.2pt}

\section*{Web Appendix M - Political vs. Non-political Fake-News Sharers}

\begin{center}
\tiny
\begin{longtable}{llllllll}
\caption{Political vs. Non Political Fake News} \\
\hline
 & No Political  & Political  & t-value & Random  & t-value  & Matched  & Fact \\
  & Fake News  & Fake News  & (p-value) & Group & (p-value) &  Group & Checkers \\ 
    & Sharers (1) & Sharers (2) & 1 vs. 2 & (3) & 3 vs. 2 &  &  \\ 
\hline
\endfirsthead
\multicolumn{8}{c}%
{{\bfseries \tablename\ \thetable{} -- Continued from previous page}} \\
\hline
 & No Political  & Political  & t-value & Random  & t-value  & Matched  & Fact \\
  & Fake News  & Fake News  & (p-value) & Group & (p-value) &  Group & Checkers \\ 
    & Sharers (1) & Sharers (2) & 1 vs. 2 & (3) & 3 vs. 2 &  &  \\ 
\hline
\endhead
\hline
\multicolumn{8}{r}{{Continued on next page}} \\
\endfoot
\hline
\endlastfoot
Word Count & 21423.407 & 20559.312 &  0.826 ( 0.410 ) & 18401.074 &   7.809 ( 0.000 ) & 22328.980 & 23477.387 \\ 
  Analytic &    80.425 &    82.245 & -1.732 ( 0.085 ) &    69.411 &  20.724 ( 0.000 ) &    75.493 &    76.995 \\ 
  Clout &    69.600 &    73.290 & -5.141 ( 0.000 ) &    63.849 &  13.405 ( 0.000 ) &    66.450 &    69.547 \\ 
  Authentic &    26.672 &    19.344 &  6.032 ( 0.000 ) &    42.021 & -24.844 ( 0.000 ) &    37.226 &    22.447 \\ 
  Emotional Tone &    52.912 &    39.025 &  5.839 ( 0.000 ) &    71.868 & -22.100 ( 0.000 ) &    71.198 &    47.859 \\ 
  Words $>$ 6 letters &    23.989 &    25.852 & -4.983 ( 0.000 ) &    20.866 &  17.529 ( 0.000 ) &    22.279 &    24.552 \\ 
  Dictionary &    71.998 &    69.364 &  5.513 ( 0.000 ) &    75.331 & -14.511 ( 0.000 ) &    73.861 &    71.562 \\ 
  Function Words &    39.186 &    37.932 &  2.801 ( 0.006 ) &    41.335 & -10.777 ( 0.000 ) &    40.323 &    40.416 \\ 
  Pronoun &    10.198 &     9.387 &  3.051 ( 0.003 ) &    12.597 & -20.187 ( 0.000 ) &    11.381 &    10.825 \\ 
  Personal Pronouns &     6.588 &     6.012 &  3.124 ( 0.002 ) &     8.893 & -24.549 ( 0.000 ) &     7.773 &     6.908 \\ 
  1st Person Singluar &     2.703 &     1.817 &  7.864 ( 0.000 ) &     4.752 & -26.804 ( 0.000 ) &     3.814 &     2.709 \\ 
  1st Person Plural &     0.654 &     0.709 & -1.887 ( 0.061 ) &     0.670 &  -1.028 ( 0.304 ) &     0.702 &     0.678 \\ 
  2nd Person &     1.759 &     1.664 &  1.132 ( 0.259 ) &     2.440 & -18.089 ( 0.000 ) &     2.194 &     1.840 \\ 
  3rd Personl Singular &     0.897 &     1.162 & -4.980 ( 0.000 ) &     0.609 &  16.710 ( 0.000 ) &     0.612 &     1.037 \\ 
  3rd Person Plural &     0.575 &     0.661 & -2.482 ( 0.014 ) &     0.428 &  15.025 ( 0.000 ) &     0.457 &     0.644 \\ 
  Impersonal Pronouns &     3.605 &     3.370 &  2.287 ( 0.023 ) &     3.698 &  -2.437 ( 0.015 ) &     3.603 &     3.912 \\ 
  Articles &     5.223 &     4.842 &  3.837 ( 0.000 ) &     4.750 &  11.009 ( 0.000 ) &     5.059 &     5.295 \\ 
  Prepositions &    11.608 &    11.937 & -1.726 ( 0.086 ) &    10.712 &  11.878 ( 0.000 ) &    11.187 &    11.194 \\ 
  Auxiliary Verbs &     5.713 &     5.684 &  0.216 ( 0.829 ) &     6.125 &  -7.686 ( 0.000 ) &     5.872 &     6.122 \\ 
  Adverbs &     3.731 &     3.518 &  2.420 ( 0.016 ) &     4.424 & -16.058 ( 0.000 ) &     4.134 &     3.985 \\ 
  Conjunctions &     3.535 &     3.299 &  2.859 ( 0.005 ) &     3.814 &  -7.704 ( 0.000 ) &     3.685 &     3.838 \\ 
  Negate &     0.892 &     0.944 & -1.601 ( 0.111 ) &     0.940 &  -3.484 ( 0.000 ) &     0.852 &     1.007 \\ 
  Verbs &    12.263 &    11.892 &  1.947 ( 0.053 ) &    13.520 & -14.631 ( 0.000 ) &    12.860 &    12.620 \\ 
  Adjectives &     4.418 &     4.145 &  3.857 ( 0.000 ) &     4.517 &  -2.965 ( 0.003 ) &     4.556 &     4.155 \\ 
  Compare &     1.724 &     1.661 &  2.028 ( 0.044 ) &     1.702 &   1.037 ( 0.300 ) &     1.734 &     1.745 \\ 
  Interrogatives &     1.274 &     1.285 & -0.348 ( 0.728 ) &     1.288 &  -0.992 ( 0.321 ) &     1.224 &     1.356 \\ 
  Number &     0.685 &     0.655 &  1.688 ( 0.093 ) &     0.657 &   2.494 ( 0.013 ) &     0.671 &     0.677 \\ 
  Quantifiers &     1.573 &     1.454 &  2.865 ( 0.005 ) &     1.661 &  -4.299 ( 0.000 ) &     1.692 &     1.628 \\ 
  Affect &     6.471 &     6.311 &  1.245 ( 0.215 ) &     7.275 & -14.377 ( 0.000 ) &     6.829 &     6.263 \\ 
  Positive Emotions &     4.005 &     3.472 &  3.807 ( 0.000 ) &     5.096 & -20.800 ( 0.000 ) &     4.860 &     3.729 \\ 
  Negative Emotions &     2.418 &     2.793 & -4.558 ( 0.000 ) &     2.128 &   9.155 ( 0.000 ) &     1.917 &     2.483 \\ 
  Anxiety &     0.280 &     0.335 & -4.145 ( 0.000 ) &     0.217 &  13.457 ( 0.000 ) &     0.209 &     0.283 \\ 
  Anger &     0.992 &     1.234 & -4.664 ( 0.000 ) &     0.852 &   6.927 ( 0.000 ) &     0.736 &     1.033 \\ 
  Sadness &     0.419 &     0.426 & -0.571 ( 0.568 ) &     0.435 &  -2.331 ( 0.020 ) &     0.385 &     0.424 \\ 
  Social &     8.759 &     8.930 & -1.035 ( 0.302 ) &     9.088 &  -4.459 ( 0.000 ) &     8.765 &     8.978 \\ 
  Family &     0.409 &     0.350 &  3.244 ( 0.001 ) &     0.479 &  -5.908 ( 0.000 ) &     0.423 &     0.360 \\ 
  Friend &     0.331 &     0.268 &  4.415 ( 0.000 ) &     0.454 & -11.814 ( 0.000 ) &     0.403 &     0.304 \\ 
  Female Words &     0.612 &     0.629 & -0.622 ( 0.534 ) &     0.580 &   2.293 ( 0.022 ) &     0.512 &     0.603 \\ 
  Male Words &     1.080 &     1.224 & -3.022 ( 0.003 ) &     0.893 &  10.417 ( 0.000 ) &     0.867 &     1.182 \\ 
  Cognitive Processes &     8.128 &     7.876 &  1.758 ( 0.080 ) &     8.342 &  -3.037 ( 0.002 ) &     8.146 &     8.753 \\ 
  Insight &     1.693 &     1.659 &  1.013 ( 0.312 ) &     1.619 &   3.924 ( 0.000 ) &     1.636 &     1.798 \\ 
  Cause &     1.341 &     1.329 &  0.461 ( 0.645 ) &     1.281 &   4.020 ( 0.000 ) &     1.308 &     1.395 \\ 
  Discrepancies &     1.257 &     1.204 &  1.538 ( 0.126 ) &     1.382 &  -7.910 ( 0.000 ) &     1.279 &     1.345 \\ 
  Tentative &     1.777 &     1.640 &  2.670 ( 0.008 ) &     1.886 &  -4.972 ( 0.000 ) &     1.865 &     1.975 \\ 
  Certain &     1.315 &     1.336 & -0.504 ( 0.615 ) &     1.355 &  -2.549 ( 0.011 ) &     1.292 &     1.367 \\ 
  Differentiation &     1.823 &     1.725 &  2.062 ( 0.040 ) &     1.873 &  -2.167 ( 0.030 ) &     1.817 &     2.065 \\ 
  Perceptional Processes &     2.773 &     2.435 &  6.315 ( 0.000 ) &     2.838 &  -1.870 ( 0.062 ) &     2.725 &     2.482 \\ 
  See &     1.424 &     1.207 &  5.073 ( 0.000 ) &     1.393 &   1.113 ( 0.266 ) &     1.360 &     1.191 \\ 
  Hear &     0.803 &     0.797 &  0.266 ( 0.790 ) &     0.746 &   3.637 ( 0.000 ) &     0.722 &     0.786 \\ 
  Feel &     0.410 &     0.334 &  6.254 ( 0.000 ) &     0.530 & -14.513 ( 0.000 ) &     0.476 &     0.380 \\ 
  Bio &     2.044 &     1.713 &  5.668 ( 0.000 ) &     2.616 & -14.992 ( 0.000 ) &     2.356 &     1.799 \\ 
  Body &     0.567 &     0.495 &  3.382 ( 0.001 ) &     0.764 & -14.195 ( 0.000 ) &     0.636 &     0.517 \\ 
  Health &     0.600 &     0.522 &  3.526 ( 0.000 ) &     0.592 &   0.485 ( 0.627 ) &     0.573 &     0.530 \\ 
  Sexual &     0.223 &     0.246 & -1.179 ( 0.240 ) &     0.232 &  -1.081 ( 0.280 ) &     0.194 &     0.229 \\ 
  Ingestion &     0.571 &     0.402 &  5.318 ( 0.000 ) &     0.763 &  -8.742 ( 0.000 ) &     0.762 &     0.452 \\ 
  Drives &     7.854 &     8.274 & -3.397 ( 0.001 ) &     7.616 &   4.265 ( 0.000 ) &     7.676 &     7.575 \\ 
  Affiliation &     2.158 &     2.010 &  3.035 ( 0.003 ) &     2.553 & -11.920 ( 0.000 ) &     2.490 &     2.026 \\ 
  Achieve &     1.358 &     1.365 & -0.106 ( 0.916 ) &     1.510 &  -7.590 ( 0.000 ) &     1.574 &     1.283 \\ 
  Power &     3.262 &     3.966 & -6.068 ( 0.000 ) &     2.311 &  25.272 ( 0.000 ) &     2.468 &     3.180 \\ 
  Reward &     1.443 &     1.348 &  0.950 ( 0.343 ) &     1.905 & -22.906 ( 0.000 ) &     1.828 &     1.353 \\ 
  Risk &     0.607 &     0.677 & -3.358 ( 0.001 ) &     0.438 &  17.566 ( 0.000 ) &     0.452 &     0.619 \\ 
  Past Focus &     2.471 &     2.409 &  1.077 ( 0.283 ) &     2.463 &   0.287 ( 0.774 ) &     2.380 &     2.579 \\ 
  Present Focus &     8.420 &     8.080 &  2.255 ( 0.025 ) &     9.756 & -19.688 ( 0.000 ) &     9.233 &     8.574 \\ 
  Future Focus &     1.019 &     0.998 &  0.915 ( 0.361 ) &     1.312 & -21.390 ( 0.000 ) &     1.199 &     0.980 \\ 
  Relativity &    12.675 &    12.282 &  1.796 ( 0.074 ) &    13.325 &  -7.134 ( 0.000 ) &    13.381 &    11.645 \\ 
  Motion &     1.702 &     1.644 &  1.591 ( 0.113 ) &     1.963 & -12.889 ( 0.000 ) &     1.899 &     1.552 \\ 
  Space &     6.479 &     6.689 & -1.304 ( 0.194 ) &     6.002 &   7.726 ( 0.000 ) &     6.312 &     6.005 \\ 
  Time &     4.622 &     4.066 &  5.594 ( 0.000 ) &     5.566 & -18.587 ( 0.000 ) &     5.355 &     4.200 \\ 
  Work &     2.662 &     3.001 & -3.677 ( 0.000 ) &     2.082 &  11.646 ( 0.000 ) &     2.430 &     2.611 \\ 
  Leisure &     2.064 &     1.582 &  4.036 ( 0.000 ) &     2.208 &  -2.751 ( 0.006 ) &     2.165 &     1.595 \\ 
  Home &     0.399 &     0.377 &  1.225 ( 0.222 ) &     0.467 &  -4.595 ( 0.000 ) &     0.460 &     0.349 \\ 
  Money &     0.968 &     1.048 & -1.787 ( 0.075 ) &     0.748 &   8.384 ( 0.000 ) &     0.917 &     0.920 \\ 
  Religion &     0.530 &     0.523 &  0.125 ( 0.901 ) &     0.446 &   3.412 ( 0.001 ) &     0.431 &     0.414 \\ 
  Death &     0.338 &     0.354 & -0.666 ( 0.506 ) &     0.181 &  21.210 ( 0.000 ) &     0.179 &     0.279 \\ 
  Informal &     1.929 &     1.626 &  3.812 ( 0.000 ) &     3.262 & -20.548 ( 0.000 ) &     2.636 &     1.865 \\ 
  Swear &     0.392 &     0.388 &  0.106 ( 0.916 ) &     0.683 & -12.519 ( 0.000 ) &     0.509 &     0.403 \\ 
  Netspeak &     0.971 &     0.748 &  5.364 ( 0.000 ) &     1.865 & -20.035 ( 0.000 ) &     1.453 &     0.853 \\ 
  Assent &     0.390 &     0.332 &  2.587 ( 0.010 ) &     0.571 & -12.644 ( 0.000 ) &     0.540 &     0.400 \\ 
  Nonfluencies &     0.221 &     0.188 &  2.467 ( 0.014 ) &     0.281 &  -9.265 ( 0.000 ) &     0.262 &     0.254 \\ 
  Filler &     0.021 &     0.018 &  1.548 ( 0.123 ) &     0.039 & -13.928 ( 0.000 ) &     0.031 &     0.023 \\ 
  Openness &     0.620 &     0.622 & -0.354 ( 0.724 ) &     0.554 &  20.836 ( 0.000 ) &     0.570 &     0.623 \\ 
  Conscientuousness &     0.520 &     0.521 & -0.130 ( 0.897 ) &     0.539 &  -6.862 ( 0.000 ) &     0.558 &     0.512 \\ 
  Extraversion &     0.416 &     0.404 &  1.641 ( 0.103 ) &     0.498 & -23.494 ( 0.000 ) &     0.475 &     0.396 \\ 
  Agreeableness &     0.448 &     0.403 &  6.749 ( 0.000 ) &     0.501 & -15.732 ( 0.000 ) &     0.502 &     0.426 \\ 
  Neuroticism &     0.419 &     0.441 & -3.435 ( 0.001 ) &     0.410 &   2.956 ( 0.003 ) &     0.392 &     0.439 \\ 
  Negative Low Arousal &     0.379 &     0.379 & -0.006 ( 0.996 ) &     0.430 &  -8.464 ( 0.000 ) &     0.382 &     0.394 \\ 
  Negative High Arousal &     0.853 &     1.031 & -4.251 ( 0.000 ) &     0.859 &  -0.340 ( 0.734 ) &     0.735 &     0.899 \\ 
  Positive Low Arousal &     1.498 &     1.223 &  6.230 ( 0.000 ) &     2.055 & -23.119 ( 0.000 ) &     1.861 &     1.422 \\ 
  Positive High Arousal &     0.847 &     0.642 &  5.484 ( 0.000 ) &     1.297 & -21.510 ( 0.000 ) &     1.164 &     0.711 \\ 
   \hline
\end{longtable}
\end{center} 
\setcounter{table}{0}
\setcounter{figure}{0}
\renewcommand{\thetable}{WN-\arabic{table}}
\renewcommand{\thefigure}{WN-\arabic{figure}}

\section*{Web Appendix N - Details on Fake-News Sharer Predictions}

\begin{singlespace*}

\subsection{ROC curves and Youden Index}

\noindent The ROC curve for the BART model with all variables is in Figure \ref{fig:ROCall}.\\

\begin{figure}[htbp]
     \centering
     \caption{ROC Curves}
     \footnotesize
     \begin{subfigure}[b]{0.40\textwidth}
         \centering
         \includegraphics[width=\textwidth]{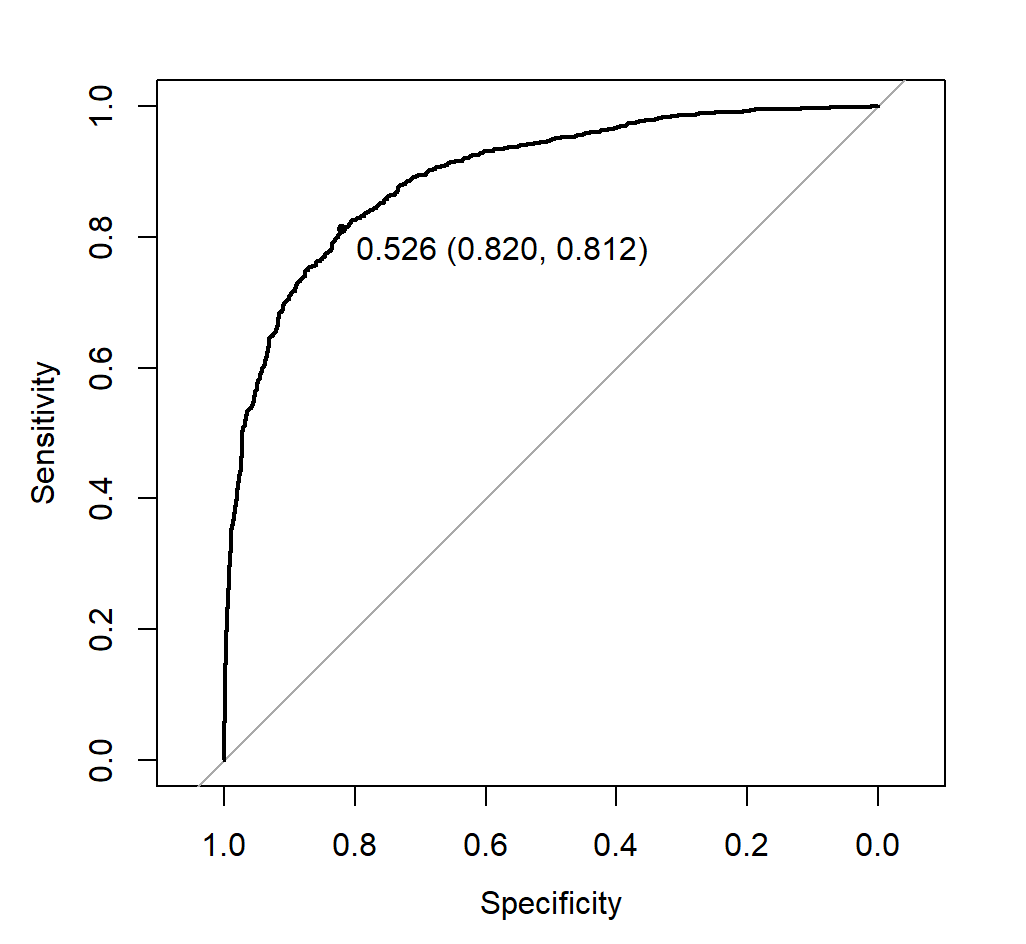}
         \caption{BART: Baseline Only (AUC=0.8946)}
         \label{fig:ROCBARTbaseline}
     \end{subfigure}
     \hfill
      \begin{subfigure}[b]{0.40\textwidth}
         \centering
         \includegraphics[width=\textwidth]{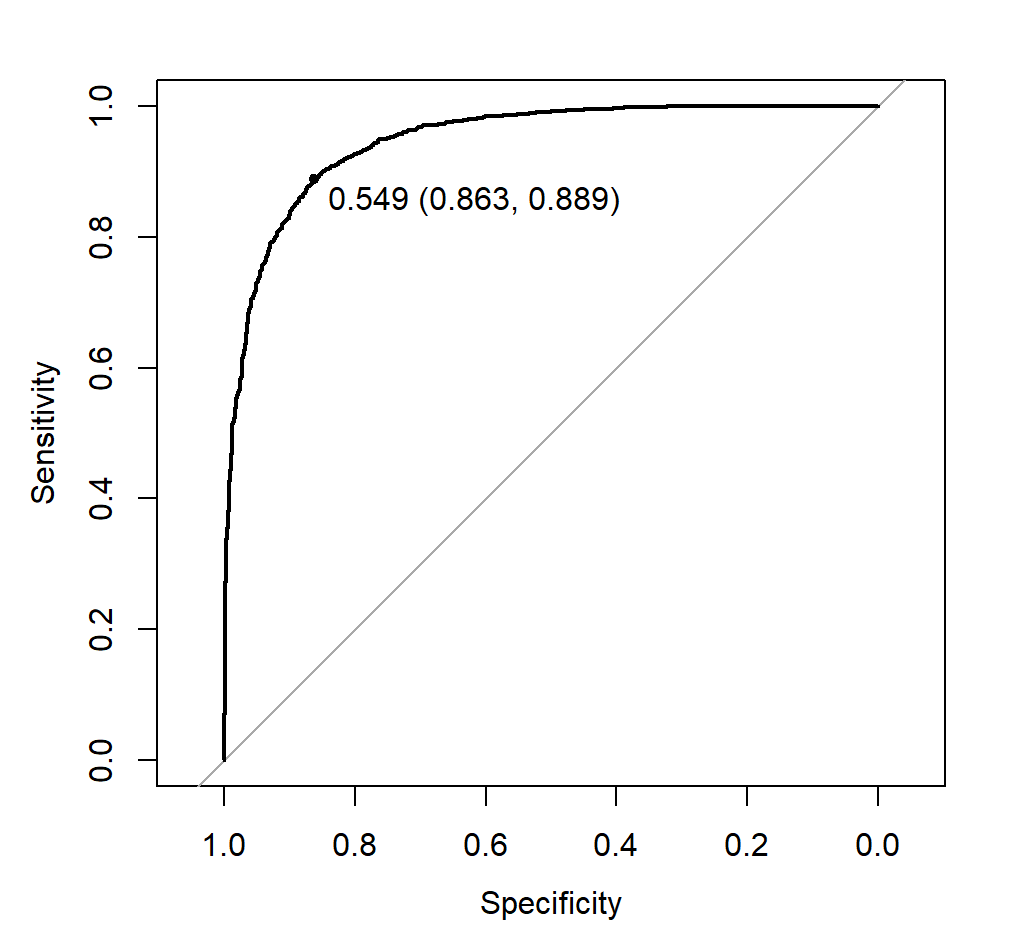}
         \caption{BART: Baseline \& Textual (AUC=0.9477)}
         \label{fig:ROCBARTbaselinetextual}
     \end{subfigure}
     \hfill
     \begin{subfigure}[b]{0.43\textwidth}
         \centering
         \includegraphics[width=\textwidth]{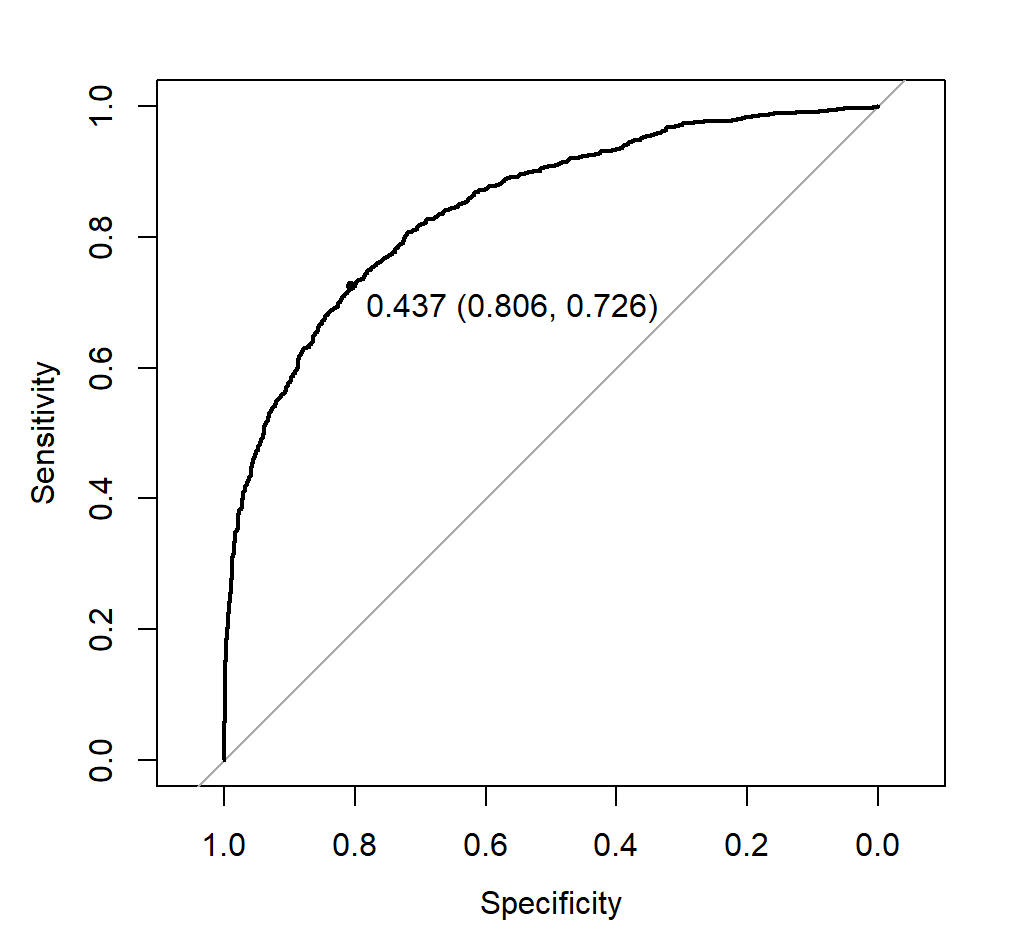}
         \caption{LR: Baseline Only (AUC=0.8462)}
         \label{fig:ROCLRbaseline}
     \end{subfigure}
     \hfill
     \begin{subfigure}[b]{0.43\textwidth}
         \centering
         \includegraphics[width=\textwidth]{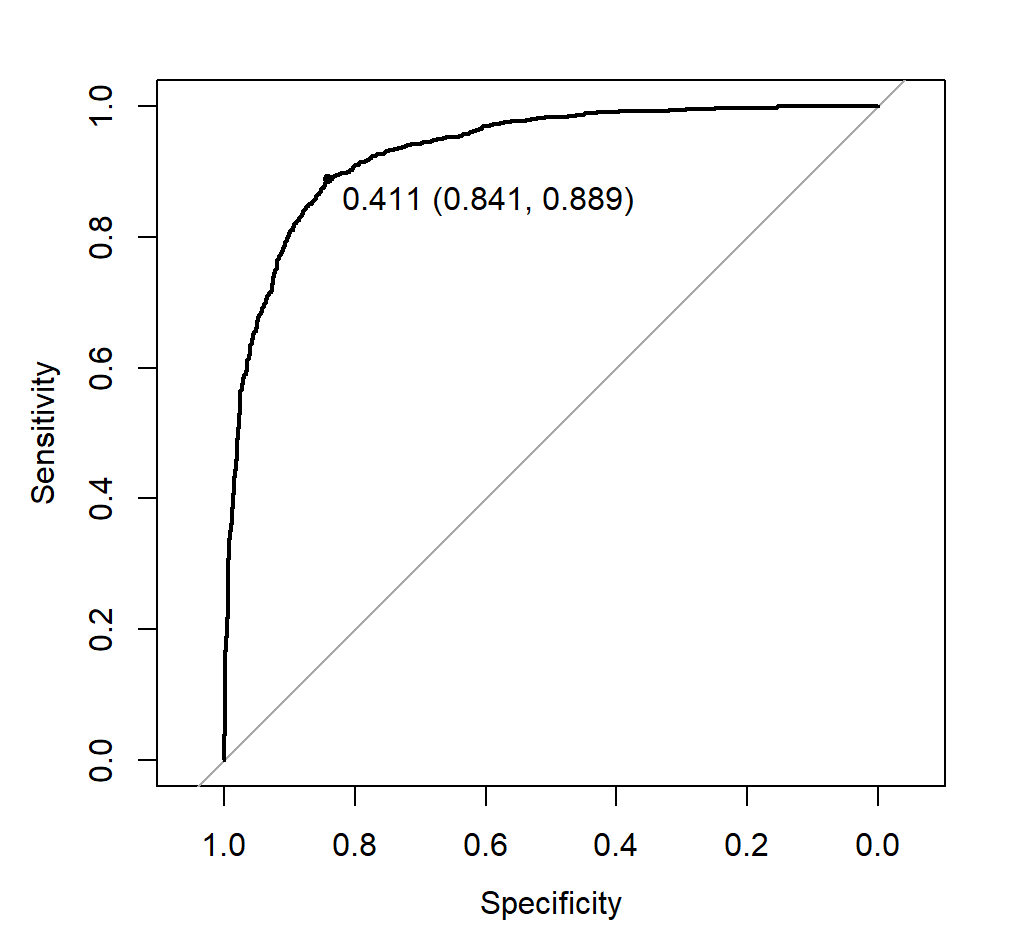}
         \caption{LR: Baseline \& Textual (AUC=0.9321)}
         \label{fig:five over x}
     \end{subfigure}        
      \label{fig:ROCall}

    \caption*{\footnotesize Note: In each subfigure, in $X (Y,Z)$, $X$ denotes the in-sample threshold that maximizes the ROC (Youden Index) while $Y$ denotes true-negative rate  (specificity) and $Z$ denotes true-positive rate (sensitivity).}
      
\end{figure}

\newpage

\subsection*{Post-hoc classification of fact-check sharers} 

We trained our model predicting fake-news sharers against a specific group of non-sharers of fake-news - random Twitter users. However, as our descriptive analyses showed, not all non-sharers of fake-news are alike. In particular, fact-check sharers may be difficult to distinguish from fake-news sharers due to their (in some respects) similar socio-demographics, social media activity, and even emotions. Our models did not use fact-check sharers for training or validation. Therefore, it is important to ask: how likely is it that a model trained to predict fake-news sharers using random Twitter users would incorrectly assign the label of fake-news to fact-check sharers? 

We used both cross-validated BART models to investigate how the models compare in their classification of fact-check sharers. We predicted the probability of each fact-check sharer being a fake-news sharer. The densities are presented in Figure \ref{fig:factcheckpred}. For the model with only baseline variables, the median predicted probability of a fact-check sharer sharing fake news is 24.31\% (mean = 30.34\%), which can be considered a particularly harmful misprediction. In contrast, the model with textual cues predicted a lower median probability for a fact-check sharer to share fake-news at 15.85\% (a reduction of 35\%; mean = 25.22\%). That is, the model with textual cues is better at labeling fact-check sharers as non-sharers of fake-news. 

\begin{figure}[htbp]
    \caption{Snopes dataset: Predictions of Fake-news sharing for Fact-check Sharers}
    \centering
    \includegraphics[width=0.7\columnwidth]{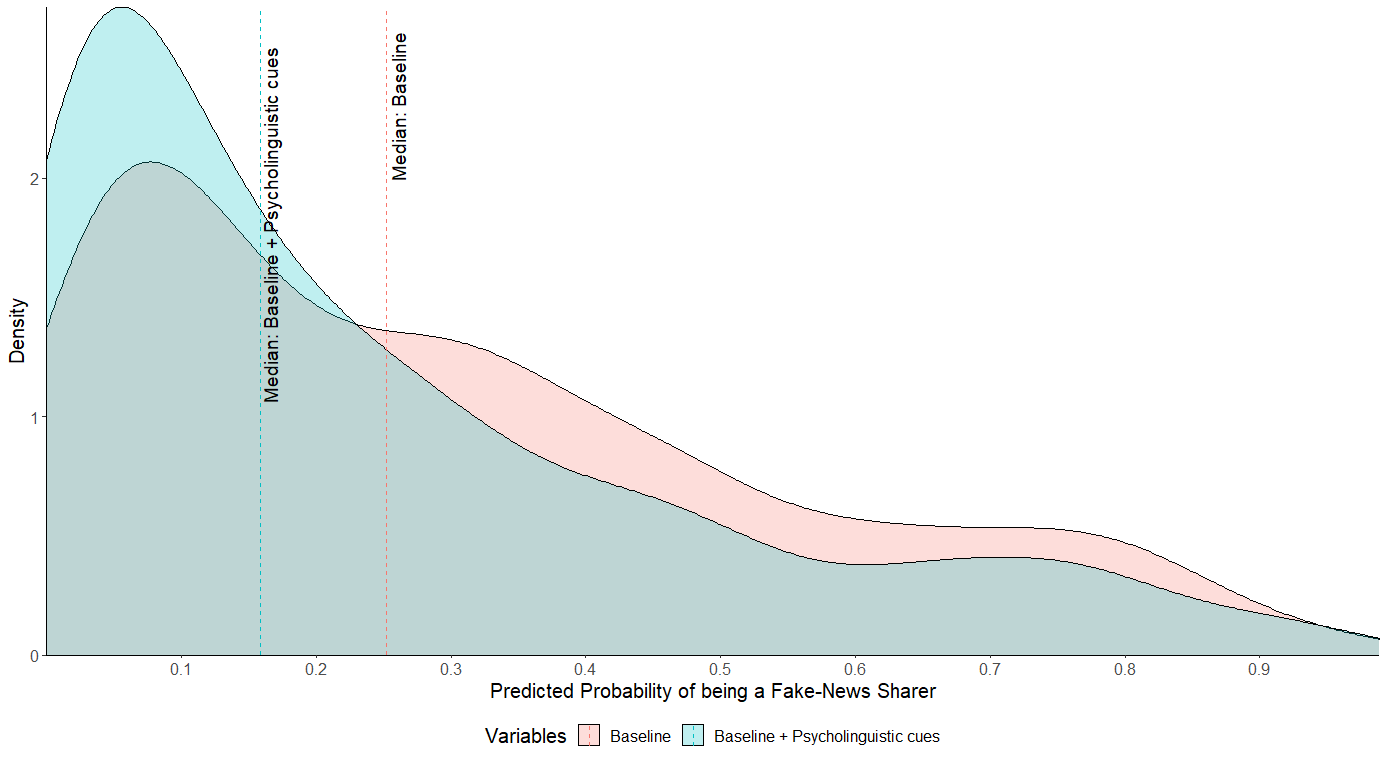}
    \caption*{\footnotesize Note: Median/Mean for Baseline model 0.2431/0.3034. Median/Mean for Baseline + Textual cues: 0.1586/0.2522}
    \label{fig:factcheckpred}
\end{figure}

\end{singlespace*}
\setcounter{table}{0}
\setcounter{figure}{0}
\renewcommand{\thetable}{WO\arabic{table}}
\renewcommand{\thefigure}{WO-\arabic{figure}}

\section*{Web Appendix O - Moderation of Content Characteristics Study 3}

\begin{singlespace*}

In the manuscript, we report that one's trait anger moderates the relationship between anger reduction and sharing through reducing the importance to share surprising news. In Table \ref{tab:moderations_anger}, we present the moderation analyses where trait anger moderates the effect of the anger reduction manipulation for each of the five content characteristics.

\begin{landscape}

 \begin{table}[!htbp] \centering 
 \tiny
   \caption{Regression Moderation of Content Characteristics} 
   \label{tab:moderations_anger} 
 \begin{tabular}{@{\extracolsep{5pt}}lD{.}{.}{-3} D{.}{.}{-3} D{.}{.}{-3} D{.}{.}{-3} D{.}{.}{-3} D{.}{.}{-3} D{.}{.}{-3} D{.}{.}{-3} } 
 \\[-1.8ex]\hline 
 \hline \\[-1.8ex] 
  & \multicolumn{8}{c}{\textit{Dependent variable:}} \\ 
 \cline{2-9} 
  & \multicolumn{1}{c}{Accurate} & \multicolumn{1}{c}{Surprising} & \multicolumn{1}{c}{Interesting} & \multicolumn{1}{c}{Aligned with Politics} & \multicolumn{1}{c}{Funny} & \multicolumn{1}{c}{Sharing Intentions} & \multicolumn{1}{c}{Believed True Shared} & \multicolumn{1}{c}{Believed False Shared} \\ 
 \\[-1.8ex] & \multicolumn{1}{c}{(1)} & \multicolumn{1}{c}{(2)} & \multicolumn{1}{c}{(3)} & \multicolumn{1}{c}{(4)} & \multicolumn{1}{c}{(5)} & \multicolumn{1}{c}{(6)} & \multicolumn{1}{c}{(7)} & \multicolumn{1}{c}{(8)}\\ 
 \hline \\[-1.8ex] 
   (Intercept) & 4.191^{***} & 2.778^{***} & 3.771^{***} & 3.411^{***} & 3.359^{***} & 2.267^{***} & 1.427 & 1.899^{***} \\ 
   & (0.238) & (0.291) & (0.241) & (0.310) & (0.321) & (0.376) & (1.067) & (0.621) \\ 
   & & & & & & & & \\ 
  Reduced Situational Anger Condition & 0.104 & -0.199^{*} & 0.067 & -0.016 & 0.018 & 0.105 & 0.012 & 0.179 \\ 
   & (0.099) & (0.121) & (0.100) & (0.129) & (0.133) & (0.103) & (0.293) & (0.171) \\ 
   & & & & & & & & \\ 
  Trait Anger & -0.100^{*} & 0.260^{***} & 0.046 & 0.251^{***} & 0.046 & 0.224^{***} & 0.477^{***} & 0.389^{***} \\ 
   & (0.055) & (0.067) & (0.056) & (0.072) & (0.074) & (0.059) & (0.169) & (0.098) \\ 
   & & & & & & & & \\ 
     Reduced Situational Anger Condition x Trait Anger & -0.141 & -0.268^{**} & -0.175 & -0.137 & 0.039 & -0.050 & 0.096 & -0.086 \\ 
   & (0.109) & (0.133) & (0.110) & (0.141) & (0.147) & (0.114) & (0.323) & (0.188) \\ 
   & & & & & & & & \\ 
  Religion & -0.018 & 0.082^{*} & 0.007 & 0.061 & 0.017 & 0.024 & 0.127 & 0.021 \\ 
   & (0.039) & (0.047) & (0.039) & (0.050) & (0.052) & (0.040) & (0.115) & (0.067) \\ 
   & & & & & & & & \\ 
  Age & 0.005 & -0.0001 & 0.005 & -0.009 & -0.005 & -0.008^{*} & -0.013 & -0.014^{*} \\ 
   & (0.005) & (0.006) & (0.005) & (0.006) & (0.006) & (0.005) & (0.013) & (0.008) \\ 
   & & & & & & & & \\ 
  Male & -0.127 & -0.052 & -0.026 & -0.327^{**} & -0.020 & 0.244^{**} & 0.564^{*} & 0.356^{**} \\ 
   & (0.101) & (0.123) & (0.102) & (0.131) & (0.136) & (0.105) & (0.299) & (0.174) \\ 
   & & & & & & & & \\ 
  Asian Pacific & -0.052 & 0.485^{**} & -0.172 & -0.305 & 0.186 & -0.006 & 0.109 & -0.411 \\ 
   & (0.165) & (0.201) & (0.167) & (0.214) & (0.222) & (0.173) & (0.493) & (0.286) \\ 
   & & & & & & & & \\ 
  Black & -0.096 & 0.448^{**} & 0.201 & -0.189 & 0.210 & 0.131 & 0.214 & 0.263 \\ 
   & (0.154) & (0.187) & (0.155) & (0.199) & (0.207) & (0.161) & (0.456) & (0.265) \\ 
   & & & & & & & & \\ 
  Hispanic & -0.191 & 0.336 & 0.036 & -0.078 & 0.052 & -0.193 & -0.194 & -0.407 \\ 
   & (0.188) & (0.229) & (0.190) & (0.244) & (0.253) & (0.195) & (0.554) & (0.322) \\ 
   & & & & & & & & \\ 
  Democrat & 0.056 & -0.237 & 0.011 & -0.017 & -0.079 & -0.129 & -0.233 & -0.035 \\ 
   & (0.144) & (0.175) & (0.145) & (0.187) & (0.193) & (0.149) & (0.423) & (0.246) \\ 
   & & & & & & & & \\ 
  Independent & 0.029 & -0.119 & 0.148 & -0.452^{**} & -0.008 & -0.081 & -0.162 & -0.139 \\ 
   & (0.150) & (0.183) & (0.151) & (0.194) & (0.202) & (0.157) & (0.445) & (0.259) \\ 
   & & & & & & & & \\ 
  Accurate &  &  &  &  &  & -0.042 & -0.089 & -0.102 \\ 
   &  &  &  &  &  & (0.057) & (0.162) & (0.094) \\ 
   & & & & & & & & \\ 
  Surprising &  &  &  &  &  & 0.189^{***} & 0.312^{**} & 0.232^{***} \\ 
   &  &  &  &  &  & (0.051) & (0.144) & (0.084) \\ 
   & & & & & & & & \\ 
  Interesting &  &  &  &  &  & 0.068 & 0.258 & 0.053 \\ 
   &  &  &  &  &  & (0.060) & (0.171) & (0.100) \\ 
   & & & & & & & & \\ 
  Aligned With Politics &  &  &  &  &  & 0.099^{**} & 0.367^{***} & -0.055 \\ 
   &  &  &  &  &  & (0.044) & (0.124) & (0.072) \\ 
   & & & & & & & & \\ 
  Funny &  &  &  &  &  & -0.111^{**} & -0.255^{**} & -0.153^{**} \\ 
   &  &  &  &  &  & (0.045) & (0.127) & (0.074) \\ 
   & & & & & & & & \\ 
 \hline \\[-1.8ex] 
 \hline 
 \hline \\[-1.8ex] 
 \textit{Note:}  & \multicolumn{8}{r}{$^{*}$p$<$0.1; $^{**}$p$<$0.05; $^{***}$p$<$0.01} \\ 
 \end{tabular} 
 \end{table}

 \end{landscape}
\setcounter{table}{0}
\setcounter{figure}{0}
\renewcommand{\thetable}{WP\arabic{table}}
\renewcommand{\thefigure}{WP-\arabic{figure}}
\setlength{\baselineskip}{21pt plus.2pt}

\section*{Web Appendix P - Exclusion Field Study}

\begin{singlespace*}

\vspace{8pt}

Below we describe the exclusions of participants from our Qualtrics study in detail and also run multiple sample comparisons on the key variables.

Of the 1,947 participants that took the Qualtrics survey, all indicated that they frequently post on social media. 1,799 participants indicated that they had not received the AstraZeneca vaccine and 148 participants chose the option “not sure/do not prefer to say”. Thus, no participant indicated to have received the AstraZeneca vaccine. All participants indicated that they had either a public or private Twitter account. In total, of the 1,947 participants, no participant was excluded from the survey due to screener questions. 

Of the 1,947 participants, 1,301 participants refused to provide their Twitter credentials using the Twitter API. Another possibility for not signing up for our API could be that they indicated owning a Twitter account during the screener question, but actually did not have a Twitter account. We compare the sample of participants that authenticated themselves via the Twitter API to those participants who did not authenticate themselves regarding the demographic information that we assessed from all 1,947 participants before the authentication. Table \ref{tab:Comparison Demographics Participants} below shows the mean differences across the four demographics between the two participant samples. The results of the comparison show no significant differences between the two samples in terms of gender and age. Regarding political affiliation, we see that participants who authenticated their Twitter account via the API are slightly less conservative on average. They also tend to be also slightly more religious.

\begin{table}[H]
\caption{Demographics Participants that Authenticated their Twitter Account vs. Not Authenticated their Twitter Account}
\label{tab:Comparison Demographics Participants}
\begin{footnotesize}
\begin{tabular}{lccc}
\hline
 & \textbf{Authenticated } & \textbf{Not authenticated} &  \\
  & \textbf{via API (n=646)} & \textbf{via API (n=1,301)} &  \textbf{P-value}\\
  \hline
Gender & \multicolumn{1}{l}{\begin{tabular}[c]{@{}l@{}}Male: 264\\ Female: 381\\ Other: 1\\ Prefer not to answer: 0\end{tabular}} & \multicolumn{1}{l}{\begin{tabular}[c]{@{}l@{}}Male: 460\\ Female: 808\\ Other: 2\\ Prefer not to answer: 1\end{tabular}} & 0.4806 \\ \hline
Age & 43.47 & 42.63 & 0.2408 \\ \hline
\begin{tabular}[c]{@{}l@{}}Political Affiliation \\ (1 = very liberal; \\ 5 = very conservative)\end{tabular} & 3.35 & 3.51 & 0.0406 \\ \hline
\begin{tabular}[c]{@{}l@{}}Religion \\ (1 = I am not religious; \\ 5 = center of my life)\end{tabular} & 3 & 2.77 & 0.0004 \\ \hline
\end{tabular}
\end{footnotesize}
\end{table}

From the remaining 646 participants that authenticated their Twitter account, Qualtrics flagged 87 participants who failed 2 or more attention checks or were marked by Qualtrics as speedsters. An additional 4 participants were removed whose answers to the last question about the study's purpose were fully unrelated to the study they had taken. This led to a sample of 555 participants.

From the 555 participants of whom we collected all tweets, we removed in line with our restrictions of the Snopes sample, participants who have less than 100 words to extract textual cues meaningfully. This leads to the exclusion of 74 participants and a final sample of 481 participants. 

For 171 users gender guesser was not able to predict the gender. The algorithm classified these participants as “unknown”. For these users, we thus imputed the gender with the gender indicated in the survey. Our final sample thus includes 481 participants in total, of which 38 participants have a protected profile. The average duration of the survey was 623 seconds (sd = 435, min = 208, max = 4,567). 

Finally, we compare the final sample of Twitter users from our field experiment to the fake-news sharers and the random sample of Twitter users in our Snopes study (see Table \ref{tab:datadescription}). Specifically, we compare the two samples' different characteristics accessed from their text (socio-demographics, media followership, fake and act-check outlet followership, and textual cues assessed via LIWC). Table \ref{tab:datadescription} below demonstrates that there exist differences in terms of demographics between the participants of the field study and the random group, but we find no systematic differences concerning our focal variables - the textual cues extracted from the users' tweets. We find that Twitter users in our field experiment tend to be about three years older on average, have a higher propensity to be male, and show similar activity on Twitter. We find no difference in the followership of fake and fact-news outlets between the survey participants and the random sample. Finally, for our key variables the textual cues assessed via LIWC, we replicate the mean differences between the survey participants and the fake-news sharer group which we found between the fake-news sharers and the random sample. 

\begin{table}[ht]
\tiny
\centering
  \caption{Study 4 vs. Random Group and Fake-Sharers Snopes' Sample} 
  \label{tab:datadescription} 
\begin{tabular}{lccccccccc}
  \hline
 & Mean/Prop & Mean/Prop & D/H& t/z-value & p-value & Mean& D/H& t/z-value & p-value\\ 
  & Survey & Random & & & & Fake & &  & \\ 
  \hline
Word Count & 2055.472 & 18401.074 & 1.692 & -62.496 & 0.000 & 21280.361 & 1.870 & -67.237 & 0.000 \\ 
  Analytic & 74.715 & 69.411 & 0.301 & 6.137 & 0.000 & 80.682 & 0.436 & -7.314 & 0.000 \\ 
  Clout & 73.871 & 63.849 & 0.676 & 12.795 & 0.000 & 70.011 & 0.339 & 5.190 & 0.000 \\ 
  Authentic & 26.354 & 42.021 & 0.819 & -16.329 & 0.000 & 25.840 & 0.030 & 0.550 & 0.583 \\ 
  Emotional Tone & 68.977 & 71.868 & 0.120 & -2.096 & 0.036 & 51.260 & 0.656 & 12.444 & 0.000 \\ 
  Words $>$ 6 letters & 23.167 & 20.866 & 0.427 & 7.803 & 0.000 & 24.198 & 0.197 & -3.504 & 0.000 \\ 
  Dictionary & 72.790 & 75.331 & 0.349 & -5.803 & 0.000 & 71.690 & 0.154 & 2.515 & 0.012 \\ 
  Function Words & 38.722 & 41.335 & 0.415 & -6.739 & 0.000 & 39.017 & 0.047 & -0.760 & 0.448 \\ 
  Pronoun & 11.326 & 12.597 & 0.346 & -6.708 & 0.000 & 10.094 & 0.369 & 6.624 & 0.000 \\ 
  Personal Pronouns & 7.806 & 8.893 & 0.359 & -7.292 & 0.000 & 6.512 & 0.523 & 9.090 & 0.000 \\ 
  1st Person Singular & 3.691 & 4.752 & 0.416 & -8.738 & 0.000 & 2.600 & 0.573 & 9.635 & 0.000 \\ 
  1st Person Plural & 0.687 & 0.670 & 0.030 & 0.574 & 0.566 & 0.660 & 0.063 & 0.964 & 0.335 \\ 
  2nd Person & 2.226 & 2.440 & 0.170 & -3.164 & 0.002 & 1.743 & 0.462 & 7.418 & 0.000 \\ 
  3rd Person Singular & 0.674 & 0.609 & 0.127 & 1.858 & 0.064 & 0.924 & 0.414 & -6.923 & 0.000 \\ 
  3rd Person Plural & 0.528 & 0.428 & 0.323 & 4.232 & 0.000 & 0.584 & 0.150 & -2.321 & 0.021 \\ 
  Impersonal Pronouns & 3.517 & 3.698 & 0.160 & -2.615 & 0.009 & 3.576 & 0.048 & -0.839 & 0.402 \\ 
  Articles & 4.669 & 4.750 & 0.060 & -1.040 & 0.299 & 5.175 & 0.388 & -6.523 & 0.000 \\ 
  Prepositions & 10.491 & 10.712 & 0.095 & -1.662 & 0.097 & 11.648 & 0.509 & -8.725 & 0.000 \\ 
  Auxiliary Verbs & 5.608 & 6.125 & 0.303 & -4.843 & 0.000 & 5.703 & 0.056 & -0.889 & 0.374 \\ 
  Adverbs & 3.607 & 4.424 & 0.611 & -10.897 & 0.000 & 3.702 & 0.075 & -1.277 & 0.202 \\ 
  Conjunctions & 3.867 & 3.814 & 0.045 & 0.696 & 0.486 & 3.507 & 0.303 & 4.653 & 0.000 \\ 
  Negate & 0.830 & 0.940 & 0.228 & -3.567 & 0.000 & 0.896 & 0.147 & -2.154 & 0.032 \\ 
  Verbs & 12.248 & 13.520 & 0.465 & -8.016 & 0.000 & 12.215 & 0.013 & 0.214 & 0.830 \\ 
  Adjectives & 4.071 & 4.517 & 0.395 & -5.940 & 0.000 & 4.390 & 0.288 & -4.254 & 0.000 \\ 
  Compare & 1.488 & 1.702 & 0.302 & -5.139 & 0.000 & 1.719 & 0.375 & -5.679 & 0.000 \\ 
  Interrogatives & 1.157 & 1.288 & 0.250 & -3.845 & 0.000 & 1.273 & 0.241 & -3.442 & 0.001 \\ 
  Number & 0.652 & 0.657 & 0.011 & -0.155 & 0.877 & 0.682 & 0.075 & -0.890 & 0.374 \\ 
  Quantifiers & 1.643 & 1.661 & 0.026 & -0.438 & 0.662 & 1.560 & 0.137 & 2.093 & 0.037 \\ 
  Affect & 6.572 & 7.275 & 0.366 & -6.138 & 0.000 & 6.452 & 0.071 & 1.069 & 0.286 \\ 
  Positive Emotions & 4.830 & 5.096 & 0.148 & -2.302 & 0.022 & 3.941 & 0.520 & 7.748 & 0.000 \\ 
  Negative Emotions & 1.697 & 2.128 & 0.418 & -7.841 & 0.000 & 2.464 & 0.846 & -14.312 & 0.000 \\ 
  Anxiety & 0.197 & 0.217 & 0.126 & -1.702 & 0.089 & 0.286 & 0.509 & -7.466 & 0.000 \\ 
  Anger & 0.700 & 0.852 & 0.222 & -4.277 & 0.000 & 1.021 & 0.572 & -9.449 & 0.000 \\ 
  Sadness & 0.328 & 0.435 & 0.445 & -7.259 & 0.000 & 0.421 & 0.416 & -6.369 & 0.000 \\ 
  Social & 10.945 & 9.088 & 0.711 & 11.061 & 0.000 & 8.775 & 0.914 & 13.102 & 0.000 \\ 
  Family & 0.449 & 0.479 & 0.072 & -1.308 & 0.191 & 0.402 & 0.145 & 2.160 & 0.031 \\ 
  Friend & 0.360 & 0.454 & 0.243 & -4.074 & 0.000 & 0.325 & 0.115 & 1.569 & 0.117 \\ 
  Female Words & 0.273 & 0.580 & 0.776 & -17.987 & 0.000 & 0.615 & 0.952 & -20.533 & 0.000 \\ 
  Male Words & 0.862 & 0.893 & 0.051 & -0.786 & 0.432 & 1.094 & 0.391 & -5.899 & 0.000 \\ 
  Cognitive Processes & 7.806 & 8.342 & 0.235 & -3.897 & 0.000 & 8.096 & 0.134 & -2.128 & 0.034 \\ 
  Insight & 1.422 & 1.619 & 0.302 & -4.646 & 0.000 & 1.688 & 0.439 & -6.347 & 0.000 \\ 
  Cause & 1.309 & 1.281 & 0.046 & 0.624 & 0.533 & 1.340 & 0.055 & -0.680 & 0.497 \\ 
  Discrepancies & 1.263 & 1.382 & 0.210 & -3.123 & 0.002 & 1.251 & 0.023 & 0.329 & 0.742 \\ 
  Tentative & 1.863 & 1.886 & 0.032 & -0.506 & 0.613 & 1.761 & 0.143 & 2.229 & 0.026 \\ 
  Certain & 1.303 & 1.355 & 0.097 & -1.380 & 0.168 & 1.316 & 0.024 & -0.343 & 0.732 \\ 
  Differentiation & 1.701 & 1.873 & 0.230 & -3.585 & 0.000 & 1.811 & 0.148 & -2.293 & 0.022 \\ 
  Perceptional Processes & 2.521 & 2.838 & 0.281 & -4.449 & 0.000 & 2.737 & 0.197 & -3.042 & 0.002 \\ 
  See & 1.235 & 1.393 & 0.184 & -3.357 & 0.001 & 1.402 & 0.212 & -3.593 & 0.000 \\ 
  Hear & 0.704 & 0.746 & 0.076 & -1.058 & 0.290 & 0.802 & 0.178 & -2.470 & 0.014 \\ 
  Feel & 0.397 & 0.530 & 0.528 & -8.291 & 0.000 & 0.401 & 0.015 & -0.243 & 0.808 \\ 
  Bio & 2.194 & 2.616 & 0.294 & -4.728 & 0.000 & 2.010 & 0.159 & 2.122 & 0.034 \\ 
  Body & 0.621 & 0.764 & 0.233 & -3.317 & 0.001 & 0.559 & 0.126 & 1.476 & 0.141 \\ 
  Health & 0.539 & 0.592 & 0.103 & -1.997 & 0.046 & 0.593 & 0.125 & -2.100 & 0.036 \\ 
  Sexual & 0.264 & 0.232 & 0.072 & 0.890 & 0.374 & 0.226 & 0.094 & 1.075 & 0.283 \\ 
  Ingestion & 0.629 & 0.763 & 0.159 & -2.507 & 0.012 & 0.554 & 0.111 & 1.449 & 0.148 \\ 
  Drives & 8.238 & 7.616 & 0.318 & 4.929 & 0.000 & 7.907 & 0.184 & 2.653 & 0.008 \\ 
  Affiliation & 2.764 & 2.553 & 0.173 & 2.923 & 0.004 & 2.143 & 0.648 & 8.958 & 0.000 \\ 
  Achieve & 1.880 & 1.510 & 0.402 & 5.403 & 0.000 & 1.358 & 0.647 & 7.731 & 0.000 \\ 
  Power & 2.821 & 2.311 & 0.479 & 7.177 & 0.000 & 3.345 & 0.399 & -7.062 & 0.000 \\ 
  Reward & 2.207 & 1.905 & 0.320 & 3.966 & 0.000 & 1.430 & 0.824 & 10.196 & 0.000 \\ 
  Risk & 0.457 & 0.438 & 0.066 & 0.971 & 0.332 & 0.616 & 0.485 & -8.094 & 0.000 \\ 
  Past Focus & 2.351 & 2.463 & 0.122 & -1.897 & 0.058 & 2.465 & 0.132 & -1.953 & 0.051 \\ 
  Present Focus & 8.774 & 9.756 & 0.455 & -7.898 & 0.000 & 8.376 & 0.196 & 3.231 & 0.001 \\ 
  Future Focus & 1.046 & 1.312 & 0.513 & -7.988 & 0.000 & 1.015 & 0.072 & 0.952 & 0.341 \\ 
  Relativity & 11.595 & 13.325 & 0.643 & -11.485 & 0.000 & 12.637 & 0.381 & -6.860 & 0.000 \\ 
  Motion & 1.893 & 1.963 & 0.094 & -1.381 & 0.168 & 1.695 & 0.290 & 3.964 & 0.000 \\ 
  Space & 5.359 & 6.002 & 0.366 & -6.548 & 0.000 & 6.507 & 0.609 & -11.444 & 0.000 \\ 
  Time & 4.539 & 5.566 & 0.643 & -10.785 & 0.000 & 4.562 & 0.014 & -0.234 & 0.815 \\ 
  Work & 2.056 & 2.082 & 0.017 & -0.313 & 0.754 & 2.702 & 0.463 & -7.834 & 0.000 \\ 
  Leisure & 2.087 & 2.208 & 0.076 & -1.199 & 0.231 & 2.013 & 0.044 & 0.728 & 0.467 \\ 
  Home & 0.487 & 0.467 & 0.034 & 0.589 & 0.556 & 0.397 & 0.224 & 2.757 & 0.006 \\ 
  Money & 1.429 & 0.748 & 0.648 & 8.698 & 0.000 & 0.977 & 0.458 & 5.821 & 0.000 \\ 
  Religion & 0.424 & 0.446 & 0.030 & -0.517 & 0.605 & 0.528 & 0.139 & -2.388 & 0.017 \\ 
  Death & 0.239 & 0.181 & 0.288 & 3.703 & 0.000 & 0.341 & 0.367 & -6.181 & 0.000 \\ 
  Informal & 4.243 & 3.262 & 0.381 & 6.257 & 0.000 & 1.895 & 1.249 & 15.683 & 0.000 \\ 
   \hline
\end{tabular}
\end{table}

\begin{table}[ht]
\tiny
\centering
\begin{tabular}{lccccccccc}
  \hline
 & Mean/Prop & Mean/Prop & D/H& t/z-value & p-value & Mean/Prop& D& t/z-value & p-value\\ 
  & Survey & Random & & & & Fake & &  & \\ 
  \hline
  Swear & 0.490 & 0.683 & 0.216 & -3.993 & 0.000 & 0.392 & 0.166 & 2.164 & 0.031 \\ 
  Netspeak & 3.239 & 1.865 & 0.697 & 10.221 & 0.000 & 0.949 & 1.535 & 17.621 & 0.000 \\ 
  Assent & 2.327 & 0.571 & 1.539 & 16.392 & 0.000 & 0.382 & 1.709 & 18.180 & 0.000 \\ 
  Nonfluencies & 0.167 & 0.281 & 0.561 & -10.471 & 0.000 & 0.216 & 0.267 & -4.620 & 0.000 \\ 
  Filler & 0.023 & 0.039 & 0.316 & -5.046 & 0.000 & 0.020 & 0.076 & 1.009 & 0.314 \\ 
  Openness & 0.558 & 0.554 & 0.035 & 0.587 & 0.558 & 0.620 & 0.675 & -9.897 & 0.000 \\ 
  Conscientiousness & 0.548 & 0.539 & 0.091 & 1.756 & 0.080 & 0.520 & 0.379 & 5.926 & 0.000 \\ 
  Extraversion & 0.464 & 0.498 & 0.280 & -4.978 & 0.000 & 0.414 & 0.503 & 7.597 & 0.000 \\ 
  Agreeableness & 0.463 & 0.501 & 0.344 & -6.307 & 0.000 & 0.443 & 0.215 & 3.571 & 0.000 \\ 
  Neuroticism & 0.396 & 0.410 & 0.122 & -2.135 & 0.033 & 0.421 & 0.279 & -4.117 & 0.000 \\ 
  Age & 33.868 & 30.417 & 0.516 & 9.225 & 0.000 & 32.596 & 0.236 & 3.535 & 0.000 \\ 
  Gender (Female) & 0.559 & 0.443 & 0.257 & 20.046 & 0.000 & 0.367 & 0.413 & 56.669 & 0.000 \\ 
  DEM & 0.162 & 0.113 & 0.144 & 8.198 & 0.004 & 0.169 & 0.019 & 0.084 & 0.772 \\ 
  GOP & 0.146 & 0.161 & 0.043 & 0.573 & 0.449 & 0.340 & 0.462 & 67.072 & 0.000 \\ 
  DEM \& GOP & 0.058 & 0.059 & 0.002 & 0.000 & 1.000 & 0.144 & 0.290 & 24.306 & 0.000 \\ 
  No Party Followed & 0.634 & 0.668 & 0.070 & 1.756 & 0.185 & 0.347 & 0.582 & 126.642 & 0.000 \\ 
  Salon & 0.015 & 0.014 & 0.002 & 0.000 & 1.000 & 0.111 & 0.438 & 41.661 & 0.000 \\ 
  Wikileaks & 0.025 & 0.051 & 0.136 & 5.200 & 0.023 & 0.277 & 0.792 & 137.157 & 0.000 \\ 
  Dailybeast & 0.033 & 0.027 & 0.034 & 0.274 & 0.601 & 0.153 & 0.438 & 47.848 & 0.000 \\ 
  New Yorker & 0.048 & 0.087 & 0.157 & 7.454 & 0.006 & 0.198 & 0.481 & 60.440 & 0.000 \\ 
  Economist & 0.025 & 0.059 & 0.173 & 8.209 & 0.004 & 0.164 & 0.518 & 61.871 & 0.000 \\ 
  Wall Street Journal & 0.035 & 0.105 & 0.283 & 21.831 & 0.000 & 0.247 & 0.663 & 103.843 & 0.000 \\ 
  Washington Post & 0.083 & 0.098 & 0.053 & 0.851 & 0.356 & 0.259 & 0.482 & 66.516 & 0.000 \\ 
  New York Times & 0.119 & 0.169 & 0.145 & 6.960 & 0.008 & 0.316 & 0.490 & 72.689 & 0.000 \\ 
  CNN & 0.146 & 0.140 & 0.015 & 0.053 & 0.818 & 0.261 & 0.290 & 27.257 & 0.000 \\ 
  LA Times & 0.033 & 0.031 & 0.011 & 0.005 & 0.944 & 0.111 & 0.311 & 25.579 & 0.000 \\ 
  USA Today & 0.048 & 0.048 & 0.002 & 0.000 & 1.000 & 0.170 & 0.408 & 44.554 & 0.000 \\ 
  Buzz Feed & 0.050 & 0.064 & 0.060 & 1.045 & 0.307 & 0.131 & 0.290 & 23.764 & 0.000 \\ 
  Aljazeera English & 0.027 & 0.021 & 0.040 & 0.410 & 0.522 & 0.080 & 0.245 & 15.950 & 0.000 \\ 
  Fox News & 0.091 & 0.083 & 0.030 & 0.254 & 0.614 & 0.276 & 0.492 & 70.188 & 0.000 \\ 
  Drudge Report & 0.000 & 0.024 & 0.312 & 10.601 & 0.001 & 0.216 & 0.967 & 123.471 & 0.000 \\ 
  Friends Count & 691.898 & 763.677 & 0.029 & -0.696 & 0.487 & 2293.471 & 0.322 & -10.193 & 0.000 \\ 
  Followers Count & 488.638 & 1708.484 & 0.083 & -3.068 & 0.002 & 4449.825 & 0.124 & -4.531 & 0.000 \\ 
  Status Count & 9430.952 & 7239.498 & 0.105 & 1.211 & 0.227 & 30071.969 & 0.480 & -9.933 & 0.000 \\ 
  Negative Low Arousal & 0.334 & 0.430 & 0.421 & -6.391 & 0.000 & 0.379 & 0.228 & -3.070 & 0.002 \\ 
  Negative High Arousal & 0.664 & 0.859 & 0.310 & -5.594 & 0.000 & 0.873 & 0.429 & -6.257 & 0.000 \\ 
  Positive Low Arousal & 1.683 & 2.055 & 0.462 & -7.770 & 0.000 & 1.467 & 0.298 & 4.587 & 0.000 \\ 
  Positive High Arousal & 0.933 & 1.297 & 0.518 & -9.347 & 0.000 & 0.823 & 0.187 & 2.927 & 0.004 \\ 
Tweets Per Day & 3.025 & 2.452 & 0.100 & 1.265 & 0.207 & 10.547 & 0.529 & -13.082 & 0.000 \\
  Number Fake Outlets Followed & 0.183 & 0.238 & 0.079 & -1.748 & 0.081 & 2.010 & 0.668 & -23.208 & 0.000 \\ 
  Number Fact Outlets Followed & 0.040 & 0.043 & 0.016 & -0.285 & 0.776 & 0.241 & 0.371 & -10.579 & 0.000 \\ 
\\ 
   \hline
\end{tabular}
\end{table}

\end{singlespace*} 
\setcounter{table}{0}
\setcounter{figure}{0}
\renewcommand{\thetable}{WQ-\arabic{table}}
\renewcommand{\thefigure}{WQ-\arabic{figure}}
\doublespacing

\section*{Web Appendix Q - Heterogeneous Effect Analyses}

\begin{singlespace*}

In the manuscript, we provide the results of two different approaches to estimate how the power manipulation affects intentions to click or download the browser extension and intentions to join a fact-checking group (i.e., outcome unrelated to the power manipulation). 

In the first analysis, we use a linear regression that includes the manipulation and controls for socio-demographics (e.g., age, gender, religiosity, political affiliation) that are self-reported in the questionnaire (i.e., variables that would not be available for scoring based only using tweet data). Although the treatment assignment (use of power-related words in ad copy) is exogenous, unanticipated differences in pre-treatment covariates (i.e., textual cues from post histories) could simultaneously explain why the use of power-related words in ad copy affects them more or less. 

In the second analysis, we report the results of a two-step approach recommended by Chen, Sridhar, and Mittal (2021) to handle unexpected heterogeneity in treatment effects by accounting for pre-treatment covariates. Specifically, we use Causal Forests to estimate individual-level treatment effects of the power manipulation while accounting for our textual cues extracted from post histories. Then, in a second step, we use the best linear projection of the conditional average treatment effects using cluster and hetero-skedastic robust standard errors (HEC3) as the dependent variable with the same self-report survey measures (e.g., age, gender, religiosity, political affiliation) as control moderators along with the desire for control and personal sense of power trait scales from the post survey. The results are similar, as reported in the main text. Table \ref{tab:ComparisonPowerModels} shows the linear regression results (Table 9) with and without controls and the Average Treatment Effects from the Causal Forest. Next, we detail unexpected heterogeneity, followed by model specifications and robustness tests. 

\begin{table}[htbp]
\caption{Comparison linear regression with and without controls to causal forest}
\label{tab:ComparisonPowerModels}
\begin{tabular}{llll}
 & Without control & With controls (Table 9) & ATE \\
 \hline\hline
Click & 1.0433 p\textless{}.01 & 0.943 p\textless{}.01 & 0.974 {[}0.49, 1.46{]} \\
Download & 1.2169 p\textless{}.01 & 1.131 p\textless{}.01 & 1.153 {[}0.66, 1.64{]} \\
Join (unrelated) & 0.1591 p=.14 & 0.129 p=.23 & 0.145 {[}-0.07, 0.36{]}\\
\hline
\end{tabular}
\end{table}

\subsubsection*{Unexpected Heterogeneity}

In Figure \ref{fig:covbalancepower}, we show the "love-plot" (standardized mean-differences) of the pre-treatment covariates across the two conditions. An absolute standardized mean difference of more than 0.1 indicates a small effect size difference between the groups on the pre-treatment covariates. Regardless of the cause, there exist numerous differences between the two groups. For example, participants allocated to the high power condition use fewer anger-related words, fewer words overall, and are more agreeable (from the Big 5). Even if we find no difference in the conditions in terms of words related to power-related words in their own text, it is important to control for such differences in our estimation. 

\begin{figure}[htbp]
    \caption{Covariate balance between the power (high use of power-related words) and control conditions}
    \centering
    \includegraphics[width=\textwidth]{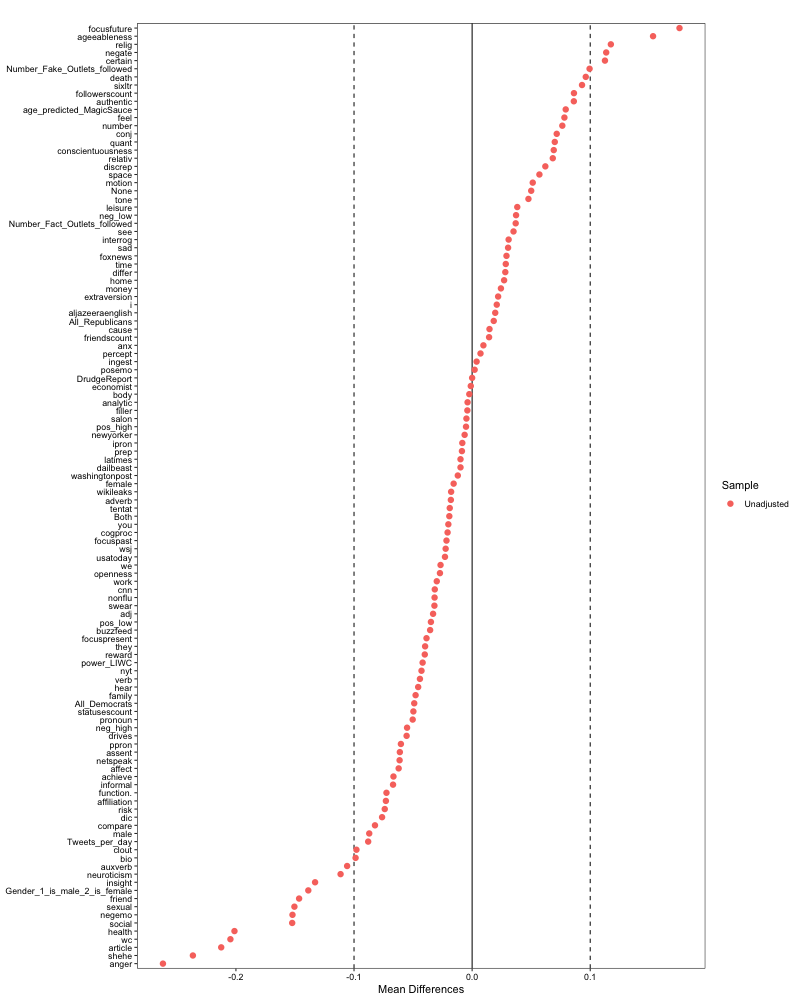}
    \label{fig:covbalancepower}
\end{figure}

\subsubsection*{Causal Forest Specification}

As previously stated for our predictive model, $X_i$ is a vector of pre-treatment covariates (i.e., user-level Twitter data and extracted information) for consumer $i$. Now, let $Y_i \in \mathbb{R}$ be our response (e.g., intentions to click or download the browser extension) and $W_i$ be an indicator set to 1 if the consumer received the ad-copy with power-related words message (0 otherwise). 

If we assume that treatment assignment $W_i$ is independent of the outcome $Y_i$ conditional on pre-treatment covariates $X_i$, causal forests aim to estimate $\tau(x)=\mathbb{E}[y_i^{1}-y_i^{0}|X_i=x]$, the treatment effect at value $x$, where $Y_i^{1}$ is the outcome if consumer $i$ is exposed to the ad copy with power-related words, and $Y_i^{0}$ if they are not. They do so by splitting the data to maximize the difference across observation splits in how the treatment affects the outcome. 

To estimate this function, we use the implementation by Tibshirani et al. (2022) (GRF package) which separately first estimates two regression forests for out-of-bag main effects and selection propensity, before using them to grow a causal forest. As causal forests have tuning parameters, we take two approaches to address robustness. First, we use cross-validation to select the tuning parameters. Second, we set the parameters to similar values used in several applications (Chen, Sridhar, and Mittal 2021). Specifically, we set the fraction of the data for each bootstrap sample to 50\% (s=.5), limited the number of variables tried at each split to 55, the minimum node size to 10, and the number of trees per forest to 4000. With the forest, we can then estimate $\hat{\tau_i}$, individual-level treatment effects, conditional on each individual's covariate values. We find very high similarity in the CATE between the specifications. We present the results with the cross-validated tuning parameters next.   

\subsubsection*{Heterogeneity in treatment effects}

In Table \ref{tab:featimppower}, we provide each of the feature importance scores for the pre-treatment covariates used to build the causal forest. The feature with the highest importance score 3.59\% is the friends' count, followed by the number of status counts, tweets per day, and mentions of words related to discrepancy. As such scores do not imply any directionality, some interpretation, and visual inspection is needed. To illustrate, friends count shows that the treatment effect is considerably weaker among those who have more friends (see Figure \ref{fig:powerfriendscount}). In contrast, 1.76\% for LIWC:discrepancy shows that the treatment effect is slowly but steadily increasing among those who use more words related to discrepancy (see Figure \ref{fig:discrepancy}). We note several things. First, we find that although friends count and statuses count (number of posts) matter a lot on the intentions to click to learn more about the browser extension, they are much less important when it comes to intentions to install (download) the browser or to join a fact-checking group. Second, followership patterns (e.g., whether people follow certain political parties or news outlets) seem to have little influence on the treatment effects.

\begin{table}[]
\caption{Study 3A: Top 40 Feature Importances (click, download, join)}
\label{tab:featimppower}
\small
\begin{tabular}{llllll}
\multicolumn{2}{c}{\textbf{(1) Click}} & \multicolumn{2}{c}{\textbf{(2) Download}} & \multicolumn{2}{c}{\textbf{(3) Join}}   \\
\textbf{Name}    & \textbf{Importance} & \textbf{Name}      & \textbf{Importance}  & \textbf{Name}     & \textbf{Importance} \\
\hline
friendscount               & 3.59\% & pronoun      & 1.47\% & social            & 1.61\% \\
statusescount              & 2.16\% & ppron        & 1.40\% & quant             & 1.58\% \\
Tweets\_per\_day           & 1.90\% & auxverb      & 1.37\% & see               & 1.56\% \\
discrep                    & 1.76\% & function.    & 1.35\% & percept           & 1.48\% \\
neg\_high                  & 1.71\% & neg\_high    & 1.35\% & body              & 1.43\% \\
focuspresent               & 1.62\% & negate       & 1.24\% & tone              & 1.40\% \\
followerscount             & 1.62\% & netspeak     & 1.24\% & anger             & 1.39\% \\
interrog                   & 1.62\% & they         & 1.21\% & negate            & 1.36\% \\
cause                      & 1.58\% & you          & 1.20\% & article           & 1.36\% \\
negemo                     & 1.57\% & anger        & 1.16\% & certain           & 1.35\% \\
space                      & 1.49\% & social       & 1.16\% & feel              & 1.32\% \\
compare                    & 1.48\% & friendscount & 1.15\% & pos\_high         & 1.28\% \\
auxverb                    & 1.42\% & sixltr       & 1.14\% & verb              & 1.27\% \\
achieve                    & 1.28\% & differ       & 1.13\% & focuspast         & 1.26\% \\
anger                      & 1.26\% & leisure      & 1.13\% & affiliation       & 1.26\% \\
verb                       & 1.25\% & pos\_low     & 1.10\% & negemo            & 1.26\% \\
power                      & 1.23\% & achieve      & 1.09\% & conscientuousness & 1.25\% \\
money                      & 1.22\% & cause        & 1.09\% & they              & 1.23\% \\
conj                       & 1.22\% & discrep      & 1.09\% & time              & 1.23\% \\
negate                     & 1.18\% & percept      & 1.08\% & friend            & 1.20\% \\
you                        & 1.18\% & adj          & 1.08\% & neuroticism       & 1.20\% \\
number                     & 1.18\% & affiliation  & 1.04\% & adj               & 1.20\% \\
they                       & 1.14\% & dic          & 1.03\% & friendscount      & 1.20\% \\
family                     & 1.14\% & money        & 1.03\% & affect            & 1.20\% \\
netspeak                   & 1.13\% & space        & 1.03\% & clout             & 1.19\% \\
age         & 1.13\% & focusfuture  & 1.02\% & i                 & 1.18\% \\
work                       & 1.11\% & negemo       & 1.02\% & focusfuture       & 1.18\% \\
focusfuture                & 1.11\% & posemo       & 1.00\% & posemo            & 1.17\% \\
reward                     & 1.06\% & openness     & 0.98\% & we                & 1.17\% \\
home                       & 1.06\% & compare      & 0.97\% & neg\_high         & 1.16\% \\
function.                  & 1.05\% & body         & 0.96\% & netspeak          & 1.15\% \\
pos\_low                   & 1.02\% & conj         & 0.96\% & pos\_low          & 1.14\% \\
focuspast                  & 1.02\% & see          & 0.96\% & assent            & 1.12\% \\
relig                      & 0.99\% & wc           & 0.95\% & home              & 1.11\% \\
dic                        & 0.97\% & analytic     & 0.95\% & reward            & 1.10\% \\
sixltr                     & 0.95\% & time         & 0.94\% & sexual            & 1.09\% \\
tentat                     & 0.95\% & authentic    & 0.93\% & discrep           & 1.08\% \\
analytic                   & 0.95\% & article      & 0.92\% & compare           & 1.07\% \\
tone                       & 0.90\% & risk         & 0.92\% & male              & 1.07\% \\
risk                       & 0.90\% & verb         & 0.92\% & cause             & 1.07\%
  \\ \hline
\end{tabular}
\end{table}

\begin{figure}[htbp]
     \centering
     \caption{Conditional Average Treatment Effects Plotted Against Two Features (examples on intention to click)}
     \begin{subfigure}[b]{0.45\textwidth}
         \centering
         \includegraphics[width=\textwidth]{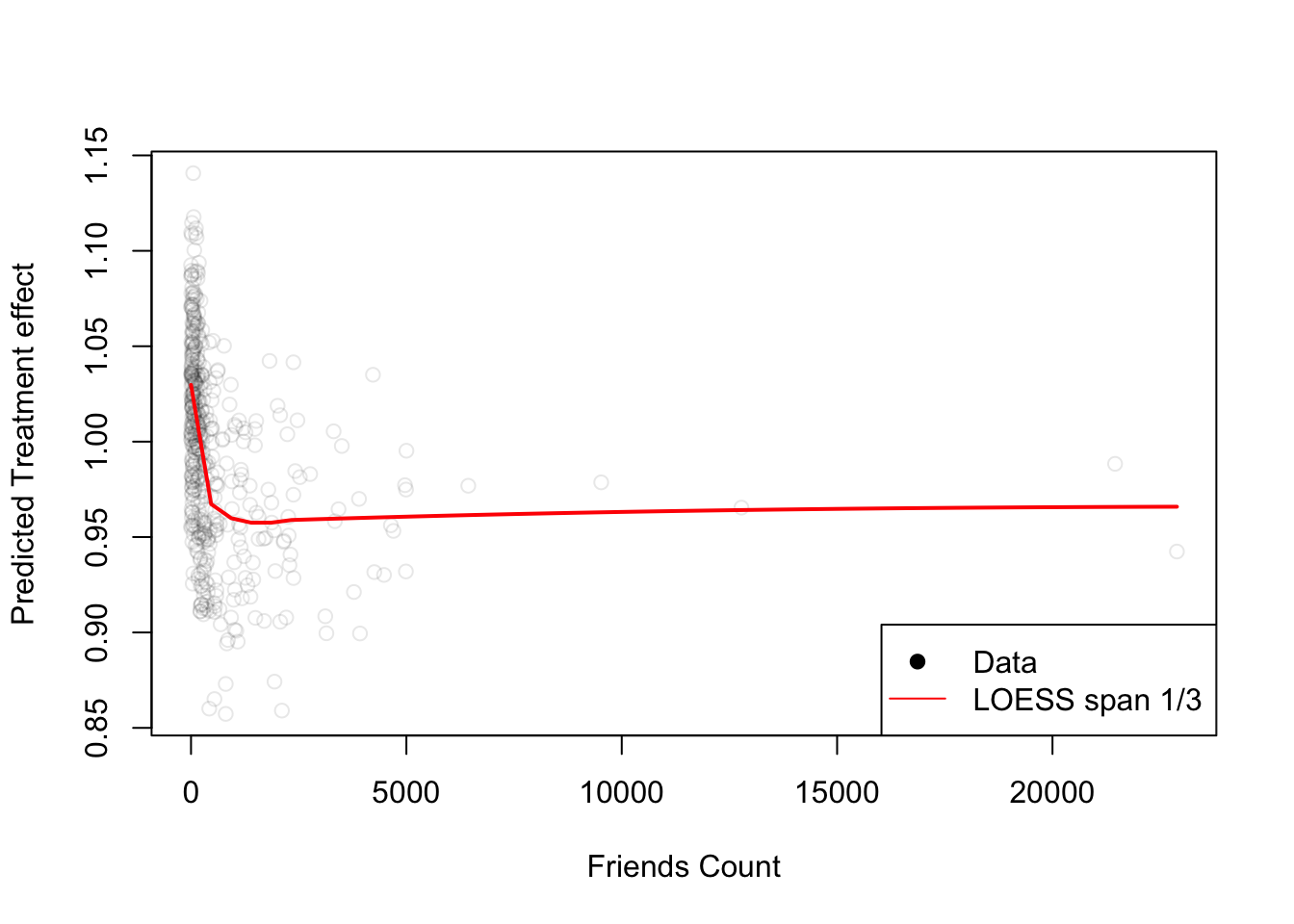}
         \caption{Friends count (3.59\% importance)}
         \label{fig:powerfriendscount}
     \end{subfigure}
     \hfill
     \begin{subfigure}[b]{0.45\textwidth}
         \centering
         \includegraphics[width=\textwidth]{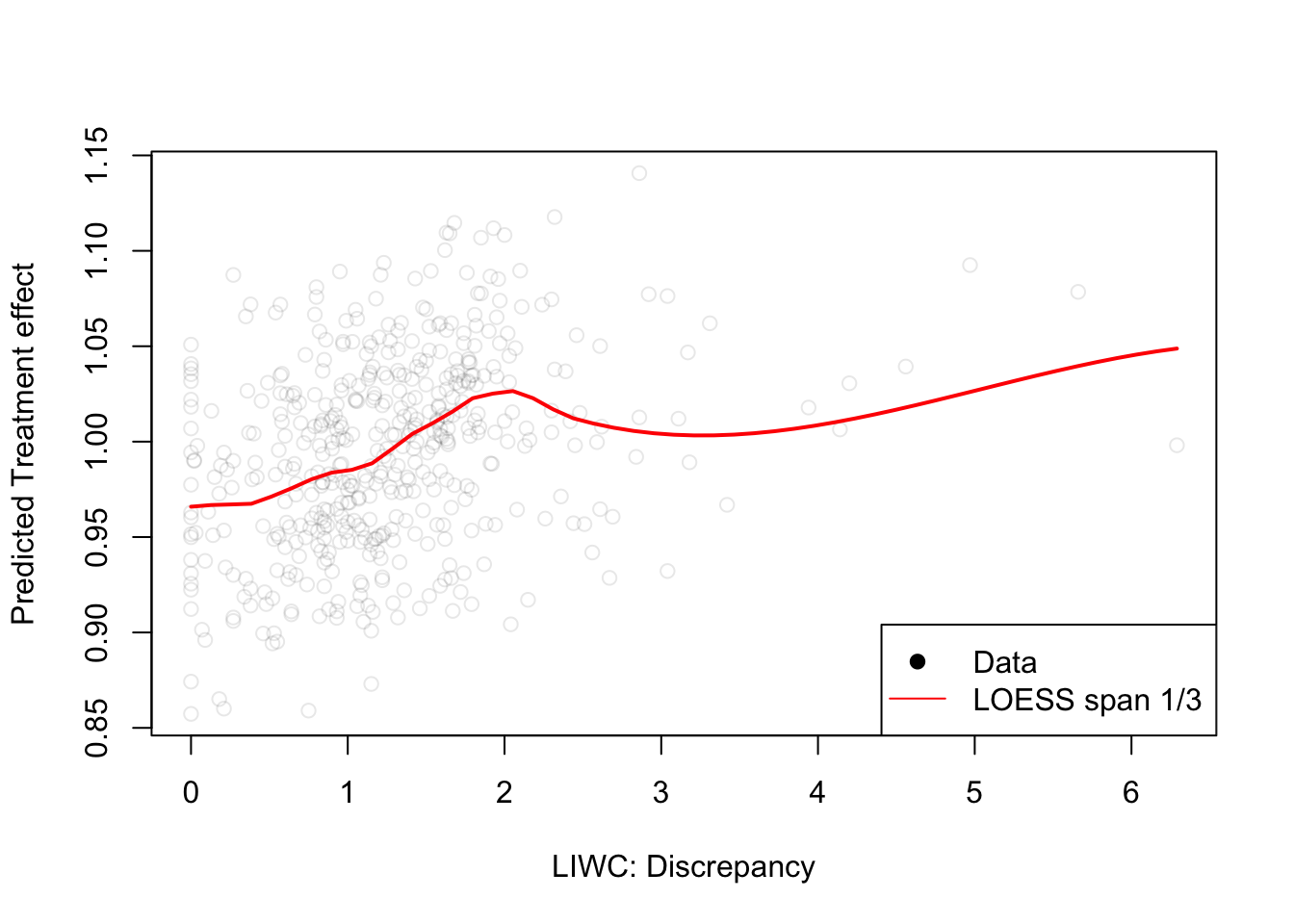}
        \caption{LIWC:discrepancy (1.76\% importance)}
         \label{fig:discrepancy}
     \end{subfigure}
     \label{fig:featimppower}
\end{figure}

\subsubsection*{Second-stage Analysis}

In addition to exploring unexpected heterogeneity in treatment effects based on pre-treatment variation in textual information, we can use the $\hat{\tau_i}$ as the dependent variable in a regression to see whether theoretically relevant predictors are moderators of the individual-level treatment effects (Athey and Wager 2021, Chen et al. 2020) even when we control for possibly endogenous covariates (i.e., self-reported gender, age, religiosity, political affiliation, and trait desire for control and sense of power). To do so, we follow a recent development in the literature and use doubly robust estimates instead of $\hat{\tau_i}$ as the latter is biased if either the propensity score or the expected outcome is misspecified (Chen, Sridhar, and Mittal 2021, Semenova and Chernozhukov 2021).

\begin{table}[!htbp] \centering 
  \caption{Second-stage models based on survey measures} 
  \label{tab:Second-stage-model} 
\begin{tabular}{@{\extracolsep{5pt}}lD{.}{.}{-3} D{.}{.}{-3} D{.}{.}{-3} } 
\\[-1.8ex]\hline 
\hline \\[-1.8ex] 
 & \multicolumn{3}{c}{\textit{Outcome:}} \\ 
\cline{2-4} 
\\[-1.8ex] & \multicolumn{3}{c}{ } \\ 
\\[-1.8ex] & \multicolumn{1}{c}{Click} & \multicolumn{1}{c}{Download} & \multicolumn{1}{c}{Join}\\ 
\hline \\[-1.8ex] 
 Religiosity & 0.372^{*} & 0.189 & 0.037 \\ 
  & (0.198) & (0.205) & (0.089) \\ 
  & & & \\ 
 Age & -0.009 & -0.017 & -0.003 \\ 
  & (0.018) & (0.018) & (0.008) \\ 
  & & & \\ 
 Gender: Female & -0.664 & 0.192 & -0.244 \\ 
  & (0.528) & (0.529) & (0.230) \\ 
  & & & \\ 
 Political (Very Liberal) & -0.040 & 0.091 & 0.019 \\ 
  & (0.781) & (0.772) & (0.331) \\ 
  & & & \\ 
 Political (Liberal) & 0.144 & 0.916 & 0.756^{***} \\ 
  & (0.640) & (0.644) & (0.274) \\ 
  & & & \\ 
 Political (Conservative) & -0.804 & -0.825 & 0.205 \\ 
  & (0.752) & (0.802) & (0.360) \\ 
  & & & \\ 
 Political (Very Conservative) & -0.705 & -1.507 & -0.126 \\ 
  & (1.048) & (1.041) & (0.419) \\ 
  & & & \\ 
 Trait: Desire for Control & -0.809 & 0.130 & -0.135 \\ 
  & (0.574) & (0.585) & (0.270) \\ 
  & & & \\ 
 Trait: Personal Sense of Power & 0.381 & -0.134 & 0.098 \\ 
  & (0.375) & (0.378) & (0.179) \\ 
  & & & \\ 
 Constant & 2.307 & 1.360 & 0.265 \\ 
  & (1.925) & (1.911) & (0.915) \\ 
  & & & \\ 
\hline \\[-1.8ex] 
\hline 
\hline \\[-1.8ex] 
\multicolumn{4}{l}{\textit{Notes:}} \\ 
\multicolumn{4}{l}{$^{*}$p$<$0.1; $^{**}$p$<$0.05; $^{***}$p$<$0.01} \\ 
\multicolumn{4}{l}{Baseline category for political orientation is middle.} \\ 
\end{tabular} 
\end{table}

As in the analyses in the main text, we continue to find strong effects of power-related words on intentions to click and download the browser extension, while failing to find evidence that it encourages joining a fact-checking group. Moreover, we still do not find evidence that the treatment differed based on the desire for control or sense of power. Instead, we find that the treatment effect is fairly homogeneous, with only some marginal evidence (p$<$.10) that those with self-reported high religiosity have a greater treatment effect of the power manipulation ($p<.10$). Table \ref{tab:Second-stage-model} summarizes the results.

\end{singlespace*}


\end{singlespace*}

\end{document}